%% file: mOGM2017.tex
\documentclass[ twocolumn]{aastex61}

\newcommand\aastex{AAS\TeX}

\received{\today}
\revised{\today}
\accepted{\today}
\submitjournal{ApJ}

\shorttitle{\aastex\ sample article}
\shortauthors{Gonz\'alez-Mart\'in et al.}

\begin{document}

\title{Hints on the gradual re-sizing of the torus in AGN \\ by decomposing IRS/\emph{Spitzer} spectra}

\correspondingauthor{Omaira Gonz\'alez-Mart\'in, tenure track}
\email{o.gonzalez@crya.unam.mx}

\author{Omaira Gonz\'alez-Mart\'in}
\affil{Instituto de Radioastronom\'ia y Astrof\'isica (IRyA-UNAM), 3-72 (Xangari), 8701, Morelia, Mexico}

\author{Josefa Masegosa}
\affil{Instituto de Astrof\'isica de Andaluc\'ia, CSIC, Glorieta de la Astronom\'ia s/n 18008, Granada, Spain}

\author{Antonio Hern\'an-Caballero}
\affil{Departamento de Astrof\'isica, Facultad de CC. F\'isicas, Universidad Complutense de Madrid, E-28040 Madrid, Spain}

\author{Isabel M\'arquez}
\affil{Instituto de Astrof\'isica de Andaluc\'ia, CSIC, Glorieta de la Astronom\'ia s/n 18008, Granada, Spain}

\author{Cristina Ramos Almeida} 
\affil{Instituto de Astrof\'isica de Canarias (IAC), C/V\'ia L\'actea, s/n, E-38205 La Laguna, Spain}
\affil{Departamento de Astrof\'isica, Universidad de La Laguna (ULL), E-38205 La Laguna, Spain}

\author{Almudena Alonso-Herrero}
\affil{Centro de Astrobiolog\'ia (CAB, CSIC-INTA), ESAC Campus,  E-28692 Villanueva de la Ca\~nada, Madrid, Spain}
\affil{Visiting professor, Department of Physics and Astronomy, University of Texas at San Antonio, San Antonio, TX 78249, USA}

\author{Itziar Aretxaga}
\affil{Instituto Nacional de Astrof\'isica, \'Optica y Electr\'onica (INAOE), Luis Enrique Erro 1, Sta. Ma. C.P. 72840 Tonantzintla, Puebla, Mexico}

\author{Jos\'e Miguel Rodr\'iguez-Espinosa}
\affil{Instituto de Astrof\'isica de Canarias (IAC), C/V\'ia L\'actea, s/n, E-38205 La Laguna, Spain}
\affil{Departamento de Astrof\'isica, Universidad de La Laguna (ULL), E-38205 La Laguna, Spain}

\author{Jose Antonio Acosta-Pulido} 
\affil{Instituto de Astrof\'isica de Canarias (IAC), C/V\'ia L\'actea, s/n, E-38205 La Laguna, Spain}
\affil{Departamento de Astrof\'isica, Universidad de La Laguna (ULL), E-38205 La Laguna, Spain}

\author{Lorena Hern\'andez-Garc\'ia}
\affil{INAF - Istituto di Astrofisica e Planetologia Spaziali di Roma (IAPS), Via del Fosso del Cavaliere 100, I-00133 Roma, Italy}

\author{Donaji Esparza-Arredondo}
\affil{Instituto de Radioastronom\'ia y Astrof\'isica (IRyA-UNAM), 3-72 (Xangari), 8701, Morelia, Mexico}

\author{Mariela Mart\'inez-Paredes}
\affil{Instituto de Radioastronom\'ia y Astrof\'isica (IRyA-UNAM), 3-72 (Xangari), 8701, Morelia, Mexico}

\author{Paolo Bonfini} 
\affil{Instituto de Radioastronom\'ia y Astrof\'isica (IRyA-UNAM), 3-72 (Xangari), 8701, Morelia, Mexico}

\author{Alice Pasetto}
\affil{Instituto de Radioastronom\'ia y Astrof\'isica (IRyA-UNAM), 3-72 (Xangari), 8701, Morelia, Mexico}

\author{Deborah Dultzin}
\affil{Instituto de Astronom\'ia, Universidad Nacional Aut\'onoma de M\'exico, Apartado Postal 70-264, 04510, M\'exico DF, Mexico}



\begin{abstract}
Several authors have claimed that the less luminous active galactic nuclei (AGN) are not capable of sustaining the dusty torus structure. Thus, a gradual re-sizing of the torus is expected when the AGN luminosity decreases.
Our aim is to confront mid-infrared observations of local AGN of different luminosities with the gradual re-sizing and disappearance of the torus.
We applied the decomposition method described by \citet{Hernan-Caballero15} to a sample of about $\rm{\sim 100}$ IRS/\emph{Spitzer} spectra of LLAGN and powerful Seyferts in order to decontaminate the torus component from other contributors. We have also included Starburst objects to ensure a secure decomposition of the IRS/\emph{Spitzer} spectra. We have used the affinity propagation (AP) method to cluster the data into five groups within the sample according to torus contribution to the 5-15$\rm{\mu}$m range($\rm{C_{torus}}$) and bolometric luminosity ($\rm{L_{bol}}$).
The AP groups show a progressively higher torus contribution and an increase of the bolometric luminosity, from Group 1 ($\rm{C_{torus}\sim 0\%}$ and $\rm{log(L_{bol})\sim 41}$) and up to Group 5 ($\rm{C_{torus}\sim 80\%}$ and $\rm{log(L_{bol})\sim 44}$). We have fitted the average spectra of each of the AP groups to clumpy models. The torus is no longer present in Group 1, supporting the disappearance at low-luminosities. We were able to fit the average spectra for the torus component in Groups 3 ($\rm{C_{torus}\sim 40\%}$ and $\rm{log(L_{bol})\sim 42.6}$), 4 ($\rm{C_{torus}\sim 60\%}$ and $\rm{log(L_{bol})\sim 43.7}$), and 5 to Clumpy torus models. We did not find a good fitting to Clumpy torus models for Group 2 ($\rm{C_{torus}\sim 18\%}$ and $\rm{log(L_{bol})\sim 42}$). This might suggest a different configuration and/or composition of the clouds for Group 2, which is consistent with a different gas content seen in Groups 1, 2, and 3, according to the detections of $\rm{H_{2}}$ molecular lines. Groups 3, 4, and 5 show a trend to decrease of the width of the torus (which yields to a likely decrease of the geometrical covering factor), although we cannot confirm it with the present data. Finally, Groups 3, 4, and 5 show an increase on the outer radius of the torus for higher luminosities, consistent with a re-sizing of the torus according to the AGN luminosity.
\end{abstract}




\section{Introduction} \label{sec:intro}

According to the unification of active galactic nuclei (AGN), the central engine is surrounded by a dusty, optically thick structure responsible for partially blocking its view (the so-called dusty torus). The AGN is powered by a supermassive black-hole (SMBH) that is fed by its accretion disk. Low- and high-velocity clouds are located at the narrow and broad line regions (NLR and BLR, respectively), the latter being located inside the dusty structure. Much of the observed diversity of AGN families is simply explained as the result of the line of sight toward this asymmetric torus structure. Reviews on the unification schemes for AGN have been presented by \citet{Antonucci93} and \citet{Urry95}. Still open questions need to be settled on the nature and geometry of the torus \citep[see e.g.][and references therein]{Netzer15}.

It was realised quite early that the nuclear dust could be distributed in clumps \citep{Krolik88}. AGN tori have a range of properties (e.g. width, size, composition, number of clouds, distribution of clouds, etc.), where the covering factor is also a key parameter to classify the object as a type-1 or type-2 AGN \citep{Elitzur12,Ramos-Almeida11,Alonso-Herrero11,Mateos16}. Although there is considerable observational support for the unified model, some observations and AGN classes have caused doubt on the most extreme form of the unified model, in which the viewing angle is the only responsable for the AGN classification. For instance, \citet{Ricci11} showed that the X-ray reflection component (associated to the torus) was intrinsically stronger for type-2 than for type-1 AGN. \citet{Ramos-Almeida11} found that type-2 AGN have larger covering factor tori than Type-1 AGN using clumpy torus models.  \citet{Mendoza-Castrejon15} (and references therein) have shown that the structure of the torus might depend even from the nearby environment of the host galaxy. An example of an AGN family not easy to be explained with the AGN ingredients are low ionisation nuclear emission-line regions \citep[LINERs,][]{Heckman80}. Their spectral energy distribution (SED) is clearly different from those of other AGN \citep{Ho08,Mason13}. 

Using mass conservation arguments, \citet{Elitzur06} showed that the dusty torus cannot be sustained under certain AGN bolometric luminosity, claiming its disappearance. \citet{Honig07} studied the balance between gravity and radiation pressure from the central source for the torus, and found that the torus changes its characteristics and obscuration becoming insufficient for luminosities of the order of $\rm{\sim 10^{42}}$\,erg/s. \citet{Elitzur09} showed that indeed this limit on the AGN bolometric luminosity depends on the SMBH mass. Recently, \citet{Elitzur16} realized that, beside this luminosity limit, there is a range on the bolometric luminosities in which the torus still might disappear depending on the combination of some of the parameters of the wind. From the observational point of view, the lack of the infrared bump in low-luminosity AGN (LLAGN) associated to dust obscuration in other AGN provides evidence for unobscured nuclei \citep{Ho08}. UV variability also invokes an unobstructed view of the accretion disk \citep{Maoz05,Hernandez-Garcia14,Hernandez-Garcia16}. 

Very little has been said about the dependence of size of the torus (i.e. outer radius of the torus) with the AGN luminosity. \citet{Mason13} already suggested a different torus than in Seyferts, in light of their low dust-to-gas ratio, although a large diversity of contributions where also found. \citet{Muller-Sanchez13} showed evidence in favor of the gradual disappearance of the torus, finding that the molecular gas in some LINERs is almost ten times more concentrated towards the center and with column densities $\rm{\sim 3}$ times smaller than in Seyfert galaxies. Indeed, \citet{Gonzalez-Martin15} showed that a large fraction of LLAGN with 2-10 keV X-ray luminosities $\rm{L_{X}<10^{41}erg/s}$ may lack torus signatures at mid-infrared. Although they excluded objects with large contamination of the ISM, these results might be somehow contaminated by the host galaxies due to the low spatial resolution inherent to the \emph{Spitzer} data they used \citep{Alonso-Herrero06,Mason13,Sturm06}. Indeed, the nuclear emission of LLAGN is affected by the host galaxy contribution even at X-rays \citep{Gonzalez-Martin14}. 

Here we take advantage of the spectral decomposition method developed by \citet[][hereafter HC15]{Hernan-Caballero15} to isolate the AGN component, and analyse the disappearance of the torus in the same sample used by \citet{Gonzalez-Martin15}. Furthermore, a full analysis of AGN with a wide range of bolometric luminosities (more than six orders of magnitude) also allows us to test the gradual disappearance of the torus. 

The paper is organised as follows. In Sect.~\ref{sec:sample} we describe the sample. The data are presented in Sect.~\ref{sec:data} and the analysis is presented in Sect~\ref{sec:analysis}. In Sect.~\ref{sec:torusvsluminosity} we analyse the torus contribution at mid-infrared (i.e. torus contribution since mid-infrared is fully dominated by the torus) as a function of the bolometric luminosity of the AGN. A full discussion of the results is presented in Sect.~\ref{sec:discussion}. Throughout this paper we use the Hubble constant $\rm{H_{0}=70 km/s/Mpc}$.

\section{Sample}\label{sec:sample}

The sample was originally presented by \citet{Gonzalez-Martin15}. The LINER sample is selected as those objects with reported X-ray luminosities from \citet{Gonzalez-Martin09A} with full coverage of the 5-30$\rm{\mu m}$ range with the InfraRed Spectrograph (IRS/\emph{Spitzer}) spectra. This guarantees that all the LINERs have $\rm{L_{X}(2-10~ keV)}$ measurements. Among the 48 LINERs with IRS/\emph{Spitzer} spectra, 40 mid-infrared spectra have been taken from the CASSIS atlas \citep{Lebouteiller11} and 8 from the SINGS database \citep{Kennicutt03}. The main properties of the LINERs sample are presented in Table~\ref{tab:sample}, where we split the sample into Type-1 (all of them Type 1.9) and Type-2 LINERs according to the optical classification done by \citet{Veron10} and \citet{Ho97}.

It is well known that the LINER classification, based only on the preponderance of the low-ionisation emission lines at optical wavelengths, yields a mixture of different types of objects \citep[][and references therein]{Ho08}. Our sample is based on available X-rays observations. Hence it could be biased towards an AGN nature since these observations might have been done because of the known AGN component in these sources. Indeed, 90\% of the X-ray sample was classified as AGN dominated based on multi-wavelength evidences such as the existence of a point-like source at hard X-rays (28 out of the 48 LINERs), the presence of iron K$\rm{\alpha}$ emission lines at 6.4 keV (26 out of the 48 LINERs), radio compact sources or radio-jet detections (31 out of the 48 LINERs), UV variability (5 out of the 48 sources), and broad $\rm{H\alpha}$ emission lines (8 out of the 48 LINERs). We refer the reader to Tables 11 and 12 in \citet{Gonzalez-Martin09A} for further details. Adding all these evidence together, among the 48 LINERs selected for the present analysis, only two sources lack evidence of the AGN (UGC4881 and NGC4676A), consistent with the idea that our sample contains AGN-like LINERs.

For consistency with our previous work, the Seyfert and Starburst samples are the same as those studied in \citet{Gonzalez-Martin15}. The Seyfert sample are all the Type-1 and Type-2 sources included in \citet{Shi06}, in the Compton-thick sample described by \citet{Goulding12}, and in the SINGS sample. In total it contains 42 Seyferts. Among them, 32 are Type-2 Seyferts (S2, including 20 Compton-thick and 12 Compton-thin) and 10 are Type-1 Seyferts (S1). The Starburst sample has been taken from \citet{Ranalli03}, \citet{Brandl06}, and \citet{Grier11}. This Starburst sample contains 19 sources. The main properties of Seyferts and Starbursts are included in Tables \ref{tab:comparisonsample1} and \ref{tab:comparisonsample2}, respectively. 

Thus, our final sample includes 48 LINERs, 42 Seyferts and 19 Starburst (109 objects). We have compared the absolute B magnitude and velocity distributions for local AGN from the Palomar sample \citep[][and references therein]{Ho03} with the sample analysed here. We found that our LINER, Seyfert, and Starburst samples have the same distributions as local AGN. Thus, although this is not a complete sample, it is representative of the population of nearby AGN. 

\begin{figure}
\begin{center}
\includegraphics[width=1.0\columnwidth]{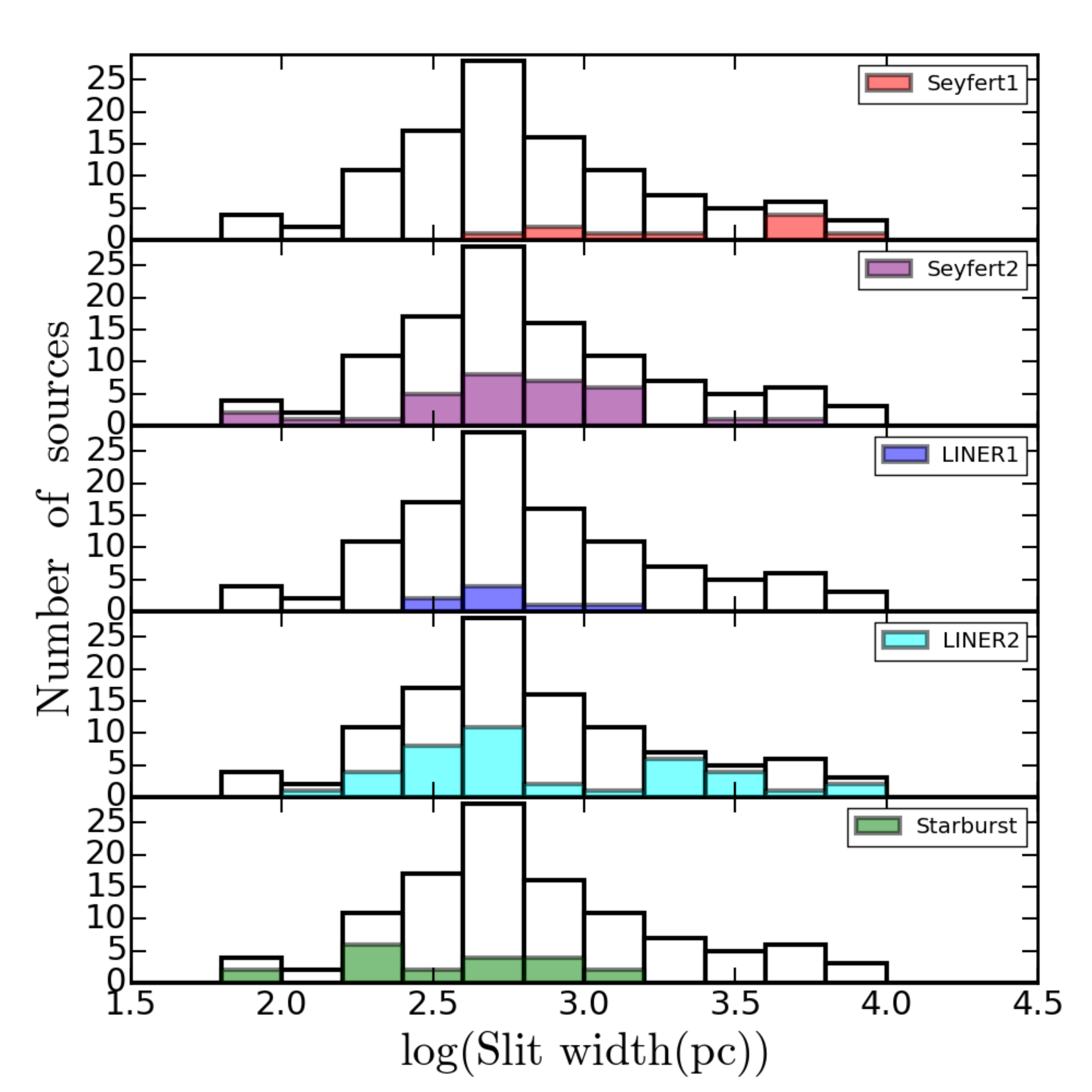}
\caption{Distributions of ``Slit sizes''  (in pc) of the IRS/\emph{Spitzer} spectra for the full sample (empty histograms in all panels) and for Seyfert 1 (red-filled histogram), Seyfert 2 (purple-filled histogram), LINER1 (cyan-filled histogram), LINER2 (blue-filled histogram), and Starbursts (green-filled histogram), from top to bottom.}
\label{fig:slitwidth}
\end{center}
\end{figure}

In order to avoid further bias in our results, we have investigated whether the physical angular resolution we obtain from IRS/\emph{Spitzer} spectra depends on the class of object. We have computed the inner portion of the galaxy (in parsec units) according to the slit width of the IRS spectra (slit width of $\rm{3.6}$ arcsec) at the distance of each object. This portion of galaxy extending out from the nucleus is recorded in Col. 5 as ``slit width'' in Tables \ref{tab:sample}, \ref{tab:comparisonsample1} and \ref{tab:comparisonsample2}. The median ``slit width'', of the sample is 520 pc, with 25-75\% percentiles of [340-1170] pc. The ``slit width'' distribution according to their optical class is shown in Fig. \ref{fig:slitwidth}. Although there is an apparent lack of Starbursts with larger ``slit widths''  while Type-1 Seyferts tend to show larger ``slit widths'', according to a K-S test we cannot reject the possibility that all optical classes come from the same parent distribution. Thus, biases due to the distance of the objects can be ruled out.

\section{Data}\label{sec:data}

\subsection{IRS/\emph{Spitzer} spectra}\label{sec:spitzerdata}

We have only included spectra observed with both the short-low (SL) and long-low (LL) modules to guarantee the full IRS/\emph{Spitzer} coverage ($\rm{\sim 5-30\mu m}$). CASSIS and SINGS provide flux and wavelength calibrated spectra. However, the observations using data from both the SL and LL spectral modules suffer from mismatches due to telescope pointing inaccuracies or due to different spatial resolution of the IRS orders. This is not corrected in the final products given by CASSIS and SINGS. We therefore scaled each spectra to the immediate prior (in wavelength range) to overcome such effects. Thus, our flux level is scaled to the level of the shortest wavelengths, which is the order with the highest spatial resolution (3.6 arcsec). This guarantees that the flux level is scaled to the best spatial resolution that \emph{Spitzer} can provide. Note, however, that this does not solve the problem related to the fact that each IRS module might be seeing a different region due to a different spatial resolutions and/or slight changes on the position of the spectrum. However, this is the best that we can do. Finally, the spectra are shifted to the rest-frame according to the redshifts of the objects (see Col. 3 in Tables~\ref{tab:sample}, \ref{tab:comparisonsample1} and \ref{tab:comparisonsample2}). 

\begin{figure*}
\begin{center}
\includegraphics[width=1.\columnwidth]{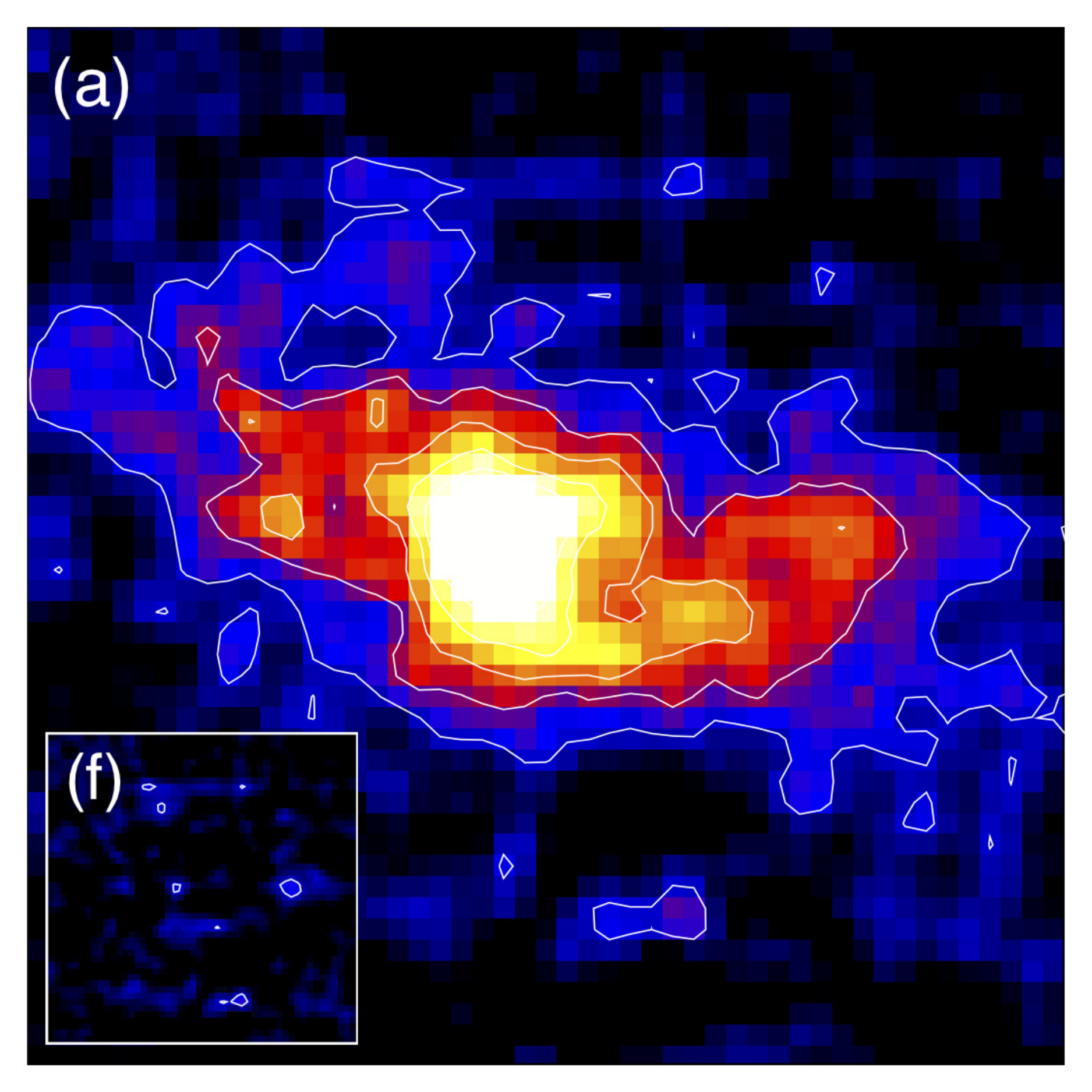}
\includegraphics[width=1.\columnwidth]{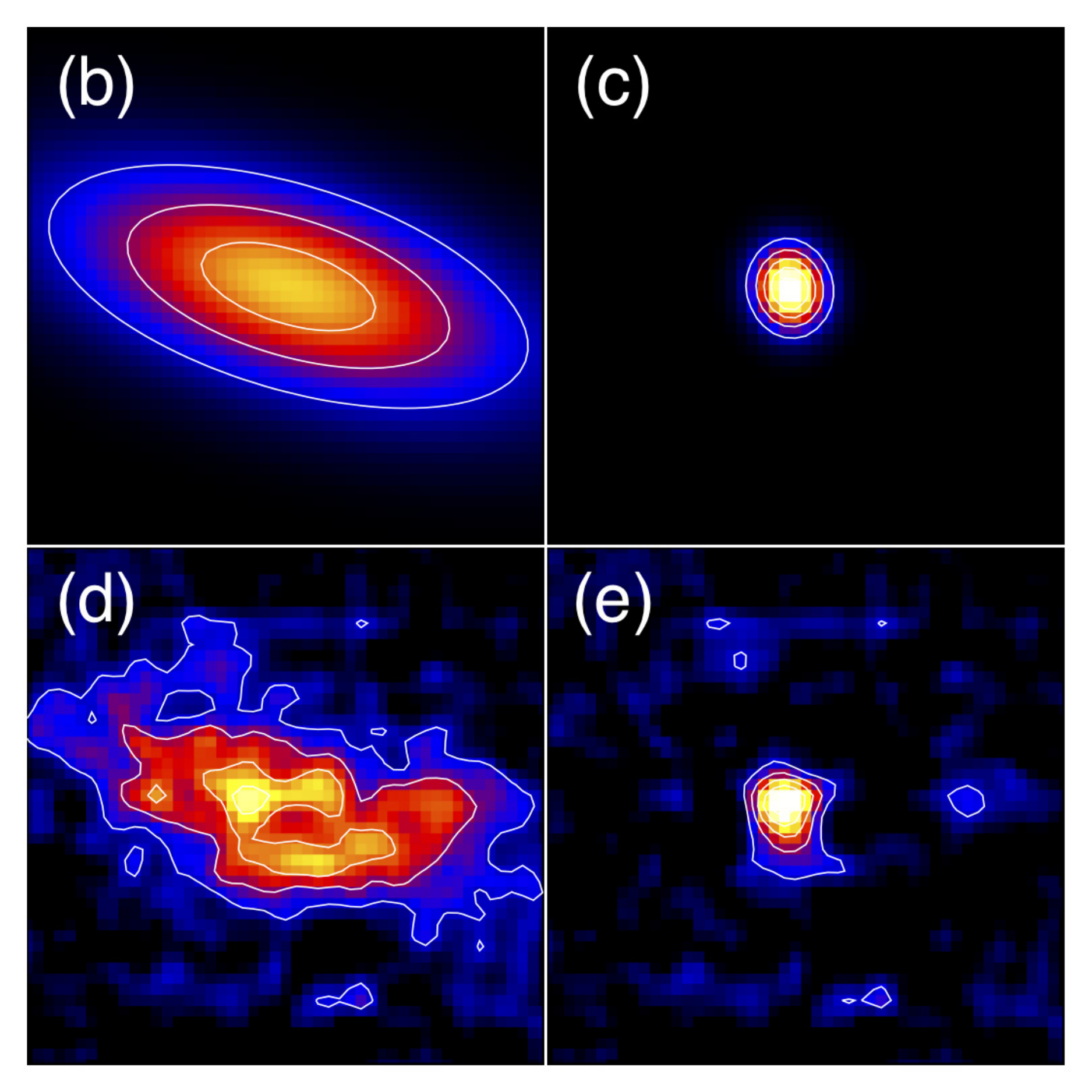}
\caption{CanariCam 11.5$\rm{\mu m}$ image of Mrk\,266SW. The size of the showed box is 4x4 arcsec. East is up and North is right. (a) Total emission; (b) Gaussian model for the extended emission; (c) Gaussian model of the nuclear emission; (d) Extended emission after subtracting the nuclear emission model to the total emission (i.e. "$\rm{a - c}$"); (e) Nuclear emission after subtracting the extended emission model to the total emission (i.e. "$\rm{a - b}$"); and (f, inset to the panel (a)) Residuals after the fitting process (i.e. "$\rm{a - b - c}$"). White contours show flux levels of 0.05, 0.1, 0.15, 0.2, and 0.25 mJy/pixel. }
\label{fig:ImageDecomp}
\end{center}
\end{figure*}

\subsection{CanariCam/GTC 11.5$\rm{\mu m}$ images}\label{sec:canaricamdata}

We have included in our analysis mid-infrared spatially resolved images taken with CanariCam/GTC using the filter `Si6' centred at 11.5$\rm{\mu m}$. These observations are part of proprietary data of a sample of faint and Compton-thick LINERs observed with CanariCam/GTC (proposal ID GTC10-14A, P.I. Gonz\'alez-Mart\'in). Some of these images were already used by \citet{Gonzalez-Martin15} to compare their nuclear fluxes with \emph{Spitzer}/IRS fluxes. Note that here we make a more sophisticated treatment of the images to better isolate the nuclear emission from its extended emission (see Section \ref{sec:analysis}). The full sample contains 19 LINERs and it will be the subject of subsequent publications focused on the nuclear \citep[Masegosa et al. in preparation, see][for preliminary results on the nuclear flux]{Masegosa13} and extended emission \citep[as an example see the analysis of the extended emission of NGC\,835 presented in][]{Gonzalez-Martin16}. Here we have used the nuclear flux taken from the CanariCam data in 12 objects in common with the IRS/\emph{Spitzer} sample of LINERs. The summary of the observations used in this paper is reported in Table~\ref{tab:CanariCam}.

CanariCam uses a Raytheon 320$\rm{\times}$240 pixels Si:As detector that covers a field of view (FOV) of 26$\rm{\times}$19 arcsec on the sky with a pixel scale of 0.0798 arcsec. Standard mid-infrared chopping-nodding techniques were used to remove the time-variable sky background, the thermal emission from the telescope, and the detector 1/f noise. The employed chopping and nodding throws and, chop and nod position angles are reported in Table~\ref{tab:CanariCam} (Col.~5).

Images of point spread function (PSF) standard stars were obtained in the same filter immediately after the science target to accurately sample the image quality and allow for flux calibration of the target observations. Table~\ref{tab:CanariCam} includes the name (Col.~6), integrating time (Col.~8), and the full width at a half maximum (FWHM) of  the standard stars associated to each target (Col.~9, representing the FWHM of the PSF at the time of the observations). To compute it we have fitted a 2D Gaussian to the standard star observations. The two numbers given in Col.~9 in Table~\ref{tab:CanariCam} show the minor and major width of this Gaussian fit. Any point-like source detected in our images should show a FWHM contained between these two values. However, as it will be discussed below, the sky at mid-infrared is highly variable and often the conditions of the sky at the time the target was observed might have slightly changed from those when the standard star was observed. Thus, although this number can give a rough estimate of the image quality, it cannot be taken as a strict limit to the FWHM of a point-like source.  

Each observing block was processed using the pipeline RedCan \citep{Gonzalez-Martin13}, which is able to produce flux-calibrated images and wavelength- and flux-calibrated spectra for CanariCam/GTC and T-ReCS/Gemini low-resolution data. The combination of the different observing blocks for the same source (when available, see Table~\ref{tab:CanariCam}) were made after flux-calibration using Python routines.

\subsection{Archival high spatial resolution images}\label{sec:otherdata}

We have compiled all the high-spatial resolution mid-infrared images associated to our IRS/\emph{Spitzer} sample. For that purpose we have used the atlas of mid-infrared observations reported in \citet{Asmus14}. It contains 895 observations of 253 AGN taken with 8-m class telescopes up to 2014.  

This complements our study in three ways: (1) mid-infrared observations centred at wavelengths other than the 11.5$\rm{\mu m}$ of the current CanariCam/GTC observed LINERs; (2) mid-infrared observations of other LINERs not observed in our campaign with CanariCam/GTC; and (3) mid-infrared observations of objects included in our comparison samples. We have retrieved 285 mid-infrared observations for 18 LINERs, 10 type-1 Seyferts, 23 type-2 Seyferts, and two Starbursts contained in the atlas by \citet{Asmus14}. Among the LINERs, only five of them (UGC\,05101, NGC\,4486, MRK\,266SW, MRK\,266NE, and NGC\,6251) are in common with the CanariCam/GTC sample, although not exactly with the same filter because CanariCam/GTC sample was selected avoiding objects already observed with 8-m class telescopes.  

\section{Analysis}\label{sec:analysis}

\subsection{Image decomposition}

Some of the high spatial resolution images in our sample show extended structures together with the nuclear point-like source \citep{Asmus15}. In order to better isolate the nuclear component we have developed a code able to decompose both emissions. This procedure is based on the idea that both emissions can be roughly fitted with 2D Gaussians. We have used a 2D Gaussian fit included in the package {\sc satrapy} within Python, which allows to vary the normalisation of the Gaussian, the width along the major and minor axis, and the angle in which the major axis is located. Firstly, we trim the image in a box of 40$\rm{\times}$40 pixels, centred at the position of the source, which is wide enough to contain all the extended emission for all the objects in our sample but sufficiently small to guarantee that the procedure avoids any artificial or real structure away from our target. Then, we followed several steps until the results converge to the final solution: (1) We fit the image using a single 2D Gaussian. At this stage the width of the Gaussian is fixed to the width of the standard star associated to the target for CanariCam data (FWHM reported in Col.~9 of Table~\ref{tab:CanariCam}) or to the major axis of the FWHM of the Gaussian fit reported by \citet{Asmus14}. This is considered as an initial guess for the nuclear component. (2) The fitted Gaussian is subtracted from the original images producing a first guess of the extended structure. This extended structure is fitted with another Gaussian which is centred at the position of the first Gaussian but now allowing the widths of the Gaussian to vary. (3) This Gaussian fit to the extended emission is now subtracted from the original image producing a new guess for the nuclear component. At this stage the process starts over in (1), using this new guess for the nuclear component as the input image. This process continues until the residuals are within three standard deviations over the background of the image. 

As an example of the result of this process, Fig.~\ref{fig:ImageDecomp} shows the case for Mrk\,266SW. Panel (a) shows the original image where both point-like and extended emission can be clearly spotted. Panels (b) and (c) correspond to the best Gaussian fit to the extended and nuclear emissions, respectively. Panels (d) and (e) show the resulting `extended' and `nuclear' images, respectively, computed as the original image minus the Gaussian best fit for the nuclear component, and the original minus the Gaussian best fit for the extended emission. Panel (f) (inset in panel (a)) shows the residuals of the final fit. This method nicely isolates nuclear from extended emission.  

The resulting minor and major axis of the FWHM are recorded in Col.~10 in Table~\ref{tab:CanariCam} for the CanariCam images and in Col.~3 in Table~\ref{tab:otherimages} for other archival data. In the case of CanariCam data we can compare the final FHWM of the nuclear component with that of the PSF (as traced by the standard star). Although in most cases the FWHM of the nuclear component is consistent with that of the PSF, it is clear that in some cases the point-like source detected in the target image shows a smaller or larger FWHM than that of the standard star (e.g. UGC\,08696 or NGC\,315). When the FWHM of the nuclear component is larger than the PSF of the standard star, it is plausible that the source is partially resolved. However, it could also be due to changes on the conditions of the observation along the night. Due to the faintness of our targets, the exposure times are rather long. Thus, a delay of one hour or more between the observation of the target and that of the standard star is usual. 

The nuclear fluxes (reported in Col.~11 of Table~\ref{tab:CanariCam} and in Col.~4 of Table~\ref{tab:otherimages}) are computed by performing aperture photometry in the nuclear images (i.e. after subtracting the extended emission), using an aperture radius 2.5 times the major width of the 2D Gaussian fit for the nuclear component. This ensures that over 97\% of the flux is contained within this aperture and, at the same time, avoids any residuals of the extended emission contributing to the nuclear flux. We computed the error as the quadratic sum of the flux calibration uncertainty plus the error due to the S/N of each observation, but the errors are fully dominated by the  former one. They are assumed to be 15\% of the flux. Note that calibration errors need to be included especially when combining datasets from different facilities. \citet{Alonso-Herrero16} computed the flux calibration in a sample of CanariCam/GTC observations of AGN, finding that the mean calibration error is 11\%, close to our estimate \citep[see also][]{Diaz-Santos10,Ramos-Almeida11}. We also compared the nuclear fluxes reported here with those reported by a Gaussian fit in \citet{Asmus14}. All of them are consistent within the errors. However, our errors are larger than those reported by \citet{Asmus14}. Note that our final purpose is to constrain the torus component when decomposing the IRS/\emph{Spitzer} spectra. A less restrictive limit could translate into a larger contribution for this component. Since we are studying the plausible disappearance of the torus, this less restrictive constraint yields to a more conservative result on the disappearance of the torus (see Section~\ref{sec:decomposition}).

\subsection{Spectral decomposition}\label{sec:decomposition}

\begin{figure*}
\begin{center}
\includegraphics[width=1.8\columnwidth]{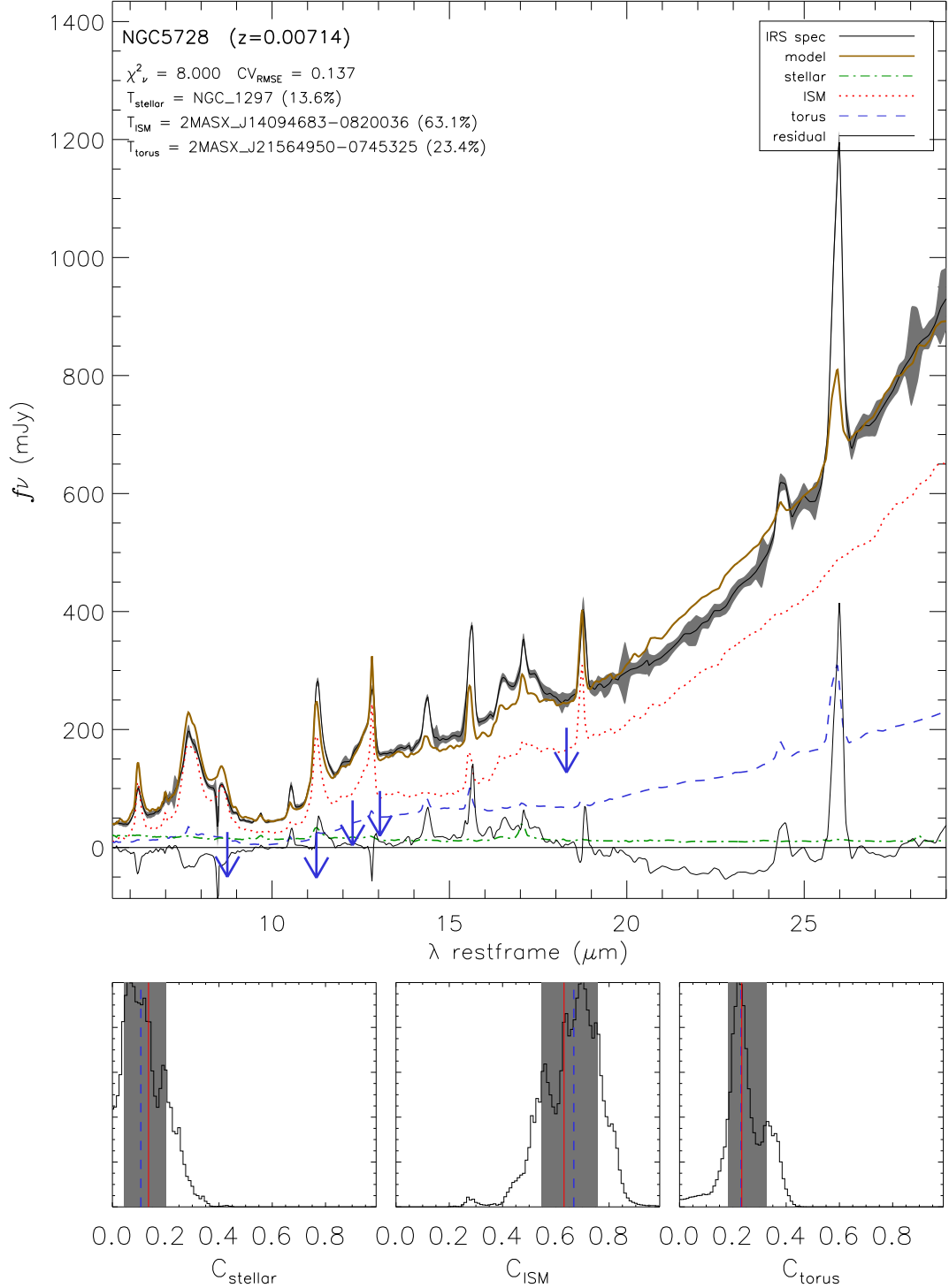}
\caption{Best-fit decomposition models for the IRS/\emph{Spitzer} spectrum of NGC\,5728 (grey shaded area and the black continuous line) using high resolution data and X-ray luminosity as constraints for the decompositions (see text). The (red) dotted, (blue) dashed, and (green) dotted-dashed lines represent the ISM, torus, and stellar components of the best-fitting model (shown in continuous yellow line), respectively. The continuous line at the bottom of each plot shows the residuals (spectrum--model). Blue arrows (in the top panel) are the high resolution data and 12$\rm{\mu m}$ flux limit (derived from the X-ray luminosity) used as constraints for the final fit. The three bottom panels show $\rm{C_{stellar}}$, $\rm{C_{ISM}}$, and $\rm{C_{torus}}$ posterior distributions, from left to right.}
\label{fig:ExampleDecomp}
\end{center}
\end{figure*}

We used the model-independent spectral decomposition called DeblendIRS presented by HC15 to decompose the IRS/\emph{Spitzer} spectra. This code uses a set of IRS spectra as templates for purely stellar, interstellar (dominated by polycyclic aromatic hydrocarbons, PAHs), and AGN dominated components. Hereinafter we call these three components \emph{stellar}, \emph{ISM}, and \emph{torus}, respectively. Note that we refer to the AGN component as torus component since the torus is the dominant source of AGN continuum in the mid-infrared. The algorithm computes the marginalised probability distribution of physical parameters, from comparison of the observed data with all the models in the library, using Bayesian inference. 

We initially have used as templates the same library presented by HC15. However, in their work they focused on the decomposition in the range between 5 and 15$\rm{\mu m}$. For that reason some of the templates did not cover the full IRS/\emph{Spitzer} wavelength range. Our purpose is to decompose the full range covered by IRS spectra as much as possible in order to study the entire \emph{Spitzer} spectra (and to compare with clumpy models). For that reason, we have removed the templates with redshift above 0.2. All together we have removed 76 sources from the torus template list and two for the \emph{ISM} list. This guarantees that all the templates can be used to decompose our local sample with spectra in the range 5.5-29$\rm{\mu m}$. We have chosen this range in order to maximise the number of templates that we are able to use, also maximising the range covered. Finally, we have also removed from the AGN template library all the sources included in our current analysis (12 AGN, mainly type-1 and type-2 Seyferts). The final template list contains 101 torus, 59 ISM, and 19 stellar dominated spectra.  

We have used the following constraints on the torus component to improve the uncertainty on the torus component as our aim is to study if the torus is present in our sample. Firstly, we have used high spatial resolution data (described in Sect.~\ref{sec:canaricamdata} and \ref{sec:otherdata}) to put an upper limit to the torus component flux at the specific observed wavelengths. Note that we are able to include these constraints because we know that the torus component is point like at both \emph{Spitzer} and high resolution imaging data. We used the flux measured in the nuclear component plus three times the error. Secondly, we have used the X-ray luminosity to set an upper limit to the torus component at 12$\rm{\mu m}$. Using the well stablished relation between the 12$\rm{\mu m}$ and 2-10 keV X-ray luminosities for AGN \citep{Asmus15}. We have used the 12$\rm{\mu m}$ to X-ray relation found by \citet{Asmus15} because it is the most recent relation with the largest number of sources. Since X-ray luminosities can show short and long term variations, that can be as high as a factor of 10 \citep[e.g.][]{Gonzalez-Martin12}, we have assumed the error on the 2-10 keV X-ray luminosity to be a factor of 10. Then, we have used this value to estimate another upper limit to the 12$\rm{\mu m}$ flux of the torus component. It is worth noticing that the X-ray to mid-infrared relation has been tested down to X-ray luminosities of $\rm{L_{X}(2-10~keV)\sim 10^{41} erg/s}$ \citep{Asmus15}. However, our current sample reaches X-ray luminosities down to $\rm{L_{X}(2-10~keV)\sim 10^{38} erg/s}$. Therefore, this relation might not apply to the very faint end of the luminosity function of AGN. Indeed, one of the goals of this work is to test the plausible disappearance of the torus, although little is known of the very low luminosity sources. In this case we expect the 12$\rm{\mu m}$ flux to be an upper limit to the actual contribution of the torus. Thus, using these limits as an upper limits does not bias our results. Furthermore, this upper limit is very useful because it is available for all the objects in our sample. 

We have used version 1.2 of the deblendIRS\footnote{http://www.denebola.org/ahc/deblendIRS} code presented by HC15. It improves over previous versions in that it allows to include flux priors to constrain the torus component. For each object we have decomposed its IRS/\emph{Spitzer} spectrum, using the upper limit on the 12$\rm{\mu m}$ flux (obtained from the 2-10 keV X-ray flux) and the mid-infrared fluxes obtained from high spatial resolution images. Fig.~\ref{fig:ExampleDecomp} shows the best fit for the spectrum of the type-2 Seyfert galaxy NGC\,5728 as an example. 

\begin{figure}
\begin{center}
\includegraphics[width=1.\columnwidth]{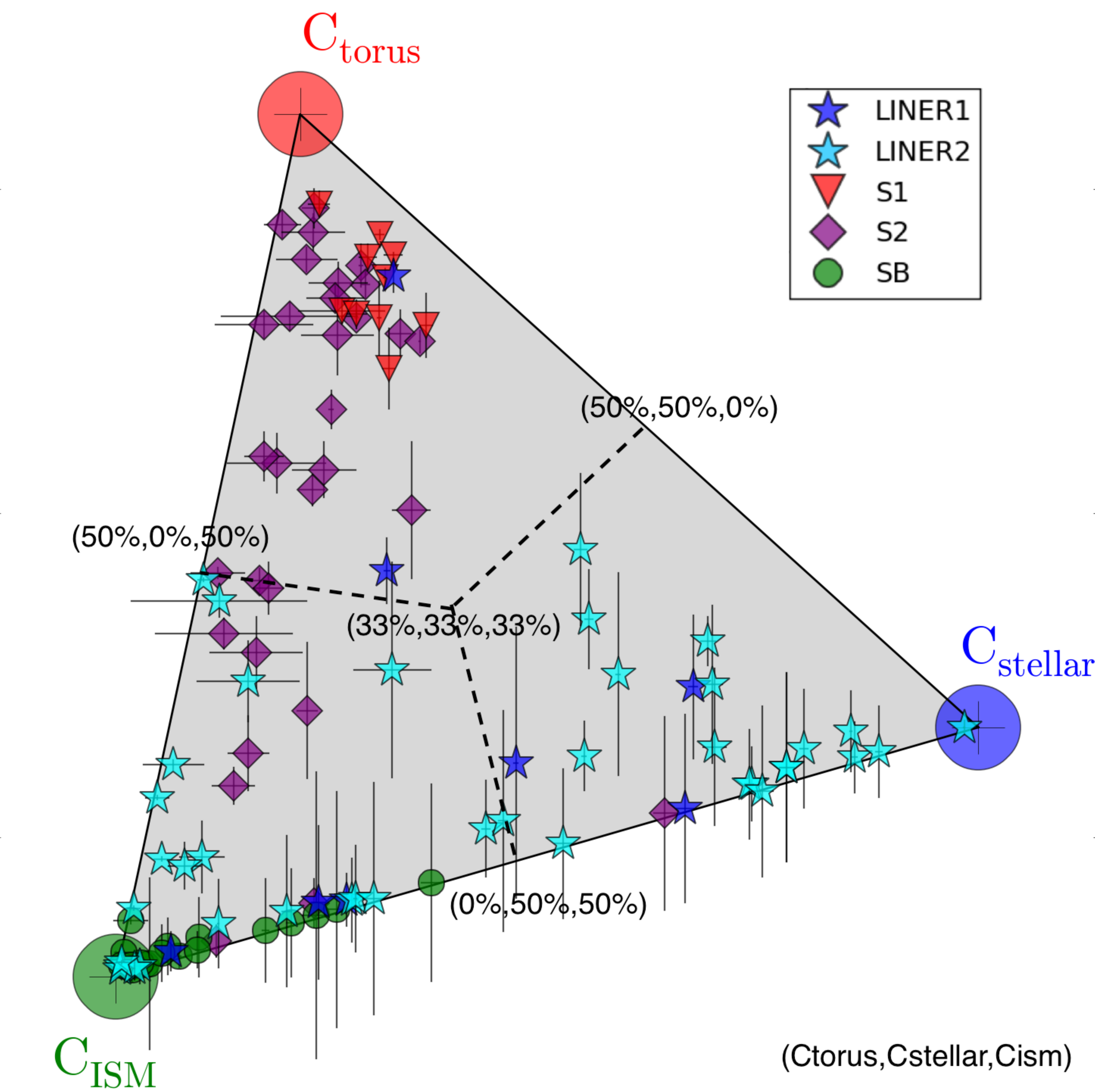}
\caption{Diagram to show the contribution of torus (top corner marked with the large red circle), ISM (bottom-left corner marked with the large green circle), and stellar (bottom-right corner marked with the large blue circle) components to the objects in our sample. Objects close to the corners of this triangle show large contributions of the corresponding component (see text). Type-1 Seyferts, type-2 Seyferts, type-1 LINERs, type-2 LINERs, and Starbursts are shown with up-side down red triangles, purple diamonds, dark blue stars, light blue stars, and green circles, respectively.}
\label{fig:CompDiagram}
\end{center}
\end{figure}


Tables \ref{tab:sample}, \ref{tab:comparisonsample1} and \ref{tab:comparisonsample2} include $\rm{C_{torus}}$ (Col.~6), $\rm{C_{stellar}}$ (Col.~7), and $\rm{C_{ISM}}$ (Col.~8) for the LINER, Seyfert, and Starburst samples, respectively. These values refer to the percentage of each component contributing to the 5-15$\rm{\mu}$m wavelength range. Appendix \ref{ape:decompose} shows the decomposition and posterior distributions for the full sample. The median values and 16-84\% percentiles for each class are listed in Table \ref{tab:contributors}. Starbursts nicely group into $\rm{C_{torus}<}$1.3\%, demonstrating the good behavior of this method to decompose \emph{Spitzer} spectra (see Table \ref{tab:contributors}). Only NGC\,3367 shows $\rm{C_{torus}>5\%}$ (8.4\%). This object might actually host an AGN \citep[see][]{Gonzalez-Martin15}. Note that we have also confirmed that the change on the spectral range used for the decomposition to 5-15$\rm{\mu m}$ has no impact in our results.

Fig.~\ref{fig:CompDiagram} shows a diagram representing the relative contribution of each component for the objects in our sample. Objects with 100\% of a single component are located at one of the corners of this diagram (marked with their names in the diagram). Objects with 0\% of one of the components would be located in the side of the triangle opposite to its corner. Interestingly, the diagram is not equally populated. There are no objects with $\rm{C_{ISM}\sim 0\%}$, indicating that all the objects in our sample show a non-negligible ISM contribution. It is worth to notice that each class can be clearly differentiated using their contributors. Starburst are mainly dominated by ISM components. Type-1 Seyferts are fully dominated by the torus component with less than 30\% of stellar component, and less than $\rm{C_{ISM}<20\%}$. Type-2 Seyferts have a wide range of torus and ISM components, with less than 30\% of stellar component. Finally, LINERs show less than 50\% of torus component with a wide range of both stellar and ISM components. The small number of type-1 LINER prevents to reach any firm conclusion on the difference between type-1 and type-2 LINERs. 

The DeblendIRS code computes the root-mean-square error (RMSE) of the final fit for each object (included in Tables \ref{tab:sample}, \ref{tab:comparisonsample1} and \ref{tab:comparisonsample2}, Col.~9). We use these numbers to investigate when the final fit is good enough to represent the data. We have selected as bad fits those with RMSE$\rm{>}$0.3 (HC15). Ten objects show RMSE above that limit; two type-2 Seyfert (NGC\,3393 and NGC\,4945), two Starbursts (NGC\,925 and NGC\,3184), a type-1 LINER (NGC\,4450), and five type-2 LINERs (IRAS\,17208-0014, IIIZW\,035, IRAS\,14348-1447, NGC\,4125, and IRAS\,12112+0305). Note that large RMSE values also imply larger error on the estimates. We marked these sources in the plots in the following analysis. According to this criterium (RMSE$\rm{<0.3}$), DeblendIRS successfully fit the spectra in 90\% of the sample. We have not found any trend on the quality of the fit according to the AGN classification. However, most of the objects with large RMSE are actually ULIRGs with deep silicate features (as already discussed by HC15).

\begin{figure*}
\begin{center}
\includegraphics[width=1.\columnwidth]{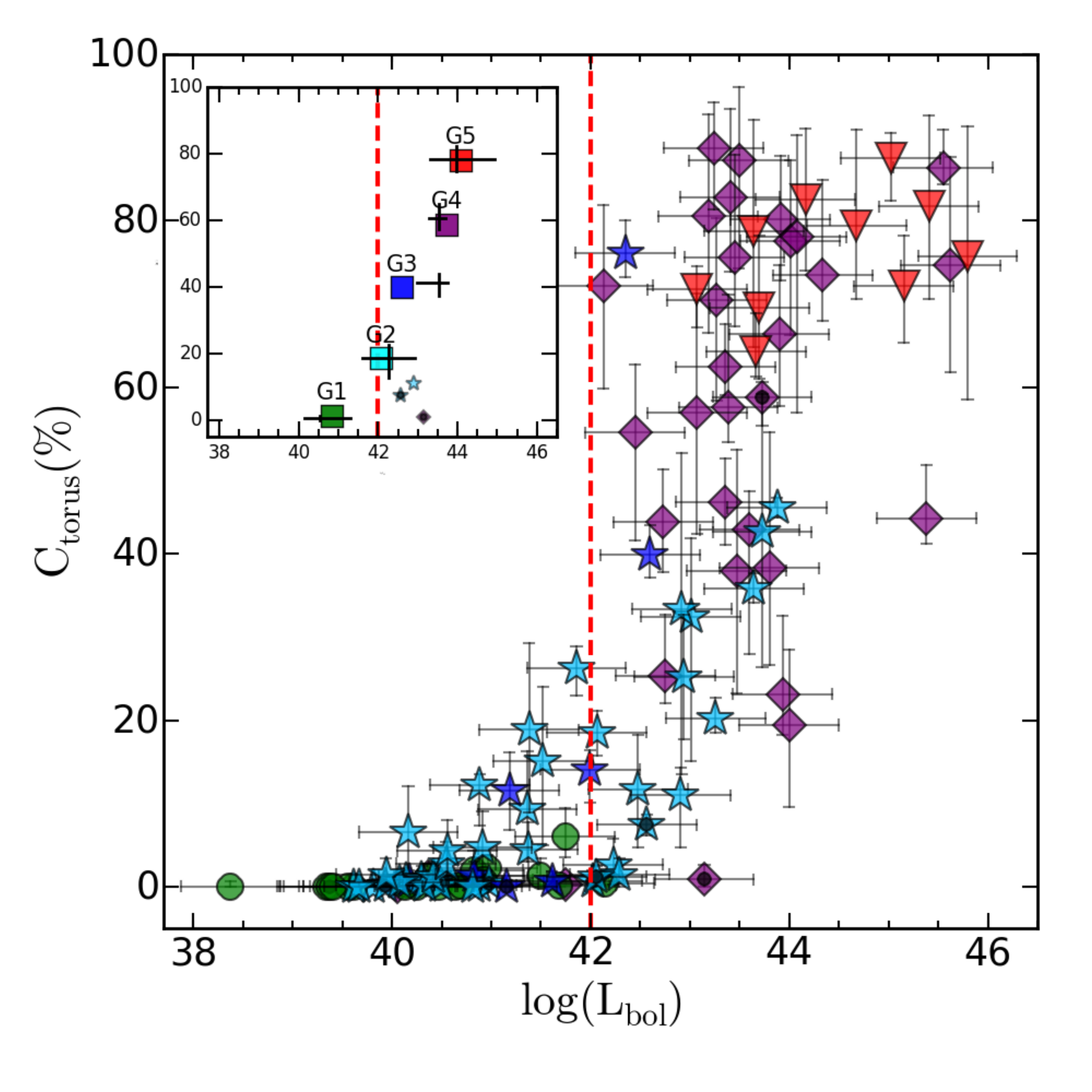}
\includegraphics[width=1.\columnwidth]{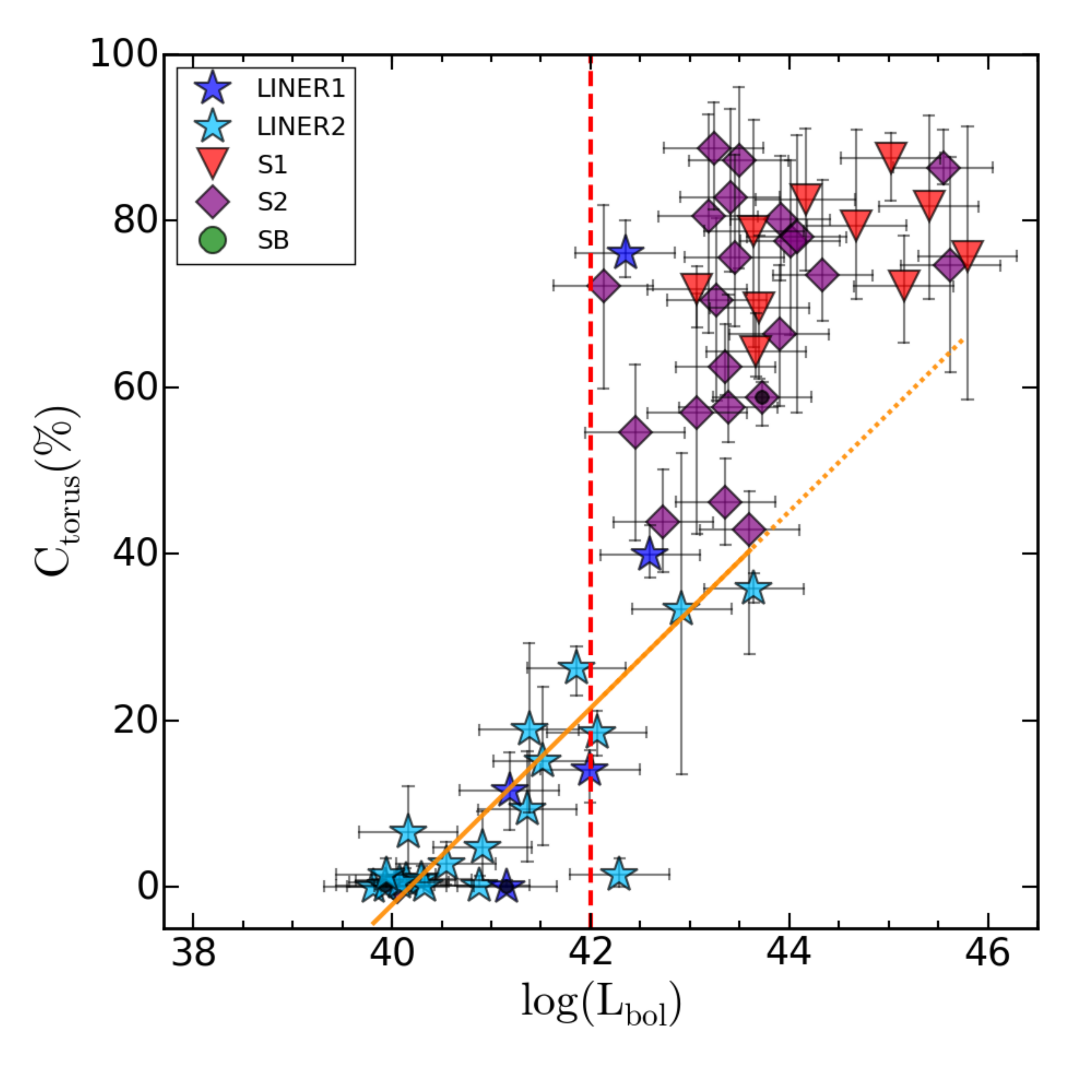}
\caption{Contribution of the torus component versus the bolometric luminosity (in units of erg/s) in logarithmic scale for the full sample (left) and for those showing $\rm{C_{ISM}<50\%}$ (right). The vertical dashed red line shows $\rm{L_{bol}=10^{42}erg/s}$, value where the torus is expected to disappear. However, note that this limit actually depends on the SMBH mass \citep[][see also text]{Elitzur09}. The small inset within the left panel shows the representative objects for the five groups found by the AP method (see text) with a square. It also shows the median values of the groups found by AP method (errors show the range using 25-75\% percentiles). Objects with $\rm{RMSE>0.3}$ are marked with a black dot in both panels. Objects shown in this small box (left panel) are those consistent with the group of negligible $\rm{C_{torus}}$ at mid-infrared and bolometric luminosities $\rm{L_{bol}>10^{42}erg/s}$ (NGC\,4945 (S2), UGC\,05101 (LINER2), and IRAS\,14348-1447 (LINER2), see text). The orange continuous line (right panel) shows the best fit linear relation for objects with $\rm{C_{torus}<50\%}$. The orange dotted line (following the orange continuous line) is the extrapolation expected for larger bolometric luminosities. }
\label{fig:LbolvsfAGN}
\end{center}
\end{figure*}

\section{Torus contribution versus AGN luminosity}\label{sec:torusvsluminosity}

We examine the relationship between $\rm{log(L_{bol})}$ and $\rm{C_{torus}}$ in Fig.~\ref{fig:LbolvsfAGN}. $\rm{C_{torus}}$ is reported in Tables \ref{tab:sample}, \ref{tab:comparisonsample1} and \ref{tab:comparisonsample2} (Col. 5). The bolometric luminosities are computed using the 2-10 keV luminosities (L(2-10 keV); reported in Tables~\ref{tab:sample}, \ref{tab:comparisonsample1} and \ref{tab:comparisonsample2}, Col.~4) using the relation $\rm{L_{bol}=}\kappa$\,L(2-10 keV), where the bolometric correction ($\kappa$) depends on the L(2-10 keV) luminosity itself with a fourth order polynomial \citep[see the prescription given by][]{Marconi04}:

\begin{equation}
log(\mathcal{L}/L(2-10~keV))=1.54+0.24\mathcal{L}+0.012\mathcal{L}^{2}-0.0015\mathcal{L}^3
\end{equation}

\noindent where $\rm{\mathcal{L}}\rm{=(logL_{bol}-12)}$ and $\rm{L_{bol}}$ is in units of $\rm{L\odot}$. 

\subsection{Affinity groups}\label{sec:affinity}

Although there is a trend showing that less luminous objects have smaller $\rm{C_{torus}}$, the relation between the bolometric luminosity and $\rm{C_{torus}}$ is not linear (coefficient of correlation $\rm{r= 0.6}$). Fig.~\ref{fig:LbolvsfAGN} (left panel) shows as a red dashed vertical line the expected luminosity below which the torus would disappear \citep{Elitzur06}. Indeed, objects below that limit tend to show small $\rm{C_{torus}}$. Above that limit there is a wide range of percentages of torus contributions $\rm{C_{torus}}$. In order to quantify this (and also to define groups within the plot for subsequent analysis), we have used the clustering Affinity Propagation (AP) method \citep{Frey07} to look for groups in this diagram. Clustering analysis is aimed at discovering the underlying clusters in the data points according to their similarities. The AP method, in particular, is based on the concept of message passing between data points. The advantage of this method compared to other clustering methods (e.g. k-means) is that AP does not require the number of clusters as an input to the algorithm. The AP method divides the sample into groups and associates one object as representative of its group. 
 
We have applied the AP method to the pair of data [$\rm{L_{bol}}$, $\rm{C_{torus}}$] for the full sample using the Python routine `AffinityPropagation' within the package {\sc scikit-learn}\footnote{http://scikit-learn.github.io}. We used `euclidean' as the affinity method, which uses the negative squared euclidean distance between points. We set the maximum number of iteration $max\_iter$ to 1000 although the actual number of iterations needed to converge was only 43. The routine iterates until the number of estimated clusters does not change for a selected number of iterations $convergence\_iter$. We set this number of iterations to $convergence\_iter$=15. Note that $convergence\_iter$ parameter is a different restriction than $max\_iter$ parameter. However, we have noticed that changes in this $convergence\_iter$ parameter do not affect the results as long as $convergence\_iter>1$. This means that the groups in the sample were found after the second iteration. 

Using the AP method, the points in Fig.~\ref{fig:LbolvsfAGN} (left panel) can be classified into five groups. Col.~10 in Tables~\ref{tab:sample}, \ref{tab:comparisonsample1} and \ref{tab:comparisonsample2} show to which group each object belongs to. Table \ref{tab:affinity} gives the pair of positions for the representative member for each group and the median (and 25-75\% percentiles as the width of the distribution) of the objects belonging to each group. The locus of the representative members and their median values is shown in the small inset within Fig.~\ref{fig:LbolvsfAGN} (left panel) as squares and black crosses, respectively. The main difference between groups is the percentage of the  torus contribution, $\rm{C_{torus}}$. Among them, only one group is below the line of $\rm{L_{bol}=10^{42}erg/s}$ and the median $\rm{C_{torus}}$ for them is consistent with zero ($\rm{C_{torus}<1.4\%}$, see Table \ref{tab:affinity}). Furthermore, only three objects are consistent with this group but showing $\rm{L_{bol}>10^{42}erg/s}$ (namely NGC\,4945, UGC\,05101, and IRAS\,14348-1447, shown in the small panel in Fig. \ref{fig:LbolvsfAGN}, left panel). Among them, all but UGC\,05101 have RMSE$\rm{>0.3}$, indicating poor decompositions. Thus, the behaviour seen in Fig.~\ref{fig:LbolvsfAGN} is consistent with a wide range of $\rm{C_{torus}}$ above $\rm{L_{bol}=10^{42}erg/s}$ and negligible $\rm{C_{torus}}$ below that limit. 

It is also worth mentioning the results of this classification method compared to the optical classes included in this analysis (see Col.~8 in Tables \ref{tab:sample} , \ref{tab:comparisonsample1} and \ref{tab:comparisonsample2}). All the Starbursts are classified as Group 1 (i.e. $\rm{C_{torus}<1.4\%}$, see Table \ref{tab:affinity}). This is fully expected under the assumption that all these sources are non-AGN. However, Group 1 also contains four type-2 Seyferts, 4 type-1 LINERs, and 25 type-2 LINERs.  Eight out of the 10 type-1 Seyferts are classified within Group 5  (i.e. $\rm{C_{torus}\simeq80\%}$, see Table \ref{tab:affinity}). Only one LINER (NGC\,1052) and 11 type-2 Seyferts are associated to Group 5. Except NGC1052, all LINERs belong to groups below Group 3. In the case of Seyferts, there is a complete mix of groups.

\begin{figure*}
\begin{center}
\includegraphics[width=2.0\columnwidth]{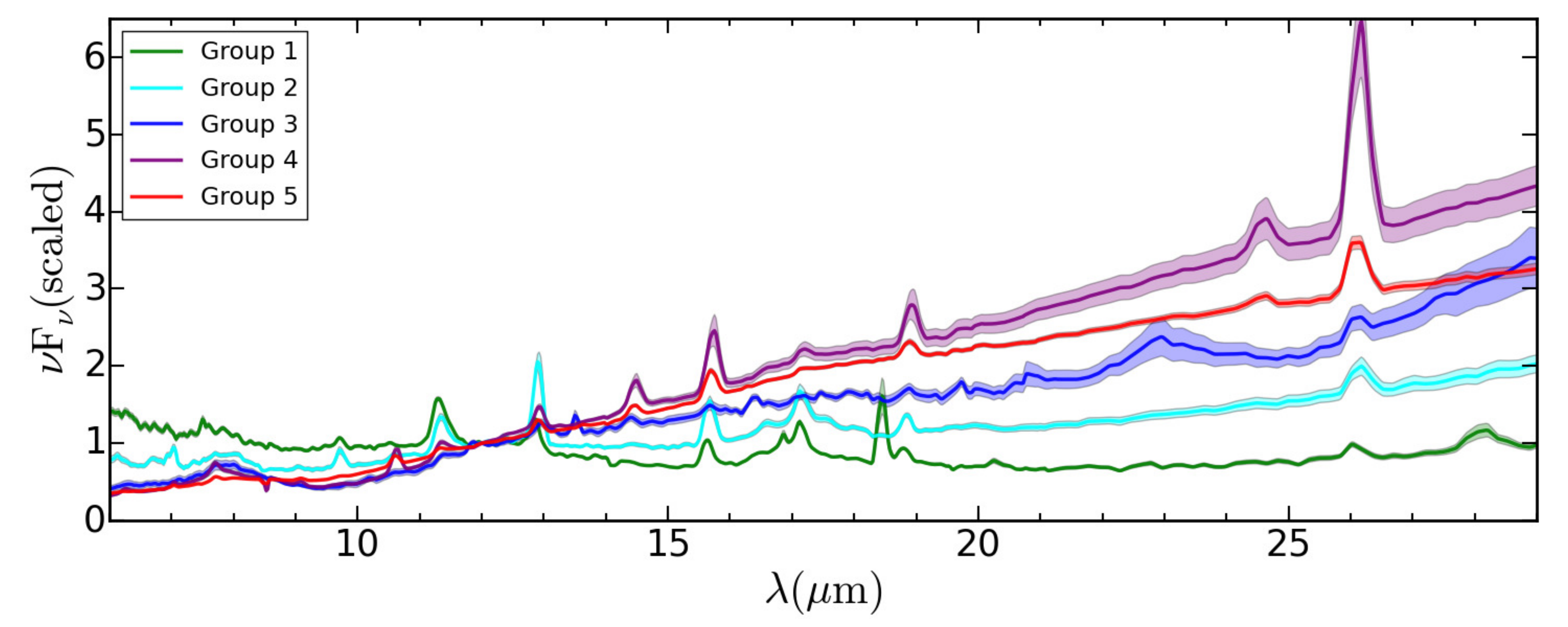}
\includegraphics[width=2.0\columnwidth]{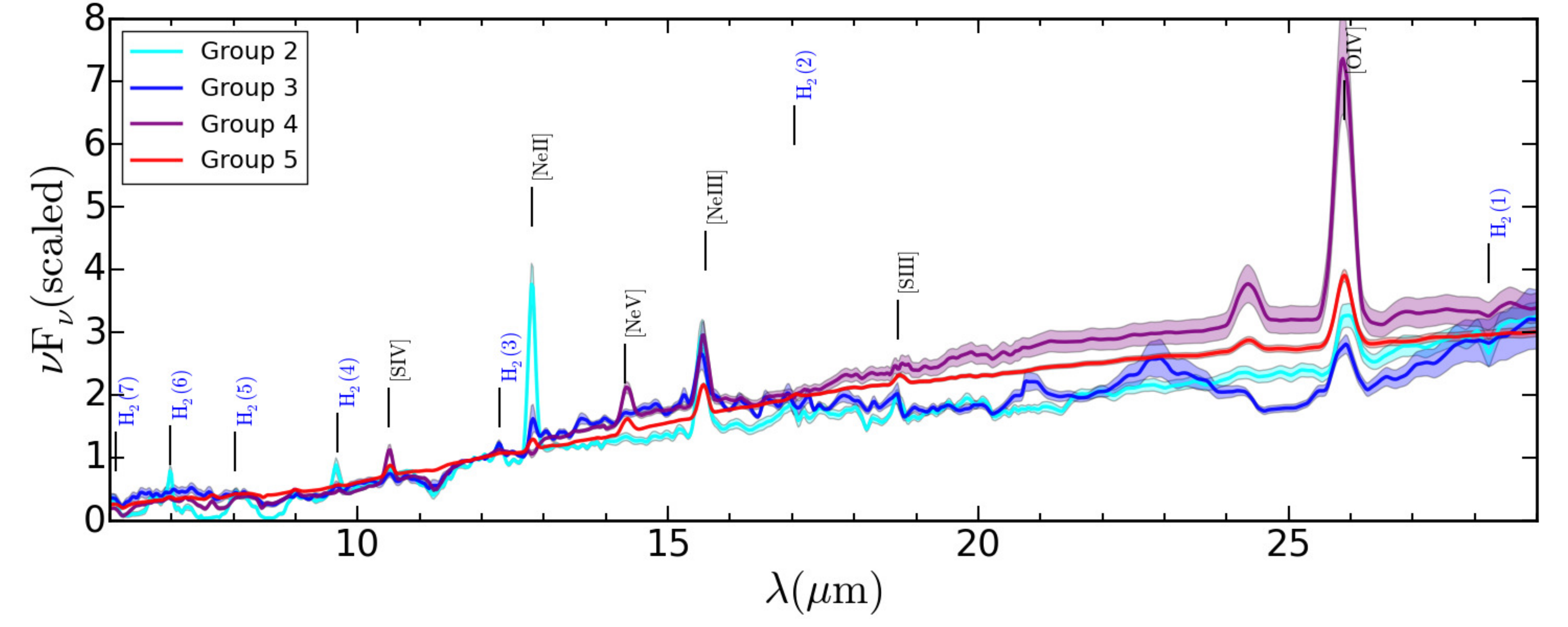}
\caption{(Top): Average IRS/\emph{Spitzer} spectra for the groups found using the AP method with $\rm{C_{ISM}<50\%}$ (see text). (Bottom):  Same average IRS/\emph{Spitzer} spectra than in the top panel but after subtracting the ISM and stellar components (i.e. the AGN component). Note that we do not show Group 1 because the AGN component is only residual (see text). $\rm{H_{2}}$ molecular lines are marked with blue letters and forbidden transitions are shown in black letters. }
\label{fig:averageAP}
\end{center}
\end{figure*}

The measurement of $\rm{C_{torus}}$ could depend on the host galaxy properties. We have collected the morphological types and B magnitudes\footnote{Note that in the case of the B magnitudes we found them for 80 sources in our sample.} for the galaxies in our sample and studied whether the distribution changes with the AP groups. We have computed the 25-75\% percentiles of each distribution for morphological types and B magnitudes (these numbers are recorded in Table~\ref{tab:affinity}). There is an increase on the B total magnitude from Group 1 to Groups 4-5, although all of them agree with each other within the percentiles. Furthermore, more distant objects could also include more dust within the slit of IRS/\emph{Spitzer}. We also computed the 25-75\% percentiles of each distribution of ``Slit widths'' (see Table~\ref{tab:affinity}). There is no correlation between the AP groups and the physical portion of the galaxy included in the IRS slit width.

We have investigated if the percentage ISM contribution ($\rm{C_{ISM}}$) might affect our results. $\rm{C_{ISM}>50\%}$ for Starbursts. This is consistent with the idea that the PAH features dominating the ISM are produced in the photo-dissociating region associated to a star-forming region. Starbursts, essentially HII galaxies, are dominated by star-forming regions all over the spectral energy distribution and, therefore, dominated by  the PAH features at mid-infrared. Among the AGN classes, both LINERs and Seyferts spread in a wide range of $\rm{C_{ISM}}$. This is consistent with the idea that $\rm{C_{ISM}}$ depends on the inner star-formation ($\rm{\sim}$1 kpc). Indeed, we have studied whether this contribution depends on AGN luminosity finding no correlation at all ($\rm{r=0.18}$). 

We have examined how Fig. \ref{fig:LbolvsfAGN} (left panel) changes if we consider only objects with $\rm{C_{ISM}<50\%}$ (see Fig.~\ref{fig:LbolvsfAGN}, right panel). Now the tendency of large $\rm{C_{torus}}$ for larger AGN bolometric luminosities is much more clear ($\rm{r=0.8}$). This correlation is still better ($\rm{r=0.9}$) if we select objects with $\rm{C_{torus}<}$50\% (see continuous orange line in Fig.~\ref{fig:LbolvsfAGN}, right panel). Furthermore, it is worth noticing that objects with $\rm{C_{torus}>}$50\% do not follow the same relation (only two objects are consistent within the error bars). In fact, they tend to have a larger $\rm{C_{torus}}$ than that expected by the extrapolation of the linear fit of objects with $\rm{C_{torus}<50\%}$ (dotted orange line in Fig.~\ref{fig:LbolvsfAGN}, right panel).  

\begin{figure}
\begin{center}
\includegraphics[width=1.0\columnwidth]{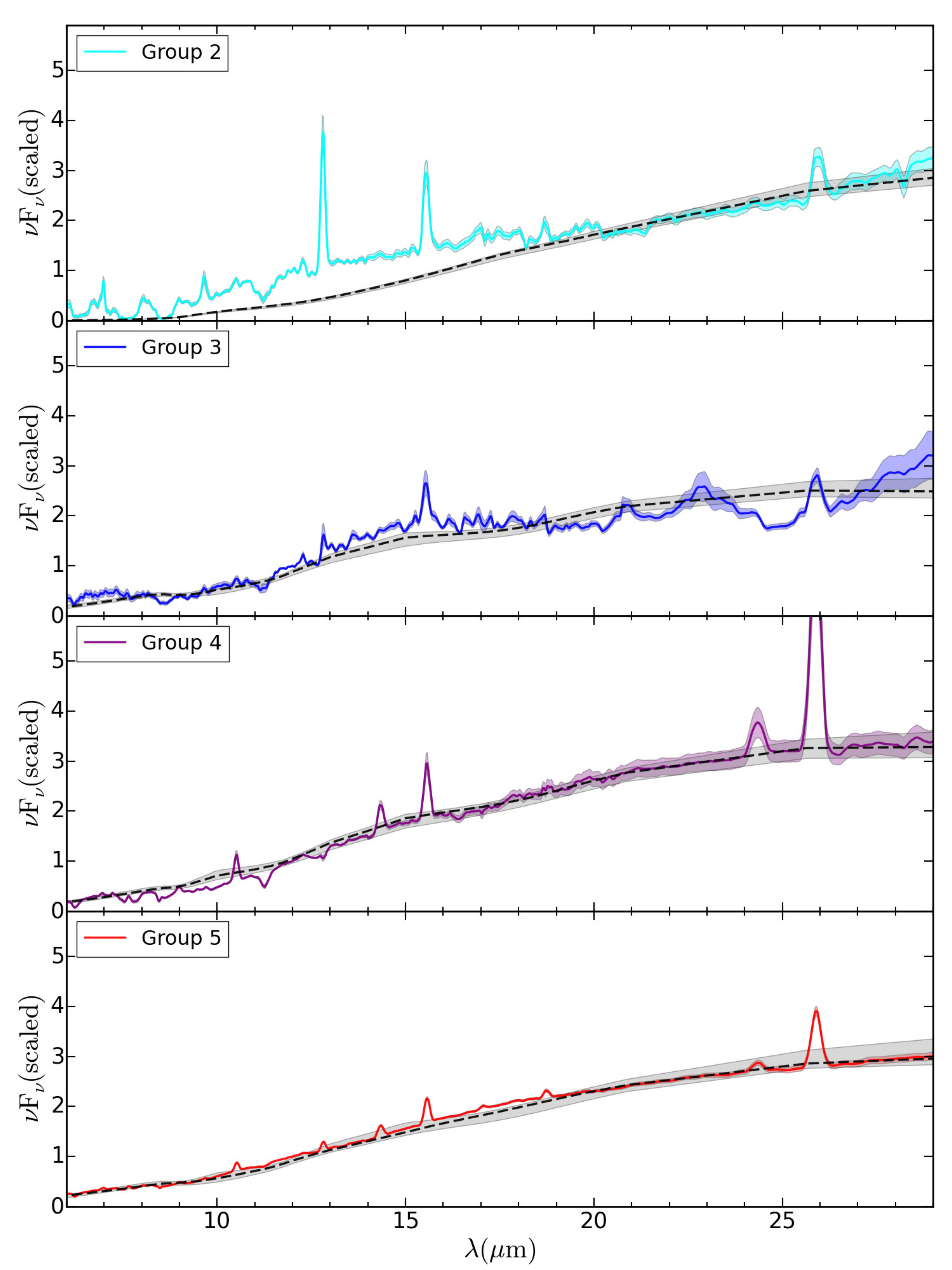}
\caption{Clumpy model fits (long dashed black lines) to the average AGN component of Group 2, 3, 4, and 5 (from the top to the bottom panel). The gray shadowed region is the lower and upper bound of the fit.}
\label{fig:averageAPmodel}
\end{center}
\end{figure}

\subsection{Average spectra}\label{sec:Averagespectra}

We constructed an average spectrum for each of the five groups found by the AP method (see Fig.~\ref{fig:averageAP}, top panel). The dispersion shown as a shadow area is the 1-$\rm{\sigma}$ uncertainty of the mean. All the spectra are scaled to their flux at 12$\rm{\mu m}$ before the average. As in Fig.~\ref{fig:LbolvsfAGN} (right panel), we have excluded objects with $\rm{C_{ISM}>50\%}$, to avoid those spectra where the ISM is dominating the observed emission\footnote{We have excluded objects with large contributions of ISM because all AGN classes show a large range of ISM contributions indicating that this component is independent on the AGN classes and entirely due to the circumnuclear conditions of each source.}. The slope of the spectra gradually becomes bluer as we move to lower $\rm{C_{torus}}$ (i.e. as we move from Group 4 and 5 to Group 1). We interpret this as a stellar dominance for the observed first AP classes ($\rm{C_{torus}<40\%}$). Our decomposition method allows us to study the average spectrum also isolating the torus component for each AP group. For that purpose, we have subtracted the stellar and ISM components to the IRS/\emph{Spitzer} spectra. The average spectra are shown in Fig.~\ref{fig:averageAP} (bottom panel). We do not show in this plot the average spectrum for Group 1 ($\rm{C_{torus}<1.4\%}$) because, after removing stellar and ISM spectra, the residuals are negligible in this group (see Table~\ref{tab:affinity}). Now the average spectra are much more similar among the four groups. The main difference is the presence of strong emission lines in Group 2. Indeed, while collisional lines such as [SIV], [NeII], [NeIII] and [OIV] are present in all the categories, it seems that there is an enhancement of the $\rm{H_{2}}$ molecular lines in Groups 2 and 3 ($\rm{C_{torus}\sim20\%}$ and $\rm{C_{torus}\sim40\%}$, respectively). 

It is worth mentioning that Groups 2 (2, 3, and 4) show the $\rm{H_{2}}$ molecular line at 28$\rm{\mu m}$ (PAH feature at 11.3 $\rm{\mu m}$) in absorption. This indicates a slight oversubtraction of those emission lines. However, note that all the other $\rm{H_{2}}$ molecular lines are shown in emission and that the other PAH features (e.g. 8.7$\rm{\mu m}$) are not seen in either emission or absorption. Thus, we believe this oversubtraction depends on the particular conditions of the lines. Perhaps this oversubtraction of the $\rm{H_{2}}$ molecular lines could indicate a different composition of the clouds. In the case of PAH features, it is even harder to evaluate since changes in the composition of PAHs could yield to a different profile of this line.

We have fitted the subtracted average spectra (Fig.~\ref{fig:averageAP}, bottom panel) with the Clumpy torus models \citep{Nenkova08A,Nenkova08B} using Bayesian inference to estimate probability distributions for the parameters of the torus model. Note that emission and absorption features have been removed from the spectral coverage since these models only account for the shape of the continuum. The BayesClumpy code uses Metropolis-Hastings Markov Chain Monte Carlo algorithm for sampling the posterior distribution function \citep[see][for more details on the code]{Asensio-Ramos09}. The parameters of the torus involved in the fitting are the half opening angle of the torus $\rm{\sigma}$, the outer radius of the torus as a function of the inner radius of the torus, $\rm{Y=R_{out}/R_{inner}}$, the number of clouds in the equatorial plane, $\rm{N_{o}}$, the slope of the power-law distribution of clouds with respect to the angle from the equatorial plane, $\rm{q}$, the optical depth of the clouds, $\rm{\tau_{\nu}}$, and the inclination between the line of sight and the equatorial plane of the torus, $\rm{i}$. Using the best fit parameters BayesClumpy marginalize over the AGN geometrical covering factor, $\rm{f_{c}}$:

\begin{equation}
f_{c} = 1 - \int_{0}^{90^{\circ}}  e^{-N_{o} e^{-(90^{\circ} - i)^{2}/\sigma^{2}}} cos (i) di
\end{equation}

The inner radius of the torus is linked to the AGN bolometric luminosity because it depends on the radius at which the dust sublimates \citep[see][]{Barvainis87,Nenkova08A,Nenkova08B}. Thus, the outer radius of the torus can be computed using the bolometric luminosity of the AGN and the parameter $\rm{Y}$. The height of the torus $\rm{H}$ can also be computed as: $\rm{H = (R_{out} + R_{in}) sin(\sigma)/2}$.

We did not fit Group 1 because after subtracting stellar and ISM components, only residuals are left. Fig. \ref{fig:averageAPmodel} shows the best fit for Groups 2, 3, 4, and 5. We failed to find a good fit for Group 2. Although the average spectrum of Group 2 resembles Groups 3, 4, and 5, below 10$\rm{\mu m}$ it decreases with a slope too steep and it increases too quickly above 25$\rm{\mu m}$ for any of the Clumpy models. We believe that the extra contribution above 25$\rm{\mu m}$ could be a residual of the ISM component. The deficit of emission compared to the model below 10$\rm{\mu m}$ could be due to extra extinction or due to a distribution and/or composition of the dusty region not predicted by the Clumpy models described by \citet{Nenkova08A,Nenkova08B}. 

We were able to successfully fit Group 3, 4, and 5, although the spectral fitting for Group 3 is visually worse than that for Groups 4 and 5. The resulting parameters are listed in Table~\ref{tab:BayesClumpy}. All the parameters are in the range found for AGN fitted to Clumpy models \citep[][and references therein]{Ramos-Almeida09,Nenkova08B}. Torus sizes recovered by the models are fully in agreement with mid-infrared interferometric observations \citep{Burtscher13}. 

Rather than focus on the actual numbers obtained for each group it is more relevant to look for differences on the measured parameters among the three Groups. The number of clouds in the equatorial plane ($\rm{N_{o}}$) and the viewing angle ($\rm{i}$) are similar for the three groups. However, when we split Group 5 into type-1 and type-2 AGN\footnote{Note that this exercise can only be made in Group 5 because it is the only one containing type-1 Seyferts.} (see the last two columns in Table \ref{tab:BayesClumpy}) we naturally recover the dependence with the viewing angle (as expected under the unified model). There is a marginal trend for an increase of the slope of the clouds distribution $q$ with the distance from the equatorial plane, the height of the torus and a decrease on the half opening angle $\rm{\sigma}$ from Group 3 to Group 5. The latter results on a marginal detection on higher geometrical covering factors for lower luminosity AGN.
 
The parameter which is changing the most among the three groups is the outer size of the torus ($\rm{R_{out}}$), increasing from Group 3 to Group 5 (i.e. for $\rm{C_{torus}}$ from 40\% to 80\%). \citet{Alonso-Herrero11} did not find a different size of the torus fitting 13 AGN to the very same Clumpy models used here. However, there are two main reasons for these results to not be in contradiction. First, the range of luminosities covered here is much larger than those reported in their sample. Indeed, the sizes of the torus and bolometric luminosities reported in their publications are consistent with all of them being mainly in Group 5. Second, we have found this tendency after averaging objects classified into the same group using the AP method while they used individual Clumpy fits to derive their conclusions. Since the torus half opening angle ($\rm{\sigma}$) is similar and the size of the torus ($\rm{R_{out}}$) is smaller when moving from Group 5 to Group 3, the covering factor (the half opening angle of the torus) of the AGN is larger (smaller) for Group 3 compared to Group 5. Note that far-infrared flux observations would be more sensitive to the torus extent independently \citep{Ramos-Almeida11}. Thus, the addition of these measurements could confirm the trend found here for the AP groups (i.e. with $\rm{C_{torus}}$).  

\subsection{Warm molecular gas}\label{sec:warmgas}

Motivated by the $\rm{H_{2}}$ molecular emission lines detected in Groups 2 and 3 as found by the AP method, we have done a search for all the $\rm{H_{2}}$ molecular lines found in the IRS/\emph{Spitzer} spectra of the sample. Due to the coverage of our spectra we have been able to look for the transitions S(0) 28.22$\rm{\mu m}$, S(1) 17.03$\rm{\mu m}$, S(2) 12.27$\rm{\mu m}$, S(3) 9.67$\rm{\mu m}$, S(4) 8.02$\rm{\mu m}$, S(5) 6.91$\rm{\mu m}$, and S(6) 6.11$\rm{\mu m}$. We have built a code using python routines (within {\sc scipy}) to automatically detect these lines. We have forced a fit to a Lorentzian profile to each of the proposed lines. We allowed to vary the amplitude, center, and width of the Lorentzian profile. We have run this fitting 200 times using Monte Carlo simulations to estimate the error of these fittings from the error of each spectrum. We consider a line to be detected if the amplitude of the line is above 3 times its error and the center of the line is consistent with the expected wavelength of the line within the width of the Lorentzian. Note that we have performed such a detection in the IRS/\emph{Spitzer} spectra and not after stellar and ISM components were subtracted. 

We have not detected S(0) in any of the objects of our sample. Furthermore, none of the $\rm{H_{2}}$ molecular lines are detected in any of the Starbursts and type-1 Seyferts included in our comparison sample. S(5) and S(6) lines are detected among both LINERs and type-2 Seyferts.  S(4) is detected only in five objects, three in Group 1 and two in Group 5. S(1), S(2), and S(3) are only detected in Groups 1, 2, and 3 (i.e. $\rm{C_{torus}<40\%}$).  

\section{Discussion}\label{sec:discussion}

An outflowing wind of material from the AGN might be responsible for the BLR, NLR, and torus componentsÊ\citep[e.g.][and references therein]{Elitzur09}. According to \citet{Emmering92}, the properties of these components depend on the AGN bolometric luminosity \citep[see also][]{Nicastro00}. Many evidences for cloud outflows indicate that instead of a hydrostatic torus, this region is part of a clumpy wind (including BLR and NLR) coming off the accretion disk \citep[][and references therein]{Emmering92,Nicastro00,Wada12}. Indeed polar dust emission is being found in some AGN, with a PA coincident with that of the NLR or the jet \citep{Mor09,Asmus16,Lopez-Gonzaga16}. Detailed fitting of the SED including dust from both NLR and torus has been presented \citep{Elitzur06,Mor09,Mor12}. The inner radius of the torus depends on the luminosity \citep{Lawrence91}, giving a dependence on the covering factor for a fixed torus half opening angle \citep[the so-called receding torus,][]{Gopal-Krishna96,Willott00,Simpson03,Arshakian05}. Furthermore, at the very low luminosity end, this torus is expected to disappear \citep[][and references therein]{Elitzur09,Elitzur12,Elitzur16}. 

We indeed find a marginal evolution of the covering factor as the AGN bolometric luminosity decreases (see Table \ref{tab:BayesClumpy}). As shown in the previous section, the geometrical covering factor depends on the equatorial number of clouds ($\rm{N_{o}}$) and on the half opening angle. In this case, this marginal increase on the geometrical covering factor comes from a marginal increase on the half opening angle towards lower luminosities. \citet{Simpson05} already proposed a slight modification of the receding torus model based on a new analysis of the fraction of type-1 versus type-2 AGN, where the height of the torus also increases when AGN ionizing luminosity increases. Indeed, we have found a marginal increase on the height of the torus when the bolometric luminosity increases. A more robust result is that the outer radius of the torus size might also depend on the AGN bolometric luminosity. Clumpy models associated to Groups 3, 4, and 5 show that the outer radius of the torus seems to increase as the bolometric luminosity of the AGN (and $\rm{C_{torus}}$) increases. The main difference in these groups is the $\rm{C_{torus}}$ which is 40\%, 60\%, and 80\% respectively. Thus, it might indicate that the dust contributing within the IRS/\emph{Spitzer} slit associated to the torus component ($\rm{C_{torus}}$), is increasing mainly because the size of the torus is increasing. The stratification of H$\rm{_2}$ molecular lines detected in this analysis is also interesting. While S(5) and S(6) transitions are seen in all the AP groups, S(1), S(2) and S(3) transitions are only detected in groups 1, 2, and 3 (i.e. $\rm{C_{torus}<40\%}$). This was already reported by \citet{Panuzzo11} in their \emph{Class-2} objects (those showing PAH features with anomalous 7.7/11.3$\rm{\mu m}$ PAH ratios). Each of these molecular lines could be tracing different densities and/or temperatures of the molecular gas content \citep[see][]{Roussel07}. Thus, independently on the origin of these lines, this result is more likely related to changes in the density and temperatures of the gas clouds. The gas could also be changing its content and distribution, coupled with the dust, if the dust is changing its morphological distribution. Supporting this, a more concentrated molecular gas distribution toward the center was reported for three LLAGN by \citet{Muller-Sanchez13}, compared to Seyferts, using integral field spectroscopy in the near infrared. They argued that this change of the concentration of molecular gas is related with the progressive disappearance of the torus. Alternatively, the change of the configuration of the dust or the AGN power is allowing the AGN to heat the gas to further distances, resulting in an enhancement of some particular transitions of the H$\rm{_2}$ molecular line. Finally, clumpy torus models applied to the different groups found by the AP method show that Group 2 (i.e. $\rm{C_{torus} \sim 20\%}$) is no longer reproduced (although it visually resembles) by clumpy torus models, suggesting also a different composition or structure of clumps for this group. As an speculation, these results might suggest a smooth transition from high to lower AGN luminosities with changes in temperature, density and/or location, of dust and gas surrounding the AGN, until the disappearance of the torus. These changes on the torus characteristics are predicted by \citet{Honig07}, based on stability arguments (i.e. gravity versus radiation pressure produced by the accretion disk). They showed that at $\rm{L_{bol}<10^{42}erg/s}$ and according to the clumpy torus model, the torus collapses to a geometrically thin disk. Thus, the mid-infrared emission below that limit is not produced by the geometrically thick torus.

Note here that we are not excluding the viewing angle as a cause for some of the AGN classes. Indeed, when we split Group 5 into type-1 and type-2 AGN the viewing angle takes an important role (see Table \ref{tab:BayesClumpy}). However, as long as we move towards low luminosities, intrinsic differences as the outer radius of the torus or the molecular gas distribution must be taken into account to produce a clear picture of AGN and their evolution from (or to) a non-active galaxy. Note also that these results are found assuming that the torus is clumpy. A detailed analysis on the clumpy versus smooth distribution of the torus, perhaps throughout the relative strength of the 10$\rm{\mu m}$ and 18$\rm{\mu m}$ silicate features strength, is also needed \citep[][]{Sirocky08,Hatziminaoglou15,Mendoza-Castrejon15}. However, the large ISM contributions in our IRS/\emph{Spitzer} spectra (see Section \ref{sec:decomposition}) prevent us from analysing  the silicate features because they are highly contaminated from this contribution. 

An expected dependence of the torus with luminosity is the disappearance of the torus below a certain bolometric luminosity \citep{Elitzur06}. The reason is that the accretion onto the SMBH can not longer sustain the required cloud outflow rate. \citet{Gonzalez-Martin15} found some evidence in favor of the disappearance of the torus below the bolometric luminosity $\rm{L_{bol}\sim10^{42} erg/s}$. The mid-infrared spectrum of AGN with $\rm{L_{X}<10^{41} erg/s}$ showed a completely different shape compared to brighter AGN. However, the main concern for the latter study is that IRS/\emph{Spitzer} spectra have low-spatial resolution. Thus, many ingredients other than the AGN can contribute to the mid-infrared emission. Therefore, galaxy dilution could still play a role. The idea is that the torus does not disappear but it gets diluted due to the lower torus contribution compared to the host galaxy. This, naturally explains the linear relation between $\rm{C_{torus}}$ and bolometric luminosities for intermediate luminosities (orange continuous line in the left panel of Fig. \ref{fig:LbolvsfAGN}; see Section \ref{sec:torusvsluminosity}). Although we cannot rule out this scenario, we believe it is less plausible because we would expect this relation to continue toward higher bolometric luminosities. Instead, we find that high luminosity AGN tend to show larger $\rm{C_{torus}}$ than expected for the linear relation. Alternatively, this relation is also expected under the evolutionary scenario in which the star formation increases when the AGN bolometric luminosity decreases.
In this scenario the contribution of the torus decreases due to an enhancement on the ISM contribution. However, the increase on the star formation for LLAGN is in contradiction with other  results. Most of the LLAGN are hosted in elliptical galaxies \citep{Carrillo99}, where there is a lack of young stars. Furthermore, young stellar populations are almost negligible in the circumnuclear environment of LLAGN, mainly constituted by old stars \citep{Cid-Fernandes04, Gonzalez-Delgado04, Sarzi05, Gonzalez-Delgado08}.  Indeed, \citet{Krongold03} discussed the possibility that  strong winds of Wolf Rayet and O stars wipe out the circumnuclear material left over from an initial circumnuclear star formation in LINERs. Thus,  this is a less preferred scenario to explain the $\rm{C_{torus}}$ versus $\rm{L_{bol}}$ relation.

\begin{figure}
\begin{center}
\includegraphics[width=1.0\columnwidth]{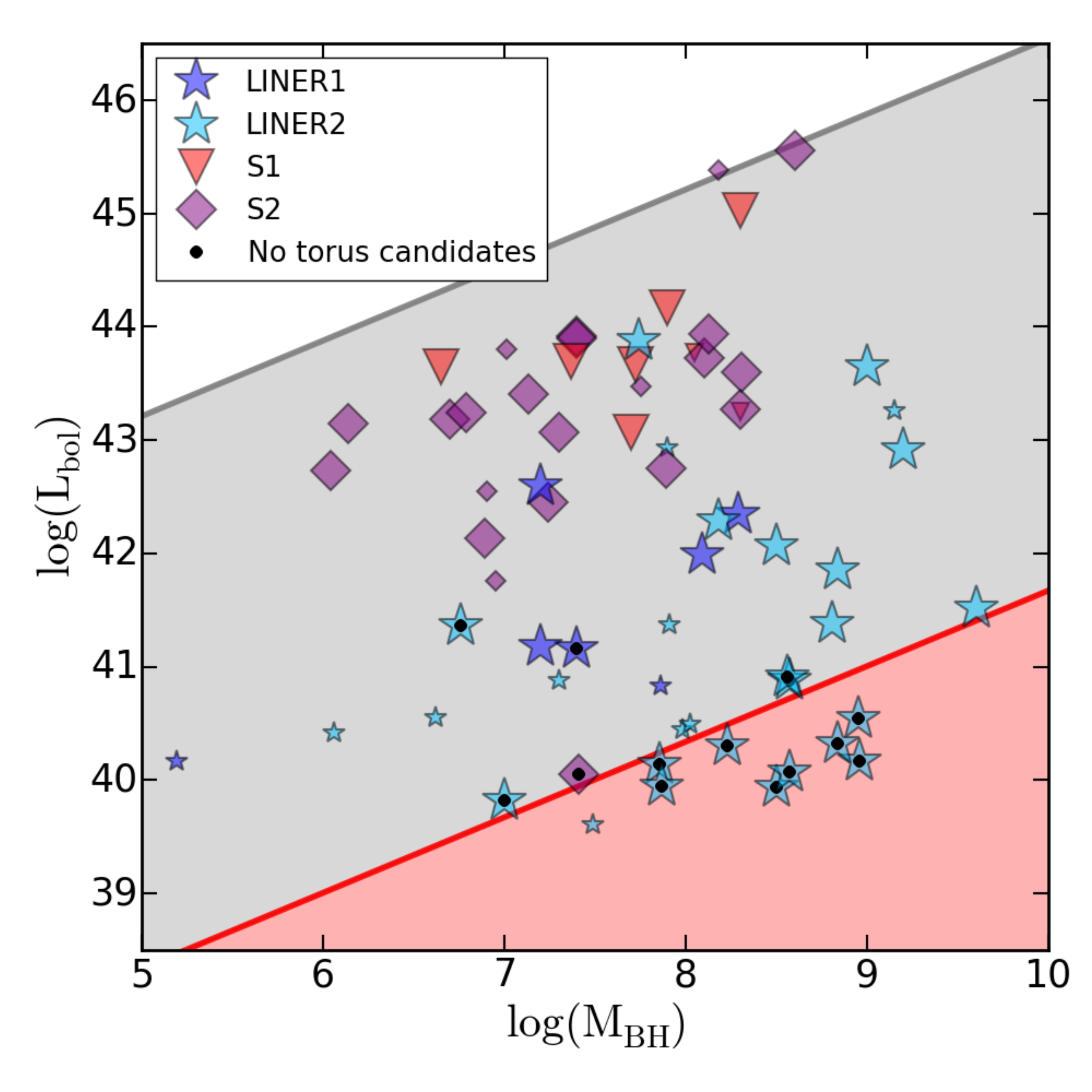}
\caption{AGN bolometric luminosity $\rm{Lbol}$ versus the SMBH masses, both in logarithmic scales. Smaller (larger) symbols show objects with $\rm{C_{ISM}>50\%}$ ($\rm{C_{ISM}<50\%}$). Red and gray continuous lines show the lower and upper limit on the AGN bolometric luminosity for the disappearance of the torus, according to \citet{Elitzur09}. The torus is not longer present in the red-shaded area and the torus might disappear only for certain combinations of the wind parameters in the gray-shaded area (see text).}
\label{fig:LbolBHM}
\end{center}
\end{figure}

On the other hand, if the torus disappears, we would expect $\rm{C_{torus}\sim 0\%}$ below a certain luminosity. This is consistent indeed with our findings, where Group 1 shows $\rm{log(L_{bol})<41.3}$ and $\rm{C_{torus}< 1.4\%}$ (see Table \ref{tab:affinity}). Another way to compute the luminosity limit where the torus is disappearing is to make a linear fit to $\rm{L_{bol}}$ versus $\rm{C_{torus}}$ for Groups 3, 4, and 5, extrapolating the $\rm{L_{bol}}$ limit when $\rm{C_{torus}=0\%}$. This is, according to our observations, $\rm{log(L_{bol} (erg/s))=41.2}$, which is fully consistent with Group 1. This luminosity limit is below that inferred by \citet{Gonzalez-Martin15}, i.e. $\rm{log(L_{bol})\simeq 42}$. This discrepancy is due to the fact that their limit was imposed to compute average spectra while the current limit is computed from observations. This limit is well below that predicted by \citet{Elitzur06}; i.e. $\rm{log(L_{bol} (erg/s))\sim42}$. However, \citet{Elitzur09} updated it, finding that it depends on the SMBH mass as $\rm{L_{bol}=5\times 10^{39} (M/10^7 M_{\odot})^{2/3}}$ erg/s. Assuming that the disappearance of the torus corresponds to the largest SMBH masses (LLAGN tend to have the largest SMBH masses, i.e. $\rm{M=10^{9}M_{\odot}}$), the bolometric luminosity limit is $\rm{log(L_{bol} (erg/s))\simeq 41}$, consistent with our findings. 

To explore more carefully the dependence of the minimum AGN bolometric luminosity (required to hold the outflowing structure)  on the SMBH mass, we have compiled the available SMBH masses in the literature for 85 out of the 109 objects in our sample \citep[][and references therein]{Woo02,Gonzalez-Martin09A,McKernan10,Gonzalez-Martin12,Zoghbi14,Bentz15}. Fig. \ref{fig:LbolBHM} shows the AGN bolometric luminosity versus the SMBH masses in our sample. Smaller symbols show objects with $\rm{C_{ISM}>50\%}$ and larger symbols  show objects with $\rm{C_{ISM}<50\%}$. 

\citet[][]{Elitzur16} show that the minimum AGN bolometric luminosity depends on the combination of several parameters of the wind (the radiative conversion efficiency, the ratio between the SMBH accretion rate and the wind accretion rate, and a factor \emph{I} that depends on the ratio between the outflow launch velocity to local Keplerian velocity and the radial density variations). For a certain SMBH mass, there is a limit on the bolometric luminosity below which they do not expect any BLR or torus to survive (shown as red continuous line in Fig. \ref{fig:LbolBHM}). Moreover, there is an upper limit above which the BLR and the torus must be present (shown as gray continuous line in Fig. \ref{fig:LbolBHM}). In the range between the upper and the lower limit both scenarios could happen.

The 14 LLAGN where the torus seems to have disappeared according to our analysis are marked with $\rm{\clubsuit}$ in Tables \ref{tab:sample}, \ref{tab:comparisonsample1} and \ref{tab:comparisonsample2} and are shown with small black dots in Fig. \ref{fig:LbolBHM}. Red shaded area shows the area where the wind radial column drops below the minimum required to produce detectable BLRs and dusty tori according to \citet{Elitzur16}. All the objects in this area are candidates to the disappearance of the torus according to our analysis. The only exception is NGC\,5866. It was not included in our list of candidates because it shows large contributions of ISM but still shows $\rm{C_{torus}<}$0.7\%. Two objects (and another four very close to the lower limit on the bolometric luminosity, red continuous line in Fig. \ref{fig:LbolBHM}) are in the gray-shaded area of the plot. \citet{Elitzur16} showed that only some sources in this gray shaded area might not have torus depending on the combination of parameters of the outflowing wind. Thus, our findings are in agreement with the most recent prediction of the disappearance of the torus presented by \citet{Elitzur16}.

\begin{figure*}
\begin{center}
\includegraphics[width=2.\columnwidth]{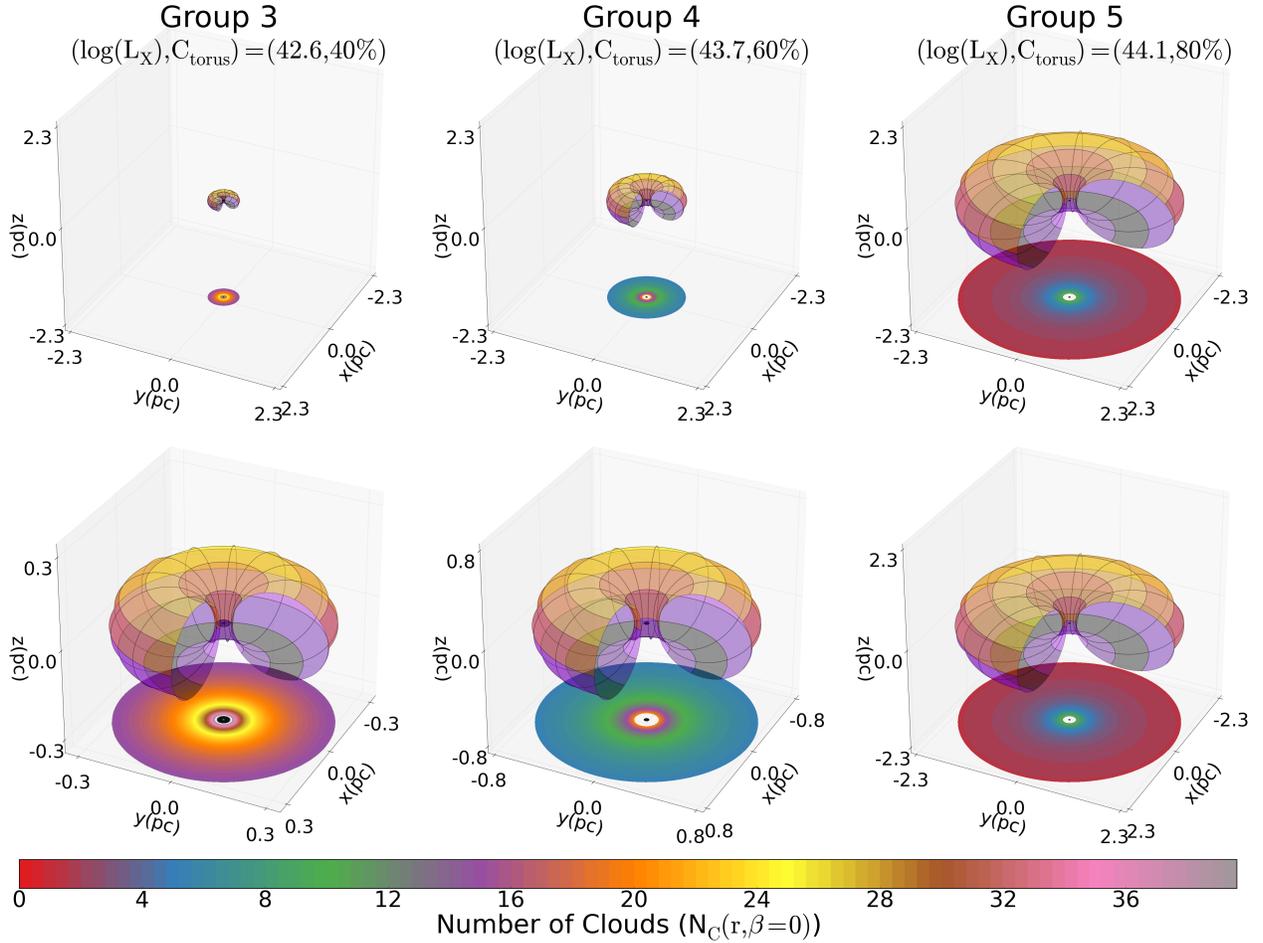}
\caption{Schematic view of the tori for Groups 3 (left), 4 (middle), and 5 (right) using the best fit parameters obtained with Clumpy models (see Table \ref{tab:BayesClumpy}). We used  a fixed size for the box of 4.6 pc in the top panels and the diameter of each torus (i.e. $\rm{\sim}$0.7, 1.6, and 4.6 pc for Groups 3, 4, and 5, respectively) in the bottom panels. We also added a disk component only for viewing purpose with a radius of 0.2 pc as a toroidal structure with half opening angle of $\rm{\sigma = 5^{\circ}}$ (i.e. this disk component has not a realistic size). The projection seen at the bottom of each box shows the radial profile of the number of clouds at the equatorial plane ($\rm{N_{C}}$, see text).}
\label{fig:torus}
\end{center}
\end{figure*}

Fig.~\ref{fig:torus} shows an schematic view of the tori for Groups 3, 4, 5 (from left to right), according to the best fit to Clumpy torus models (Table \ref{tab:BayesClumpy}). Top panels show the tori for the three groups in a box with a fixed side of 4.6~pc. This clearly shows how drastically the outer radius and height of the torus decreases towards lower-luminosities. This tendency seems to evolve to its disappearance at bolometric luminosities of $\rm{log(L_{bol} (erg/s))\simeq 41}$. The bottom panels in Fig.~\ref{fig:torus} show the same schematic view of the tori for Groups 3, 4, and 5 but optimising the size of the box to match the outer size of the torus. The half opening angle slightly increases toward lower luminosities, being responsible for the tentative increase on the geometrical covering factor. Although larger covering factors and smaller torus sizes seem to be contradictory, this figure illustrates how both can coexist.

Fig.~\ref{fig:torus} also shows, as a projection at the bottom of each panel, the radial profile of the number of clouds per unit length at the equatorial plane, $\rm{N_{C}(r,\beta=0)}$: 

\begin{equation}
N_{C}(r,\beta=0)  = C~N_{o} (R_{in}/r)^{q}
\end{equation}

\noindent where C is the normalisation to guarantee the number of clouds at the equatorial plane along any ray is $\rm{N_{o}}$ (i.e. $\rm{ \int N_{C}(r,\beta=0) dr = N_{o}  }$). The distribution of the number of clouds per unit of length also seems to be different for Groups 3, 4, and 5. Group 3 shows the largest maximum values and the wider distribution of clouds and Group 5 shows the lowest values and narrower distributions of clouds. This shows the effect on the change of the $\rm{q}$ parameter on the structure of the torus.

It is worth emphasising that our result on the plausible disappearance of the torus is not in contradiction with the 2-10 keV X-ray versus mid-infrared luminosity correlation found for AGN by several authors \citep[][]{Krabbe01,Lutz04,Horst06,Ramos-Almeida07,Horst08,Gandhi09,Asmus11,Mason12,Gonzalez-Martin13,Asmus14,Asmus15,Garcia-Bernete16}. The deepest sample ever put together to study this correlation was presented by \citet{Asmus14} and the X-ray to mid-infrared luminosity relation was analysed by \citet{Asmus15}. They found a good correlation down to 2-10 keV X-ray luminosities of $\rm{log(L_{X}(2-10 keV))\simeq 40}$ ($\rm{log(L_{MIR}(12\mu m))\simeq 39.5}$). This correlation is being interpreted as an indication that both are tracing the same mechanism, through reprocessed UV emission by the torus. Thus, the torus must be present up to these luminosities. The limit found in our paper is $\rm{log(L_{bol} (erg/s))\simeq 41.3}$ (from AP analysis). This implies $\rm{log(L_{X}(2-10 keV))\simeq 40.1}$, assuming a conversion $\rm{L_{bol}=15.8\times L_{X}(2-10 keV)}$ \citep{Ho09}. Only seven sources are below that limit in the correlation given by \citet{Asmus15} (namely Fornax, NGC\,1553, NGC\,4111, NGC\,4278, NGC\,4374, NGC\,4395, and NGC\,4594). In two cases they did not detect mid-infrared emission (NGC\,1553 and NGC\,4374), consistent with the lack of a torus for these sources. Other three sources (Fornax, NGC\,4111, and NGC\,4278) might be above $\rm{log(L_{X}(2-10 keV))\simeq 40.1}$, considering the error bars of the X-ray luminosities. Thus, only two sources are consistent with being in the correlation and below the limit where the torus seems to disappear, namely NGC\,4395 and NGC\,4594. NGC\,4395 is a strongly variable X-ray source \citep[][and references therein]{Cameron12}. Its intrinsic X-ray luminosity could change a factor of two, which already places this source above our limit. Indeed, variability is being discussed as a source of dispersion in this correlation \citep{Mayo13}. NGC\,4594 (Sombrero Galaxy) shows a dust lane within the disk of the galaxy. It is plausible, although we cannot guarantee it, that this dust lane is contributing to the mid-infrared emission, even with high resolution data. \citet{Mason12} studied the behaviour of this correlation for LLAGN using high spatial resolution mid-infrared data. Only NGC\,4736 has an X-ray luminosity below the expected limit for the disappearance of the torus (after taking into account Compton-thickness and X-ray variability). Interestingly this object appears out of the X-ray to mid-infrared relation, as expected if it lacks the torus. Finally, no point-like nuclear sources are detected in high spatial resolution CanariCam data (Table \ref{tab:CanariCam}) below that limit. We would like to remark that this is a limitation with the current ground-based mid-infrared instrumentation. Thus, new generation telescopes, going towards very low-luminosities are needed to confirm or reject the prediction of the torus disappearance with high spatial resolution imaging. 

\citet{Gonzalez-Martin09B} showed that a large percentage ($\rm{>40\%}$) of their LLAGN are Compton-thick candidates, i.e. with hydrogen column densities $\rm{N_{H}>2\times 10^{24}cm^{-2}}$. This might be in contradiction with our findings of the lack of the torus for some of them. Eleven out of these 14 LLAGN are included in \citet{Gonzalez-Martin09B}. Among them, only NGC\,4589 is classified as a Compton-thick candidate. From the 40 LINERs classified as Compton-thick candidates in \citet{Gonzalez-Martin09B}, only another two (NGC\,4314 and NGC\,4636) show X-ray luminosities below the expected value where the torus might disappear ($\rm{log(L_{X})<39.1}$ and $\rm{log(L_{X})<39.0}$, respectively). Although it is out of the scope of this paper to understand the explanation for the simultaneous occurrence of the Compton-thickness and the lack of torus for these three objects, these are not many cases. Furthermore, among our sample, NGC\,4321 does not show multi-wavelength signatures of AGN according to \citet{Gonzalez-Martin09A}. Thus, 12 out of these 14 LLAGN (excluding NGC\,4321 and NGC\,4589) are good candidates for the disappearance of the torus.

If the obscuring torus is a smooth continuation of the BLR, i.e. BLR and torus are the inner and outer bounds of a single cloud distribution \citep{Risaliti02,Suganuma06,Honig13}, the smooth disappearance of the dusty torus might be linked to changes in the structure of BLR and accretion disk \citep{Elitzur06,Elitzur09}. Several authors support the disappearance of the torus \citep[e.g.][]{Maoz05,Muller-Sanchez13} and the BLR according to the AGN luminosity \citep{Cao10,Elitzur14}. Recently, \citet{Ramos-Almeida16} showed that the BLR was present for a large portion of their AGN throughout polarised emission. However, none of the sources presented in their analysis have bolometric luminosities consistent with the disappearance of the BLR \citep{Elitzur09}. \citet{Hernandez-Garcia16} recently showed UV variability in the spatially-unresolved nuclear emission of LINERs, consistent with the idea that the torus is no longer obstructing the view of the accretion disk, in contrast with type-2 Seyferts, where no UV point-like source is found for them. Furthermore, absorption variations are not seen in most LINERs, while it appears to be more common in type-2 Seyferts \citep[with average higher luminosities than LINERs,][]{Hernandez-Garcia16}. This is interpreted as a lack of gas clouds close enough to the AGN (i.e. the BLR) to produce eclipses responsible for these absorption variations. 

All together, these results show intrinsic differences of LLAGN compared to more luminous AGN. Further studies on the candidates where the torus might disappear, and perhaps those belonging to the group which is close to the disappearance (i.e. Group 2, $\rm{C_{torus}\sim20\%}$), need to be conducted to verify if the torus is actually disappearing at those LLAGN. 

\section{Conclusions}

We have studied the torus contribution to the mid-infrared IRS/\emph{Spitzer} spectra of a sample of LINERs and Seyferts, with bolometric luminosities ranging more than six orders of magnitude and used the code called DeblendIRS (HC15) to decontaminate the torus emission from the stellar and ISM contributions. We have then compared the torus contribution with the bolometric luminosity of the sources. We have used the affinity propagation (AP) method, finding five groups defined by the torus contribution and the AGN bolometric luminosity. These groups have been studied to understand the main differences among them. The main results are:

\begin{enumerate}
\item Below a threshold luminosity ($\rm{L_{bol}\sim 10^{41}erg/s}$), we find a negligible torus contribution. This limit is fully consistent with the latest predictions for the disappearance of the torus \citep{Elitzur12}.
\item The average spectrum of the AP groups with the lowest torus contributions cannot be reproduced by the clumpy torus, even when other contributors are removed from the spectra. 
\item Clumpy torus models fitted to the AP groups show that the outer radius of the torus decreases as long as the torus contribution (and $\rm{L_{bol}}$) decreases. We also found tentative dependencies on the half opening angle of the torus, the height of the torus, and the radial profile of the cloud distributionl.
\item We have found a different molecular gas content for those groups with the lowest torus contributions. The  S(1), S(2), and S(3) transitions of the $\rm{H_{2}}$ molecular line are only seen in the groups with the lowest torus contributions while S(5) and S(6) are spread among all groups. This might also indicate that the gas content is also changing at the lowest luminosities. 
\end{enumerate}

\begin{acknowledgements}
OGM thanks to Daniel Asmus for his suggestions that have improved this manuscript significantly. This scientific publication is based on observations made with the Gran Telescopio CANARIAS (GTC), installed at the Spanish Observatorio del Roque de los Muchachos of the Instituto de Astrof\'isica de Canarias on the island of La Palma. This research has been supported by the UNAM PAPIIT grant (IA100516 PAPIIT/UNAM, 108716 PAPIIT/UNAM), CONACyT grant (CB-2011-01-167291), the Spanish Ministry of Economy and Competitiveness (MINECO, project refs. AYA 2010-15169, AYA2012-31277, AYA 2012-39168-C03-01, AYA2013-42227-P, AYA2015-70815-ERC, AYA2015-64346-C2-1-P, and AYA2016-76682-C3-1-P) and by La Junta de Andaluc\'ia (TIC 114). MM-P acknowledges support by the UNAM Postdoctoral fellowship programme and CRA acknowledges the Ram\'on y Cajal Program of the Spanish Ministry of Economy and Competitiveness through project RyC-2014-15779. LHG acknowledges the ASI/INAF agreement number 2013-023-R1.
\end{acknowledgements}

\newpage

\input{Tabla_IAcomm.tex}

\input{ApendixA.tex}



\end{document}

%% file: Tabla_IAcomm.tex
\begin{table*}[!h]
\def\arraystretch{1.1}
\caption{Details of the LINER sample and results of IRS/\emph{Spitzer} decomposition. }
\label{tab:sample}
\begin{footnotesize}
\centering
\begin{tabular}{l l r c c r r r r l}
\hline\hline
Object name     & type      & D (Mpc) & $\rm{log(L_{X})}$ &      Slit width (pc)               & $\rm{C_{torus}}$ (\%) & $\rm{C_{Stellar}}$ (\%)             & $\rm{C_{ISM}}$ (\%)            & RMSE    & AP     \\ 
      (1)       &    (2)    & (3)  & (4)  &             (5)                &       (6)                      &         (7)                   &    (8)        &    (9)  & (10) \\ \hline
NGC315          & LINER1  &  56.0 &  41.8 & 1159 & 39.8$\rm{_{  -2.7}^{+   3.6}}$ &  22.3$\rm{_{  -4.5}^{+   8.8}}$ &  36.4$\rm{_{  -8.7}^{+   4.6}}$ & 0.071 &  3  \\
NGC1052         & LINER1  &  19.7 &  41.5 & 408 & 76.2$\rm{_{  -2.9}^{+   3.9}}$ &  15.4$\rm{_{  -5.1}^{+   5.0}}$ &   7.2$\rm{_{  -3.8}^{+   4.2}}$ &  0.056 & 5  \\
NGC1097         & LINER1  &  19.6 &  40.8 &  406 & 0.6$\rm{_{  -0.6}^{+   1.3}}$ &   5.6$\rm{_{  -2.8}^{+   4.0}}$ &  92.4$\rm{_{  -3.8}^{+   3.0}}$ &  0.049 & 1  \\
NGC2639         & LINER1  &  47.7 &  40.3 & 986 & 11.6$\rm{_{  -4.7}^{+   4.6}}$ &  43$\rm{_{ -23}^{+  21}}$ &  44$\rm{_{ -22}^{+  24}}$ & 0.178  & 2  \\
NGC4438         & LINER1  &  13.4 &  39.0 &  282 & 0.6$\rm{_{  -0.6}^{+   1.3}}$ &  26.0$\rm{_{  -6.1}^{+   2.9}}$ &  71.9$\rm{_{  -2.8}^{+   5.9}}$ & 0.13  & 1  \\
NGC4450$\rm{\clubsuit}$         & LINER1  &  16.1 &  40.3 &  333 & 0.0$\rm{_{  -0.0}^{+   0.7}}$ &  66$\rm{_{  -4}^{+  24}}$ &  33$\rm{_{ -24}^{+   4}}$ & 0.48  & 1  \\
NGC4579         & LINER1  &  19.6 &  41.2 & 402 & 14.1$\rm{_{  -4.0}^{+   2.3}}$ &  63$\rm{_{ -15}^{+  10}}$ &  21$\rm{_{ -10}^{+  14}}$ & 0.101  & 2  \\
NGC5005         & LINER1  &  20.6 &  39.9 &  426 & 1.3$\rm{_{  -1.3}^{+   1.6}}$ &  23$\rm{_{ -15}^{+   10}}$ &  75$\rm{_{  -10}^{+  14}}$ &  0.104 & 1  \\ \hline
IIIZW035        & LINER2  & 117.5 &  40.0 &  2431 & 0.4$\rm{_{  -0.4}^{+   1.1}}$ &   0.0$\rm{_{  -0.0}^{+   0.7}}$ &  98.6$\rm{_{  -1.1}^{+   0.9}}$ & 0.818  & 1  \\
NGC835          & LINER2  &  34.0 &  41.4 &  703 & 2.6$\rm{_{  -2.4}^{+   3.1}}$ &  11.0$\rm{_{  -7.3}^{+   9.0}}$ &  85.7$\rm{_{  -9.0}^{+   7.2}}$ & 0.091  & 1  \\
NGC1291$\rm{\clubsuit}$         & LINER2  &   8.6 &  39.0 &  178 & 0.7$\rm{_{  -0.7}^{+   1.5}}$ &  73.1$\rm{_{  -6.6}^{+   6.2}}$ &  25.3$\rm{_{  -5.9}^{+   5.8}}$ &  0.106 & 1  \\
NGC2685$\rm{\clubsuit}$        & LINER2  &  13.3 &  39.0 &  274 & 6.6$\rm{_{  -5.1}^{+   5.5}}$ &  68$\rm{_{ -16}^{+  18}}$ &  25$\rm{_{ -15}^{+  12}}$ & 0.104  & 1  \\
NGC2655         & LINER2  &  24.4 &  41.2 & 505 & 18.5$\rm{_{  -2.8}^{+   2.5}}$ &  53$\rm{_{ -19}^{+  13}}$ &  26$\rm{_{ -13}^{+  21}}$ & 0.077  & 2  \\
UGC04881        & LINER2  & 168.3 &  38.4 &  3482 & 0.0$\rm{_{  -0.0}^{+   0.7}}$ &   1.7$\rm{_{  -1.7}^{+   2.3}}$ &  97.3$\rm{_{  -2.3}^{+   1.7}}$ &  0.053 & 1  \\
3C218           & LINER2  & 235.0 &  42.1 & 4862 & 33$\rm{_{ -20}^{+  19}}$ &  31$\rm{_{ -22}^{+  28}}$ &  48$\rm{_{ -30}^{+  27}}$ & 0.177  & 3  \\
NGC2841$\rm{\clubsuit}$         & LINER2  &  17.3 &  39.2 &  358 & 0.8$\rm{_{  -0.8}^{+   1.7}}$ &  72.9$\rm{_{  -9.2}^{+   8.7}}$ &  25.5$\rm{_{  -8.4}^{+   9.0}}$ & 0.167  & 1  \\
UGC05101        & LINER2  & 168.6 &  42.1 & 3488 & 11.0$\rm{_{  -6.2}^{+   3.3}}$ &   5.3$\rm{_{  -4.3}^{+   5.2}}$ &  83.1$\rm{_{  -7.4}^{+   6.4}}$ &  0.08 & 2  \\
NGC3079         & LINER2  &  19.3 &  42.1 & 399 & 25.2$\rm{_{  -7.4}^{+   7.5}}$ &   2.4$\rm{_{  -2.4}^{+   2.5}}$ &  74.8$\rm{_{  -8.2}^{+   4.4}}$ & 0.122  & 2  \\
NGC3185         & LINER2  &  22.9 &  39.4 &  473 & 1.2$\rm{_{  -1.2}^{+   1.7}}$ &  19$\pm 12$ &  78$\pm 12$ & 0.082  & 1  \\
NGC3190         & LINER2  &  24.3 &  39.5 &  504 & 0.2$\rm{_{  -0.2}^{+   1.1}}$ &  26.7$\rm{_{  -7.5}^{+   8.0}}$ &  70.9$\rm{_{  -8.5}^{+   8.3}}$ & 0.086  & 1  \\
NGC3627         & LINER2  &   9.8 &  39.4 &  202 & 0.6$\rm{_{  -0.6}^{+   1.3}}$ &  26.7$\rm{_{  -7.3}^{+   7.2}}$ &  71.7$\rm{_{  -5.3}^{+   7.5}}$ & 0.064  & 1  \\
NGC3628         & LINER2  &  10.9 &  39.9 & 225 & 12.2$\rm{_{  -4.9}^{+   1.3}}$ &   2.0$\rm{_{  -1.3}^{+   1.3}}$ &  84.2$\rm{_{  -1.4}^{+   2.0}}$ & 0.172  & 2  \\
NGC4125$\rm{\clubsuit}$         & LINER2  &  22.0 &  38.7 &  455 & 0.1$\rm{_{  -0.1}^{+   1.1}}$ &  88.0$\rm{_{  -7.0}^{+   7.1}}$ &  10.8$\rm{_{  -7.3}^{+   6.9}}$ &  0.322 & 1  \\
IRAS12112+0305  & LINER2  & 314.0 &  41.2 & 6496  & 1.2$\rm{_{  -1.1}^{+   1.2}}$ &   0.0$\rm{_{  -0.0}^{+   0.7}}$ &  97.8$\rm{_{  -1.2}^{+   1.1}}$ & 0.758  &  1  \\
NGC4261         & LINER2  &  29.9 &  41.0 & 619 & 26.3$\rm{_{  -3.3}^{+   2.6}}$ &  48.1$\rm{_{  -9.8}^{+   7.1}}$ &  23$\rm{_{ -10}^{+   8}}$ & 0.098  & 2  \\
NGC4321$\rm{\clubsuit}$         & LINER2  &  16.4 &  40.5 &  340 & 9.3$\rm{_{  -6.2}^{+   7.0}}$ &  51$\pm 10$ &  37$\rm{_{ -11}^{+   6}}$ & 0.116  & 1  \\
NGC4374$\rm{\clubsuit}$         & LINER2  &  16.7 &  39.5 &  346 & 2.7$\rm{_{  -2.5}^{+   2.6}}$ &  78$\pm 11$ &  17.0$\pm 10$ & 0.07  & 1  \\
NGC4486         & LINER2  &  16.7 &  40.7 & 344 & 15$\rm{_{ -10}^{+   9}}$ &  66$\rm{_{ -21}^{+  19}}$ &  19$\rm{_{ -14}^{+  18}}$ & 0.099  & 2  \\
NGC4552$\rm{\clubsuit}$         & LINER2  &  16.0 &  39.2 &  330 & 0.1$\rm{_{  -0.1}^{+   1.1}}$ &  97.9$\rm{_{  -1.6}^{+   1.4}}$ &   1.1$\rm{_{  -1.1}^{+   1.6}}$ &  0.216 & 1  \\
NGC4589$\rm{\clubsuit}$         & LINER2  &  28.1 &  38.9 &  581 & 0.6$\rm{_{  -0.6}^{+   1.3}}$ &  76$\rm{_{ -13}^{+  14}}$ &  26$\rm{_{ -16}^{+   10}}$ & 0.109  & 1  \\
NGC4594$\rm{\clubsuit}$         & LINER2  &  11.1 &  39.9 &  230 & 0.1$\rm{_{  -0.1}^{+   1.1}}$ &  85.0$\rm{_{  -5.6}^{+   5.6}}$ &  13.5$\rm{_{  -5.6}^{+   5.6}}$ &  0.075 & 1  \\
NGC4676A        & LINER2  &  94.5 &  39.9 &  1955 & 0.0$\rm{_{  -0.0}^{+   1.0}}$ &   2.2$\rm{_{  -2.2}^{+   4.6}}$ &  96.6$\rm{_{  -2.8}^{+   2.4}}$ & 0.075  & 1  \\
NGC4698$\rm{\clubsuit}$         & LINER2  &  23.3 &  38.7 &  469 & 1.5$\rm{_{  -1.5}^{+   1.9}}$ &  77$\rm{_{ -16}^{+  14}}$ &  21$\rm{_{ -14}^{+  15}}$ & 0.122  & 1  \\
NGC4696$\rm{\clubsuit}$         & LINER2  &  37.6 &  40.0 &  778 & 4.7$\rm{_{  -4.0}^{+   4.3}}$ &  84.9$\rm{_{  -8.6}^{+   9.4}}$ &  11.9$\rm{_{  -8.4}^{+   7.6}}$ & 0.166  & 1  \\
NGC4736$\rm{\clubsuit}$         & LINER2  &   5.1 &  38.6 &  105 & 0.0$\rm{_{  -0.0}^{+   0.8}}$ &  51$\rm{_{ -15}^{+   8}}$ &  47$\rm{_{  -8}^{+  15}}$ &  0.115 & 1  \\
MRK266SW        & LINER2  & 118.2 &  42.2 & 2446 & 32$\rm{_{ -17}^{+   9}}$ &   9$\rm{_{  -7}^{+  11}}$ &  60$\rm{_{ -15}^{+  17}}$ & 0.059 &  3  \\
MRK266NE        & LINER2  & 120.1 &  41.6 & 2485 & 11.7$\rm{_{  -6.9}^{+   6.6}}$ &   7.4$\rm{_{  -6.0}^{+   8.4}}$ &  81$\rm{_{ -11}^{+   10}}$ & 0.083  & 2  \\
UGC08696        & LINER2  & 161.8 &  43.0 & 3348 & 45.5$\rm{_{  -1.2}^{+   1.2}}$ &   0.0$\rm{_{  -0.0}^{+   0.7}}$ &  53.5$\rm{_{  -1.2}^{+   1.2}}$ & 0.129  & 3  \\
IRAS14348-1447  & LINER2  & 355.5 &  41.7 & 7354  & 7.4$\rm{_{  -1.3}^{+   1.2}}$ &   0.0$\rm{_{  -0.0}^{+   0.7}}$ &  91.6$\rm{_{  -1.2}^{+   1.3}}$ & 0.851  & 1  \\
NGC5866         & LINER2  &  12.2 &  38.3 & 253  & 0.0$\rm{_{  -0.0}^{+   0.7}}$ &  29$\rm{_{  -5}^{+  29}}$ &  70$\rm{_{ -29}^{+   5}}$ & 0.096  & 1  \\
NGC6251         & LINER2  &  98.2 &  42.8 & 2032 & 35.8$\rm{_{  -1.6}^{+   1.8}}$ &  46$\rm{_{ -15}^{+   10}}$ &  18$\rm{_{  -9}^{+  15}}$ & 0.149  & 3  \\
NGC6240         & LINER2  & 104.8 &  42.4 & 2169 & 20.2$\rm{_{  -1.7}^{+   2.5}}$ &   0.0$\rm{_{  -0.0}^{+   0.7}}$ &  78.8$\rm{_{  -2.5}^{+   1.7}}$ & 0.081  & 2  \\
IRAS17208-0014  & LINER2  & 183.3 &  41.2 & 3793 &  0.7$\rm{_{  -0.7}^{+   1.2}}$ &   0.0$\rm{_{  -0.0}^{+   0.7}}$ &  98.3$\rm{_{  -1.2}^{+   0.8}}$ &  1.075 & 1  \\
NGC7130         & LINER2  &  69.2 &  42.9 & 1431 & 43$\rm{_{ -16}^{+  17}}$ &   2.7$\rm{_{  -2.5}^{+   3.2}}$ &  54$\rm{_{ -15}^{+  18}}$ & 0.032  & 3  \\
NGC7331         & LINER2  &  14.2 &  40.5 & 295  & 4.5$\rm{_{  -3.5}^{+   3.2}}$ &  41$\rm{_{ -11}^{+   9}}$ &  52.8$\rm{_{  -7.5}^{+   9.9}}$ & 0.071  & 1  \\
IC1459          & LINER2  &  24.0 &  40.5 & 497 & 19$\rm {\pm10}$ &  63$\rm{_{ -15}^{+  11}}$ &  14.6$\rm{_{  -7.5}^{+   8.4}}$ &  0.083 & 2  \\
NPM1G-12.0625   & LINER2  &  23.3 &  41.5 & 469 & 25$\rm{_{ -17}^{+  18}}$ &  40$\pm 28$ &  48$\rm{_{ -33}^{+  34}}$ & 0.69 &  2  \\
NGC7743         & LINER2  &  21.1 &  39.5 & 438  & 4.3$\rm{_{  -3.5}^{+   3.6}}$ &  43$\rm{_{ -21}^{+  15}}$ &  50$\rm{_{ -15}^{+  22}}$ & 0.105  & 1  \\ \hline \hline
\end{tabular}
\newline
{ Note: AP: Affinity Propagation groups (1 to 5, see text). X-ray luminosity given in units of erg/s. $\rm{\clubsuit}$: Candidates lacking torus because they belong to Group 1 with $\rm{C_{ISM}<50\%}$ (see text).}
\end{footnotesize}
\end{table*}

\begin{table*}[!h]
\def\arraystretch{1.1}
\caption{Details of the comparison samples and results of IRS/\emph{Spitzer} decomposition. }
\label{tab:comparisonsample1}
\begin{footnotesize}
\centering
\begin{tabular}{l l r c c r r r r l}
\hline\hline
Object name     & type      & D (Mpc) & $\rm{log(L_{X})}$ & Slit width (pc) & $\rm{C_{torus}}$ (\%) & $\rm{C_{Stellar}}$ (\%)             & $\rm{C_{ISM}}$ (\%)      & RMSE          & AP \\ 
      (1)       &    (2)    & (3)  & (4)  &             (5)                &       (6)                      &         (7)                   &    (8)  &    (9)  & (10) \\ \hline
MCG-6-30-15     & S1      &  33.2 &  42.8 & 687  & 79$\pm 14$ &  17$\rm{_{ -12}^{+  15}}$ &   9.4$\rm{_{  -6.7}^{+   8.6}}$ &  0.031 & 5  \\
Fairall9        & S1      & 201.4 &  44.0 & 4166 & 87.5$\rm{_{  -5.1}^{+   3.1}}$ &   4.1$\rm{_{  -3.4}^{+   3.7}}$ &   6.7$\rm{_{  -4.1}^{+   5.0}}$ &  0.034 & 5  \\
NGC526A         & S1      &  81.8 &  43.2 & 1692 & 82.6$\rm{_{  -8.5}^{+   8.5}}$ &  13.2$\rm{_{  -7.8}^{+   9.4}}$ &   4.8$\rm{_{  -3.9}^{+   4.8}}$ & 0.06  & 5  \\
NGC3783         & S1      &  47.8 &  42.8 & 988 & 69.5$\rm{_{  -8.4}^{+   8.8}}$ &  21$\rm{_{ -14}^{+  11}}$ &   9.5$\rm{_{  -7.3}^{+   9.6}}$ & 0.038  & 4  \\
IC4329A         & S1      &  68.8 &  43.7 &  1422 & 79$\rm{_{  -9}^{+   12}}$ &  15.4$\rm{_{  -7.7}^{+   6.4}}$ &  5.6$\rm{_{  -4.4}^{+   5.4}}$ & 0.034  & 5  \\
NGC5548         & S1      & 216.8 &  42.8 & 4484 & 64.3$\rm{_{  -3.0}^{+   4.7}}$ &  16.7$\rm{_{  -7.4}^{+   8.1}}$ &  16.4$\rm{_{  -8.1}^{+   8.6}}$ &  0.081 & 4  \\
H1846-786       & S1      & 317.4 &  44.6 & 6567 & 76$\rm{_{ -17}^{+  16}}$ &  11.6$\rm{_{  -8.6}^{+   9.7}}$ &  16$\rm{_{ -12}^{+  13}}$ & 0.055  & 5  \\
MRK509          & S1      & 240.5 &  44.1 & 4976 & 72.1$\rm{_{  -6.8}^{+   6.0}}$ &  11.3$\rm{_{  -6.5}^{+   5.3}}$ &  14.0$\rm{_{  -4.9}^{+   6.7}}$ & 0.039  & 5  \\
NGC7213         & S1      &  22.0 &  42.2 & 455 & 71.8$\rm{_{  -4.6}^{+   2.8}}$ &  14.9$\rm{_{  -7.7}^{+   8.9}}$ &  12.7$\rm{_{  -6.5}^{+   8.0}}$ & 0.099  & 5  \\
MCG-2-58-22     & S1      & 200.7 &  44.3 & 4152 & 82$\pm 11 $ &  14$\pm 10 $ &   8.8$\rm{_{  -6.8}^{+   8.6}}$ & 0.028  & 5  \\ \hline
MRK1066         & S2      &  51.7 &  42.9 & 1070 & 38$\rm{_{ -12}^{+  16}}$ &   4.1$\rm{_{  -3.4}^{+   3.6}}$ &  57$\rm{_{ -15}^{+  13}}$ & 0.036  & 3  \\
NGC1386         & S2      &  16.1 &  41.6 & 334 & 55$\rm{_{ -13}^{+   8}}$ &  13$\rm{_{ -10}^{+   8}}$ &  36.1$\rm{_{  -9.6}^{+   6.8}}$ &  0.089 & 4  \\
NGC2110         & S2      &  35.6 &  42.4 & 736 & 71$\rm{_{ -12}^{+   10}}$ &  19$\rm{_{ -14}^{+  15}}$ &  13.0$\rm{_{  -8.5}^{+   7.5}}$ & 0.053  & 4  \\
ESO005-G004     & S2      &  25.6 &  41.9 & 531 & 25.3$\rm{_{  -3.4}^{+   7.3}}$ &  16$\rm{_{  -9}^{+  15}}$ &  57$\rm{_{ -16}^{+  11}}$ &  0.117 & 2  \\
MRK3            & S2      &  63.2 &  44.4 & 1308 & 86.4$\rm{_{  -2.0}^{+   4.5}}$ &   0.2$\rm{_{  -0.2}^{+   3.3}}$ &  12.0$\rm{_{  -7.0}^{+   2.5}}$ & 0.093  & 5  \\
NGC2273         & S2      &  31.0 &  42.2 &  641& 57$\rm{_{ -15}^{+  14}}$ &  14$\rm{_{ -10}^{+  13}}$ &  33$\rm{_{ -14}^{+  13}}$ & 0.03  & 4  \\
IRAS07145-2914  & S2      &  23.2 &  42.5 & 480 & 83$\rm{_{  -9}^{+  11}}$ &   5.4$\rm{_{  -4.4}^{+   6.4}}$ &  14.0$\rm{_{  -8.9}^{+   8.2}}$ & 0.075  & 5  \\
MCG-5-23-16     & S2      &  36.3 &  43.0 & 752 & 80.1$\rm{_{  -7.3}^{+   7.7}}$ &  12.4$\rm{_{  -9.1}^{+   8.9}}$ &   9.8$\rm{_{  -5.9}^{+   7.8}}$ & 0.027  & 5  \\
NGC3081         & S2      &  26.5 &  42.5 & 548 & 62.4$\rm{_{  -3.4}^{+   5.1}}$ &  11.6$\rm{_{  -6.8}^{+   4.9}}$ &  25.9$\rm{_{  -4.4}^{+   5.7}}$ & 0.106  & 4  \\
NGC3281         & S2      &  45.7 &  43.2 & 946 & 78$\rm{_{ -21}^{+  12}}$ &  10.9$\rm{_{  -8.1}^{+  15.4}}$ &  16$\rm{_{  -9}^{+  14}}$ &  0.058 & 5  \\
NGC3393         & S2      &  53.6 &  42.9 & 1108 & 58.8$\rm{_{  -3.4}^{+   1.8}}$ &   4.4$\rm{_{  -2.9}^{+   2.3}}$ &  36.3$\rm{_{  -3.6}^{+   9.1}}$ & 0.314  & 4  \\
NGC3621         & S2      &   6.9 &  39.3 &  142 & 1.2$\rm{_{  -1.2}^{+   1.5}}$ &  21.8$\rm{_{  -6.4}^{+   2.2}}$ &  75.1$\rm{_{  -1.8}^{+   6.9}}$ & 0.156  & 1  \\
NGC4388         & S2      &  20.5 &  42.5 & 417 & 58$\rm{_{ -4}^{+   14}}$ &   6.2$\rm{_{  -3.3}^{+   4.8}}$ &  36$\rm{_{ -17}^{+   7}}$ & 0.131  & 4  \\
NGC4507         & S2      &  50.5 &  43.1 & 1046 & 78$\rm{_{ -11}^{+   10}}$ &  13$\rm{_{  -9.0}^{+  11}}$ &  11.2$\rm{_{  -6.8}^{+   7.5}}$ &  0.039 & 5  \\
NGC4725$\rm{\clubsuit}$         & S2      &  13.6 &  38.9 & 281  & 0.0$\rm{_{  -0.0}^{+   0.7}}$ &  63$\rm{_{  -2}^{+  27}}$ &  36$\rm{_{ -26}^{+   2}}$ &  0.278 & 1  \\
MRK231          & S2      & 180.6 &  44.3 & 3736 & 44.2$\rm{_{  -3.0}^{+   6.4}}$ &   0.2$\rm{_{  -0.2}^{+   1.2}}$ &  50.8$\rm{_{  -3.5}^{+   6.3}}$ &  0.138 & 3  \\
NGC4941         & S2      &  17.0 &  41.3 & 351 & 72$\rm{_{ -12}^{+   10}}$ &  11$\rm{_{  -8}^{+  10}}$ &  19$\rm{_{ -13}^{+  15}}$ & 0.052  & 5  \\
NGC4939         & S2      &  38.9 &  42.6 & 805 & 87$\rm{_{ -13}^{+   9}}$ &   6.8$\rm{_{  -5.6}^{+   8.5}}$ &  11$\rm{_{  -9}^{+  11}}$ &  0.118 & 5  \\
NGC4945         & S2      &   3.9 &  42.3 &  81 & 0.9$\rm{_{  -0.9}^{+   1.7}}$ &   0.0$\rm{_{  -0.0}^{+   0.7}}$ &  98.1$\rm{_{  -1.7}^{+   0.9}}$ & 0.7  & 1  \\
NGC5135         & S2      &  58.6 &  43.1 & 1213 & 19.4$\rm{_{  -9.8}^{+   9.1}}$ &   9.5$\rm{_{  -7.0}^{+   7.6}}$ &  71.1$\rm{_{  -8.5}^{+   8.9}}$ & 0.028  & 2  \\
NGC5194         & S2      &   8.0 &  40.9 &  165 & 0.2$\rm{_{  -0.2}^{+   1.1}}$ &  11.0$\rm{_{  -2.9}^{+   2.6}}$ &  87.4$\rm{_{  -2.7}^{+   2.9}}$ & 0.059  & 1  \\
NGC5347         & S2      &  27.3 &  42.4 & 565 & 88.7$\rm{_{  -7.3}^{+   5.5}}$ &   5.0$\rm{_{  -4.2}^{+   8.4}}$ &   8.3$\rm{_{  -6.2}^{+   7.4}}$ & 0.054  & 5  \\
CircinusGalaxy  & S2      &   4.2 &  41.9 & 87 & 43.8$\rm{_{  -5.9}^{+   6.3}}$ &   7.3$\rm{_{  -6.3}^{+   4.5}}$ &  48.8$\rm{_{  -4.9}^{+   5.9}}$ &  0.276 & 3  \\
NGC5506         & S2      &  23.8 &  43.0 & 493 & 66.3$\rm{_{  -8.7}^{+   8.3}}$ &  19.7$\rm{_{  -7.9}^{+   9.7}}$ &  11.0$\rm{_{  -4.9}^{+   6.3}}$ & 0.04  & 4  \\
NGC5643         & S2      &  16.9 &  42.6 & 350 & 76$\rm{_{  -8}^{+  12}}$ &   1.4$\rm{_{  -1.4}^{+   4.7}}$ &  24$\rm{_{ -12}^{+   8}}$ & 0.06  & 5  \\
NGC5728         & S2      &  30.5 &  43.0 & 631 & 23.1$\rm{_{  -4.9}^{+   9.4}}$ &  10.7$\rm{_{  -6.5}^{+   9.4}}$ &  67$\rm{_{ -12}^{+   9}}$ & 0.137  & 2  \\
ESO138-G001     & S2      &  39.1 &  42.8 & 810 & 46.2$\rm{_{  -5.2}^{+   5.2}}$ &  24$\rm{_{ -10}^{+  11}}$ &  28$\rm{_{ -16}^{+  13}}$ & 0.141 &  3  \\
ESO103-G035     & S2      &  56.9 &  43.4 & 1177 & 74$\rm{_{  -6}^{+  11}}$ &  12.9$\rm{_{ -10.3}^{+   5.2}}$ &  15.1$\rm{_{  -4.9}^{+   4.6}}$ & 0.085 &  5  \\
IRAS19254-7245  & S2      & 264.3 &  44.5 & 5468 & 75$\pm 13$ &   3.3$\rm{_{  -2.4}^{+   2.5}}$ &  20$\rm{_{ -12}^{+  13}}$ & 0.031  & 5  \\
NGC7172         & S2      &  33.9 &  42.7 & 701 & 43$\rm{_{ -15}^{+   5}}$ &   9$\rm{_{  -7}^{+  12}}$ &  49$\rm{_{ -11}^{+   8}}$ & 0.145  & 3  \\
NGC7314         & S2      &  18.2 &  42.3 & 376 & 81$\rm{_{ -14}^{+  12}}$ &  12$\rm{_{  -8}^{+  11}}$ &  14$\rm{_{ -11}^{+  12}}$ &  0.076 & 5  \\
NGC7582         & S2      &  21.2 &  42.6 & 439 & 38$\pm 15$ &  11$\rm{_{  -9}^{+  12}}$ &  58$\rm{_{ -16}^{+  15}}$ & 0.046  & 3  \\ \hline
\end{tabular}
\newline
{Note: AP: Affinity Propagation groups (1 to 5, see text). X-ray luminosity given in units of erg/s. $\rm{\clubsuit}$: Candidates to have not torus because they belong to Group 1 with $\rm{C_{ISM}<50\%}$ (see text).}
\end{footnotesize}
\end{table*}

\begin{table*}[!h]
\def\arraystretch{1.1}
\caption{Details of the comparison samples and results of IRS/\emph{Spitzer} decomposition. }
\label{tab:comparisonsample2}
\begin{footnotesize}
\centering
\begin{tabular}{l l r c c r r r r l}
\hline\hline
Object name     & type      & D (Mpc) & $\rm{log(L_{X})}$ & Slit width (pc) & $\rm{C_{torus}}$ (\%) & $\rm{C_{Stellar}}$ (\%)             & $\rm{C_{ISM}}$ (\%)      & RMSE          & AP \\ 
      (1)       &    (2)    & (3)  & (4)  &             (5)                &       (6)                      &         (7)                   &    (8)  &    (9)  & (10) \\ \hline
NGC520          & SB      &  34.4 &  40.0 &  712 & 0.8$\rm{_{  -0.8}^{+   1.3}}$ &   0.0$\rm{_{  -0.0}^{+   0.7}}$ &  98.2$\rm{_{  -1.3}^{+   1.2}}$ & 0.085  & 1  \\
NGC0855         & SB      &   9.3 &  37.9 &  192 & 0.0$\rm{_{  -0.0}^{+   1.1}}$ &  23$\rm{_{ -16}^{+  27}}$ &  76$\rm{_{ -26}^{+  16}}$ &  0.133 & 1  \\
NGC0925         & SB      &   8.6 &  38.3 &  178 & 0.0$\rm{_{  -0.0}^{+   0.9}}$ &  25$\rm{_{ -10}^{+  25}}$ &  74$\rm{_{ -25}^{+  10}}$ & 0.456  & 1  \\
IC342           & SB      &   3.4 &  39.0 &  69 & 0.3$\rm{_{  -0.3}^{+   1.3}}$ &   0.7$\rm{_{  -0.7}^{+   1.5}}$ &  97.5$\rm{_{  -1.5}^{+   1.4}}$ & 0.07  & 1  \\
NGC1482         & SB      &  19.6 &  39.4 & 405  & 0.0$\rm{_{  -0.0}^{+   0.7}}$ &   3.1$\rm{_{  -1.3}^{+   1.4}}$ &  95.9$\rm{_{  -1.4}^{+   1.3}}$ &  0.061 & 1  \\
NGC1614         & SB      &  68.3 &  41.3 &  1412 & 0.4$\rm{_{  -0.4}^{+   1.3}}$ &   1.2$\rm{_{  -1.2}^{+   2.3}}$ &  97.3$\rm{_{  -2.6}^{+   1.8}}$ & 0.103  & 1  \\
NGC1808         & SB      &   9.8 &  39.7 &  204 & 0.1$\rm{_{  -0.1}^{+   1.1}}$ &   1.4$\rm{_{  -1.4}^{+   2.0}}$ &  97.2$\rm{_{  -2.4}^{+   1.8}}$ & 0.04  & 1  \\
NGC2146         & SB      &  21.9 &  39.0 &  453 & 0.0$\rm{_{  -0.0}^{+   0.7}}$ &   1.6$\rm{_{  -1.6}^{+   2.1}}$ &  97.7$\rm{_{  -2.3}^{+   1.8}}$ & 0.038  & 1  \\
NGC2798         & SB      &  26.4 &  39.6 &  546 & 0.0$\rm{_{  -0.0}^{+   0.7}}$ &   7.0$\rm{_{  -1.8}^{+   2.4}}$ &  92.2$\rm{_{  -2.1}^{+   1.7}}$ & 0.036  & 1  \\
NGC2903         & SB      &   8.7 &  39.9 &  181 & 1.9$\rm{_{  -1.9}^{+   2.4}}$ &   5.6$\rm{_{  -4.8}^{+   5.7}}$ &  92.5$\rm{_{  -5.8}^{+   5.1}}$ & 0.039  & 1  \\
NGC2976         & SB      &   3.9 &  36.6 &  81 & 0.0$\rm{_{  -0.0}^{+   0.7}}$ &  20$\rm{_{  -4.6}^{+  12}}$ &  79$\rm{_{ -12}^{+   5}}$ & 0.107  & 1  \\
NGC3184         & SB      &  12.0 &  38.0 &  249 & 0.0$\rm{_{  -0.0}^{+   1.0}}$ &  36$\rm{_{ -20}^{+  10}}$ &  63$\rm{_{ -10}^{+  19}}$ & 0.303  & 1  \\
NGC3198         & SB      &  13.9 &  38.2 &  288 & 0.0$\rm{_{  -0.0}^{+   0.7}}$ &   3$\rm{_{  -3}^{+  23}}$ &  96$\rm{_{ -23}^{+   3}}$ & 0.161  & 1  \\
NGC3256         & SB      &  37.4 &  40.8 &  774 & 0.1$\rm{_{  -0.1}^{+   1.1}}$ &   2.1$\rm{_{  -2.1}^{+   1.6}}$ &  96.2$\rm{_{  -1.8}^{+   2.4}}$ & 0.028  & 1  \\
NGC3310         & SB      &  18.1 &  40.0 &  374 & 2.3$\rm{_{  -2.0}^{+   1.5}}$ &   0.0$\rm{_{  -0.0}^{+   0.7}}$ &  96.7$\rm{_{  -1.5}^{+   2.0}}$ &  0.14 & 1  \\
NGC3367         & SB      &  43.6 &  40.9 &  902 & 6.0$\rm{_{  -2.5}^{+   3.4}}$ &   0.0$\rm{_{  -0.0}^{+   0.7}}$ &  93.0$\rm{_{  -3.4}^{+   2.5}}$ & 0.25  & 1  \\
M108            & SB      &  11.8 &  39.3 &  244 & 1.2$\rm{_{  -1.2}^{+   1.6}}$ &   8.6$\rm{_{  -6.6}^{+   5.9}}$ &  88.6$\rm{_{  -6.0}^{+   7.5}}$ &  0.084 & 1  \\
MRK52           & SB      &  33.5 &  38.0 &  693 & 0.0$\rm{_{  -0.0}^{+   0.7}}$ &   9.0$\rm{_{  -2.4}^{+   3.1}}$ &  90.0$\rm{_{  -3.1}^{+   2.4}}$ & 0.193  & 1  \\
NGC7252         & SB      &  58.6 &  40.6 & 1213  & 1.3$\rm{_{  -1.3}^{+   2.1}}$ &   5.2$\rm{_{  -3.9}^{+   6.7}}$ &  93.7$\rm{_{  -6.6}^{+   3.6}}$ & 0.048  & 1  \\ \hline \hline
\end{tabular}
\newline
{Note: AP: Affinity Propagation groups (1 to 5, see text). X-ray luminosity given in units of erg/s. $\rm{\clubsuit}$: Candidates to have not torus because they belong to Group 1 with $\rm{C_{ISM}<50\%}$ (see text).}
\end{footnotesize}
\end{table*}

\begin{table*}[!h]
\caption{CanariCam observations.}
\label{tab:CanariCam}
\begin{footnotesize}
\centering
\begin{tabular}{l c c c l | l c c c | c l l}
\hline\hline
 & \multicolumn{4}{c}{Target observations}         &   \multicolumn{4}{|c|}{Standard star observations}  &  \multicolumn{3}{c}{Results}   \\ 

Name            &   Date     & ObsID &  Expt. & Config. & Name &  ObsID & Expt. & FWHM$\rm{_{PSF}}$ & FWHM$\rm{_N}$  & Flux$\rm{_T}$ & Flux$\rm{_N}$ \\
                &   (Y-M-D)  &       &   (s)  &         &  (HD)&        &  (s)  & (arcsec)          &   (arcsec)      & (mJy)   &  (mJy)   \\
   (1)             &   (2)  &  (3)  &   (4)  &  (5)    &  (6)&   (7)   &  (8)  & (9)          &   (10)      & (11)   &  (12)   \\ \hline
NGC315 			& 14-09-22 & 1048 & 927 & 10/10/90/-180 & 4502   & 1045 & 66 & 0.34-0.44 & 0.24-0.29 & 28$\rm{\pm}$5  & 9$\rm{\pm}$2     \\
      			& 14-09-22 & 1050 & 927 & 10/10/90/-180 & 4502	 & 1045 & 66 & 0.34-0.44 &           &                &                  \\ 
NGC835 			& 14-09-24 & 1281 & 993 & 10/10/90/-180 & 11353  & 1287 & 66 & 0.31-0.36 & 0.25-0.39 & 32$\rm{\pm}$5  & 27$\rm{\pm}$4    \\
      			& 14-09-24 & 1283 & 993 & 10/10/90/-180 & 11353  & 1287 & 66 & 0.31-0.36 &           &                &                  \\ 
NGC2685* 		& 13-01-04 & 7636 & 661 & 10/10/90/-180 & 73108  & 7640 & 83 & 0.30-0.32 &     --    &      --        &      --          \\ 
      			& 13-01-04 & 7638 & 661 & 10/10/90/-180 & 73108  & 7640 & 83 & 0.30-0.32 &           &                &                  \\ 
NGC2655 		& 13-01-04 & 7642 & 661 & 10/10/90/-180 & 73108  & 7640 & 83 & 0.30-0.32 & 0.26-0.27 & 29$\rm{\pm}$5  & 8$\rm{\pm}$1     \\
        		& 13-01-04 & 7644 & 661 & 10/10/90/-180 & 73108  & 7640 & 83 & 0.30-0.32 &           &                &                  \\ 
UGC05101 		& 13-01-01 & 7270 & 617 & 10/10/90/-180 & 86378  & 7268 & 77 & 0.34-0.43 & 0.28-0.37 & 96$\rm{\pm}$15 & 11$\rm{\pm}$2    \\
NGC4321 		& 15-04-03 & 2700 & 927 & 10/10/90/-180 & 108381 & 2698 & 66 & 0.24-0.36 & 0.26-0.26 & 15$\rm{\pm}$3  & 3.0$\rm{\pm}$0.6 \\
      			& 15-04-04 & 2702 & 927 & 10/10/90/-180 & 108381 & 2698 & 66 & 0.24-0.36 &           &                &                  \\ 
NGC4486 		& 15-04-06 & 3117 & 927 & 16/16/0/-180  & 108985 & 3113 & 66 & 0.36-0.42 & 0.28-0.33 & 14$\rm{\pm}$2  & 6$\rm{\pm}$1     \\
      			& 15-04-06 & 3119 & 927 & 16/16/0/-180  & 108985 & 3113 & 66 & 0.36-0.42 &           &                &                  \\ 
MRK266NE 		& 13-01-01 & 7276 & 617 & 10/10/90/-180 & 120933 & 7272 & 77 & 0.32-0.35 & 0.37-0.46 & 34$\rm{\pm}$5  & 16$\rm{\pm}$3    \\
MRK266SW 		& 13-01-01 & 7276 & 617 & 10/10/90/-180 & 120933 & 7272 & 77 & 0.32-0.35 & 0.31-0.37 & 94$\rm{\pm}$15 & 11$\rm{\pm}$2    \\
UGC08696 		& 13-01-01 & 7274 & 308 & 10/10/90/-180 & 120933 & 7272 & 77 & 0.33-0.35 & 0.48-0.57 & 61$\rm{\pm}$10 & 25$\rm{\pm}$4    \\ 
NGC6251 		& 12-09-25 & 5990 & 617 & 12/12/0/-180  & 144204 & 5988 & 77 & 0.40-0.52 & 0.25-0.38 & 18$\rm{\pm}$3  & 7$\rm{\pm}$1     \\
      			& 13-08-24 & 3063 & 662 & 16/16/0/0     & 144204 & 3061 & 66 & 0.25-0.26 &           &                &                  \\ 
IRAS17208-0014 	& 13-07-19 & 1897 & 596 & 10/10/90/-180 & 153210 & 1901 & 66 & 0.31-0.36 & 0.28-0.42 & 87$\rm{\pm}$14  & 4$\rm{\pm}$1    \\
              	& 13-07-19 & 1899 & 596 & 10/10/90/-180 & 153210 & 1901 & 66 & 0.31-0.36 &           &                &                  \\ \hline 
\hline
\end{tabular}
\newline
{Note:  Object marked with an asterisk (NGC\,2685) was not detected with CanariCam. 
The column called ``Config." corresponds to the nod throw (keyword NODTHROW, in units of arcsec), chop throw (keyword CHPTHROW, in units of arcsec), 
instrument position angle (keyword INSTRPA, in degrees), and chop position angle (keyword CHPPA, in degrees) (written as NODTHROW/CHPTHROW/INSTRPA/CHPPA). }
\end{footnotesize}
\end{table*}

\clearpage

\begin{longtable}{l c c c}
\caption{Archival high resolution mid-IR imaging results.}\\
\hline\hline
Name & Wavelength     & FWHM$\rm{_N}$ & Flux$\rm{_N}$  \\
     & ($\rm{\mu m}$)  &   (arcsec)   &  (mJy)         \\ \hline
\hline
\endfirsthead
\caption{continued.}\\
\hline\hline
Name & Wavelength     & FWHM$\rm{_N}$ & Flux$\rm{_N}$  \\
     & ($\rm{\mu m}$)  &     (arcsec)  &  (mJy)         \\ \hline
\hline
\endhead
\hline
\endfoot
\label{tab:otherimages}
        NGC1052  & 10.6 &  0.5-0.4  &   133$\rm{\pm}$   24 \\
                 & 11.5 &  0.4-0.5  &   144$\rm{\pm}$   26 \\
                 & 12.5 &  0.6-0.5  &   172$\rm{\pm}$   31 \\
                 &  9.8 &  0.4-0.4  &   112$\rm{\pm}$   20 \\
                 &  7.8 &  0.6-0.4  &    65$\rm{\pm}$   11 \\
                 &  8.6 &  0.4-0.5  &    62$\rm{\pm}$   11\\
                 & 11.9 &  0.4-0.4  &   125$\rm{\pm}$   21 \\
                 & 18.7 &  0.5-0.5  &   307$\rm{\pm}$   51 \\
                 & 18.3 &  0.6-0.5  &   315$\rm{\pm}$   53 \\
                 &  8.7 &  0.8-0.6  &    63$\rm{\pm}$   10 \\
        NGC1097  & 12.3 &  0.4-0.4  &    24.9$\rm{\pm}$    4.2 \\
                 &  8.6 &  0.3-0.6  &    25.1$\rm{\pm}$    4.2 \\
                 & 11.2 &  0.6-0.7  &    25.7$\rm{\pm}$    4.5 \\
                 & 11.2 &  0.4-0.5  &    26.6$\rm{\pm}$    4.4 \\
                 & 11.9 &  0.5-0.4  &    22.1$\rm{\pm}$    3.9 \\
                 & 11.9 &  0.3-0.4  &    21.8$\rm{\pm}$    3.7 \\
                 & 18.7 &  0.5-0.5  &    47.3$\rm{\pm}$    7.9 \\
                 & 18.3 &  0.5-0.6  &    45.1$\rm{\pm}$    7.7 \\
                 & 10.8 &  0.4-0.5  &    24.5$\rm{\pm}$    4.1 \\
                 & 11.7 &  0.5-0.5  &    23.3$\rm{\pm}$    4.0 \\
        NGC4438  &  8.7 &  0.3-0.4  &     1.3$\rm{\pm}$    0.3 \\
                 &  8.7 &  0.7-0.5  &    10.2$\rm{\pm}$    1.7 \\
        NGC4579  & 12.3 &  0.4-0.4  &    51.2$\rm{\pm}$    8.9 \\
                 &  8.6 &  0.5-0.6  &    13.7$\rm{\pm}$    2.4 \\
                 & 11.2 &  0.3-0.3  &    51.3$\rm{\pm}$    9.1 \\
                 & 11.9 &  0.4-0.4  &    58$\rm{\pm}$   10 \\
                 & 18.7 &  0.5-0.4  &    93$\rm{\pm}$   16 \\
                 & 10.5 &  0.3-0.4  &    69$\rm{\pm}$   12 \\
                 & 10.8 &  0.4-0.3  &    61$\rm{\pm}$   11 \\
                 & 11.6 &  0.6-0.5  &    74$\rm{\pm}$   13 \\
                 & 12.5 &  0.5-0.4  &    64$\rm{\pm}$   11 \\
        NGC5005  & 11.2 &  0.4-0.9  &     1.3$\rm{\pm}$    0.3 \\
       IIIZW035  &  8.6 &  0.6-0.5  &    47.7$\rm{\pm}$    7.9 \\
       UGC05101  &  8.8 &  0.7-1.0  &   275$\rm{\pm}$   45 \\
                 & 17.7 &  0.7-0.8  &   227$\rm{\pm}$   39 \\
        NGC3627  &  8.6 &  1.2-0.8  &    52.2$\rm{\pm}$    8.6 \\
        NGC3628  &  8.6 &  0.8-1.0  &    34.1$\rm{\pm}$    5.3 \\
        NGC4261  & 12.3 &  0.6-0.4  &    15.0$\rm{\pm}$    2.5 \\
                 &  8.6 &  0.4-0.7  &     5.5$\rm{\pm}$    0.9 \\
                 & 11.2 &  0.3-0.5  &    12.2$\rm{\pm}$    2.1 \\
                 & 18.7 &  0.4-0.6  &    20.3$\rm{\pm}$    3.4 \\
                 & 11.9 &  0.7-0.6  &    21.6$\rm{\pm}$    3.5 \\
                 & 10.8 &  0.2-0.4  &     2.2$\rm{\pm}$    0.4 \\
                 &  8.7 &  0.4-0.6  &     4.1$\rm{\pm}$    0.7 \\
        NGC4486  & 11.7 &  1.3-1.8  &    83$\rm{\pm}$   13 \\
                 &  8.6 &  0.3-0.3  &    11.3$\rm{\pm}$    2.0 \\
                 & 11.9 &  0.3-0.4  &    19.3$\rm{\pm}$    3.3 \\
                 & 10.8 &  0.3-0.3  &    14.8$\rm{\pm}$    2.6 \\
        NGC4594  &  8.7 &  1.0-1.5  &    12.2$\rm{\pm}$    2.0 \\
                 &  8.7 &  0.9-1.9  &    14.3$\rm{\pm}$    2.3 \\
                 &  8.7 &  0.3-0.2  &     0.4$\rm{\pm}$    0.1 \\
                 &  8.7 &  0.9-0.8  &     6.9$\rm{\pm}$    1.1 \\
        NGC4736  & 11.2 &  0.7-0.5  &     6.4$\rm{\pm}$    1.1 \\
                 & 18.1 &  0.9-1.4  &   110$\rm{\pm}$   18 \\
       MRK266SW  &  8.8 &  1.2-1.1  &    49.2$\rm{\pm}$    7.9 \\
       MRK266NE  &  8.8 &  1.3-1.1  &    23.7$\rm{\pm}$    3.9 \\
        NGC6251  & 11.7 &  0.6-0.5  &    15.7$\rm{\pm}$    2.8 \\
        NGC6240  & 11.2 &  0.6-0.5  &   131$\rm{\pm}$   23 \\
                 & 11.9 &  0.9-0.7  &   200$\rm{\pm}$   34 \\
                 & 18.7 &  0.7-0.8  &   519$\rm{\pm}$   88 \\
        NGC7130  & 11.5 &  0.6-0.6  &   203$\rm{\pm}$   35 \\
                 & 12.3 &  0.5-0.6  &   229$\rm{\pm}$   40 \\
                 & 13.0 &  0.6-0.6  &   443$\rm{\pm}$   77 \\
                 & 10.4 &  0.6-0.6  &   143$\rm{\pm}$   24 \\
                 &  8.6 &  0.7-0.7  &   128$\rm{\pm}$   22 \\
                 & 10.5 &  0.5-0.5  &   101$\rm{\pm}$   18 \\
                 & 10.5 &  0.5-0.5  &   117$\rm{\pm}$   21 \\ \hline
    MCG-6-30-15  & 12.3 &  0.4-0.4  &   346$\rm{\pm}$   59 \\
                 &  8.6 &  0.3-0.3  &   220$\rm{\pm}$   38 \\
                 & 11.2 &  0.3-0.4  &   330$\rm{\pm}$   57 \\
                 & 11.9 &  0.4-0.4  &   313$\rm{\pm}$   54 \\
                 & 10.5 &  0.4-0.3  &   277$\rm{\pm}$   48 \\
       Fairall9  &  9.0 &  0.3-0.4  &   187$\rm{\pm}$   32 \\
                 & 12.3 &  0.4-0.5  &   273$\rm{\pm}$   46 \\
                 & 12.8 &  0.4-0.4  &   269$\rm{\pm}$   46 \\
                 &  8.6 &  0.6-0.5  &   195$\rm{\pm}$   32 \\
                 &  8.6 &  0.3-0.3  &   171$\rm{\pm}$   30 \\
                 &  8.6 &  0.5-0.4  &   189$\rm{\pm}$   32 \\
                 &  8.6 &  0.3-0.3  &   173$\rm{\pm}$   31 \\
                 &  8.6 &  0.3-0.3  &   169$\rm{\pm}$   30 \\
                 &  8.6 &  0.3-0.3  &   178$\rm{\pm}$   32 \\
                 &  8.6 &  0.3-0.3  &   175$\rm{\pm}$   31 \\
                 &  8.6 &  0.3-0.3  &   171$\rm{\pm}$   31 \\
                 &  8.6 &  0.3-0.3  &   184$\rm{\pm}$   32 \\
                 &  8.6 &  0.5-0.5  &   182$\rm{\pm}$   30 \\
                 &  8.6 &  0.3-0.3  &   173$\rm{\pm}$   31 \\
                 & 11.9 &  0.4-0.4  &   254$\rm{\pm}$   43 \\
                 & 10.5 &  0.4-0.4  &   263$\rm{\pm}$   45 \\
        NGC526A  & 12.3 &  0.4-0.4  &   221$\rm{\pm}$   38 \\
                 & 10.5 &  0.4-0.3  &   161$\rm{\pm}$   28 \\
        NGC3783  &  9.0 &  0.4-0.3  &   351$\rm{\pm}$   61 \\
                 & 12.3 &  0.4-0.4  &   630$\rm{\pm}$  110 \\
                 &  8.6 &  0.3-0.3  &   359$\rm{\pm}$   63 \\
                 &  8.6 &  0.3-0.3  &   368$\rm{\pm}$   65 \\
                 &  8.6 &  0.4-0.3  &   310$\rm{\pm}$   53 \\
                 &  8.6 &  0.4-0.4  &   358$\rm{\pm}$   61 \\
                 &  8.6 &  0.4-0.3  &   316$\rm{\pm}$   55 \\
                 &  8.6 &  0.4-0.5  &   334$\rm{\pm}$   56 \\
                 &  8.6 &  0.3-0.3  &   337$\rm{\pm}$   61\\
                 &  8.6 &  0.5-0.4  &   319$\rm{\pm}$   53 \\
                 &  8.6 &  0.4-0.5  &   398$\rm{\pm}$   66 \\
                 &  8.6 &  0.4-0.3  &   358$\rm{\pm}$   61 \\
                 &  8.6 &  0.4-0.4  &   371$\rm{\pm}$   64 \\
                 & 11.2 &  0.4-0.4  &   680$\rm{\pm}$  120 \\
                 & 11.9 &  0.4-0.4  &   580$\rm{\pm}$  100 \\
                 & 11.9 &  0.3-0.3  &   551$\rm{\pm}$   96 \\
                 & 11.9 &  0.4-0.4  &   550$\rm{\pm}$   95 \\
                 & 11.9 &  0.4-0.3  &   640$\rm{\pm}$  110 \\
                 & 11.9 &  0.4-0.3  &   600$\rm{\pm}$  100 \\
                 & 17.6 &  0.5-0.5  &  1200$\rm{\pm}$  200 \\
                 & 18.7 &  0.5-0.5  &  1240$\rm{\pm}$  210 \\
                 & 10.5 &  0.3-0.4  &   503$\rm{\pm}$   88 \\
        IC4329A  &  9.0 &  0.4-0.3  &   800$\rm{\pm}$  140 \\
                 & 12.3 &  0.4-0.4  &  1080$\rm{\pm}$  180 \\
                 & 11.2 &  0.3-0.4  &  1000 $\rm{\pm}$  170 \\
                 & 11.9 &  0.4-0.5  &  1060$\rm{\pm}$  180 \\
                 & 18.3 &  0.7-0.8  &  2140$\rm{\pm}$  350 \\
                 & 10.5 &  0.4-0.3  &   920$\rm{\pm}$  160 \\
                 &  8.7 &  0.8-0.9  &   700$\rm{\pm}$  110 \\
        NGC5548  &  9.0 &  0.8-0.6  &   127$\rm{\pm}$   21 \\
                 & 11.9 &  0.7-0.5  &   172$\rm{\pm}$   28 \\
         MRK509  & 12.8 &  0.4-0.4  &   232$\rm{\pm}$   40 \\
                 & 11.2 &  0.4-0.4  &   223$\rm{\pm}$   38 \\
                 & 18.3 &  0.5-0.6  &   426$\rm{\pm}$   71 \\
                 & 10.5 &  0.4-0.4  &   203$\rm{\pm}$   35 \\
                 &  8.7 &  0.4-0.4  &   143$\rm{\pm}$   25 \\
                 &  8.7 &  0.4-0.6  &   157$\rm{\pm}$   27 \\
                 & 11.7 &  0.4-0.5  &   218$\rm{\pm}$   38 \\
        NGC7213  & 12.3 &  0.4-0.4  &   238$\rm{\pm}$   41 \\
                 & 12.3 &  0.3-0.4  &   204$\rm{\pm}$   36 \\
                 & 12.3 &  0.6-0.7  &   245$\rm{\pm}$   40 \\
                 & 12.8 &  0.6-0.5  &   258$\rm{\pm}$   42 \\
                 & 11.2 &  0.4-0.4  &   270$\rm{\pm}$   46 \\
                 & 11.2 &  0.3-0.3  &   205$\rm{\pm}$   36 \\
                 & 18.7 &  0.5-0.5  &   397$\rm{\pm}$   66 \\
                 & 18.3 &  0.6-0.6  &   351$\rm{\pm}$   58 \\
                 & 10.5 &  0.3-0.4  &   211$\rm{\pm}$   37 \\
                 &  8.7 &  0.5-0.5  &    92$\rm{\pm}$   16 \\
                 &  8.7 &  0.5-0.6  &    84$\rm{\pm}$   14 \\
                 &  8.7 &  0.5-0.7  &    81$\rm{\pm}$   13 \\ \hline
        NGC1386  &  8.6 &  0.7-0.7  &   198$\rm{\pm}$   32 \\
                 & 11.9 &  0.5-0.4  &   381$\rm{\pm}$   64 \\
                 & 18.7 &  0.6-0.5  &   593$\rm{\pm}$   97 \\
                 & 18.7 &  0.7-0.6  &   740$\rm{\pm}$  120 \\
                 & 10.8 &  0.8-0.7  &   295$\rm{\pm}$   47 \\
        NGC2110  &  9.0 &  0.3-0.4  &   167$\rm{\pm}$   29 \\
                 & 12.3 &  0.4-0.4  &   324$\rm{\pm}$   56 \\
                 & 12.8 &  0.4-0.4  &   341$\rm{\pm}$   59 \\
                 & 11.2 &  0.5-0.5  &   283$\rm{\pm}$   48 \\
                 &  8.6 &  0.3-0.4  &   157$\rm{\pm}$   27 \\
                 & 11.9 &  0.3-0.4  &   272$\rm{\pm}$   47 \\
                 & 17.6 &  0.5-0.5  &   457$\rm{\pm}$   77 \\
           MRK3  & 11.6 &  0.4-0.6  &   495$\rm{\pm}$   84 \\
    MCG-5-23-16  &  9.0 &  0.3-0.3  &   320$\rm{\pm}$   57 \\
                 &  8.6 &  0.4-0.5  &   318$\rm{\pm}$   58 \\
                 & 11.9 &  0.4-0.3  &   490$\rm{\pm}$   85 \\
                 & 11.9 &  0.4-0.3  &   559$\rm{\pm}$   97 \\
                 & 17.6 &  0.5-0.5  &  1270$\rm{\pm}$  210 \\
                 & 18.7 &  0.5-0.5  &  1300$\rm{\pm}$  220 \\
        NGC3081  & 13.0 &  0.3-0.5  &   160$\rm{\pm}$   28 \\
                 &  8.6 &  0.3-0.5  &    89$\rm{\pm}$   15 \\
                 & 11.2 &  0.3-0.4  &   141$\rm{\pm}$   24 \\
                 & 18.3 &  0.7-0.6  &   331$\rm{\pm}$   55 \\
                 & 10.5 &  0.3-0.5  &   136$\rm{\pm}$   23 \\
                 &  8.7 &  0.4-0.4  &    81$\rm{\pm}$   14 \\
        NGC3281  & 13.0 &  0.4-0.4  &   900$\rm{\pm}$  150 \\
                 & 10.4 &  0.5-0.4  &   440$\rm{\pm}$   75 \\
                 &  8.6 &  0.3-0.3  &   397$\rm{\pm}$   69 \\
                 & 11.2 &  0.4-0.3  &   422$\rm{\pm}$   73 \\
                 & 18.3 &  0.7-0.7  &  1070$\rm{\pm}$  180 \\
                 & 11.9 &  0.4-0.4  &   562$\rm{\pm}$   96 \\
                 & 10.5 &  0.4-0.4  &   255$\rm{\pm}$   44 \\
                 &  8.7 &  0.5-0.6  &   366$\rm{\pm}$   62 \\
                 &  8.7 &  0.5-0.4  &   365$\rm{\pm}$   62 \\
        NGC3393  & 13.0 &  0.3-0.5  &    89$\rm{\pm}$   15 \\
                 &  8.6 &  0.3-0.4  &    17.1$\rm{\pm}$    3.1 \\
                 &  8.6 &  0.3-0.2  &    14.0$\rm{\pm}$    2.5 \\
                 & 11.2 &  0.4-0.3  &    39.3$\rm{\pm}$    6.7 \\
                 & 11.2 &  0.3-0.5  &    36.9$\rm{\pm}$    6.3 \\
                 & 10.5 &  0.5-0.3  &    43.2$\rm{\pm}$    7.4 \\
                 & 10.5 &  0.4-0.3  &    23.9$\rm{\pm}$    4.3 \\
        NGC4388  & 13.0 &  0.5-0.5  &   414$\rm{\pm}$   69 \\
                 & 11.2 &  0.4-0.5  &   222$\rm{\pm}$   38 \\
                 &  8.6 &  0.5-0.6  &   140$\rm{\pm}$   23 \\
                 & 11.2 &  0.4-0.5  &   191  $\rm{\pm}$   32 \\
                 & 18.1 &  0.6-0.6  &   860$\rm{\pm}$  140 \\
                 & 10.5 &  0.4-0.5  &   141$\rm{\pm}$   24 \\
        NGC4507  & 12.3 &  0.4-0.4  &   551$\rm{\pm}$   95 \\
                 & 11.2 &  0.3-0.4  &   522$\rm{\pm}$   91 \\
                 & 10.5 &  0.3-0.4  &   458$\rm{\pm}$   80 \\
        NGC4941  & 12.3 &  0.3-0.4  &    68$\rm{\pm}$   12 \\
                 &  8.6 &  0.3-0.3  &    34.0$\rm{\pm}$    6.1 \\
                 & 11.2 &  0.4-0.4  &    64$\rm{\pm}$   11 \\
                 & 18.7 &  0.5-0.5  &   211$\rm{\pm}$   35 \\
                 & 10.8 &  0.3-0.4  &    62$\rm{\pm}$   11 \\
        NGC4945  &  8.6 &  1.3-0.9  &   106$\rm{\pm}$   18 \\
                 & 18.7 &  1.1-1.4  &   268$\rm{\pm}$   44 \\
                 & 18.3 &  0.5-0.5  &    20.7$\rm{\pm}$    4.0 \\
        NGC5135  & 11.5 &  0.5-0.5  &    95$\rm{\pm}$   17 \\
                 & 12.3 &  0.4-0.4  &   109$\rm{\pm}$   19 \\
                 & 12.3 &  0.5-0.5  &   123$\rm{\pm}$   22 \\
                 & 12.8 &  0.5-0.4  &   138$\rm{\pm}$   25 \\
                 & 13.0 &  0.5-0.5  &   130$\rm{\pm}$   23 \\
                 & 13.0 &  0.5-0.5  &   158$\rm{\pm}$   28 \\
                 & 10.5 &  0.4-0.4  &    76$\rm{\pm}$   13 \\
                 & 10.5 &  0.5-0.5  &    60$\rm{\pm}$   11 \\
        NGC5194  & 11.6 &  0.4-0.5  &     8.2$\rm{\pm}$    1.4 \\
                 & 12.5 &  0.7-0.8  &    34.0$\rm{\pm}$    5.6 \\
        NGC5347  & 11.2 &  0.5-0.5  &   215$\rm{\pm}$   37 \\
 CircinusGalaxy  & 12.8 &  0.5-0.6  & 14700$\rm{\pm}$ 2400 \\
                 & 13.0 &  0.5-0.6  & 14600$\rm{\pm}$ 2400 \\
                 &  8.6 &  0.4-0.4  &  7600$\rm{\pm}$ 1300 \\
                 &  8.6 &  0.4-0.3  &  6300$\rm{\pm}$ 1100 \\
                 &  8.6 &  0.4-0.3  &  5320$\rm{\pm}$  920 \\
                 &  8.6 &  0.4-0.3  &  6400$\rm{\pm}$ 1100 \\
                 &  8.6 &  0.5-0.4  &  7300$\rm{\pm}$ 1200 \\
                 &  8.6 &  0.3-0.4  &  6200$\rm{\pm}$ 1100 \\
                 &  8.6 &  0.4-0.3  &  6100$\rm{\pm}$ 1100 \\
                 & 11.9 &  0.4-0.4  &  8200$\rm{\pm}$ 1400 \\
                 & 11.9 &  0.4-0.4  & 12200$\rm{\pm}$ 2100 \\
                 & 11.9 &  0.4-0.4  & 10700$\rm{\pm}$ 1800 \\
                 & 11.9 &  0.4-0.3  &  7900$\rm{\pm}$ 1400 \\
                 & 11.9 &  0.3-0.3  &  6600$\rm{\pm}$ 1200 \\
                 & 11.9 &  0.5-0.5  & 12700$\rm{\pm}$ 2100 \\
                 & 11.9 &  0.4-0.4  & 11100$\rm{\pm}$ 1900 \\
                 & 11.9 &  0.4-0.4  & 10700$\rm{\pm}$ 1800 \\
                 & 17.6 &  0.3-0.3  &  8700$\rm{\pm}$ 1600 \\
                 & 18.7 &  0.5-0.5  & 16500$\rm{\pm}$ 2700 \\
                 & 19.5 &  0.6-0.7  & 24100$\rm{\pm}$ 3900 \\
                 & 18.3 &  0.6-0.7  & 12400$\rm{\pm}$ 2000 \\
                 & 10.5 &  0.5-0.5  &  4580$\rm{\pm}$  770 \\
                 & 10.8 &  0.6-0.6  &  6100$\rm{\pm}$ 1000 \\
                 &  8.7 &  0.4-0.4  &  6100$\rm{\pm}$ 1100 \\
        NGC5506  & 12.3 &  0.4-0.4  &  1060$\rm{\pm}$  180 \\
                 & 13.0 &  0.4-0.4  &  1190$\rm{\pm}$  200 \\
                 & 11.2 &  0.5-0.4  &   810$\rm{\pm}$  140 \\
                 & 11.2 &  0.4-0.4  &   670$\rm{\pm}$  120 \\
                 & 11.2 &  0.4-0.4  &   750$\rm{\pm}$  130 \\
                 & 11.9 &  0.4-0.4  &   790$\rm{\pm}$  140 \\
                 & 11.9 &  0.4-0.4  &   880$\rm{\pm}$  150 \\
                 & 18.7 &  0.6-0.7  &  1930$\rm{\pm}$  310 \\
                 & 18.7 &  0.5-0.5  &  1530$\rm{\pm}$  250 \\
                 & 18.7 &  0.5-0.5  &  1590$\rm{\pm}$  260 \\
                 & 18.1 &  0.6-0.6  &  1880$\rm{\pm}$  320 \\
        NGC5643  &  9.0 &  0.4-0.3  &   117$\rm{\pm}$   20 \\
                 & 11.9 &  0.4-0.4  &   287$\rm{\pm}$   49 \\
        NGC5728  & 12.3 &  0.4-0.5  &    69$\rm{\pm}$   11 \\
                 & 13.0 &  0.3-0.6  &    81$\rm{\pm}$   15 \\
                 & 11.2 &  0.3-0.4  &    21.1$\rm{\pm}$    3.6 \\
                 & 18.3 &  0.7-0.8  &   174$\rm{\pm}$   29 \\
                 &  8.7 &  0.7-0.8  &    22.5$\rm{\pm}$    3.7 \\
    ESO138-G001  &  9.0 &  0.5-0.4  &   439$\rm{\pm}$   74 \\
                 & 13.0 &  0.4-0.4  &   700$\rm{\pm}$  120 \\
                 &  8.6 &  0.3-0.3  &   389$\rm{\pm}$   68 \\
                 & 11.2 &  0.3-0.3  &   620$\rm{\pm}$  110 \\
                 & 11.9 &  0.5-0.4  &   710$\rm{\pm}$  120 \\
                 & 11.9 &  0.4-0.4  &   710$\rm{\pm}$  120 \\
                 & 10.5 &  0.3-0.4  &   571$\rm{\pm}$   98 \\
    ESO103-G035  & 13.0 &  0.5-0.6  &   660$\rm{\pm}$  110 \\
                 &  8.6 &  0.4-0.5  &   255$\rm{\pm}$   42 \\
                 & 11.2 &  0.4-0.6  &   410$\rm{\pm}$   67 \\
                 & 10.5 &  0.4-0.8  &   304$\rm{\pm}$   50 \\
        NGC7172  & 12.3 &  0.4-0.5  &   141$\rm{\pm}$   24 \\
                 & 13.0 &  0.4-0.4  &   176$\rm{\pm}$   30 \\
                 & 11.2 &  0.3-0.4  &    56.7$\rm{\pm}$    9.7 \\
        NGC7314  & 11.2 &  0.6-0.7  &    67$\rm{\pm}$   11 \\
                 & 11.2 &  0.4-0.5  &    68$\rm{\pm}$   12 \\
                 & 18.3 &  0.6-0.6  &   177$\rm{\pm}$   30 \\
                 & 10.5 &  0.5-0.6  &    69$\rm{\pm}$   11 \\
                 & 10.5 &  0.5-0.5  &    78$\rm{\pm}$   13 \\
                 &  8.7 &  0.5-0.5  &    46.8$\rm{\pm}$    7.9 \\
        NGC7582  &  9.0 &  0.5-0.5  &   231$\rm{\pm}$   38 \\
                 &  9.0 &  0.4-0.3  &   231$\rm{\pm}$   39 \\
                 & 12.8 &  0.4-0.4  &   475$\rm{\pm}$   87 \\
                 & 10.4 &  0.4-0.4  &   311$\rm{\pm}$   54 \\
                 &  8.6 &  0.4-0.5  &   250$\rm{\pm}$   42 \\
                 & 11.9 &  0.3-0.4  &   348$\rm{\pm}$   61 \\
                 & 11.9 &  0.5-0.4  &   386$\rm{\pm}$   65 \\
                 & 11.9 &  0.4-0.4  &   356$\rm{\pm}$   61 \\
                 & 18.7 &  0.5-0.6  &   489$\rm{\pm}$   81 \\ \hline
        NGC1614  & 11.2 &  1.1-1.2  &   390$\rm{\pm}$   64 \\
                 & 17.7 &  1.2-1.3  &  1280$\rm{\pm}$  210 \\
                 & 18.7 &  1.2-1.2  &  1370$\rm{\pm}$  220 \\
                 &  8.7 &  1.5-1.5  &   660$\rm{\pm}$  100 \\
                 &  8.7 &  1.5-1.5  &   635$\rm{\pm}$   99 \\
        NGC1808  & 11.9 &  0.5-0.6  &   389$\rm{\pm}$   64 \\ \hline
\end{longtable}

\clearpage

\begin{table}[!h]
\def\arraystretch{1.1}
\caption{Median and percentiles 16 and 84 (in parenthesis) of the \emph{torus}, stellar, and ISM distributions.}
\label{tab:contributors}
\begin{footnotesize}
\begin{center}
\begin{tabular}{l r r r}
\hline\hline
                                 &$\rm{C_{torus}}$ & $\rm{C_{stellar}}$ & $\rm{C_{ISM}}$    \\
                                 &  (\%)                   &   (\%)                    &    (\%)                  \\ \hline
Starbursts              &  0.1 (0.0-1.3)     &   4.3 (0.7-19.8)  & 93.3 (79.2-97.3) \\
LINER1                  & 6.4 (0.6-36.7)   & 24.3 (16.2-60.4) & 40.1 (22.2-74.3) \\
LINER2                  & 3.5 (0.1-24.0)   & 36.0 (1.8-76.7) & 51.6 (17.9-85.3) \\
Seyfert1                  &77.2 (70.5-82.2)& 14.3 (11.4-16.6) & 9.4 (6.1-15.2) \\
Seyfert 2                &58.2 (23.0-80.2)& 10.9 (4.1-16.2) & 30.7 (12.0-57.9) \\ \hline
\end{tabular}
\end{center}
\end{footnotesize}
\end{table}

\begin{table*}[!h]
\def\arraystretch{1.1}
\caption{Affinity Propagation (AP) clustering method overall results. }
\label{tab:affinity}
\begin{footnotesize}
\begin{center}
\begin{tabular}{l c r r r r l l }
\hline\hline
 Group &\multicolumn{2}{c}{Representative} & &  \multicolumn{2}{c}{Median}       &            Morph.                    & Slit width        \\
	   &\multicolumn{2}{c}{member}         & &  \multicolumn{2}{c}{of the group}    &                                             &    log(pc)            \\ \cline{2-3} \cline{5-6}
	   &   $\rm{log(L_{bol})}$ & $\rm{C_{torus}}$ & & $\rm{log(L_{bol})}$   & $\rm{C_{torus}}$     &                  &                       \\ \hline
1      &   40.8           &   1.3            & &  40.5 (40.1,41.3) & 0.6 (0,1.4)  &   2 (0,5)             & 2.6 (2.4,2.8)  \\
2      &   42.1           &  18.5          & &  42.5 (41.9,43.0) & 18.8 (14.1,23.0)  &  1 (-1,3)       & 2.8 (2.6,3.1)  \\
3      &   42.6           &  39.8          & &  43.5 (43.0,43.7) & 41.2 (37.4,43.9) &   2 (-3,2)   & 3.1 (2.9,3.4)   \\
4      &   43.7           &  58.8          & &  43.5 (43.2, 43.7) & 60.6 (57.4,64.8) &  1 (0,1)              & 2.8 (2.7,3.0)  \\
5      &   44.1           &  78.0          & &  44.0 (43.3,45.0) & 78.4 (74.9,82.3) &   1(-2,4)      & 3.0 (2.7,3.2)     \\ \hline \hline
\end{tabular}
\newline
{Note: The confidence levels computed for the median values are the percentiles 25-75\%; they are not an error on the estimates. }
\end{center}
\end{footnotesize}
\end{table*}	 

\begin{table}[!h]
\def\arraystretch{1.2}
\caption{Parameters of the Clumpy models obtained using the code BayesClumpy. }
\label{tab:BayesClumpy}
\begin{footnotesize}
\begin{center}
\begin{tabular}{l r r r r r r}
\hline\hline
Param.                       &     Group 3     &     Group 4      &     \multicolumn{3}{c}{Group 5}       \\
($\rm{C_{torus}}$)             & ($\rm{40\%}$)          &   ($\rm{60\%}$)  & \multicolumn{3}{c}{($\rm{80\%}$)} \\ \cline{4-6}
                                    &                         &                         &     \multicolumn{1}{c}{All}   & \multicolumn{1}{c}{ Type 1}   &  \multicolumn{1}{c}{Type 2}     \\   \hline
 $\rm{\sigma}$     &  $\rm{52_{ -11}^{ +9}}$   & $\rm{48_{ -13}^{ +12}}$    &  $\rm{38_{ -14}^{ +19}}$       &    $\rm{36_{ -11}^{ +15}}$     &  $\rm{35_{ -7}^{ +17}}$      \\
 $\rm{Y}$          &  $\rm{13.2_{  -2.0}^{  +2.8}}$   & $\rm{14.6_{  -2.0}^{  +4.2}}$    &  $\rm{21.0_{  -3.1}^{  +7.7}}$       &   $\rm{27_{  -8}^{  +26}}$       &     $\rm{22.0_{  -2.8}^{  +5.3}}$      \\
 $\rm{R_{in}(pc)}$    &              0.025               &            0.089                 &             0.14                          &   0.14    &    0.14    \\
 $\rm{R_{out}(pc)}$    &          $\rm{0.33_{-0.05}^{+0.07}}$   &          $\rm{0.82_{-0.11}^{+0.23}}$               &         $\rm{2.36_{-0.34}^{+0.84}}$        &     $\rm{3.0_{-0.9}^{+2.8}}$    &         $\rm{2.42_{-0.58}^{+0.31}}$       \\
  $\rm{H(pc)} $        &  $\rm{0.14_{-0.05}^{+0.05}}$   &          $\rm{0.34_{-0.14}^{+0.14}}$               &         $\rm{0.76_{-0.45}^{+0.45}}$  & $\rm{\sim0.9}$  & $\rm{\sim0.7}$ \\
 $\rm{N_{o}}$      &  $\rm{6.3_{ -1.5}^{ +2.9}}$      & $\rm{6.5_{ -1.7}^{ +3.3}}$       &  $\rm{7.0_{ -2.2}^{ +3.6}}$           &   $\rm{6.7_{ -2.8}^{ +3.8}}$       &  $\rm{8.5_{ -2.5}^{ +3.5}}$       \\
 $\rm{q}$          &  $\rm{0.39_{ -0.25}^{ +0.45}}$   & $\rm{0.61_{ -0.40}^{ +0.72}}$    &  $\rm{0.72_{ -0.46}^{ +0.68}}$          &   $\rm{1.26_{ -0.56}^{ +0.46}}$     &   $\rm{0.56_{ -0.36}^{ +0.58}}$   \\
 $\rm{\tau_{\nu}}$ &  $\rm{  71_{ -18}^{ +25}}$ & $\rm{110_{ -25}^{ +21}}$   &  $\rm{64_{ -24}^{ +27}}$          &   $\rm{42_{ -16}^{ +32}}$     &  $\rm{43_{ -17}^{ +21}}$    \\
 $\rm{i}$          &  $\rm{58_{ -32}^{ +18}}$   & $\rm{60_{ -24}^{ +16}}$    &  $\rm{65_{ -26}^{ +13}}$           &    $\rm{35_{ -21}^{ +22}}$      &  $\rm{73_{ -14}^{ +9}}$      \\
 $\rm{f_{c}}$      &           $\rm{0.74_{-0.24}^{+0.13}}$                   &     $\rm{0.67_{-0.24}^{+0.16}}$   &     $\rm{0.48_{-0.29}^{+0.26}}$        &   $\rm{0.42_{-0.18}^{+0.22}}$          &   $\rm{0.45_{-0.14}^{+0.27}}$           \\  \hline \hline	
\end{tabular}
\newline
{Note:  {\bf $\rm{R_{in}}$ is estimated from the bolometric luminosity, $\rm{R_{out}}$ is estimated from Y  $\rm{R_{in}}$,   $\rm{H(pc)} $ is estimated using  $\rm{R_{out}}$, $\rm{R_{in}}$, and  $\rm{\sigma}$, and  $\rm{f_{c}}$ is estimated using the others parameters using equation 2.} Group 1 was not fitted. We did not find a good fit for Group 2 (see text). We have also fitted the average spectrum of type-1 and type-2 Seyferts included in Group 5 (see text).}
\end{center}
\end{footnotesize}
\end{table}

%% file: ApendixA.tex
\clearpage

\begin{appendix} 
\section{Catalog of spectral decompositions} \label{ape:decompose}

\begin{figure*}[!h]
\begin{center}
\includegraphics[width=0.45\columnwidth]{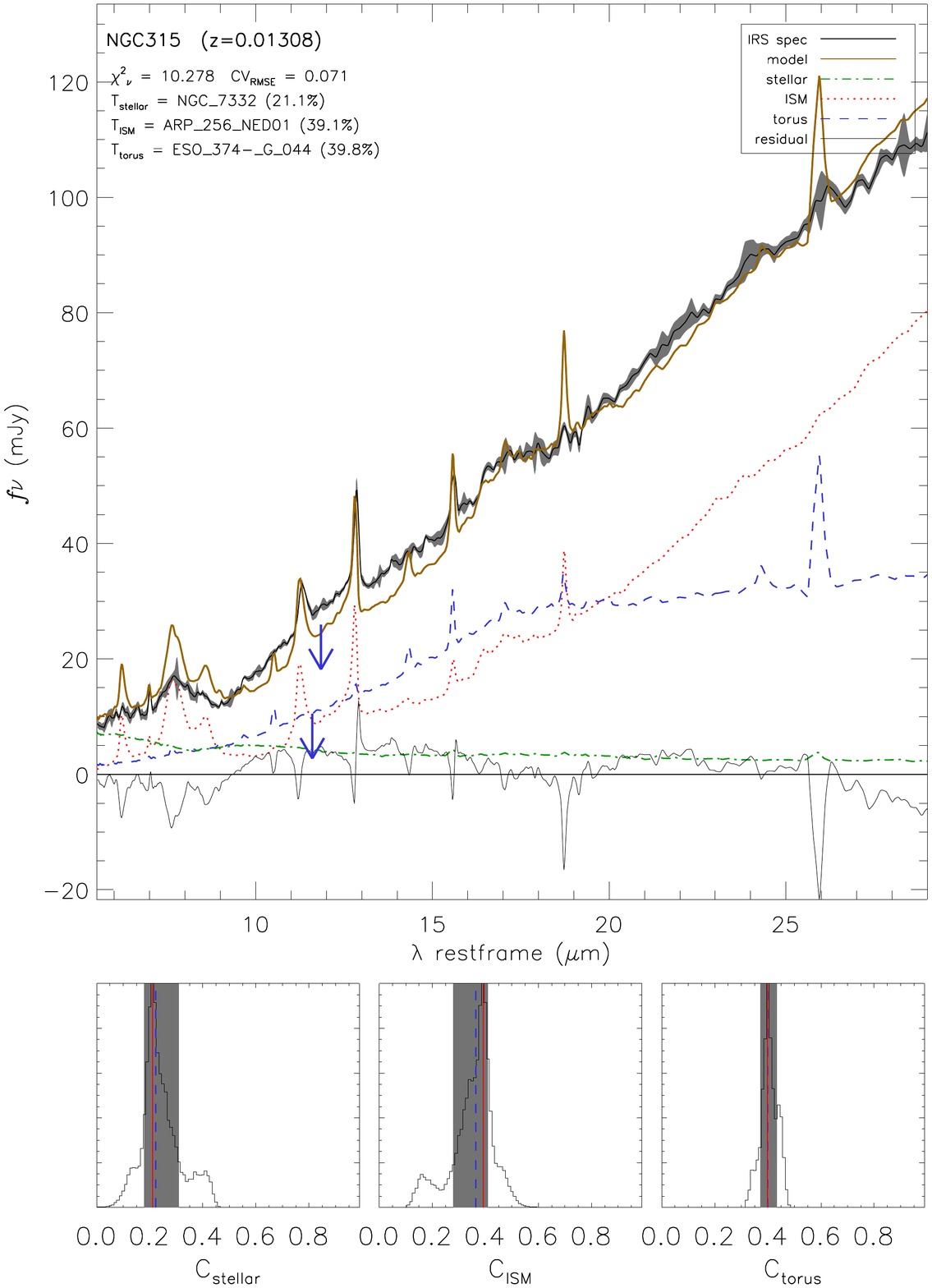}
\includegraphics[width=0.45\columnwidth]{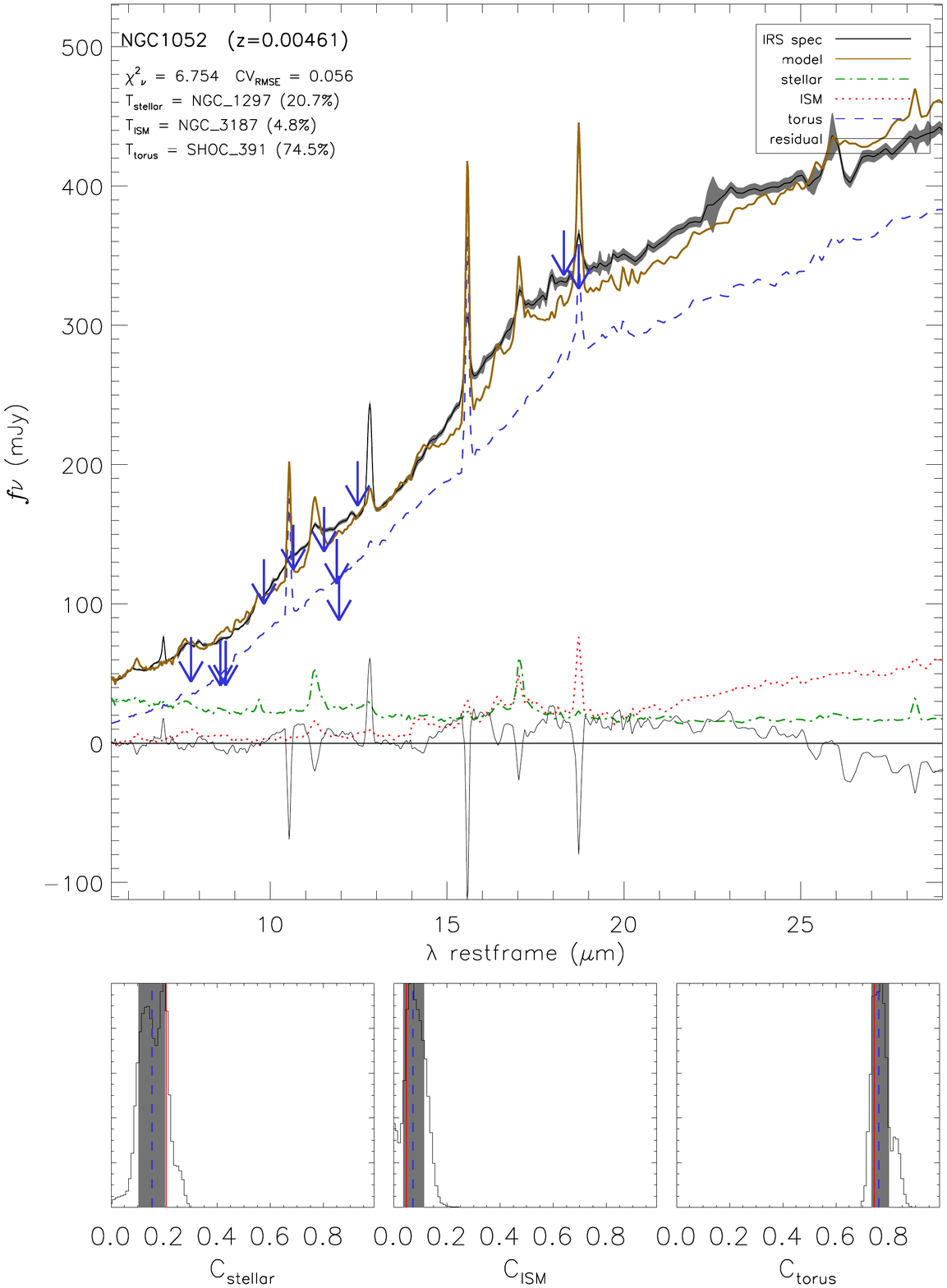}
\includegraphics[width=0.45\columnwidth]{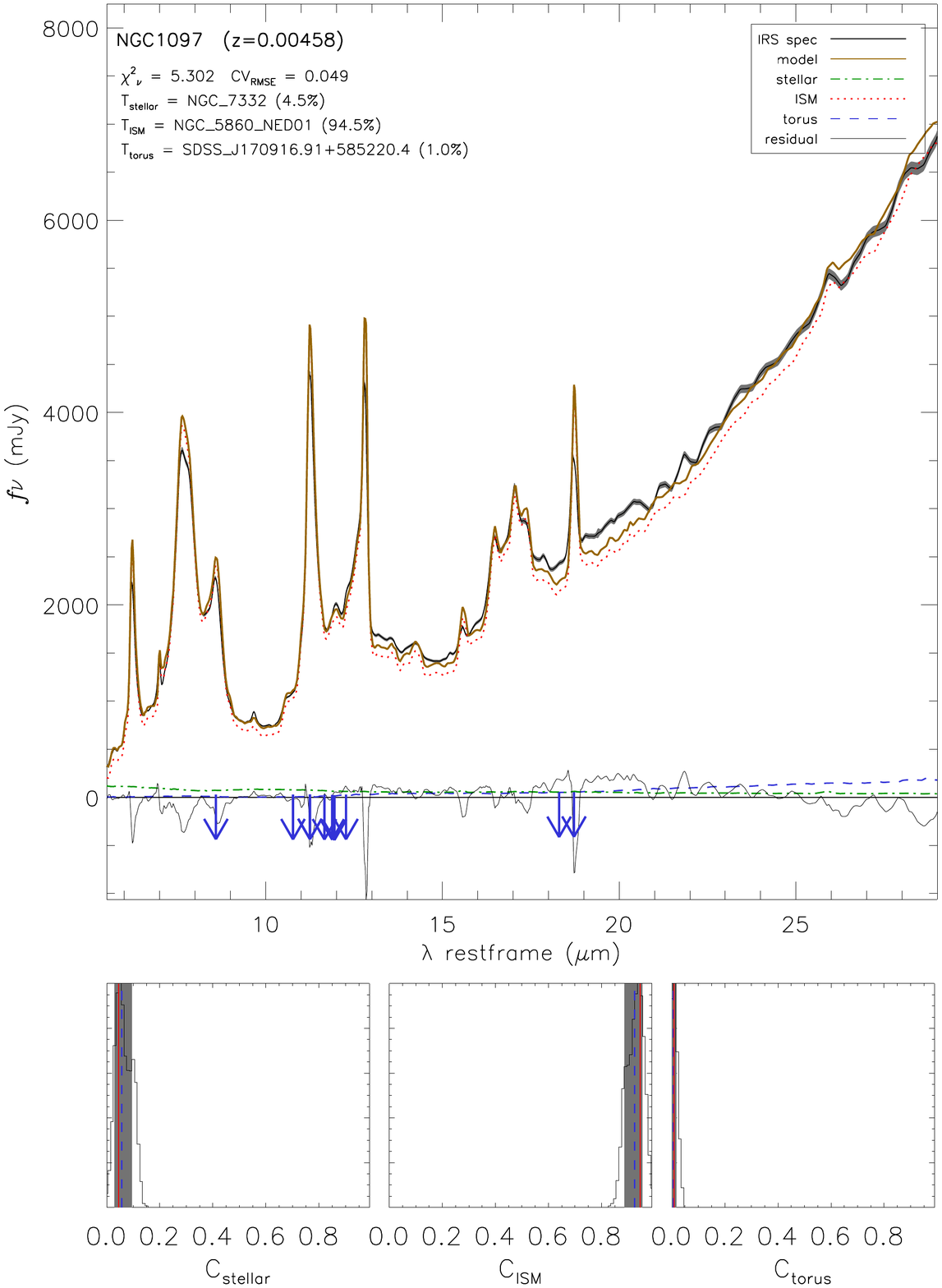}
\includegraphics[width=0.45\columnwidth]{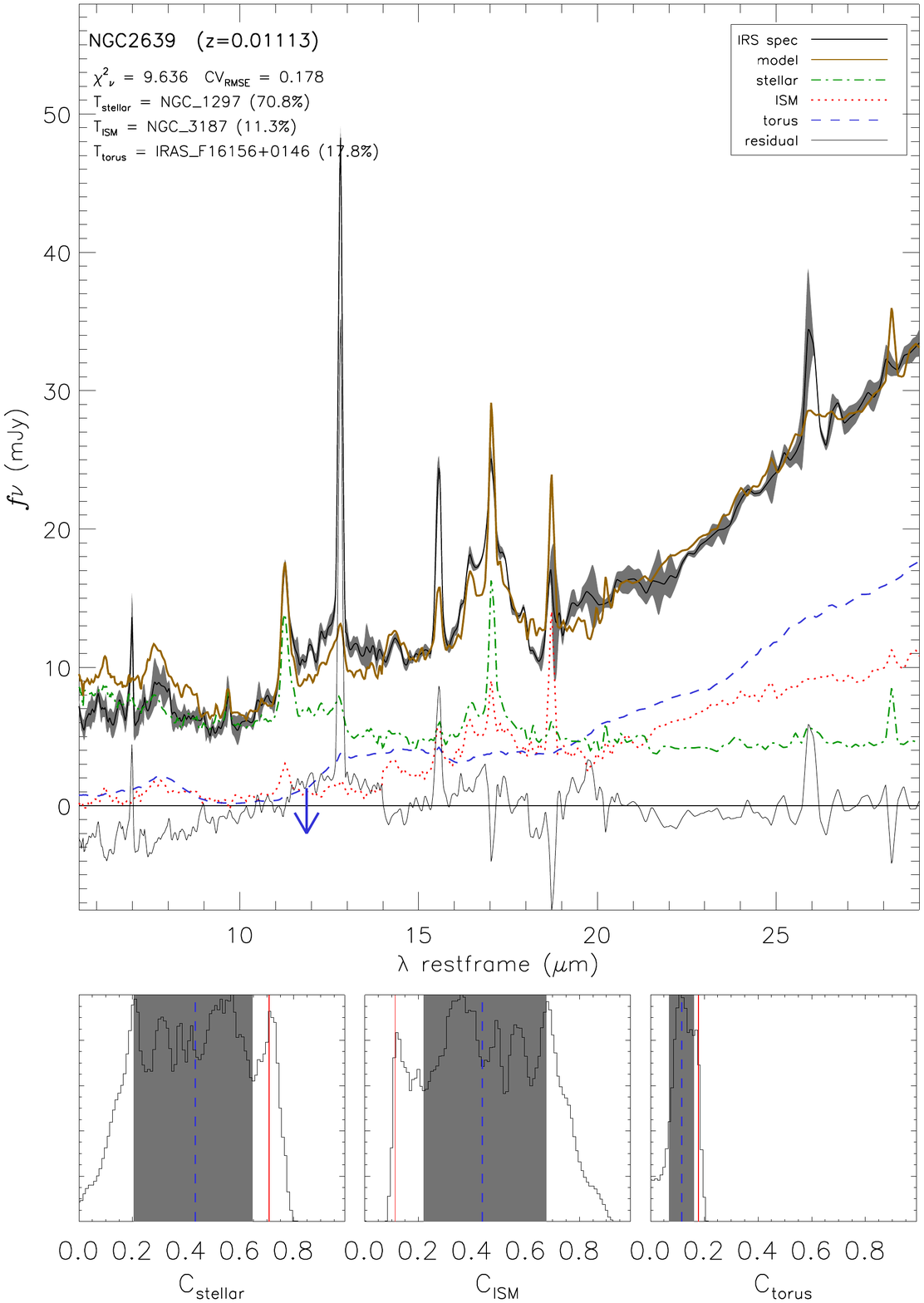}
\caption{Plots of the spectral decomposition of the IRS/\emph{Spitzer} spectra, as reported by \citet{Hernan-Caballero15}. Note that blue arrows are upper-limits derived from the 2-10 keV X-ray luminosity and blue bars are constraints from ground-based high spatial resolution images (see text). The small panels show the posterior distributions for contributions to the 5-15$\rm{\mu m}$ of the stellar, ISM, and torus components. }
\label{fig:CatSpectra}
\end{center}
\end{figure*}

\begin{figure*}
\begin{center}
\includegraphics[width=0.45\columnwidth]{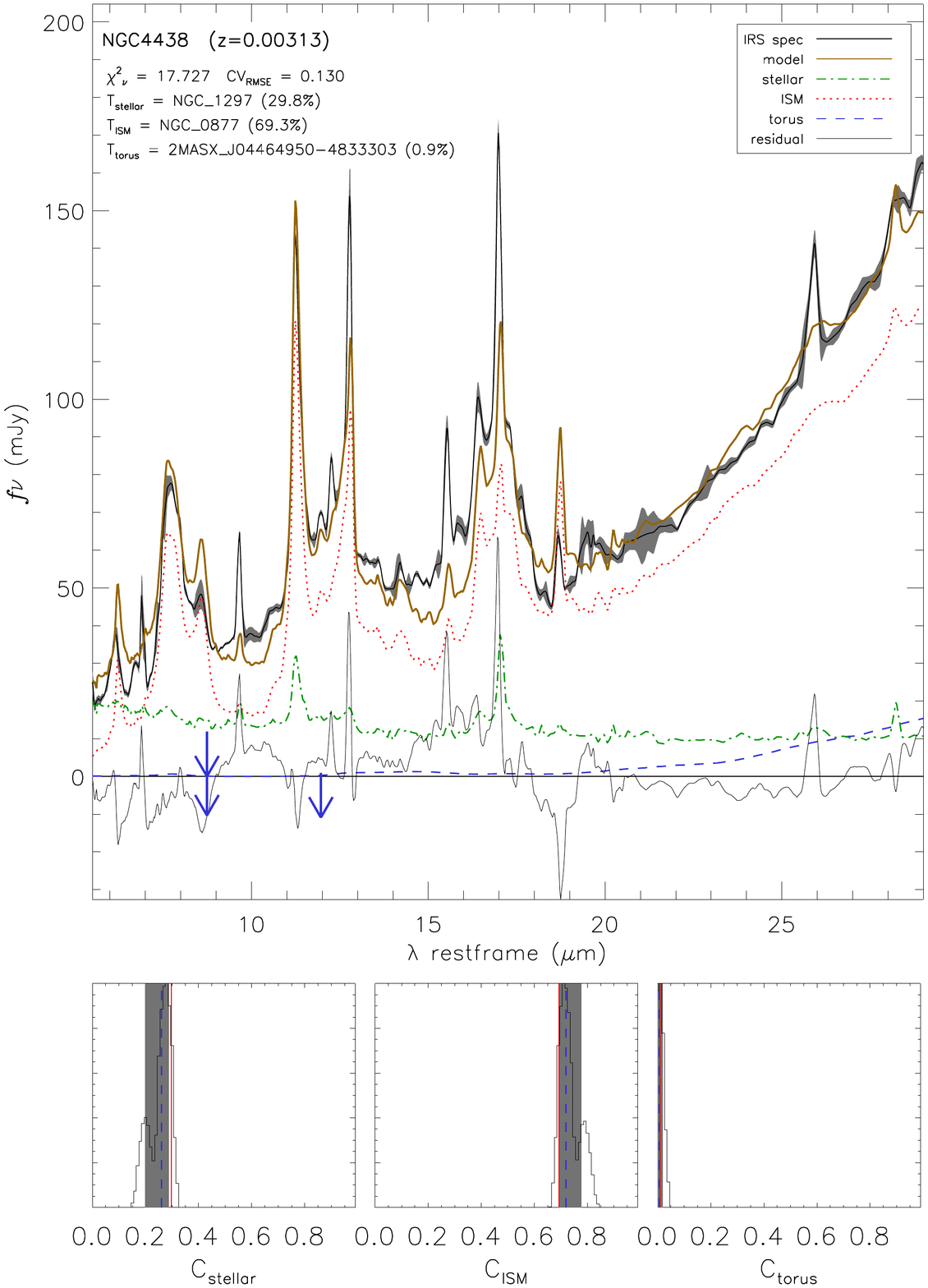}
\includegraphics[width=0.45\columnwidth]{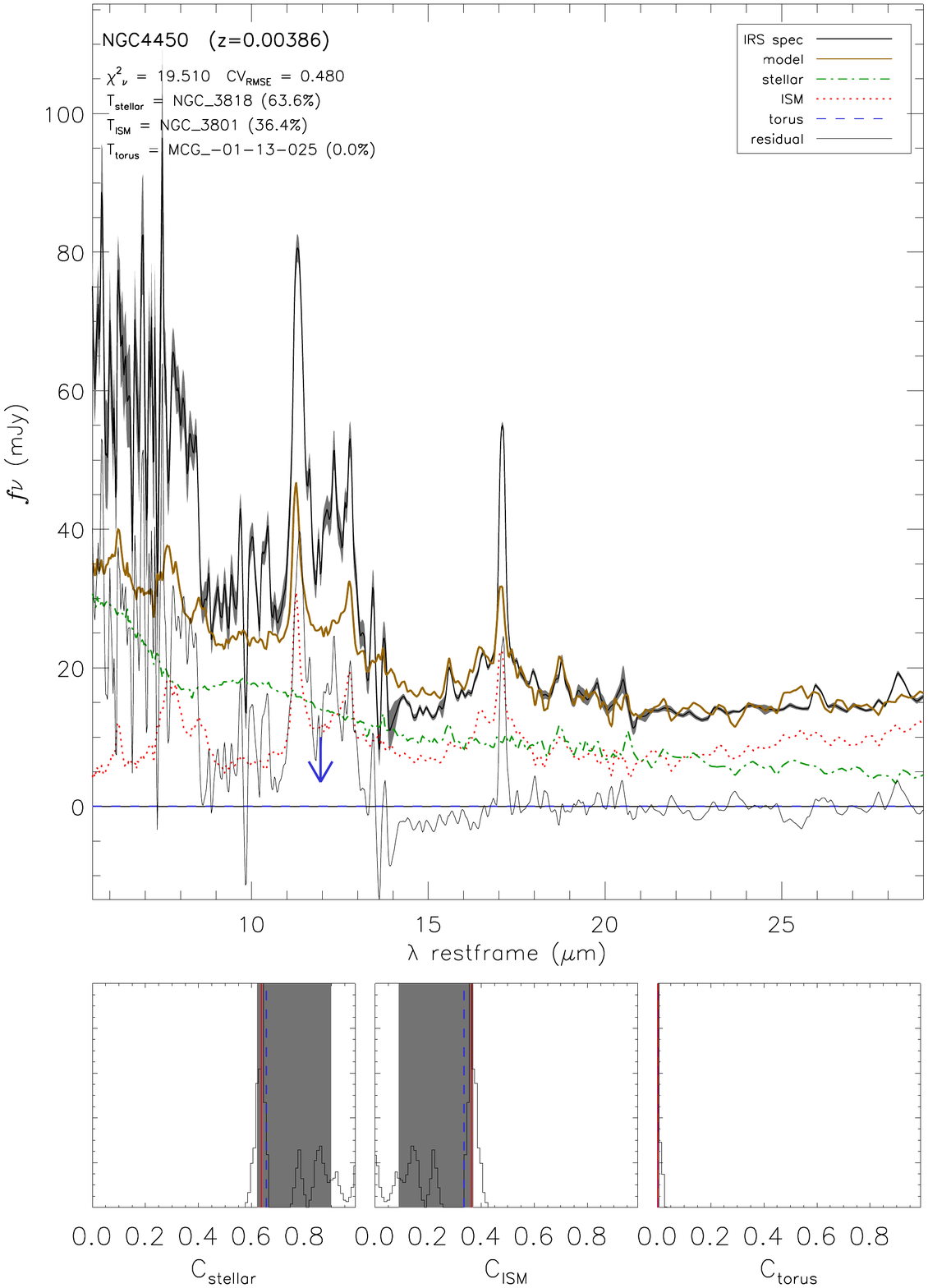}
\includegraphics[width=0.45\columnwidth]{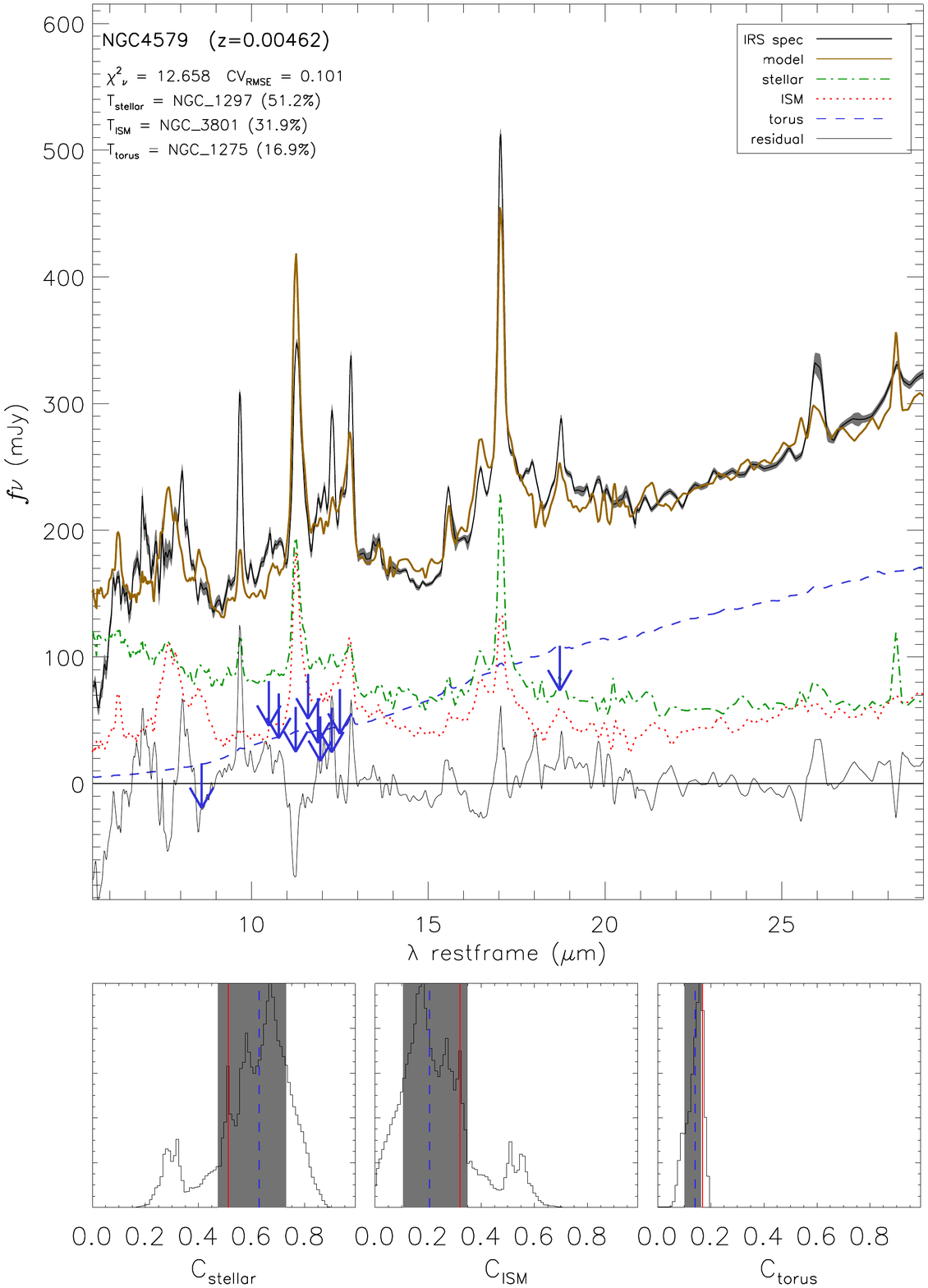}
\includegraphics[width=0.45\columnwidth]{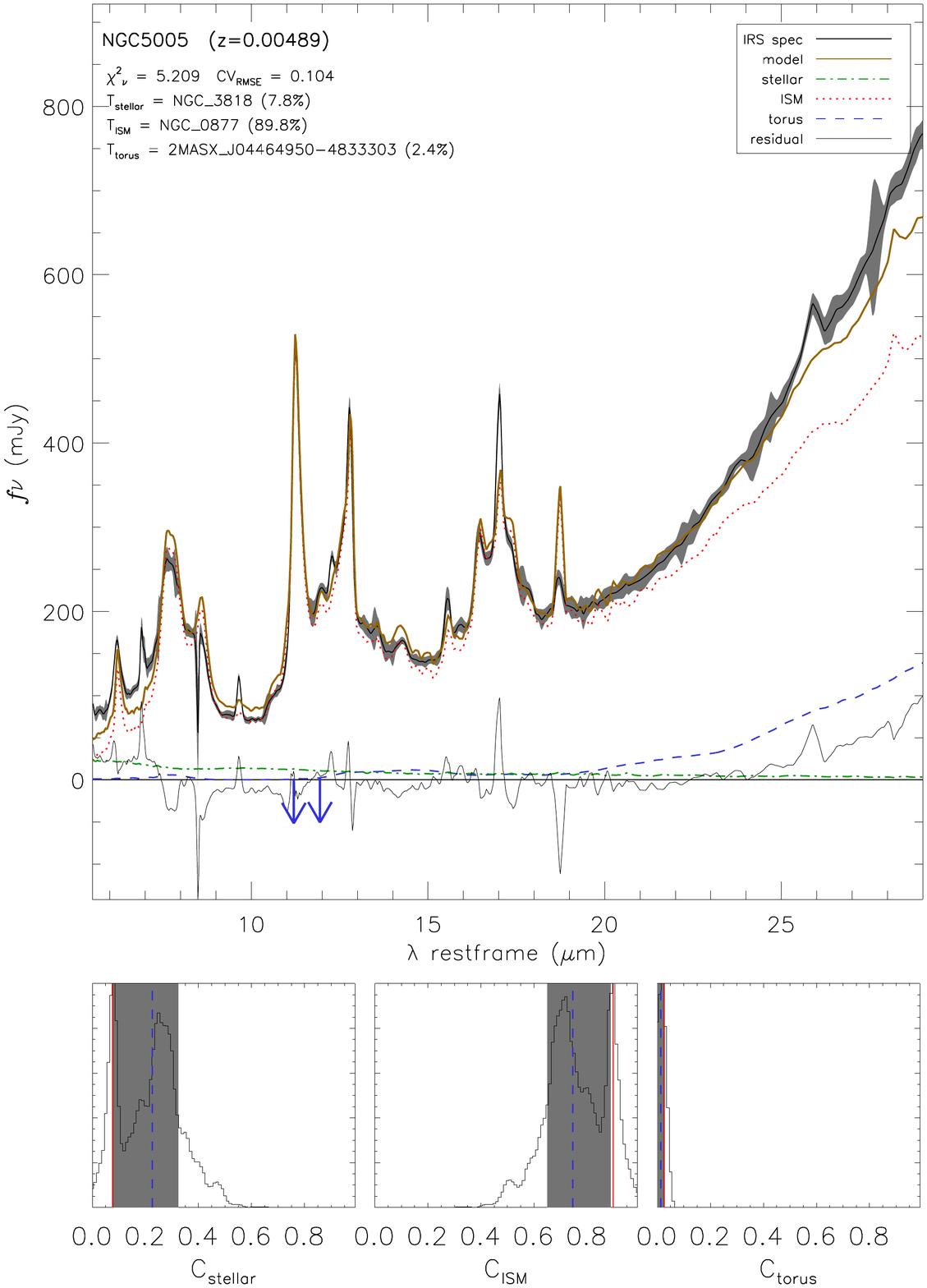}
\caption{...continued.}
\label{fig:CatSpectra}
\end{center}
\end{figure*}

\begin{figure*}
\begin{center}
\includegraphics[width=0.45\columnwidth]{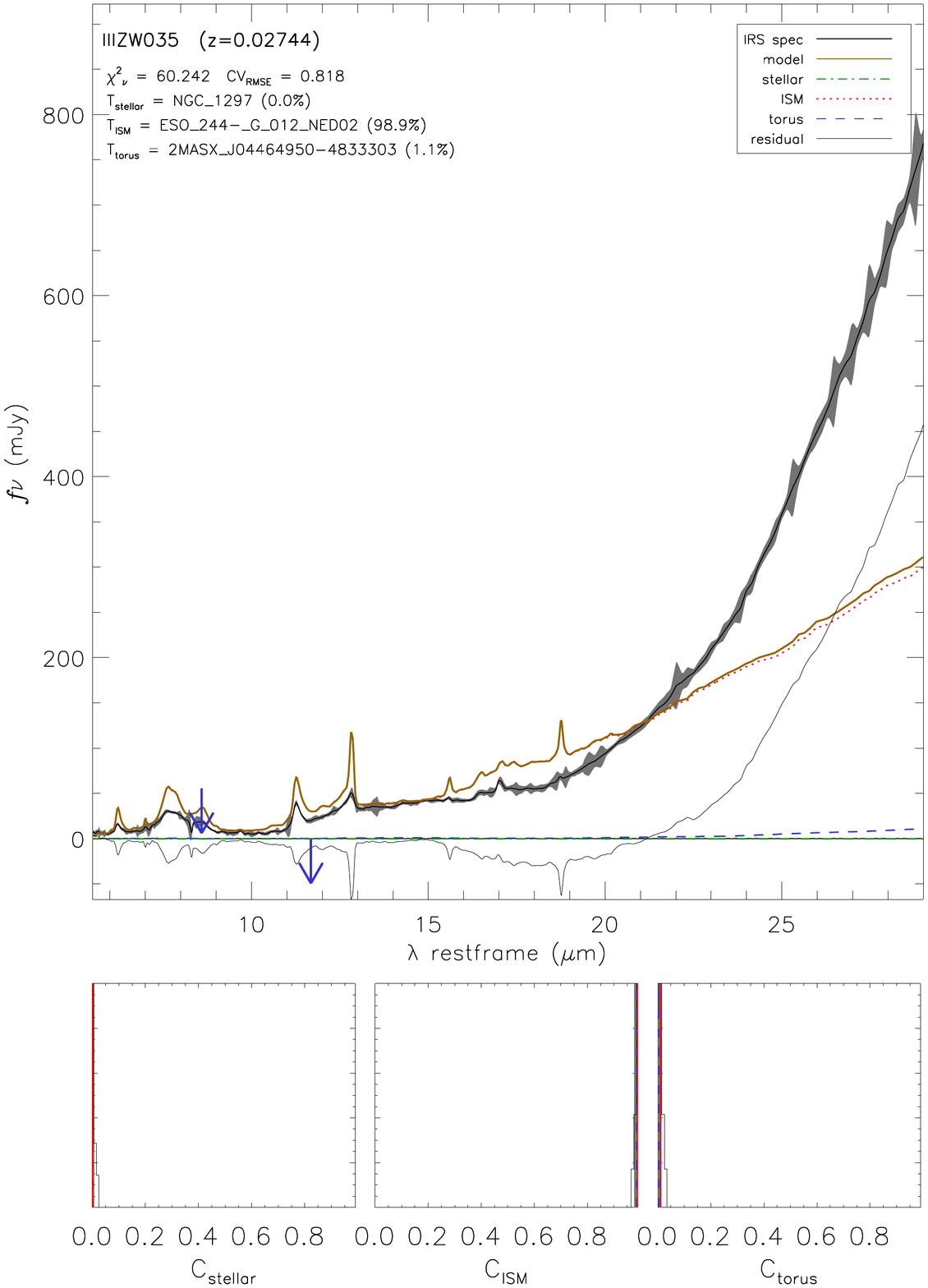}
\includegraphics[width=0.45\columnwidth]{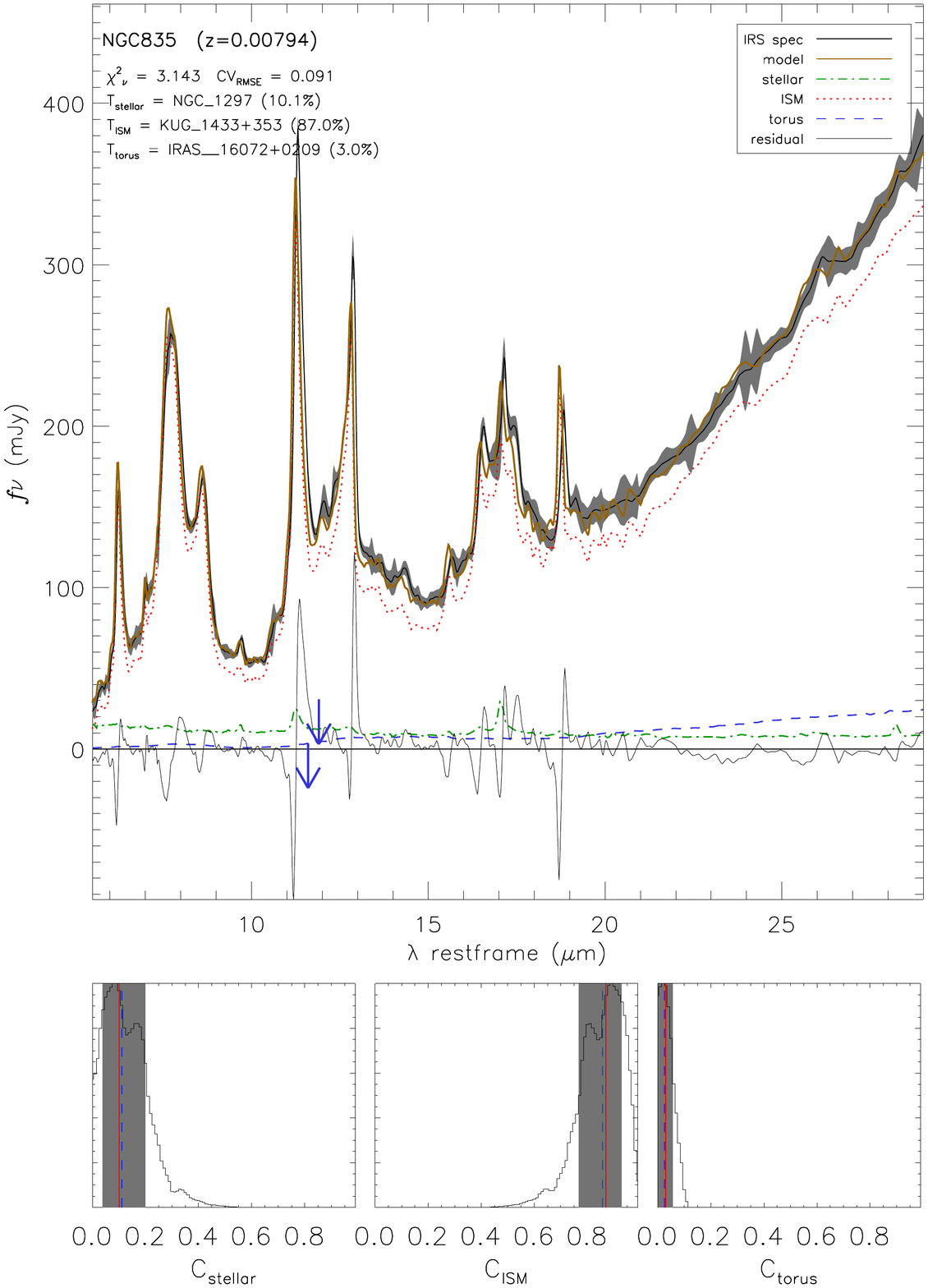}
\includegraphics[width=0.45\columnwidth]{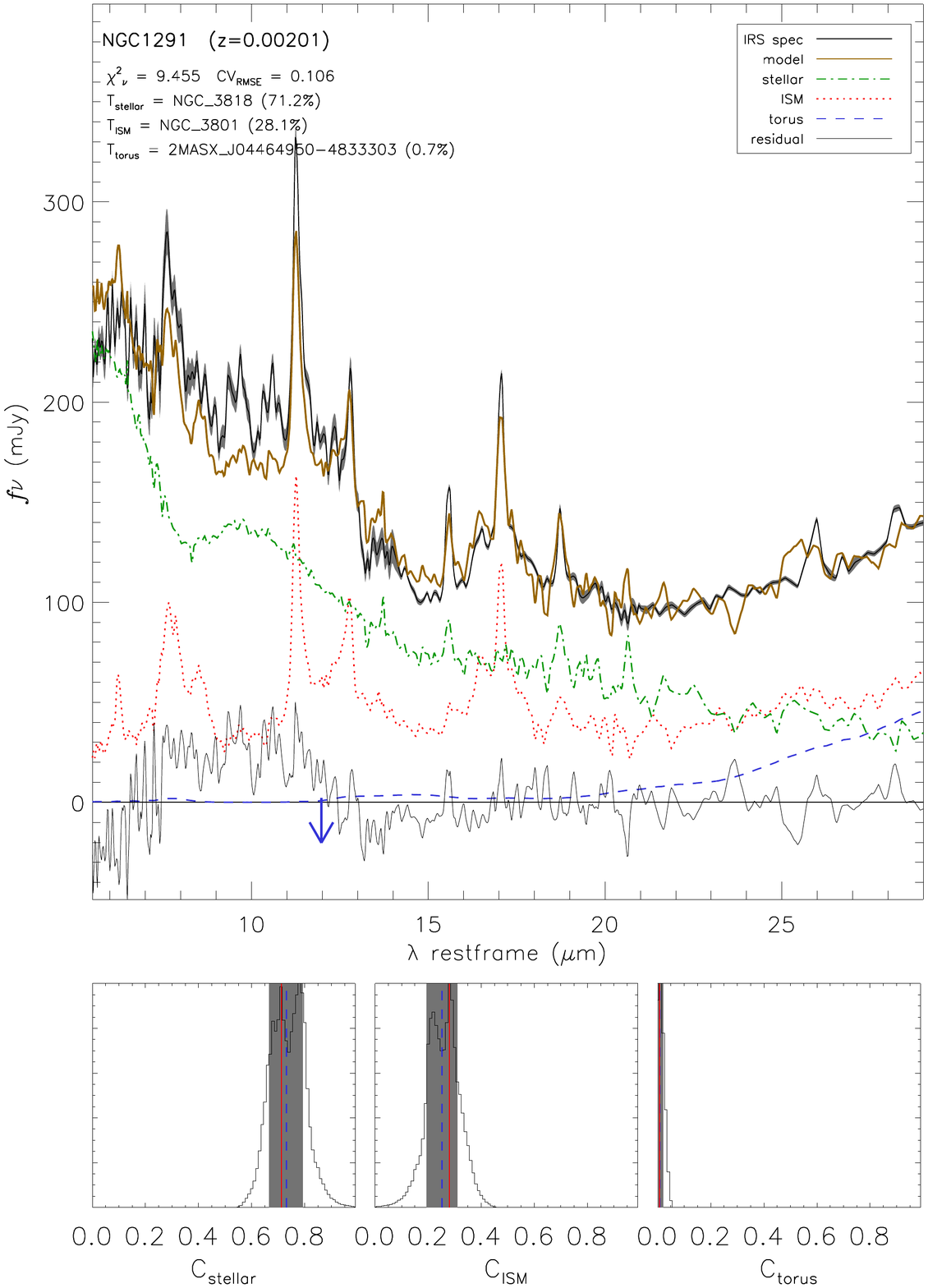}
\includegraphics[width=0.45\columnwidth]{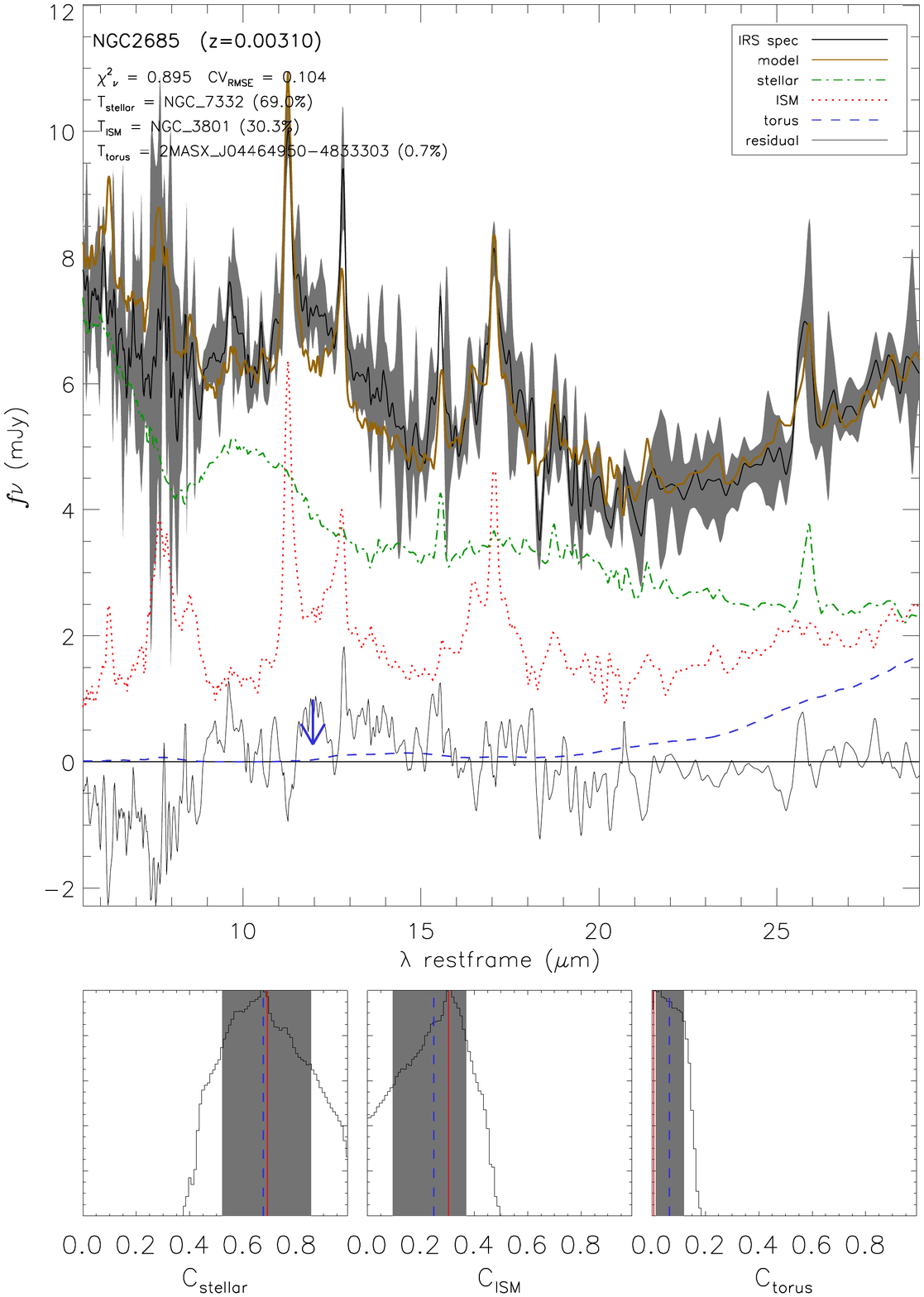}
\caption{...continued.}
\label{fig:CatSpectra}
\end{center}
\end{figure*}

\begin{figure*}
\begin{center}
\includegraphics[width=0.45\columnwidth]{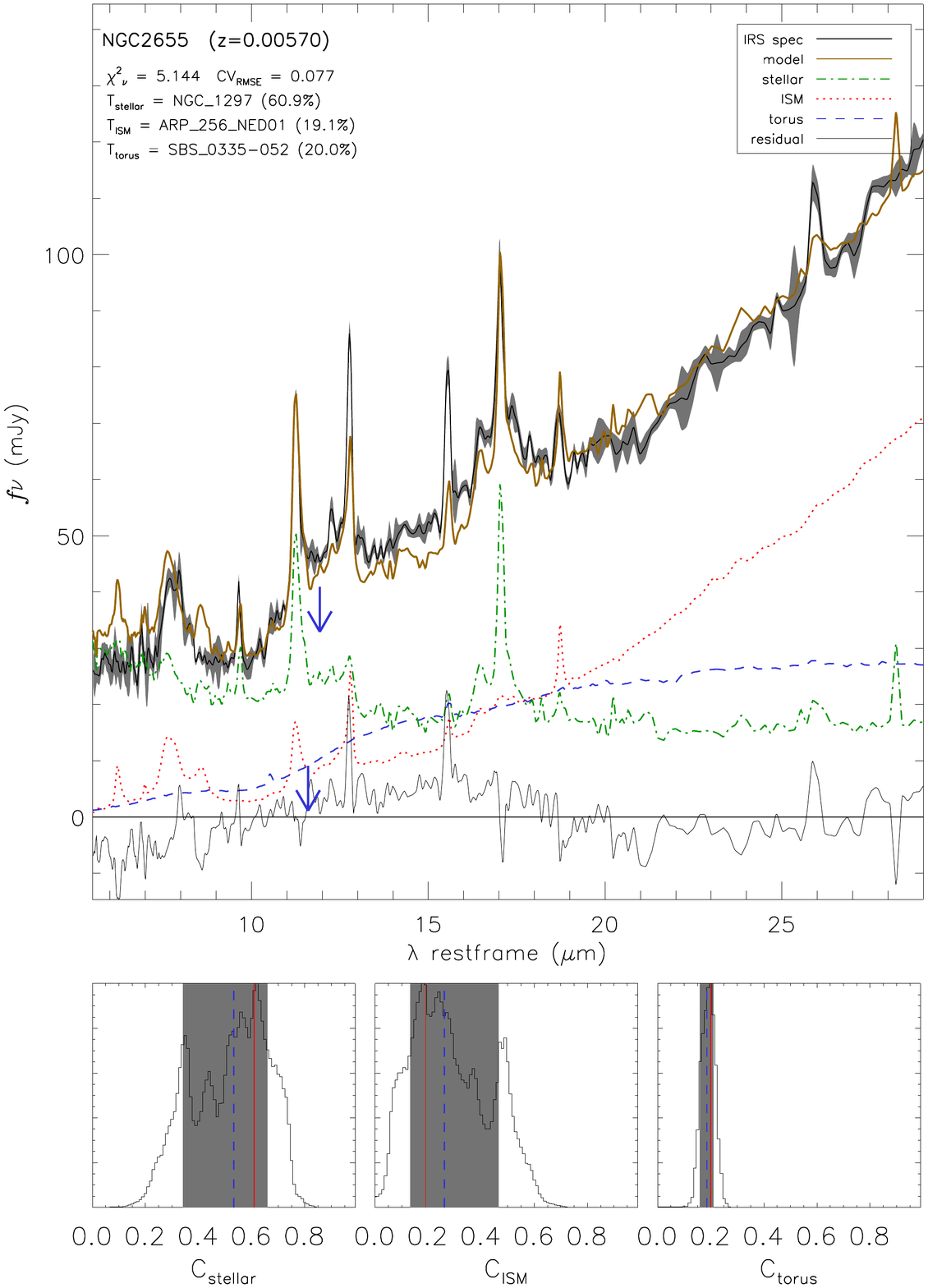}
\includegraphics[width=0.45\columnwidth]{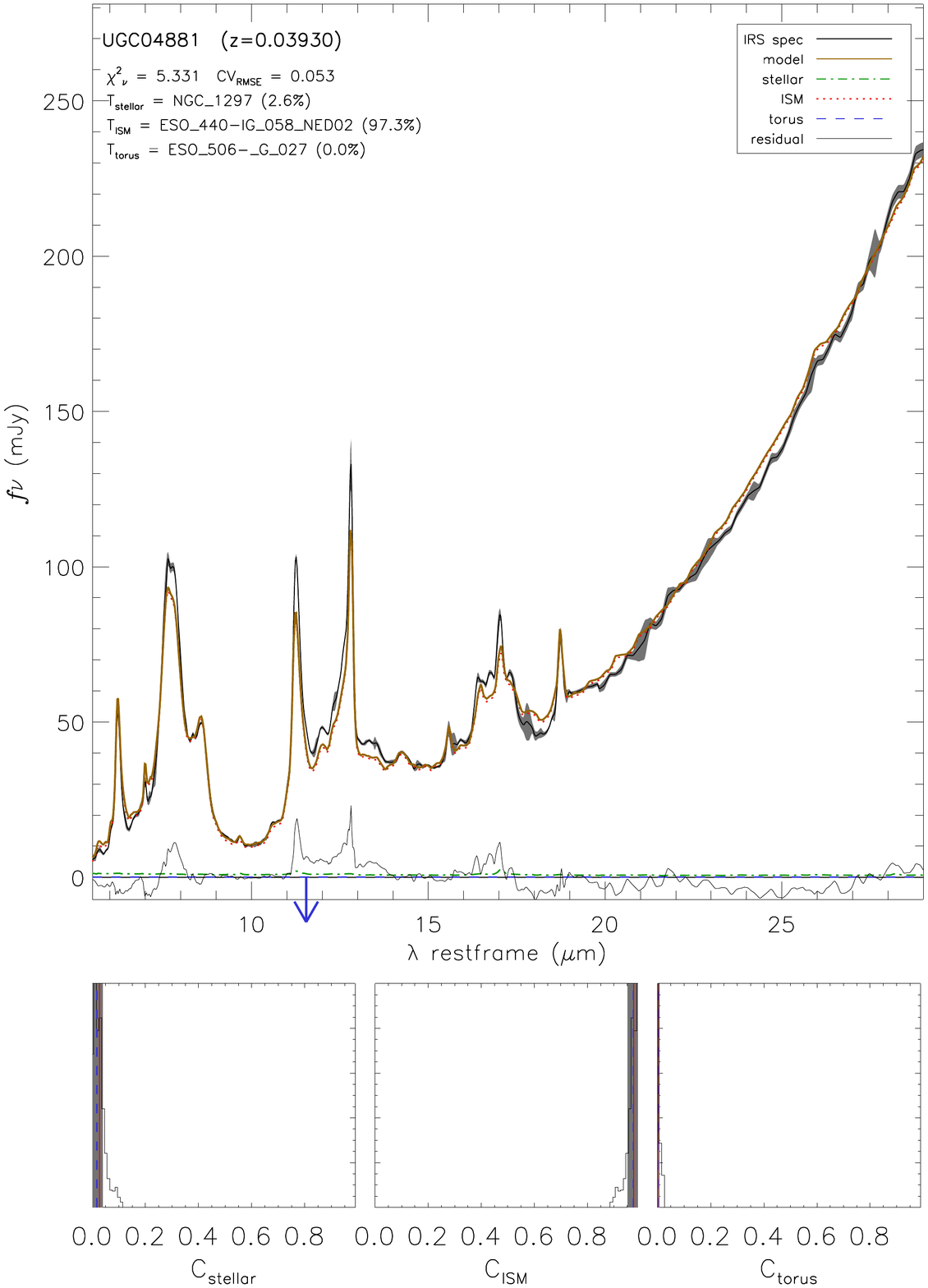}
\includegraphics[width=0.45\columnwidth]{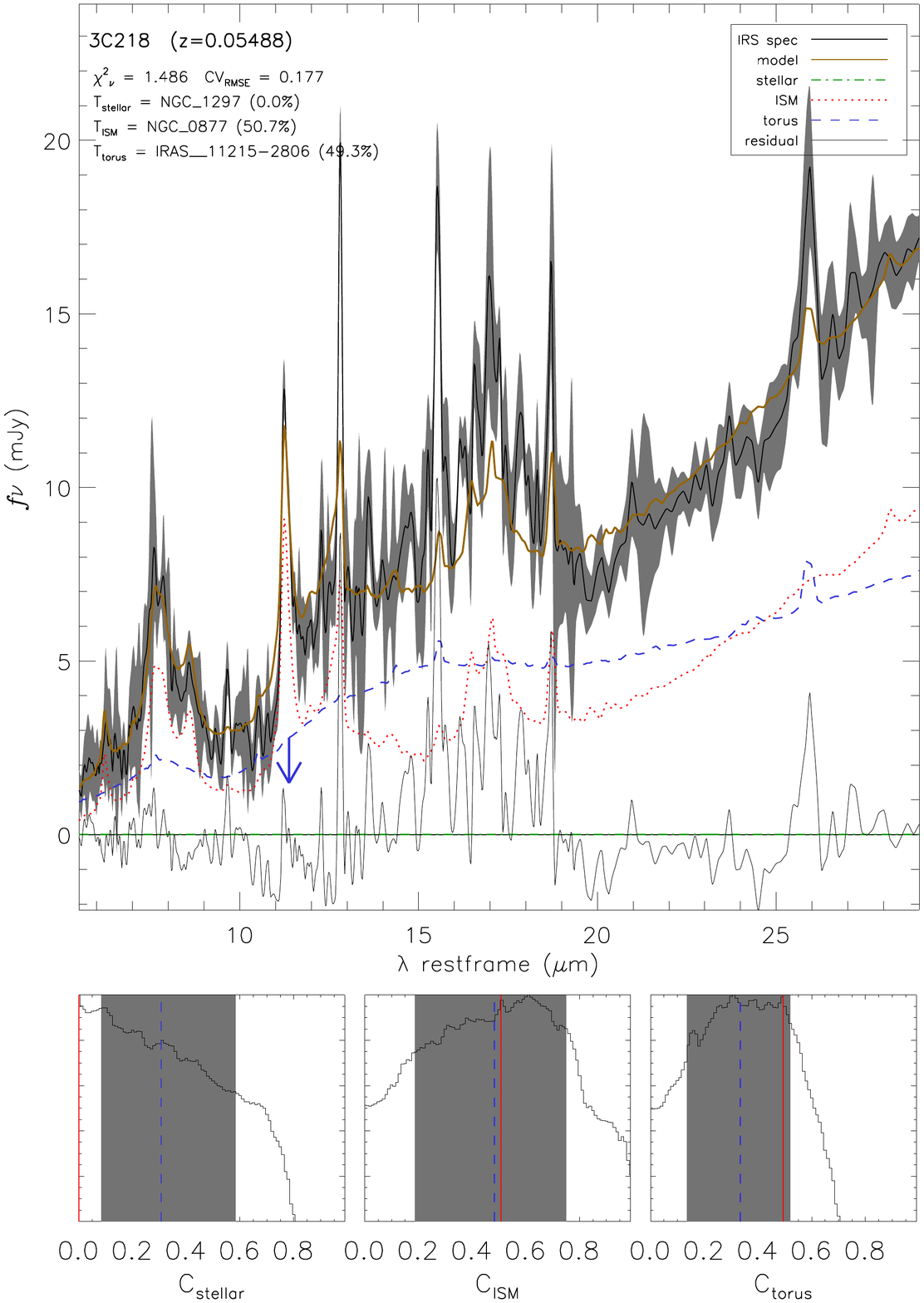}
\includegraphics[width=0.45\columnwidth]{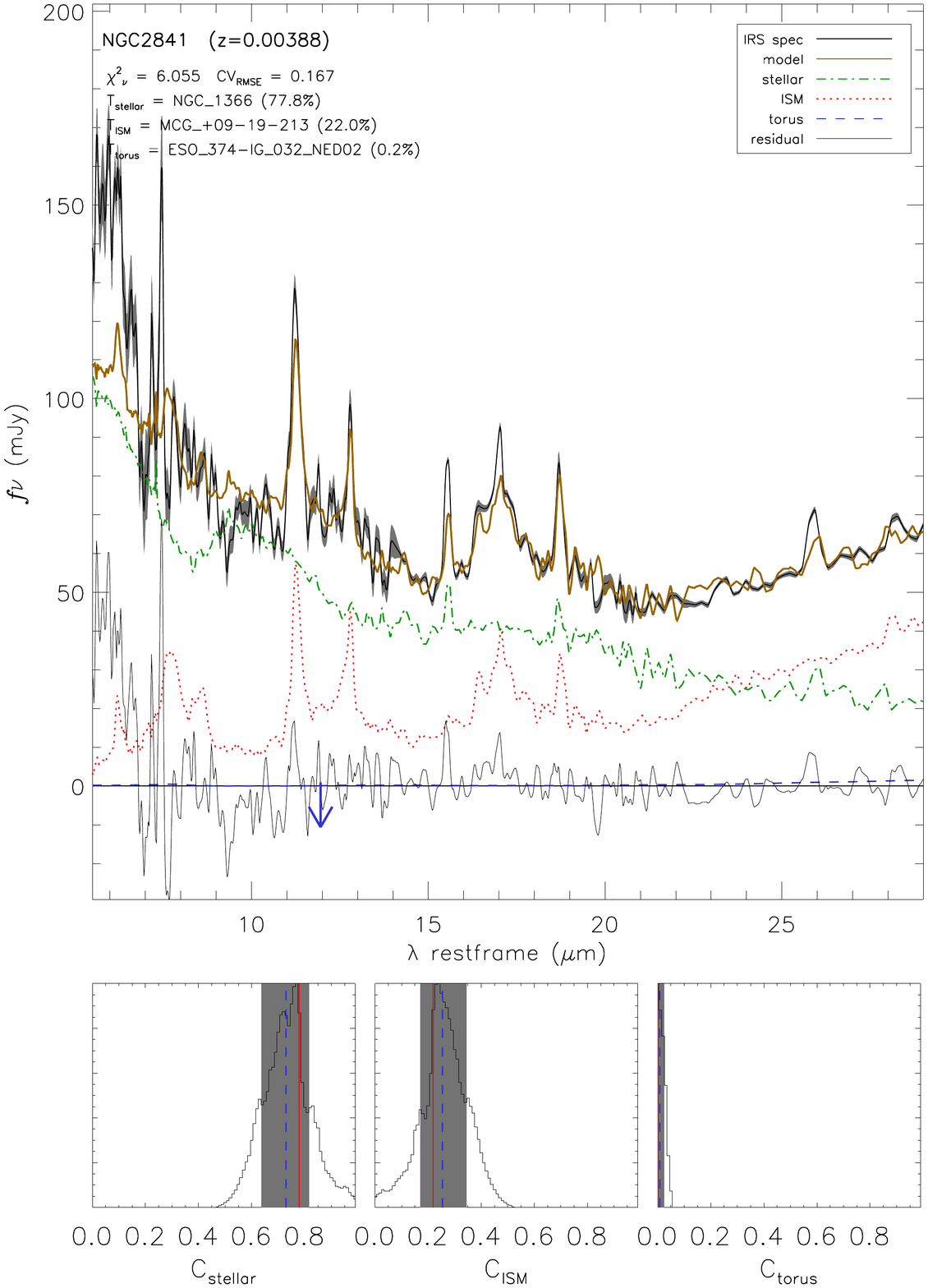}
\caption{...continued.}
\label{fig:CatSpectra}
\end{center}
\end{figure*}

\begin{figure*}
\begin{center}
\includegraphics[width=0.45\columnwidth]{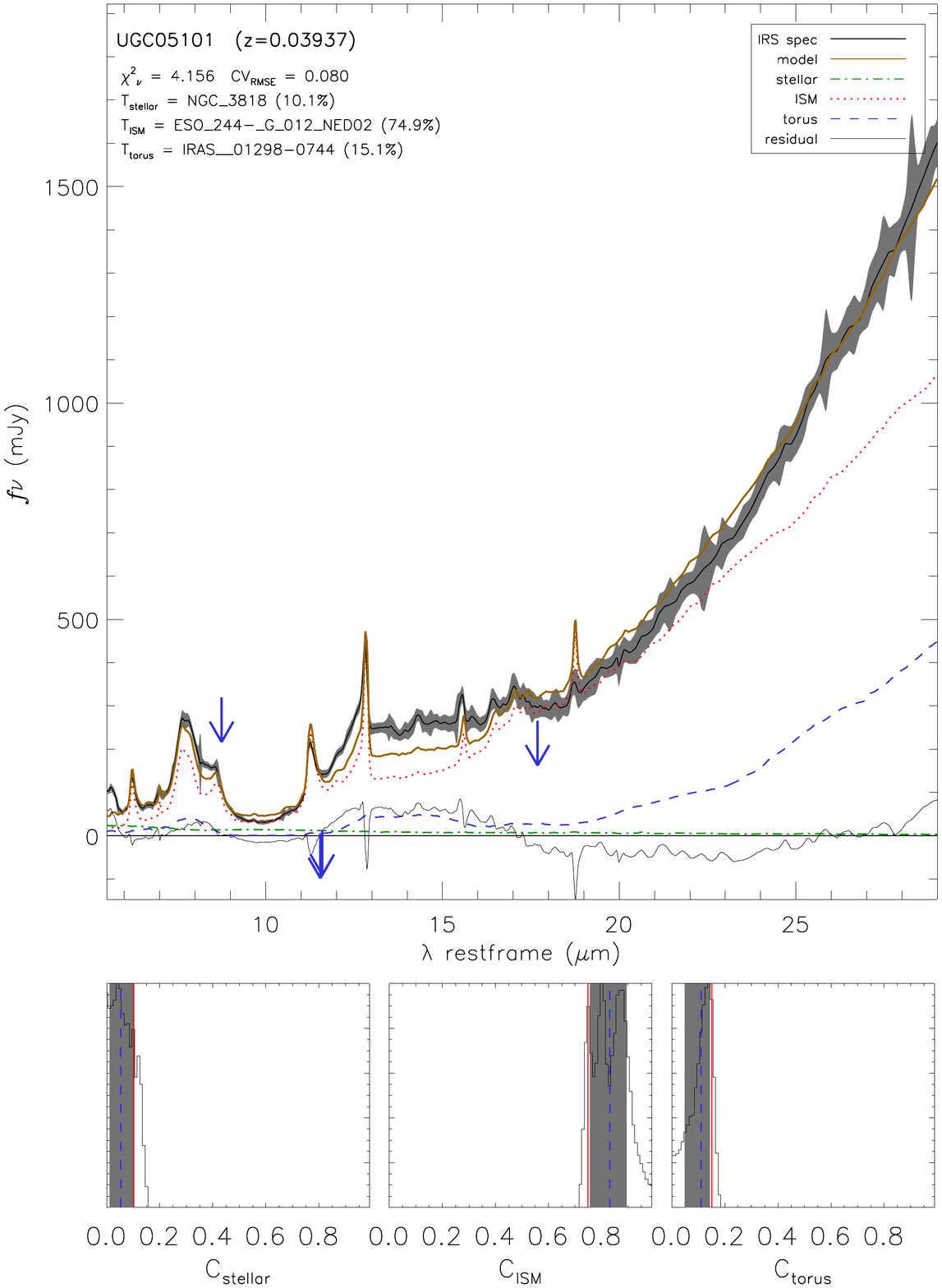}
\includegraphics[width=0.45\columnwidth]{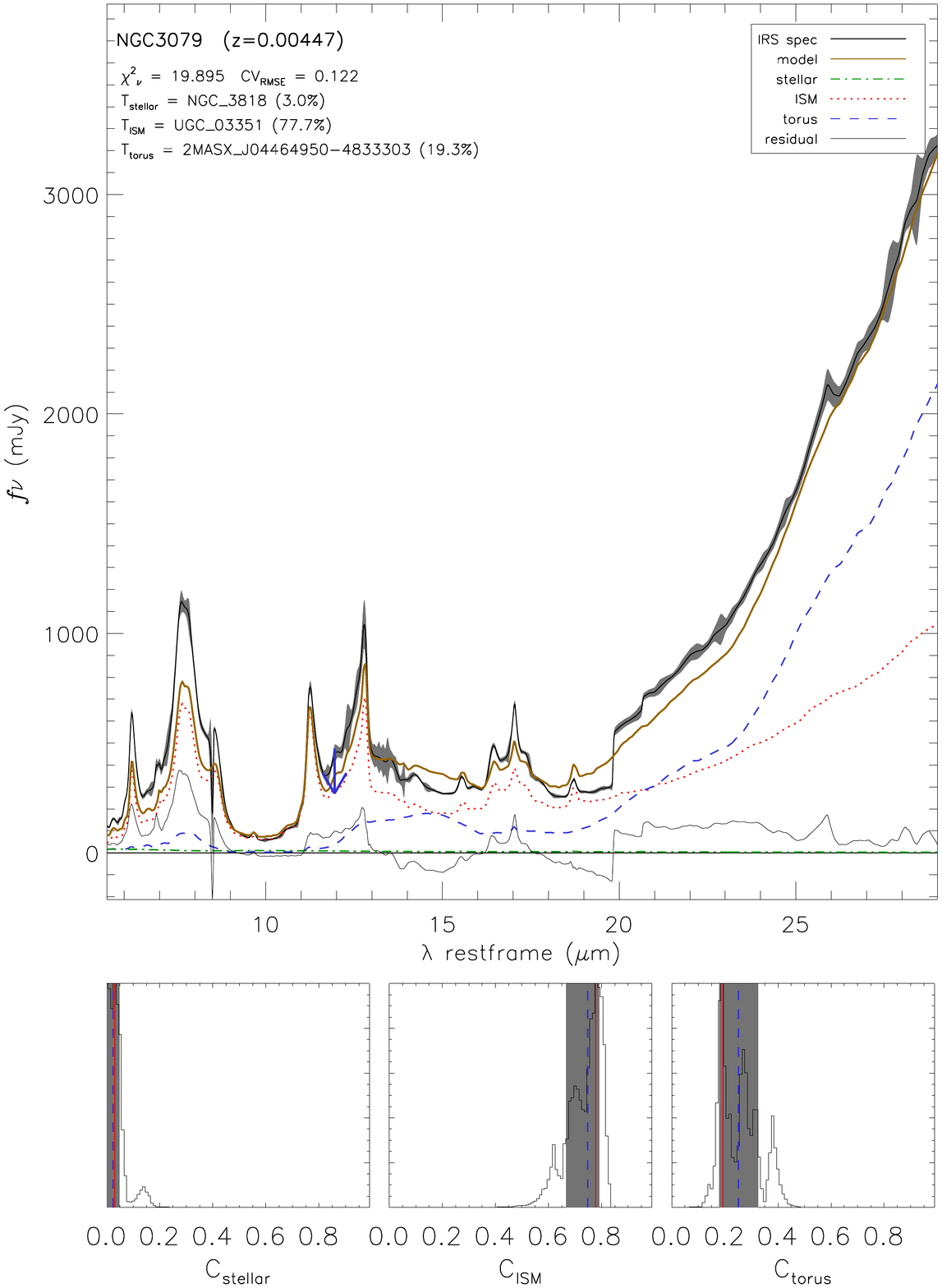}
\includegraphics[width=0.45\columnwidth]{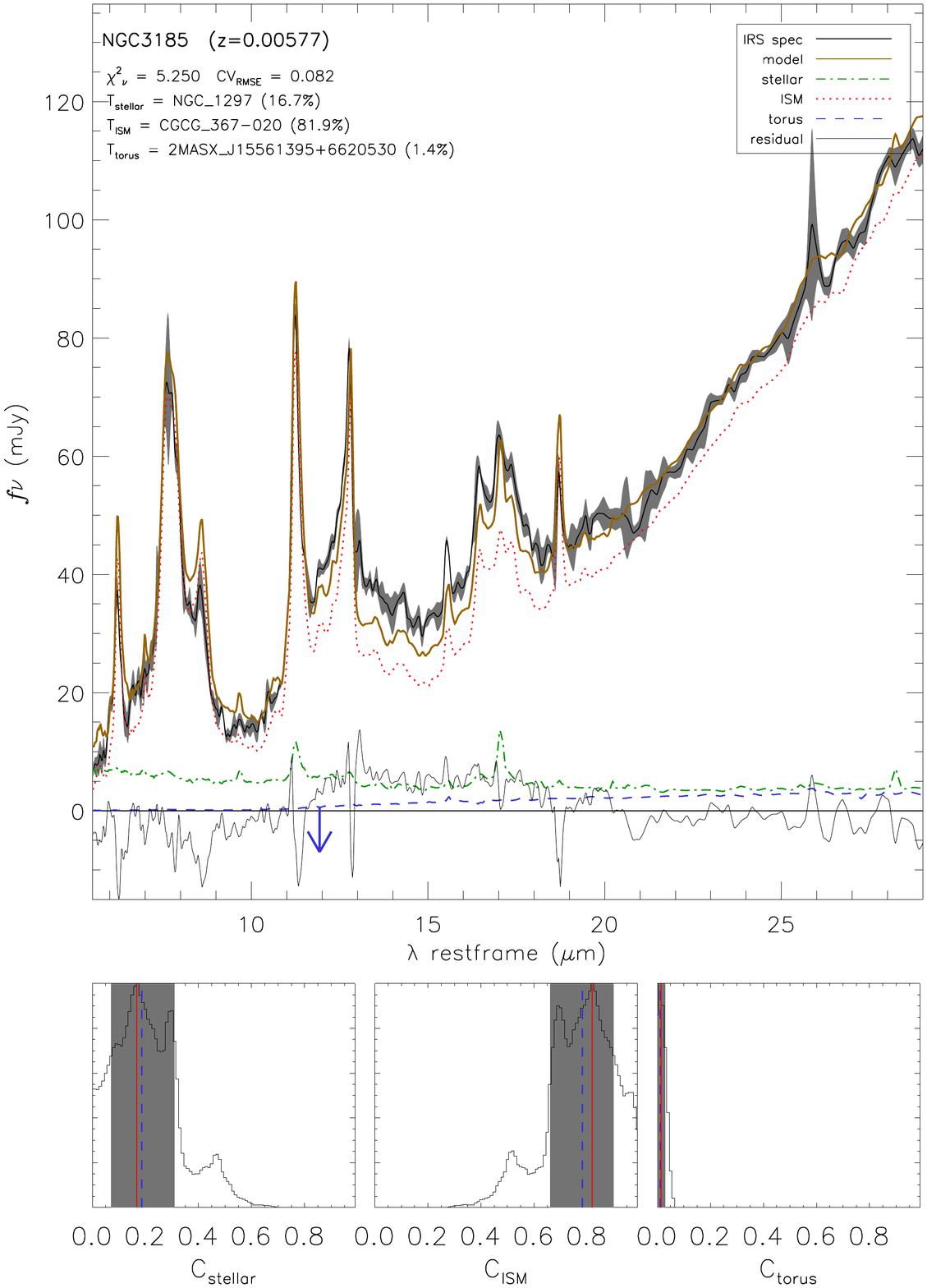}
\includegraphics[width=0.45\columnwidth]{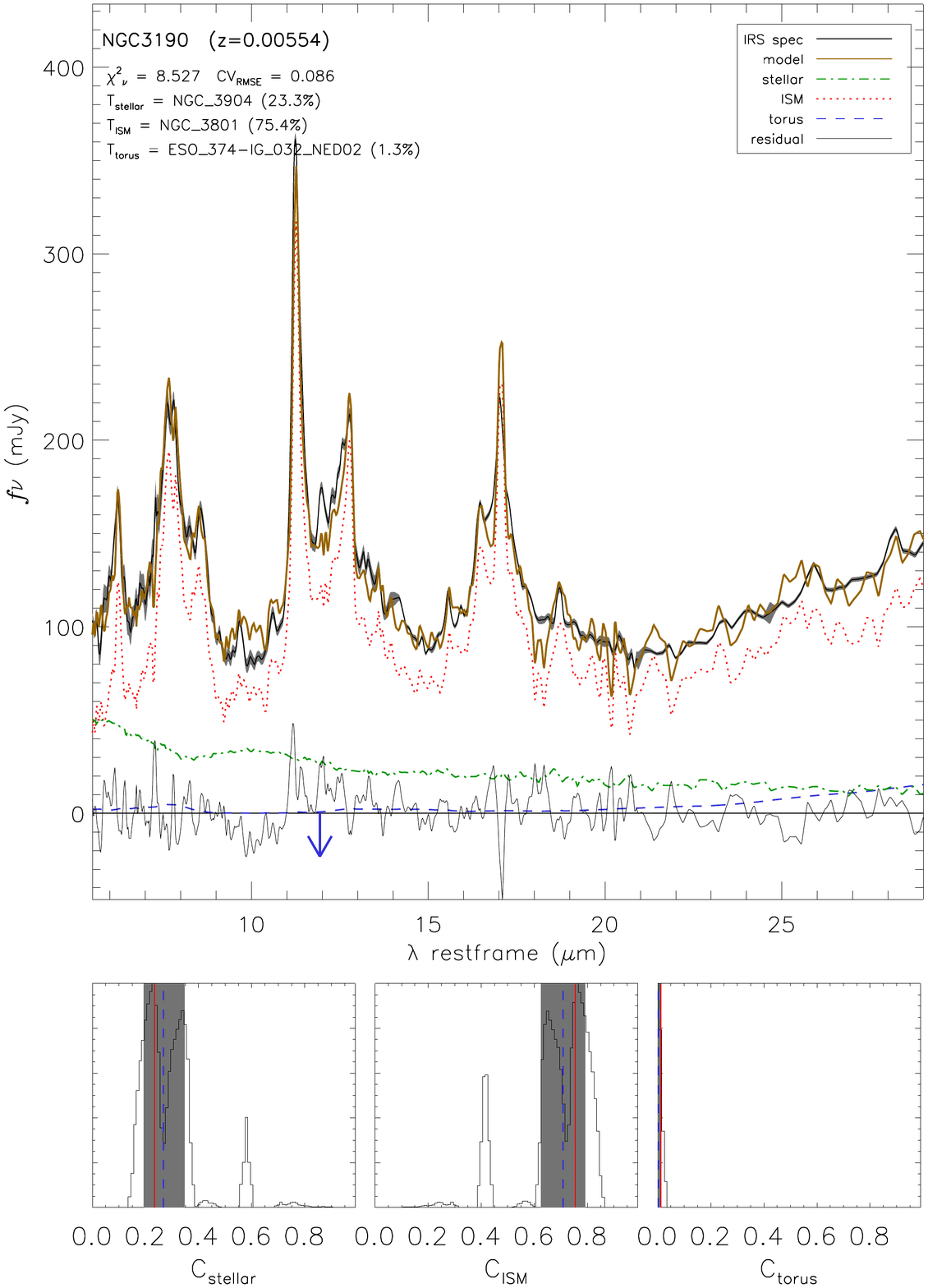}
\caption{...continued.}
\label{fig:CatSpectra}
\end{center}
\end{figure*}

\begin{figure*}
\begin{center}
\includegraphics[width=0.45\columnwidth]{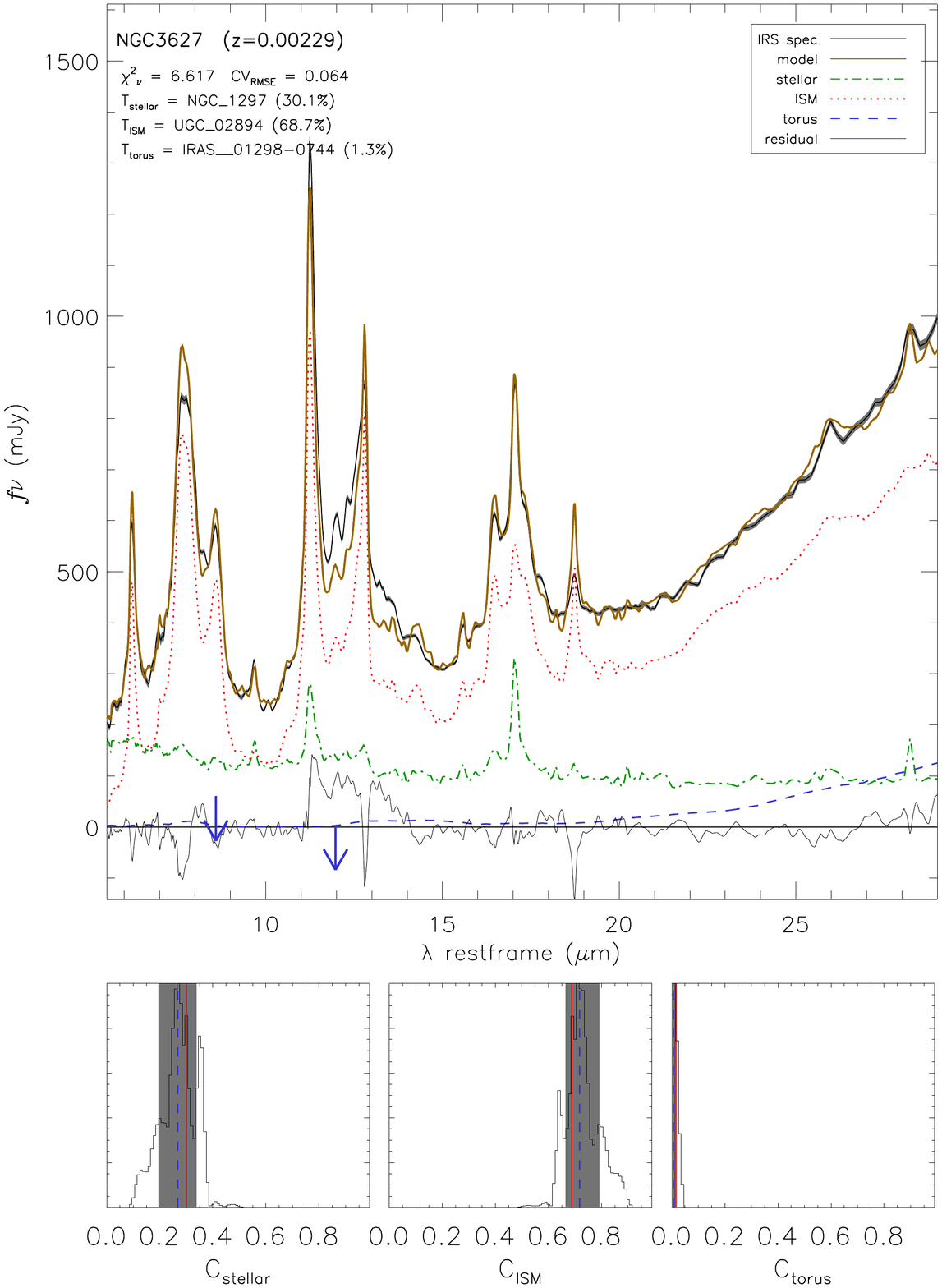}
\includegraphics[width=0.45\columnwidth]{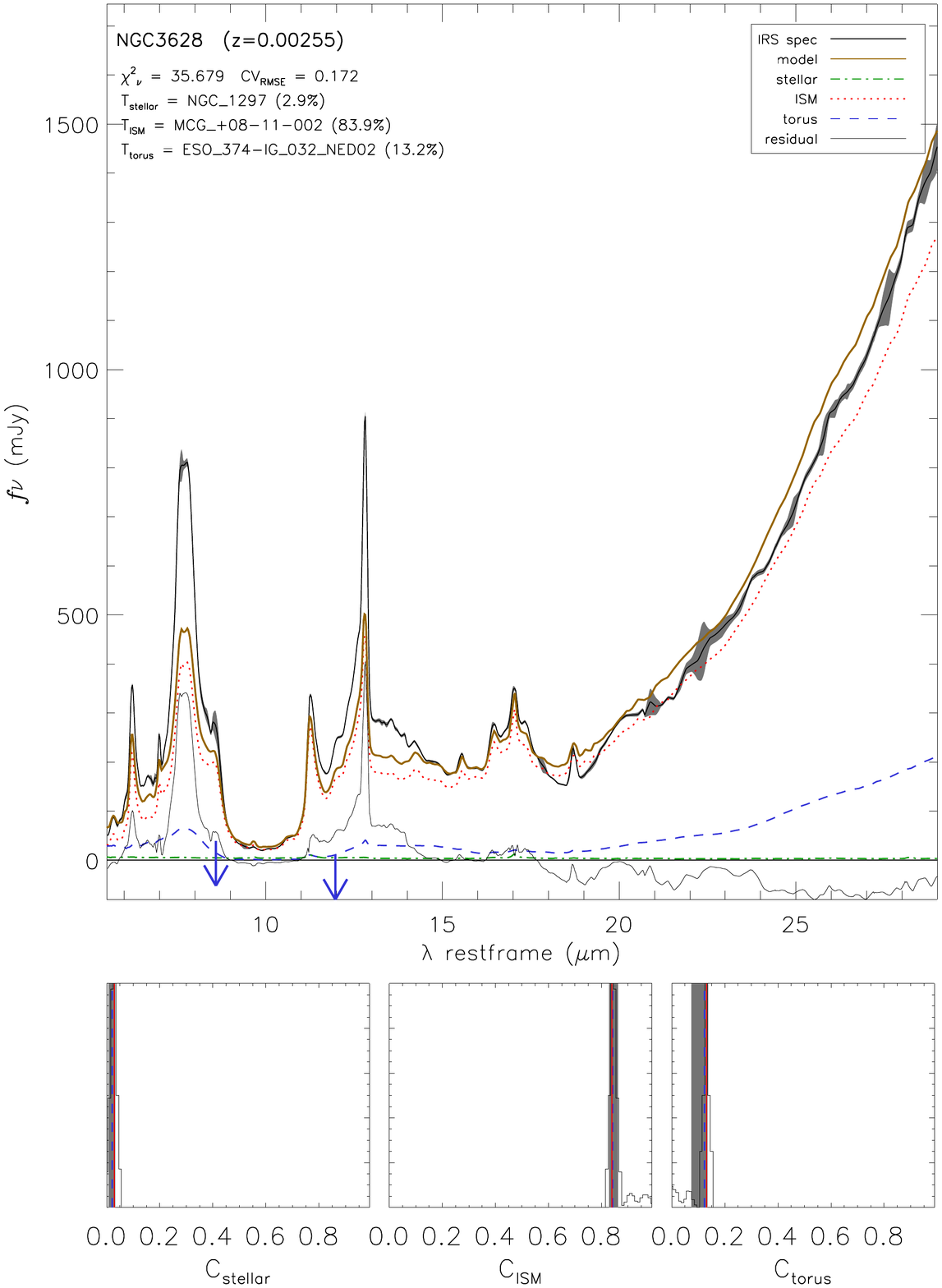}
\includegraphics[width=0.45\columnwidth]{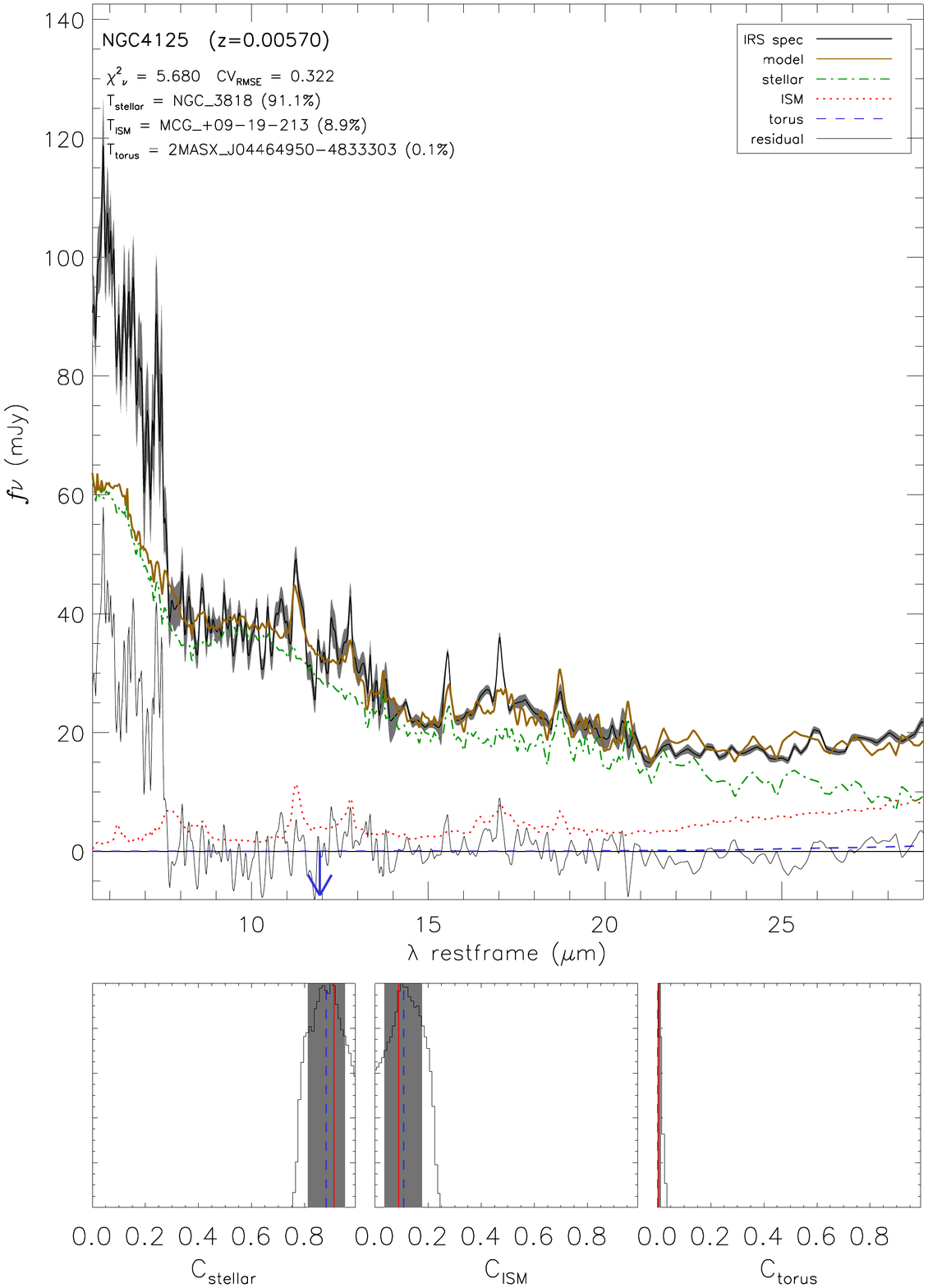}
\includegraphics[width=0.45\columnwidth]{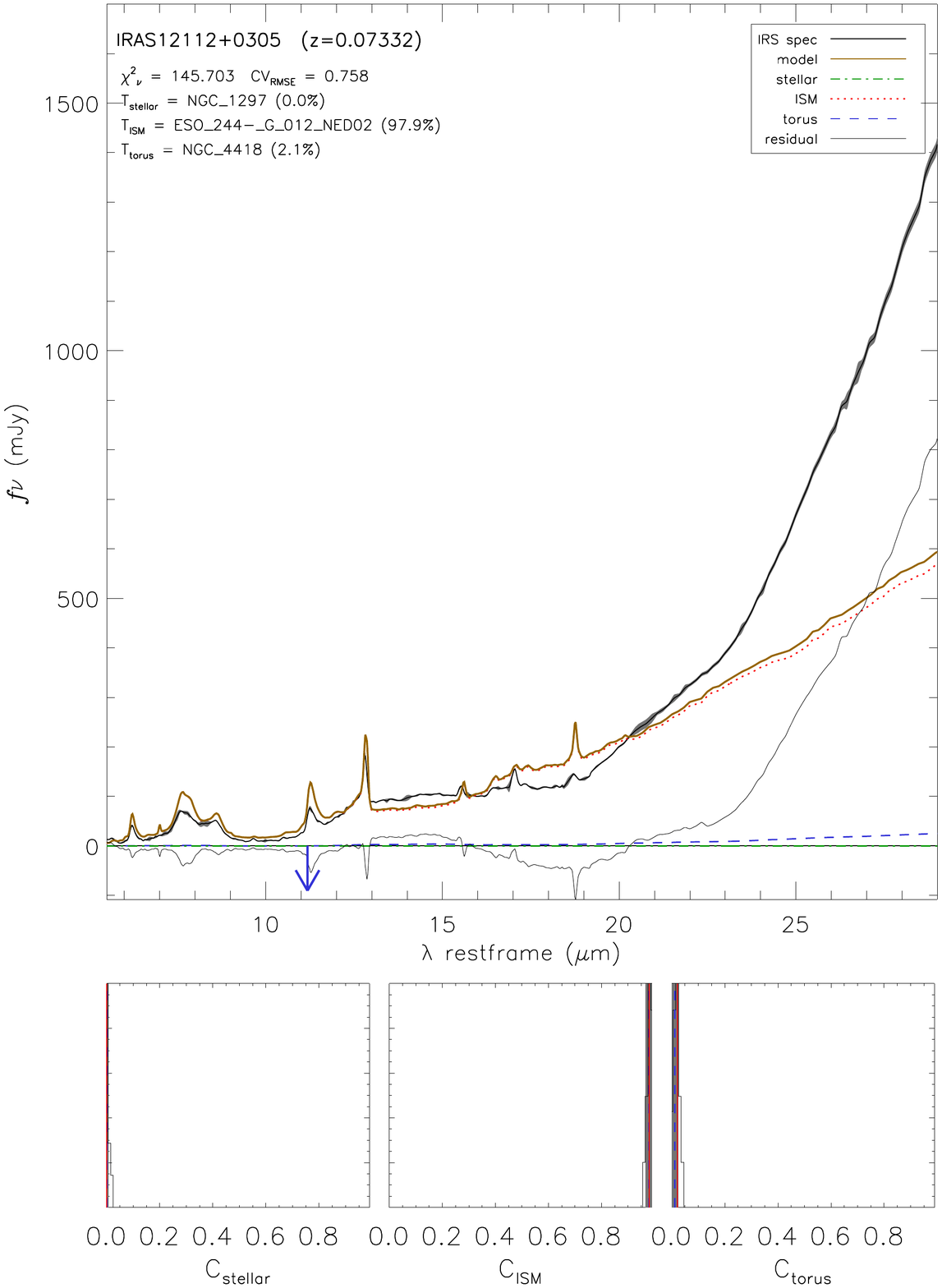}
\caption{...continued.}
\label{fig:CatSpectra}
\end{center}
\end{figure*}

\begin{figure*}
\begin{center}
\includegraphics[width=0.45\columnwidth]{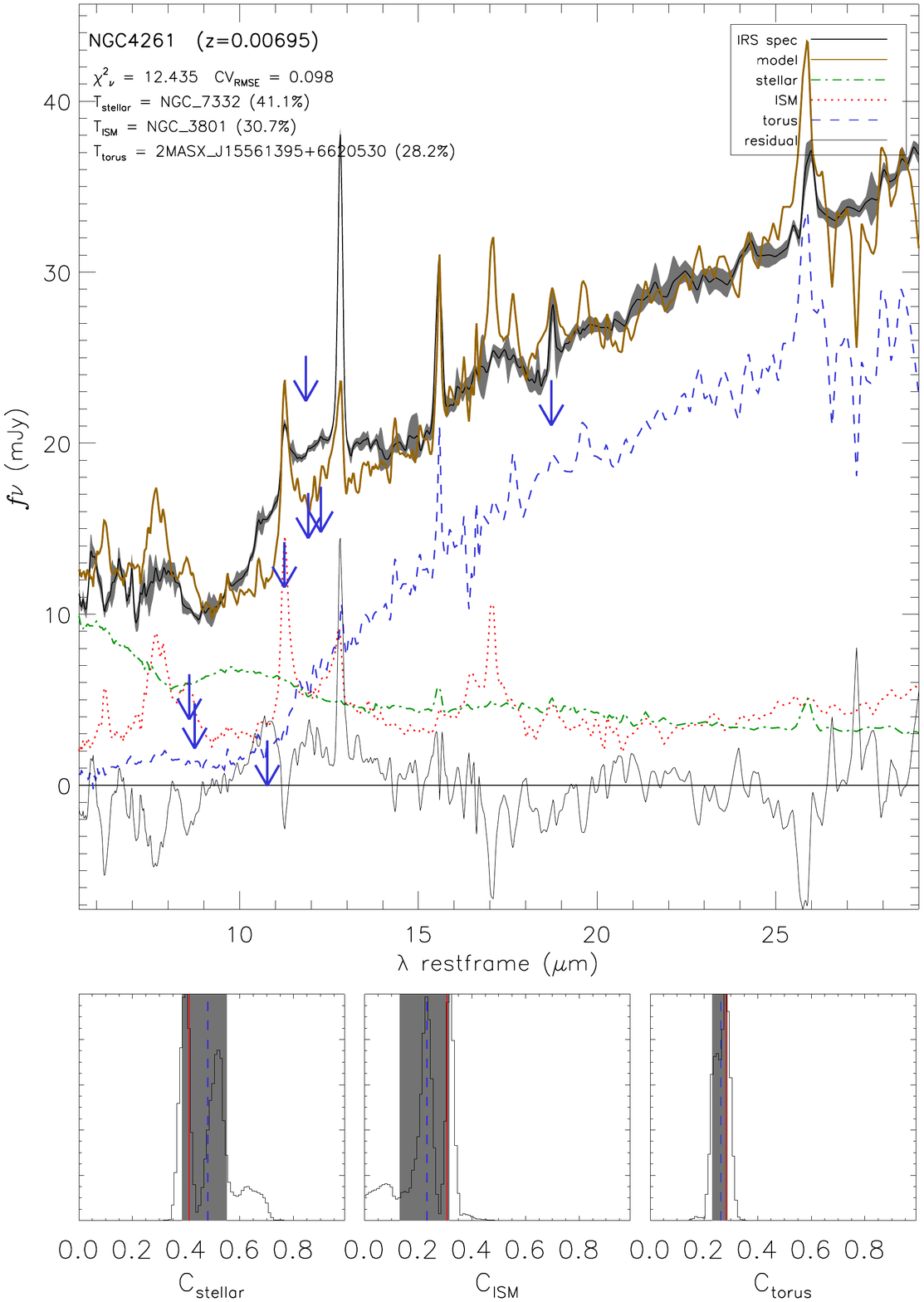}
\includegraphics[width=0.45\columnwidth]{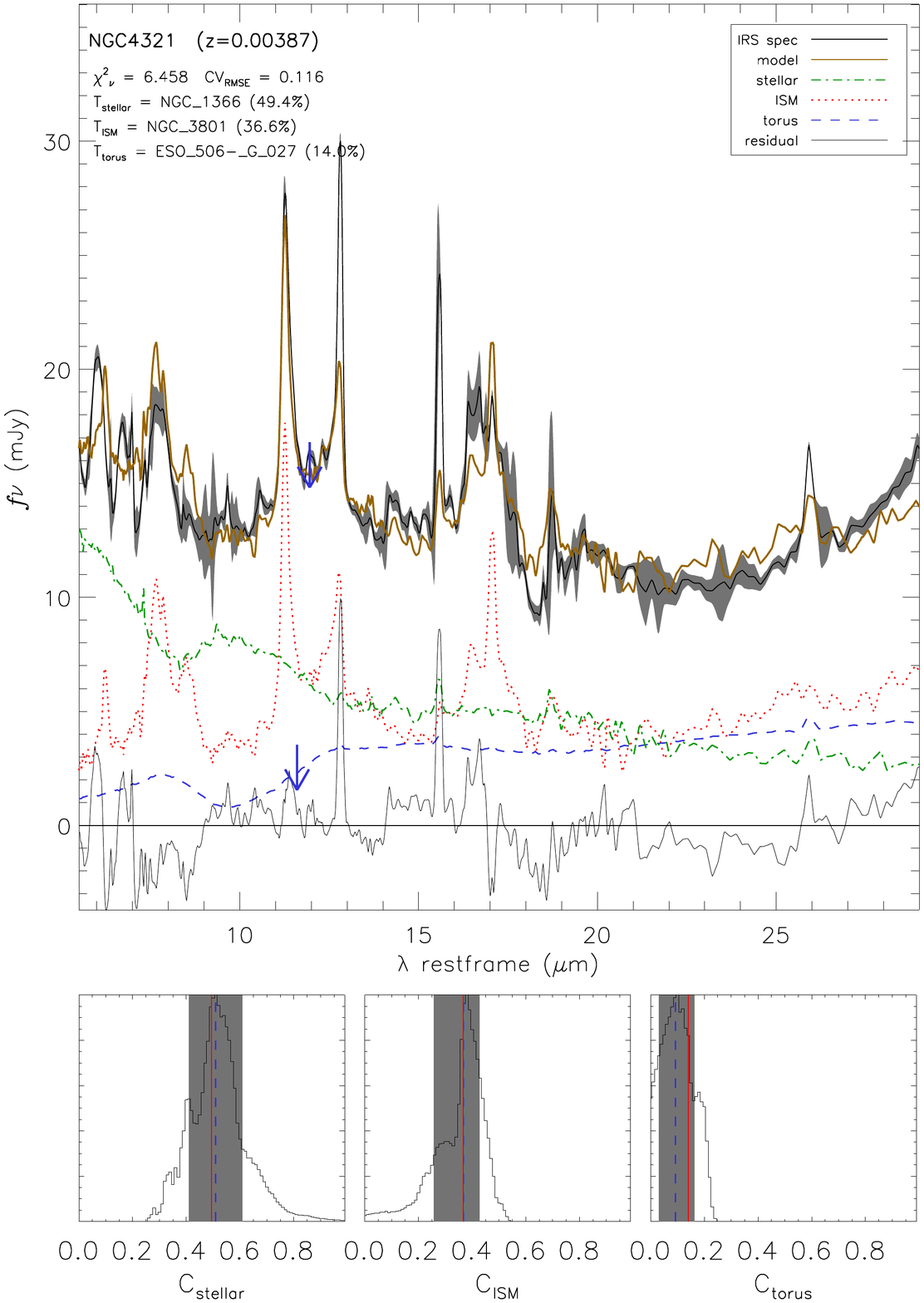}
\includegraphics[width=0.45\columnwidth]{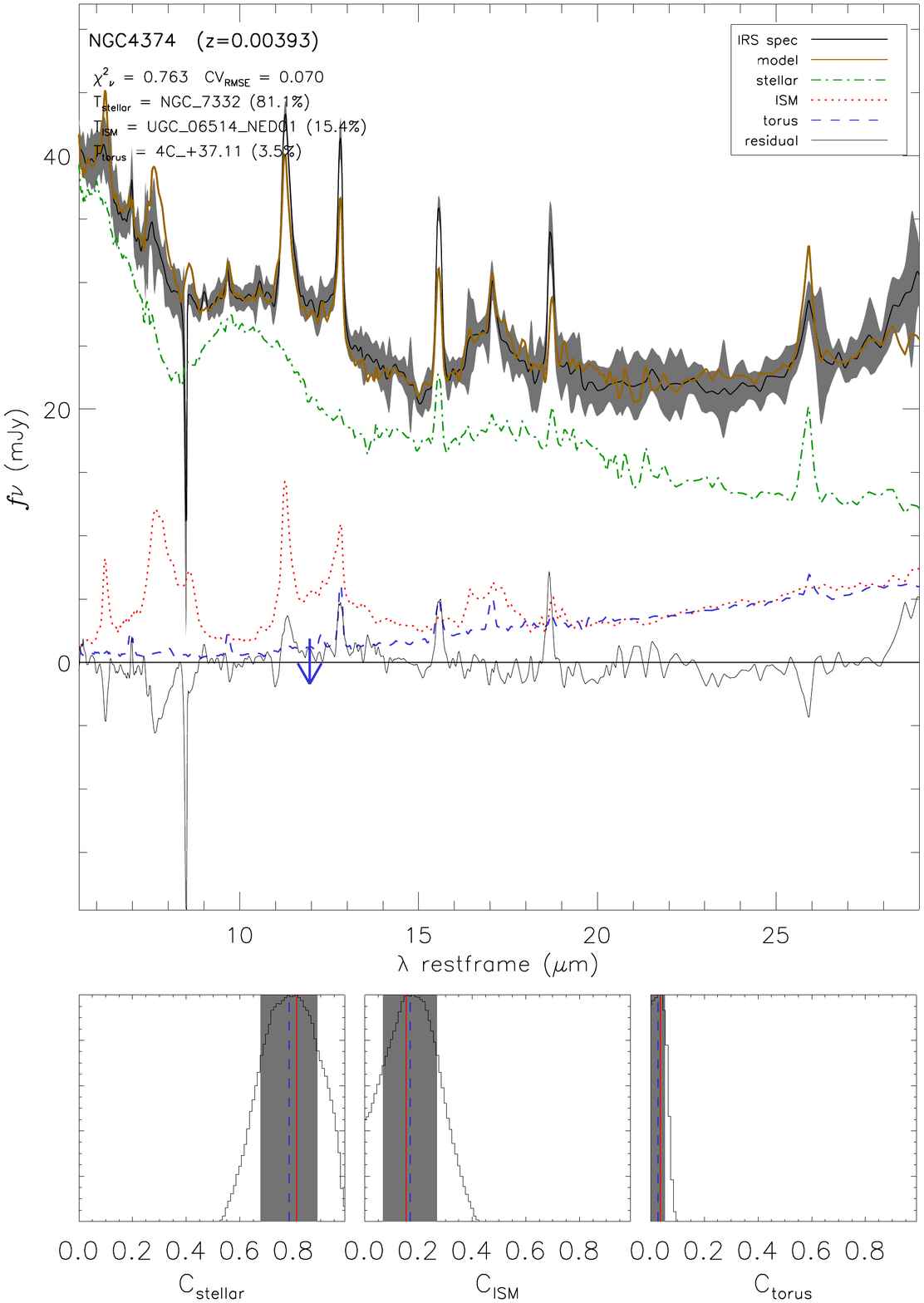}
\includegraphics[width=0.45\columnwidth]{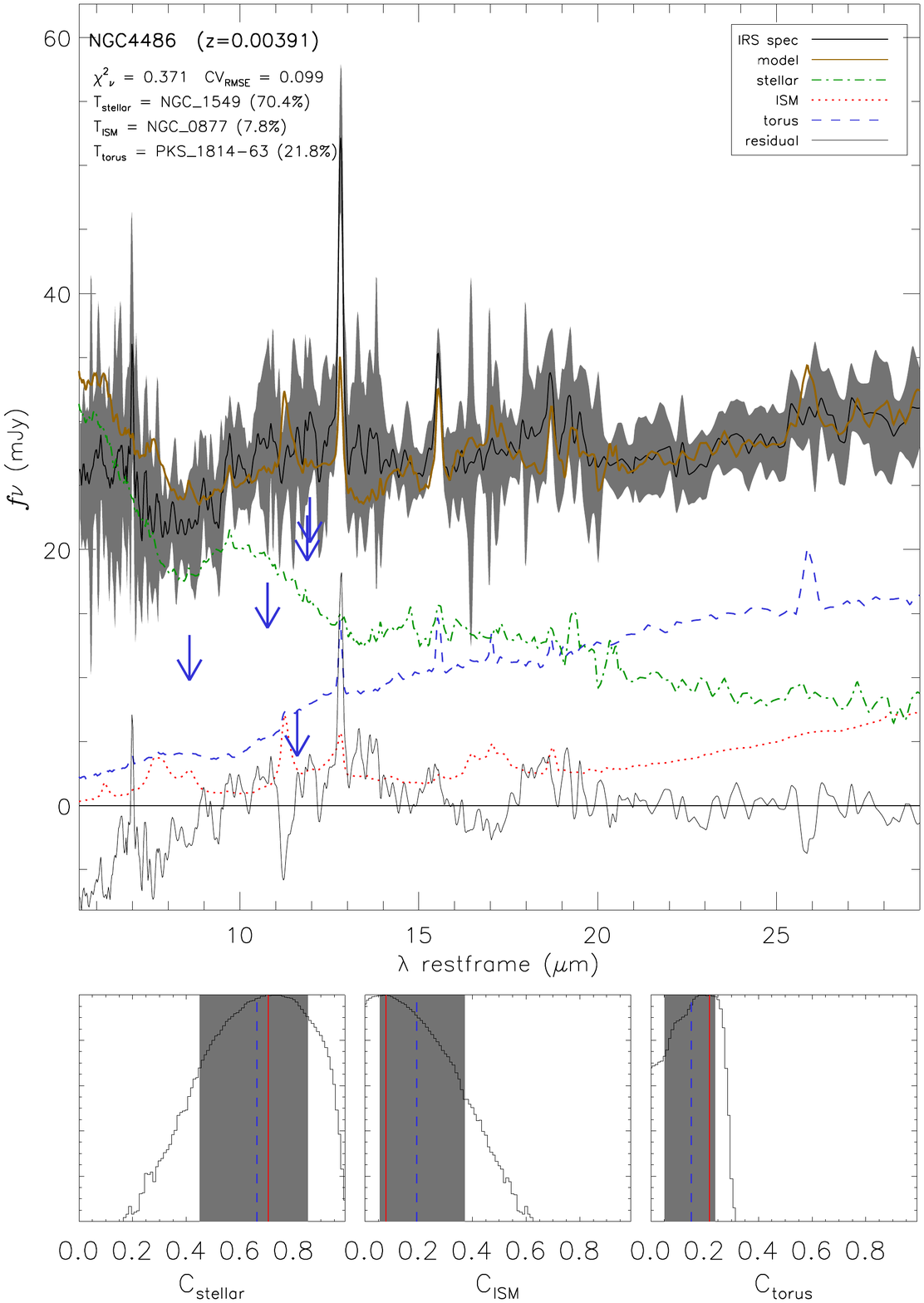}
\caption{...continued.}
\label{fig:CatSpectra}
\end{center}
\end{figure*}

\begin{figure*}
\begin{center}
\includegraphics[width=0.45\columnwidth]{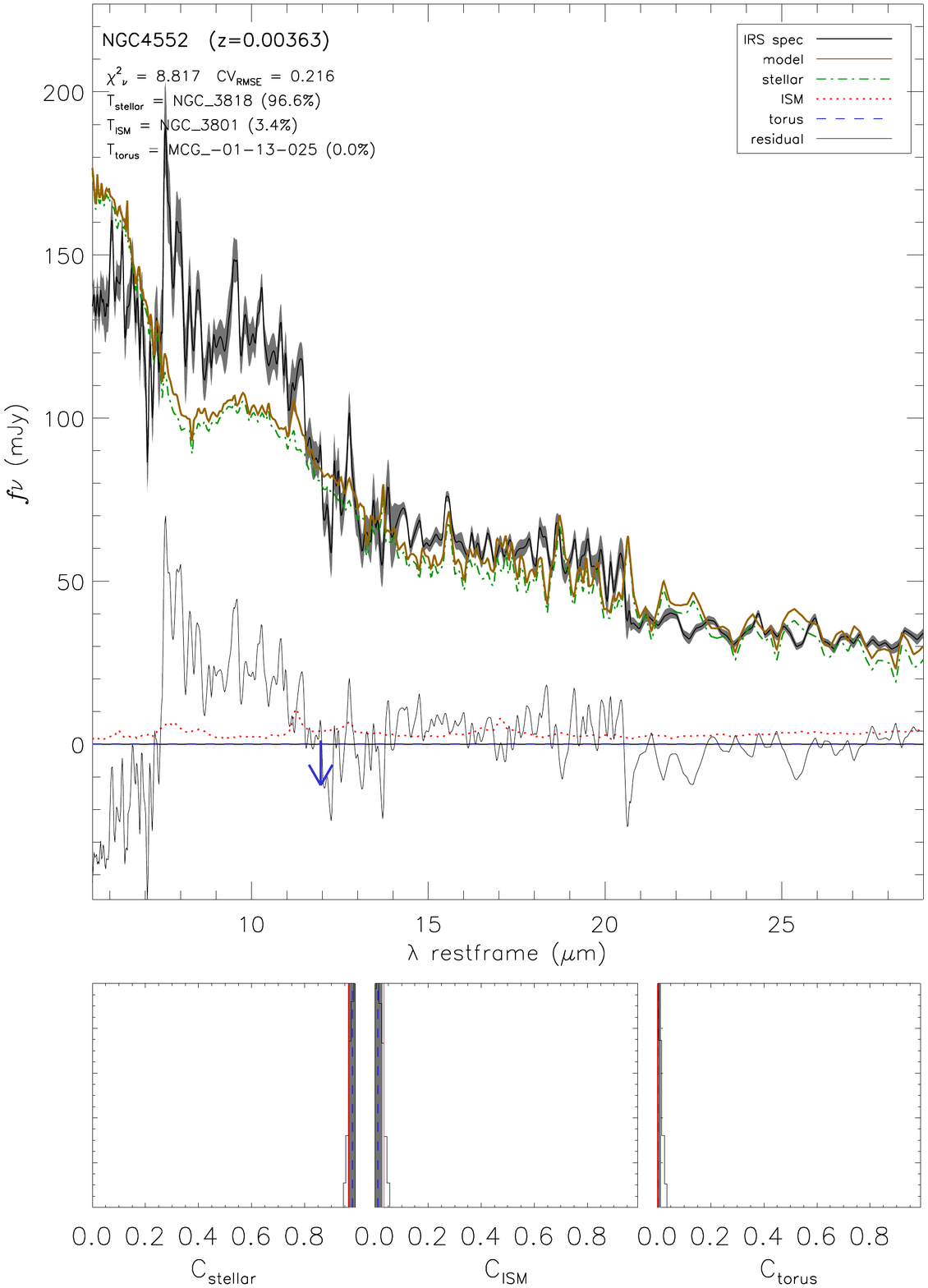}
\includegraphics[width=0.45\columnwidth]{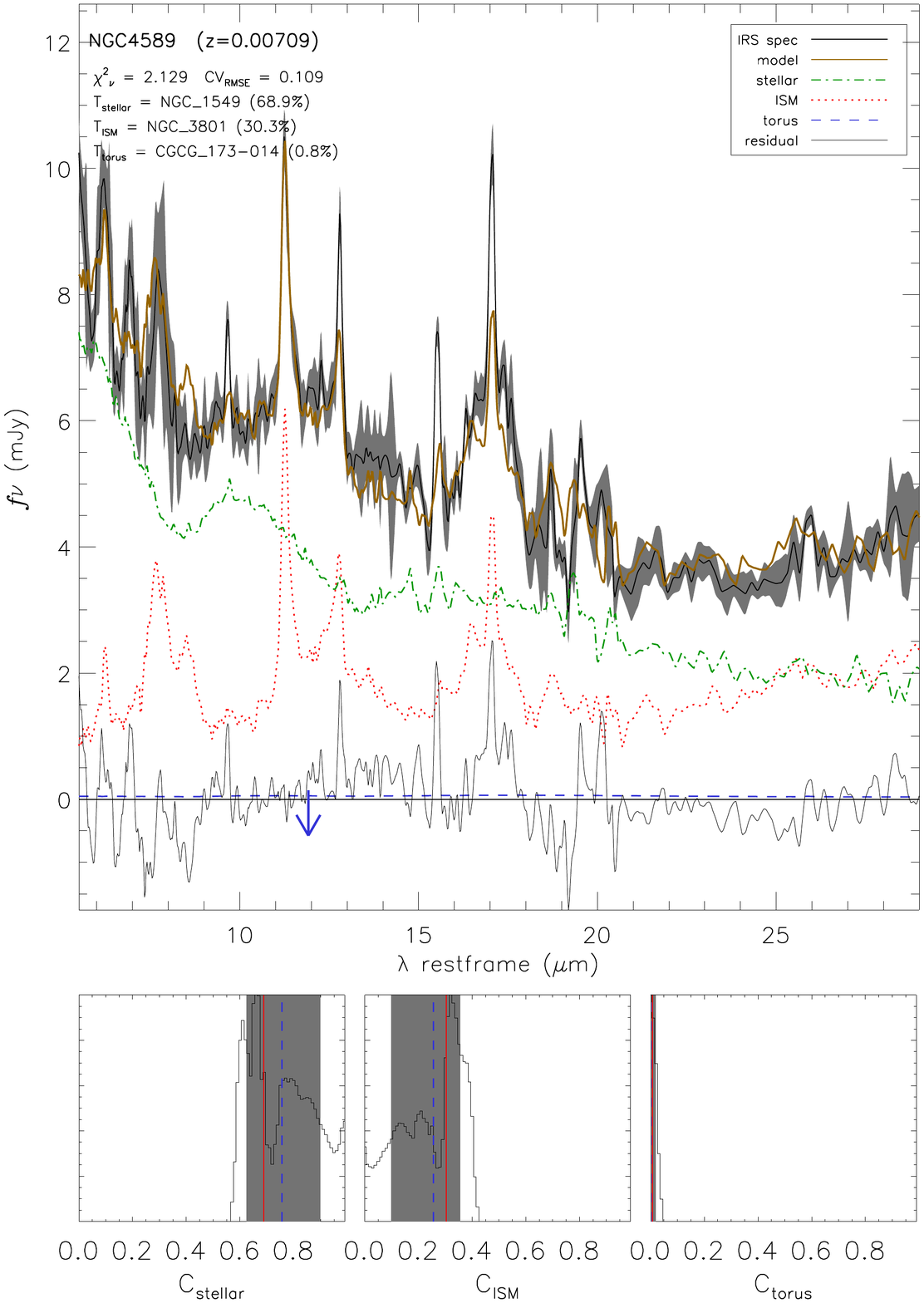}
\includegraphics[width=0.45\columnwidth]{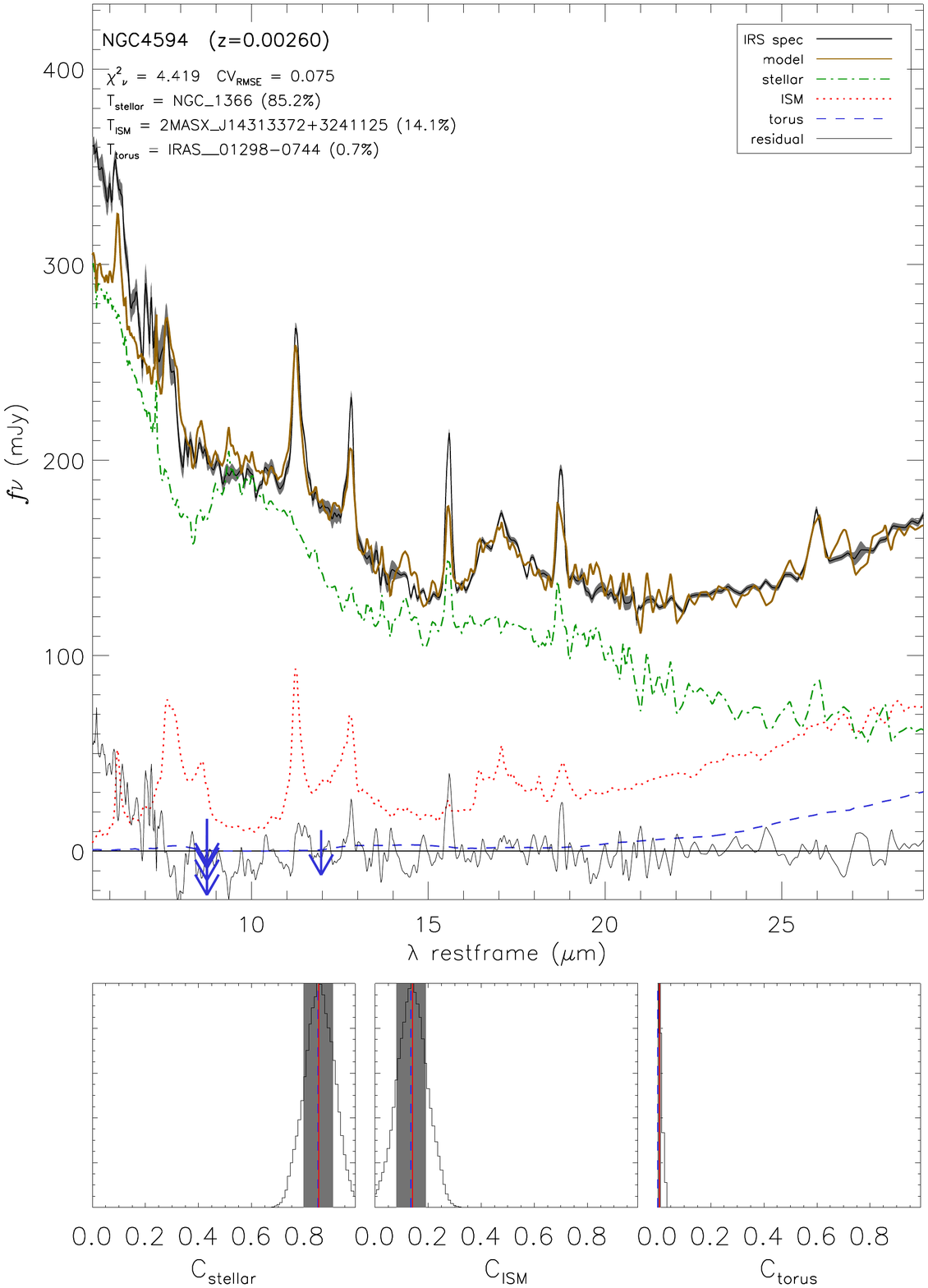}
\includegraphics[width=0.45\columnwidth]{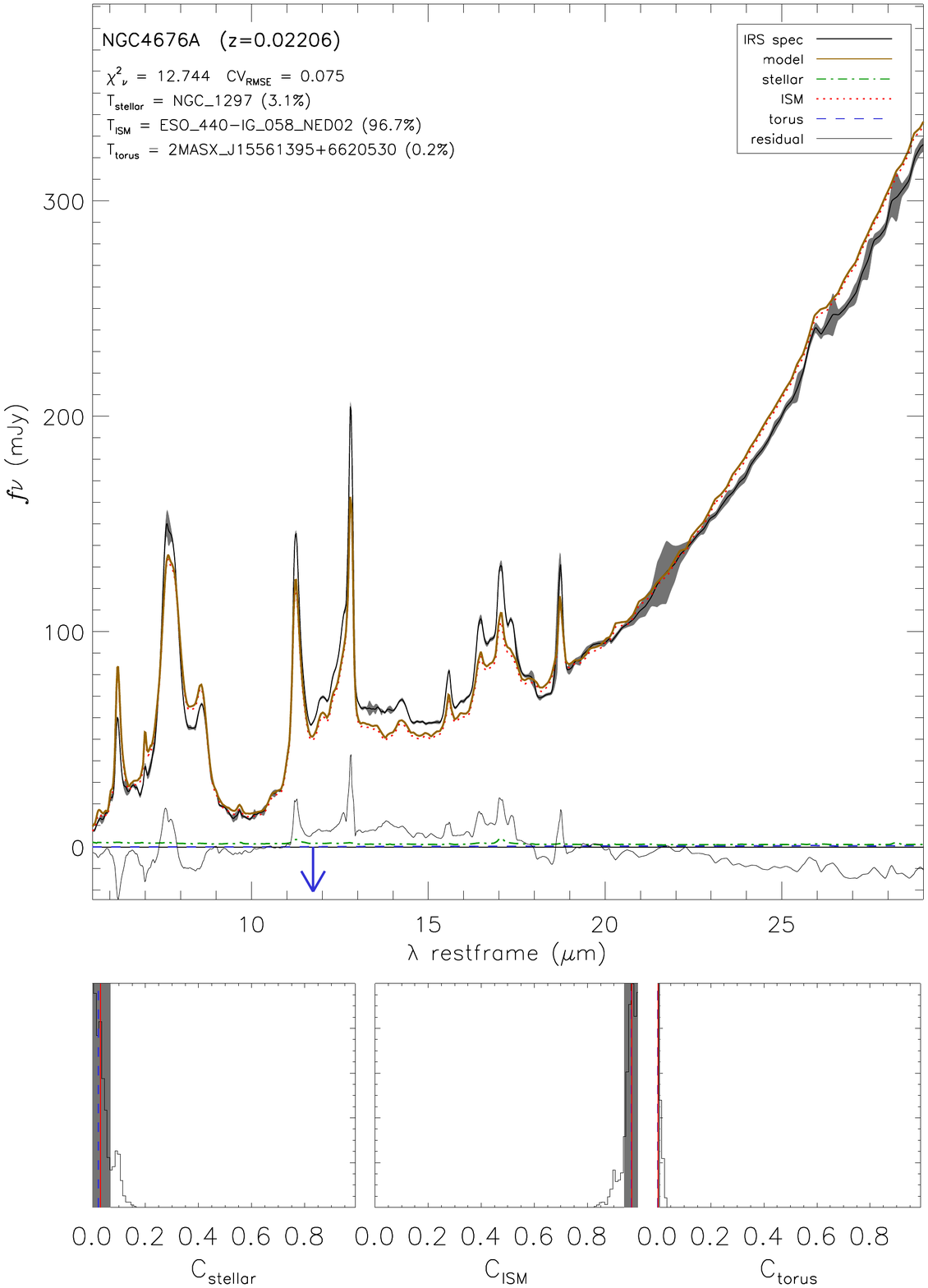}
\caption{...continued.}
\label{fig:CatSpectra}
\end{center}
\end{figure*}

\begin{figure*}
\begin{center}
\includegraphics[width=0.45\columnwidth]{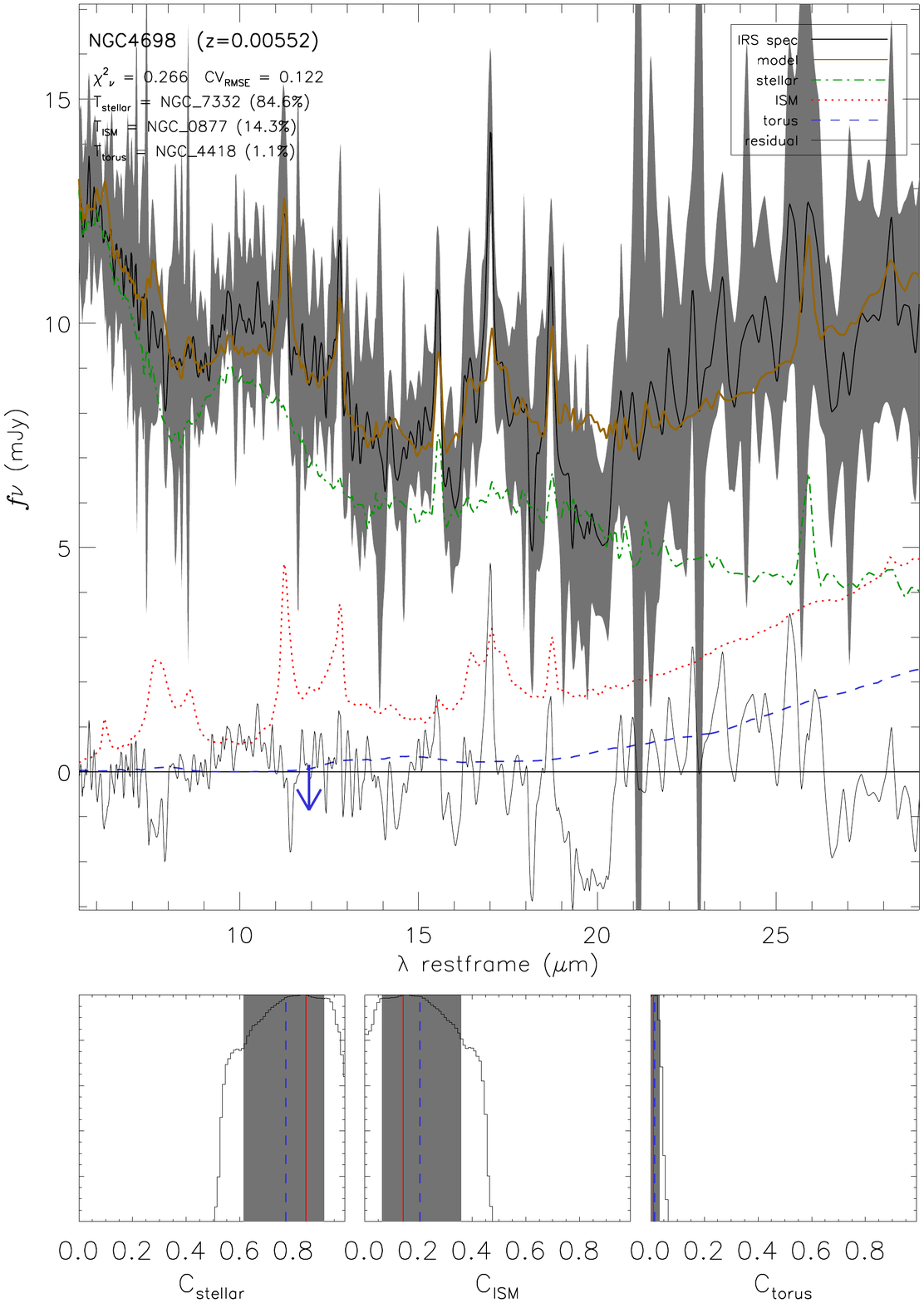}
\includegraphics[width=0.45\columnwidth]{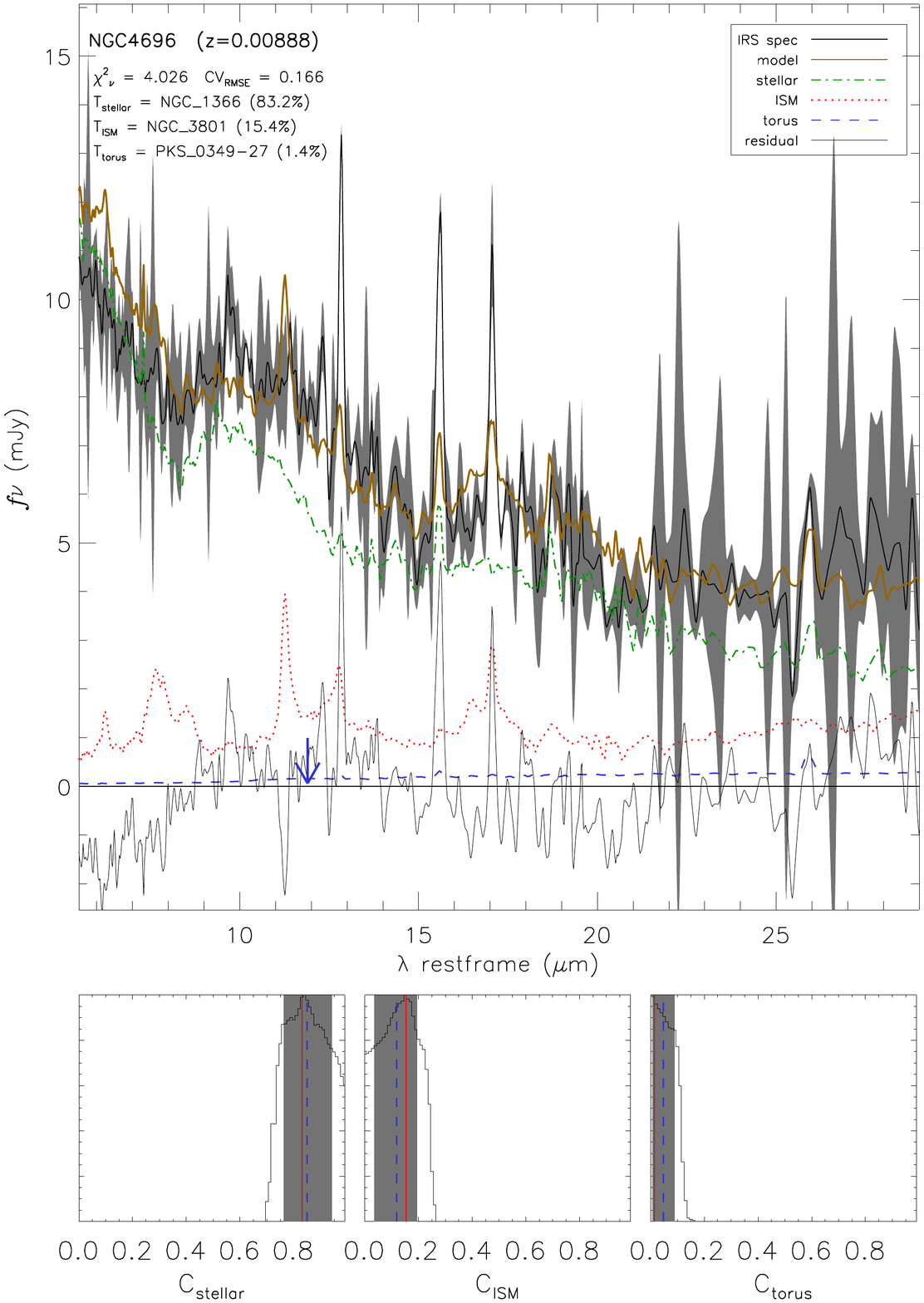}
\includegraphics[width=0.45\columnwidth]{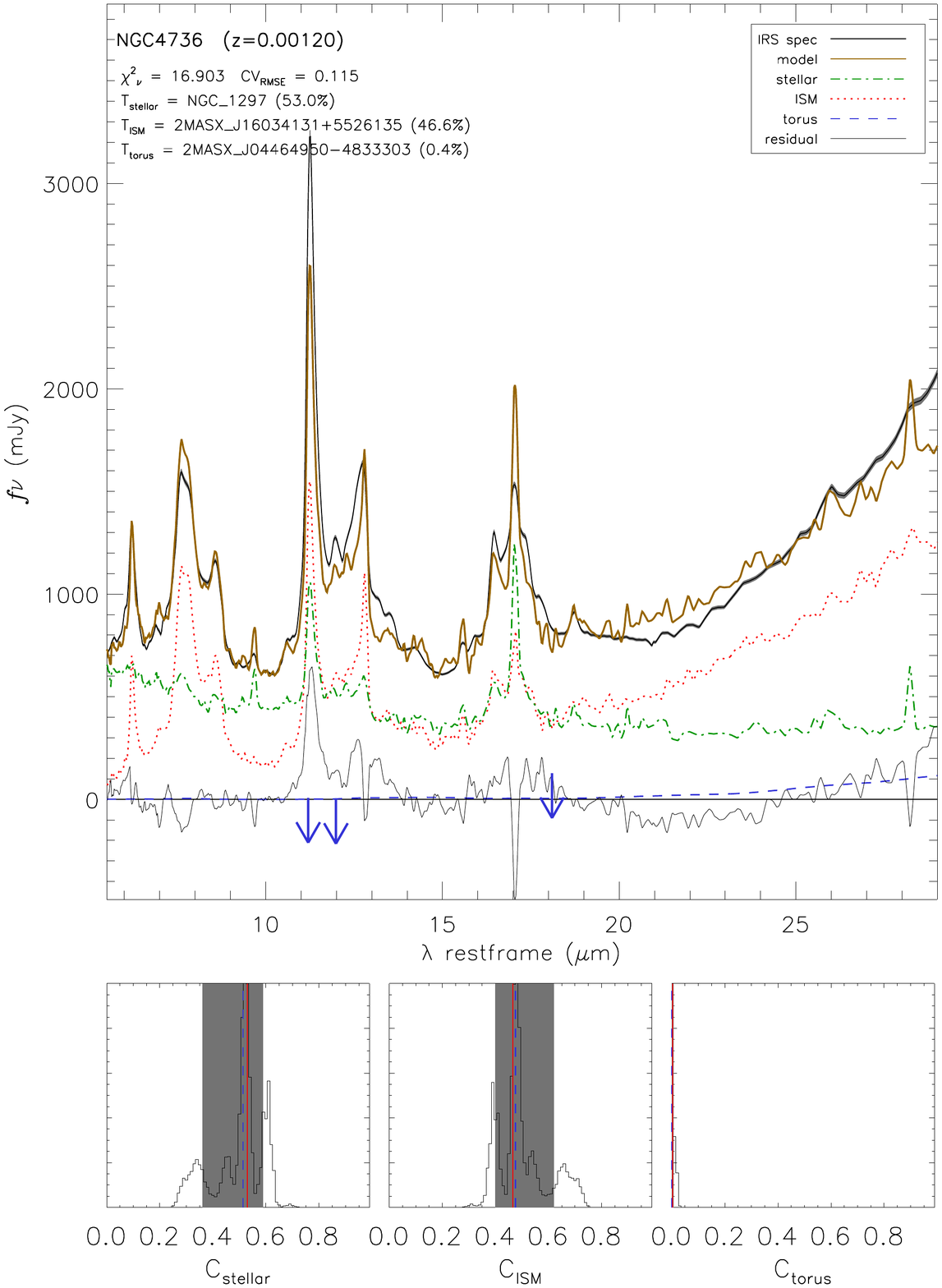}
\includegraphics[width=0.45\columnwidth]{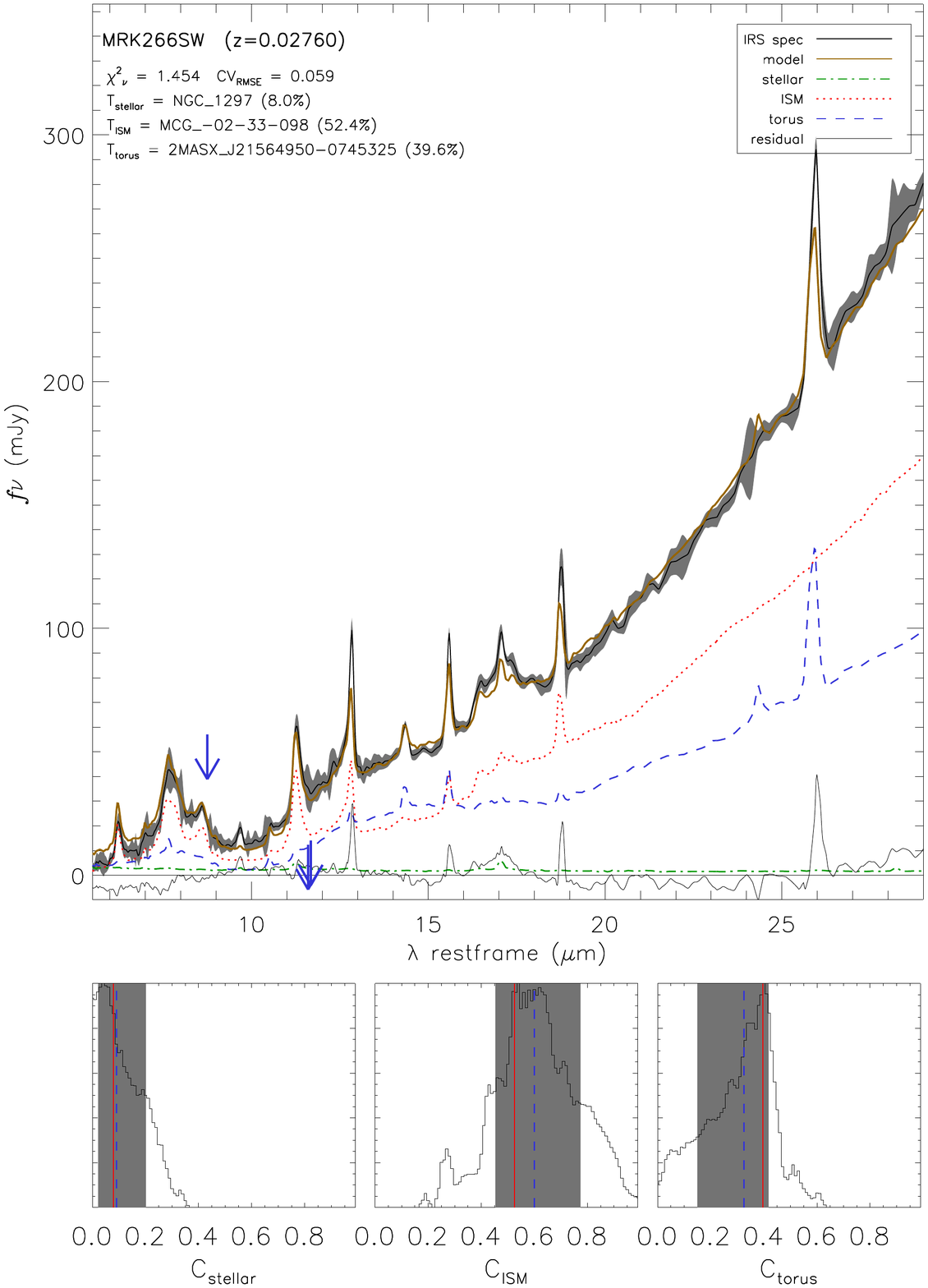}
\caption{...continued.}
\label{fig:CatSpectra}
\end{center}
\end{figure*}

\begin{figure*}
\begin{center}
\includegraphics[width=0.45\columnwidth]{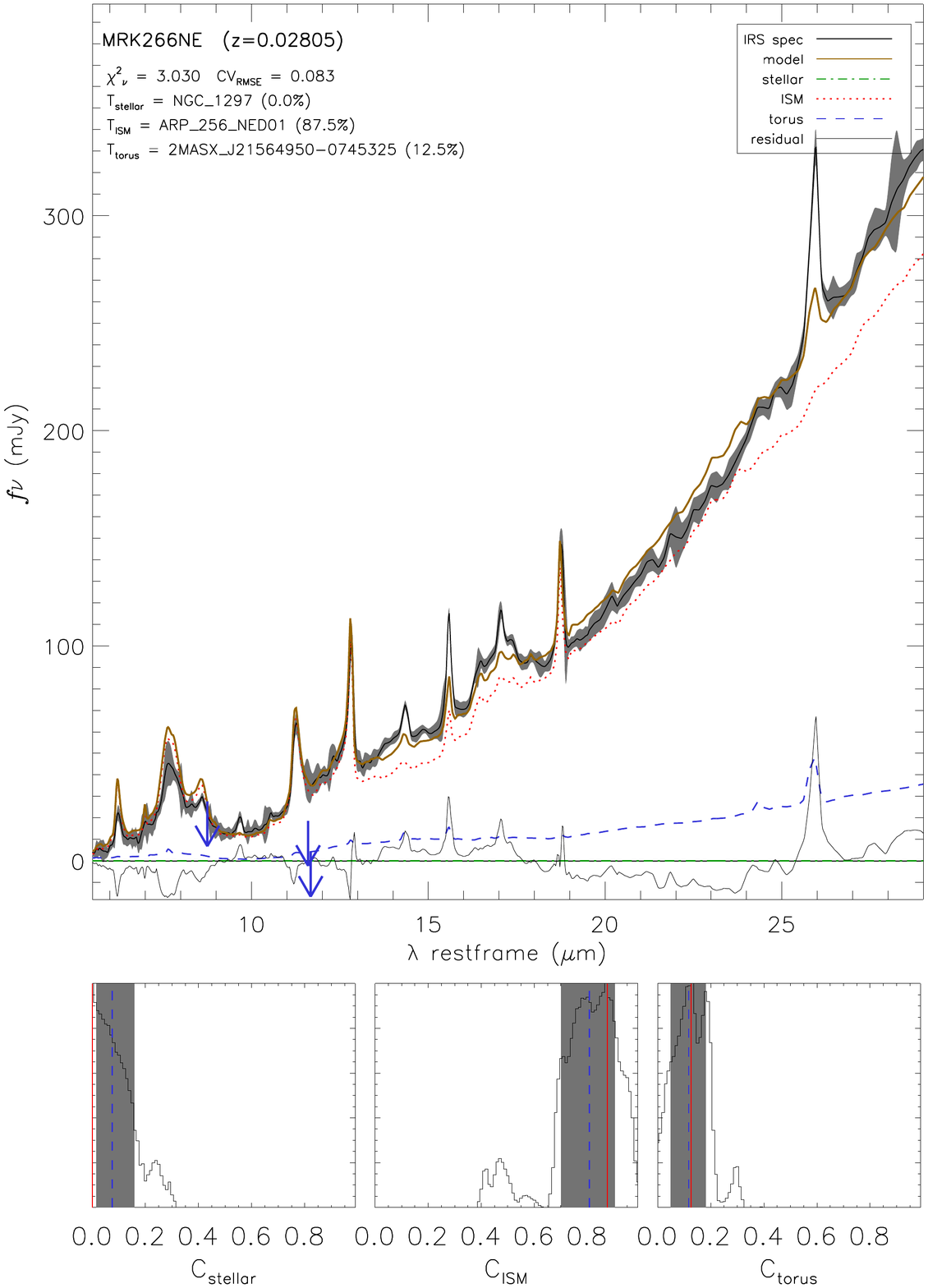}
\includegraphics[width=0.45\columnwidth]{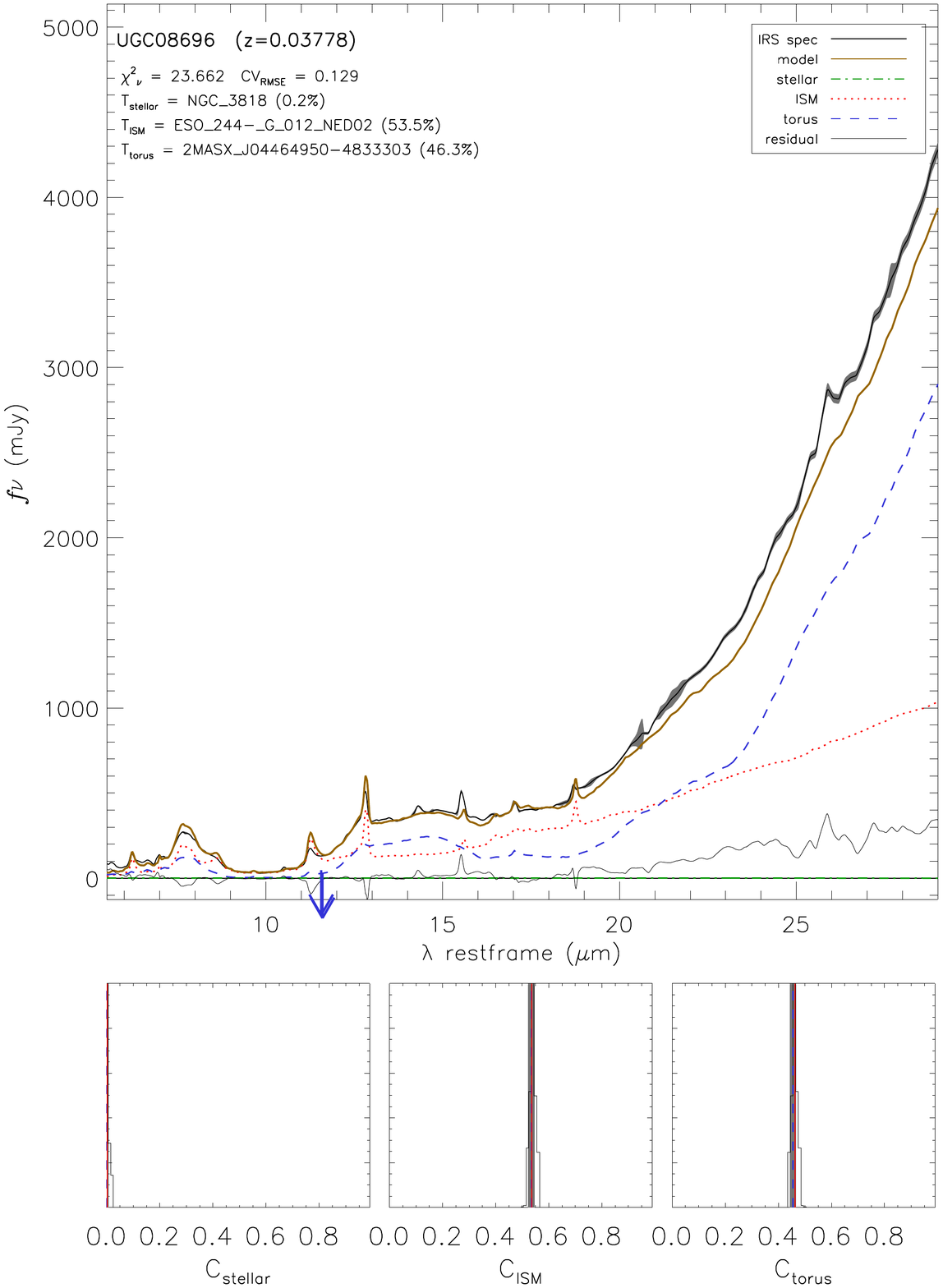}
\includegraphics[width=0.45\columnwidth]{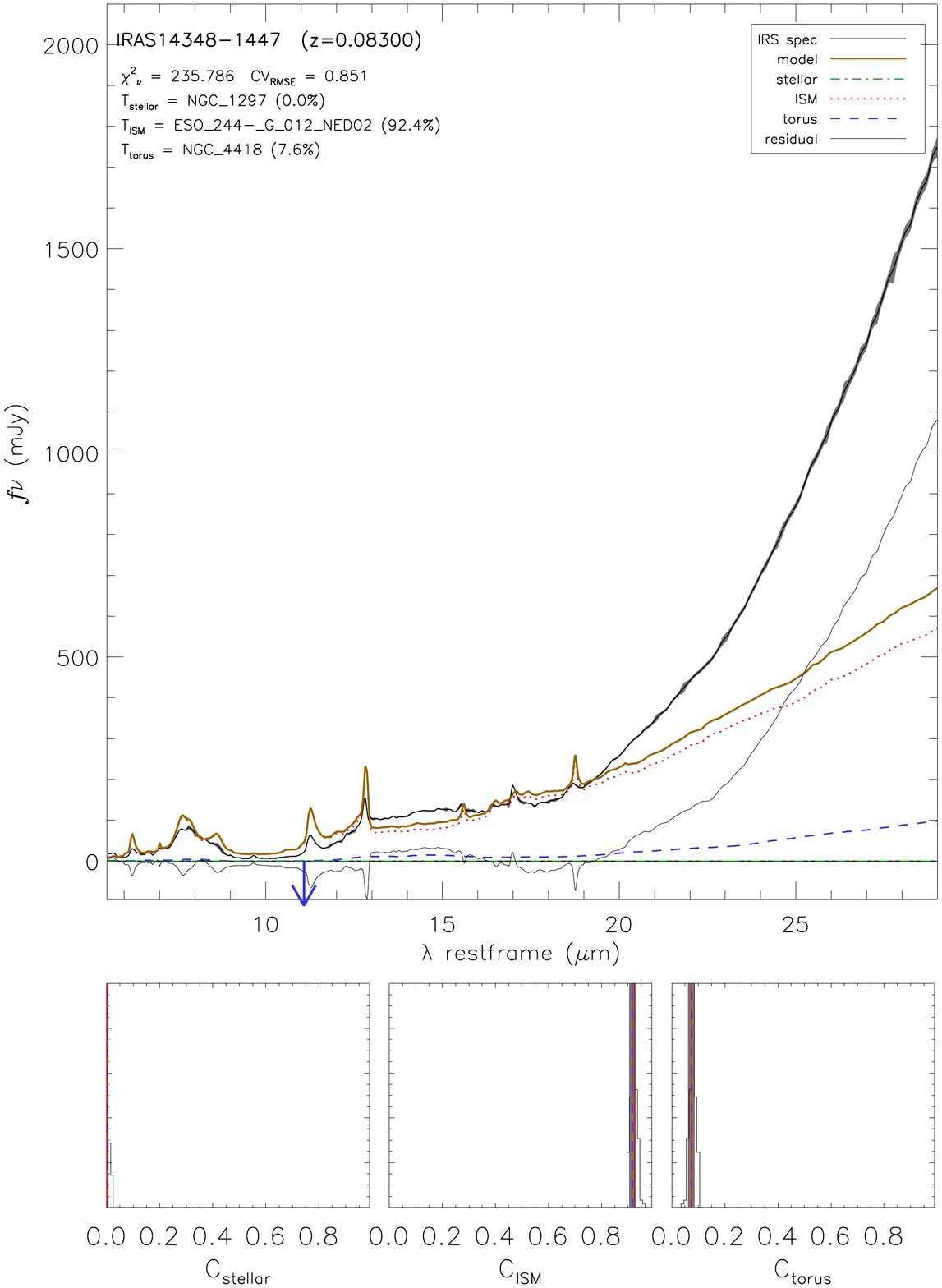}
\includegraphics[width=0.45\columnwidth]{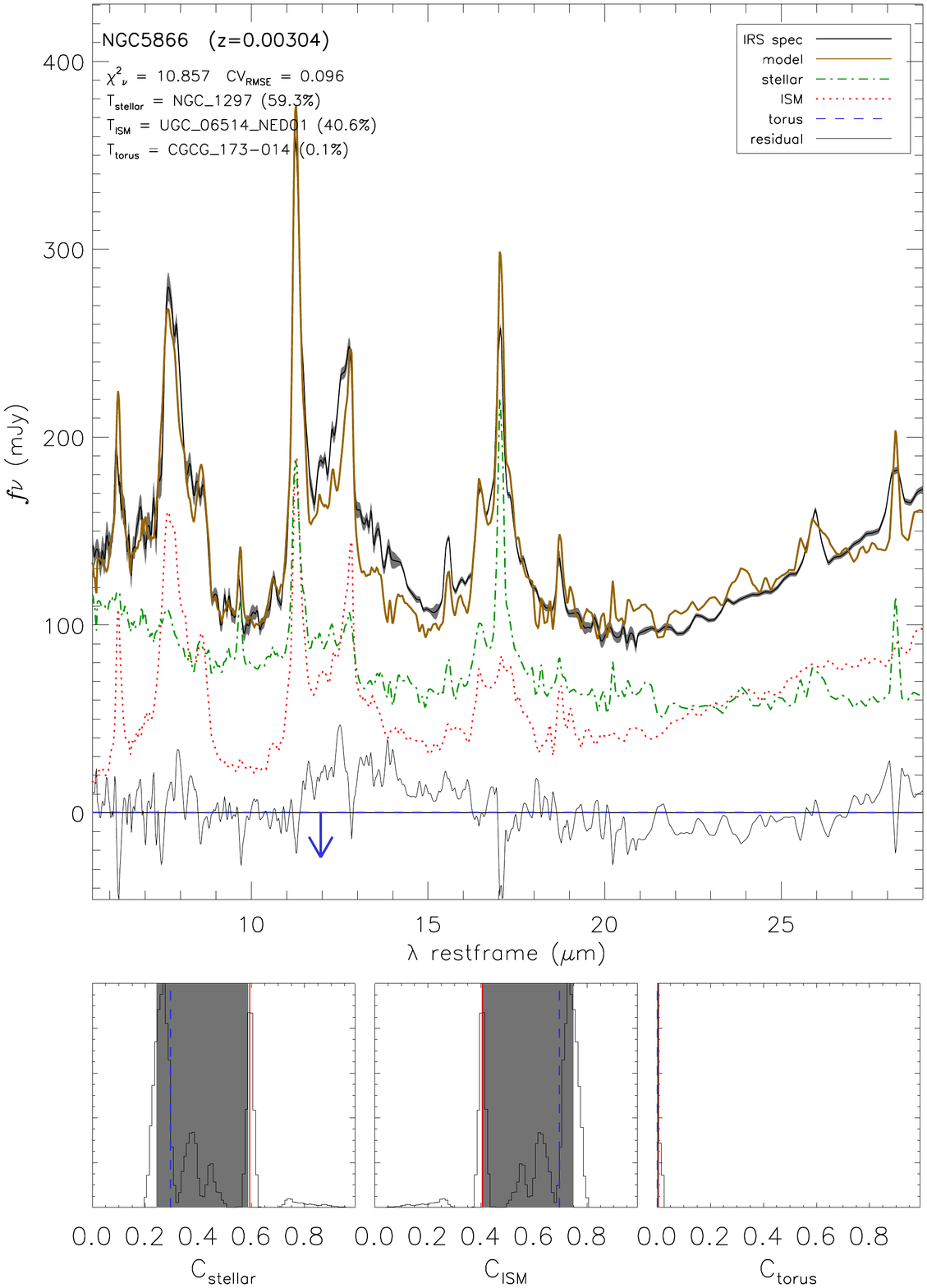}
\caption{...continued.}
\label{fig:CatSpectra}
\end{center}
\end{figure*}

\begin{figure*}
\begin{center}
\includegraphics[width=0.45\columnwidth]{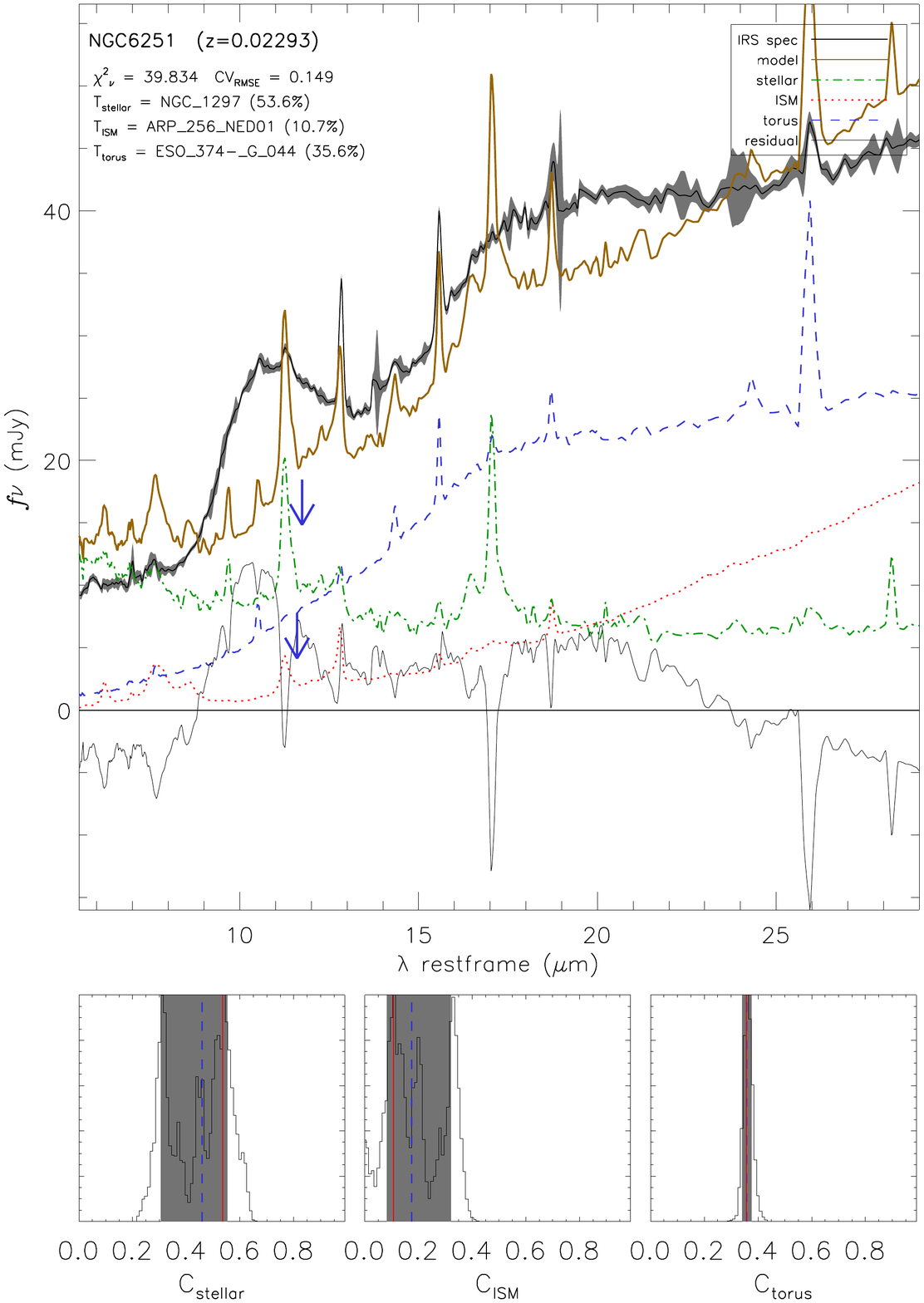}
\includegraphics[width=0.45\columnwidth]{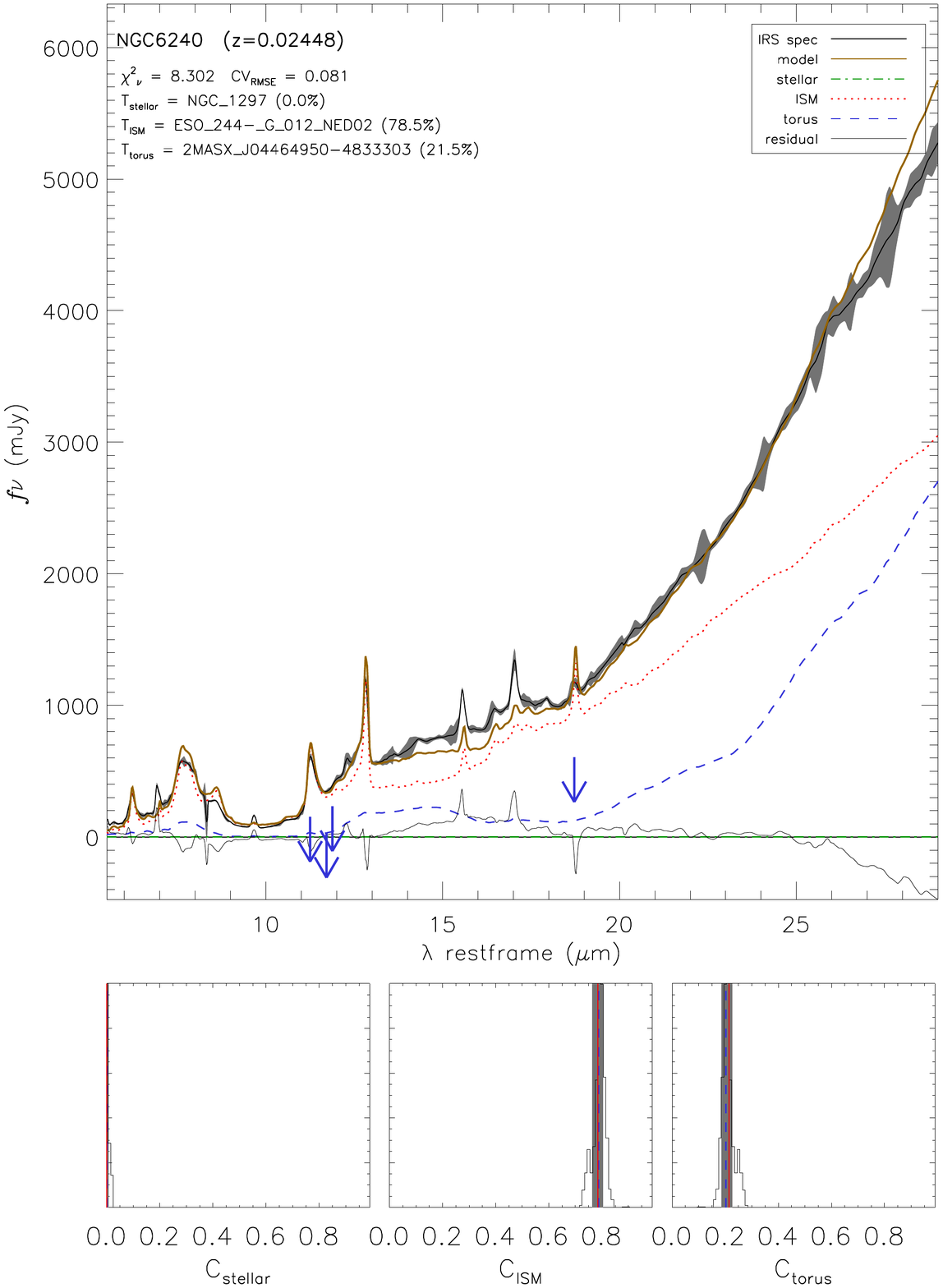}
\includegraphics[width=0.45\columnwidth]{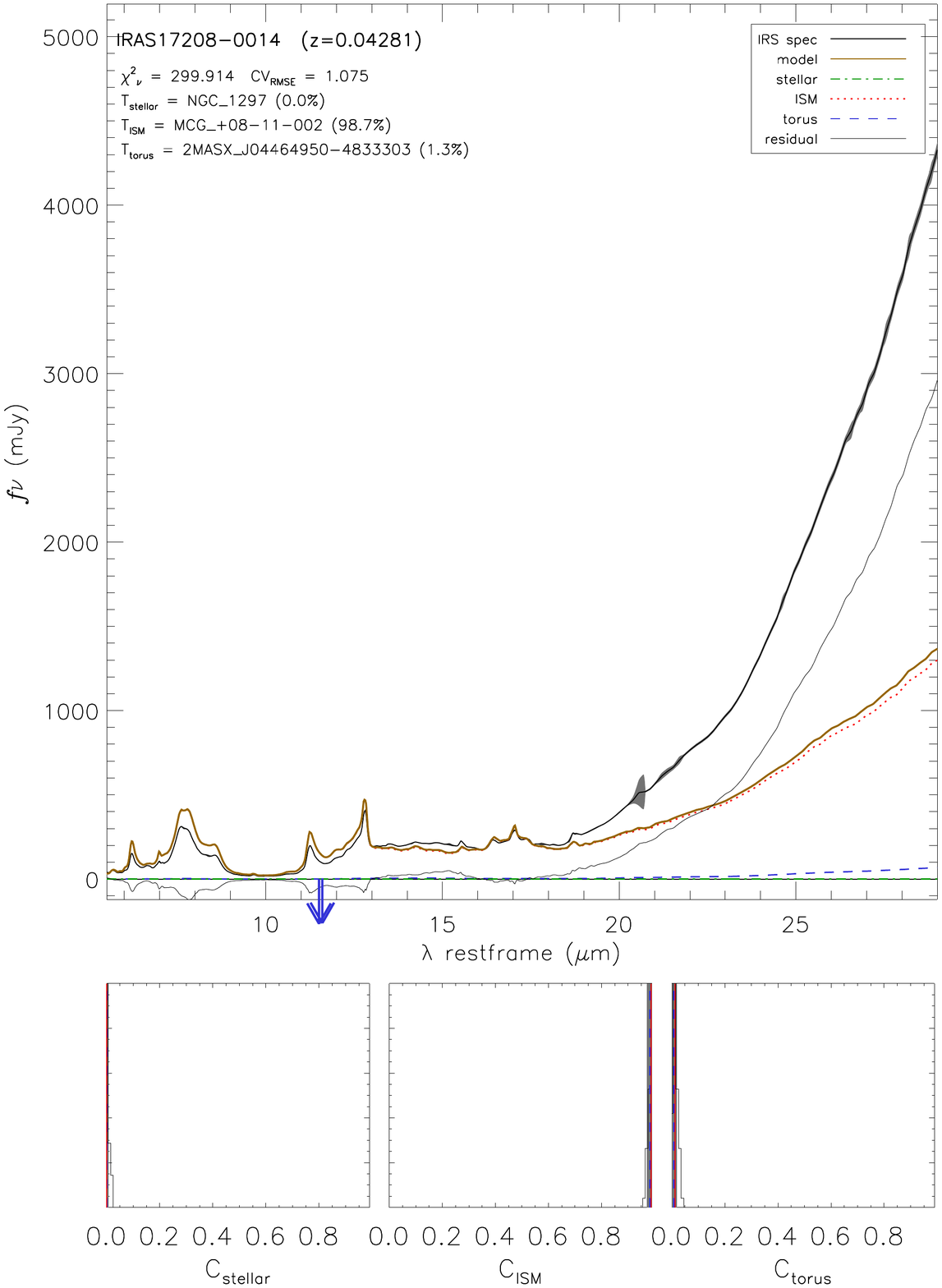}
\includegraphics[width=0.45\columnwidth]{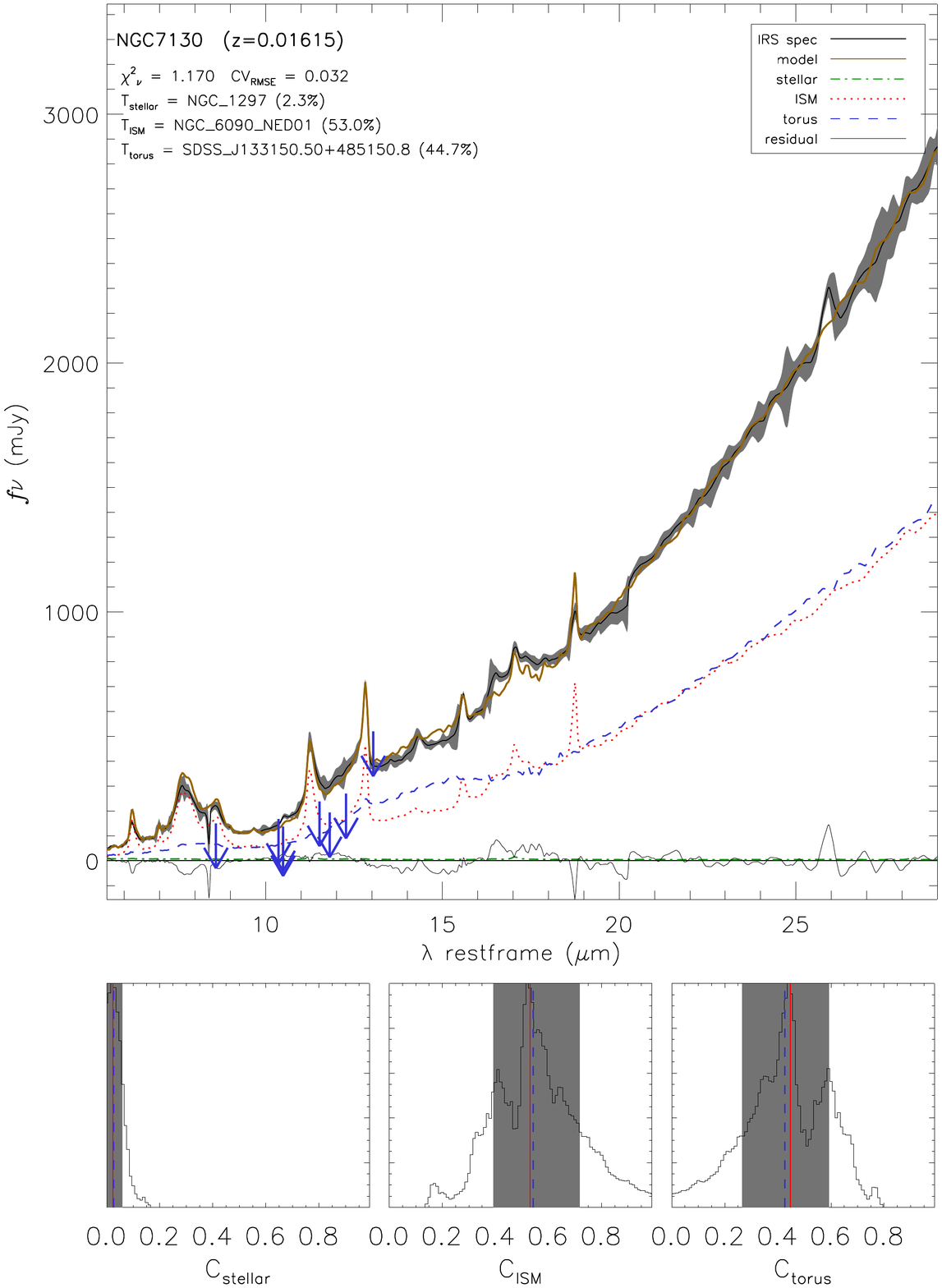}
\caption{...continued.}
\label{fig:CatSpectra}
\end{center}
\end{figure*}

\begin{figure*}
\begin{center}
\includegraphics[width=0.45\columnwidth]{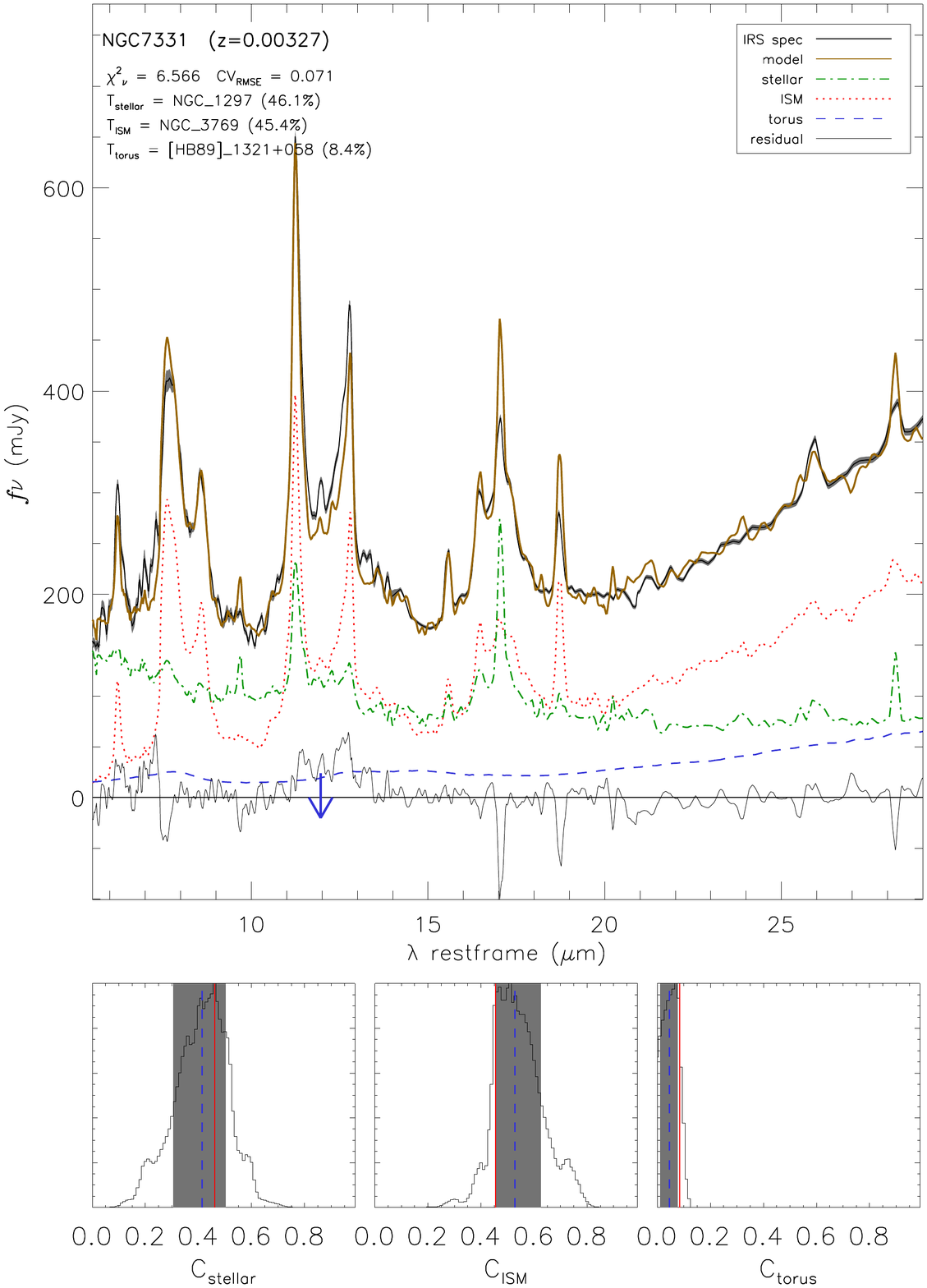}
\includegraphics[width=0.45\columnwidth]{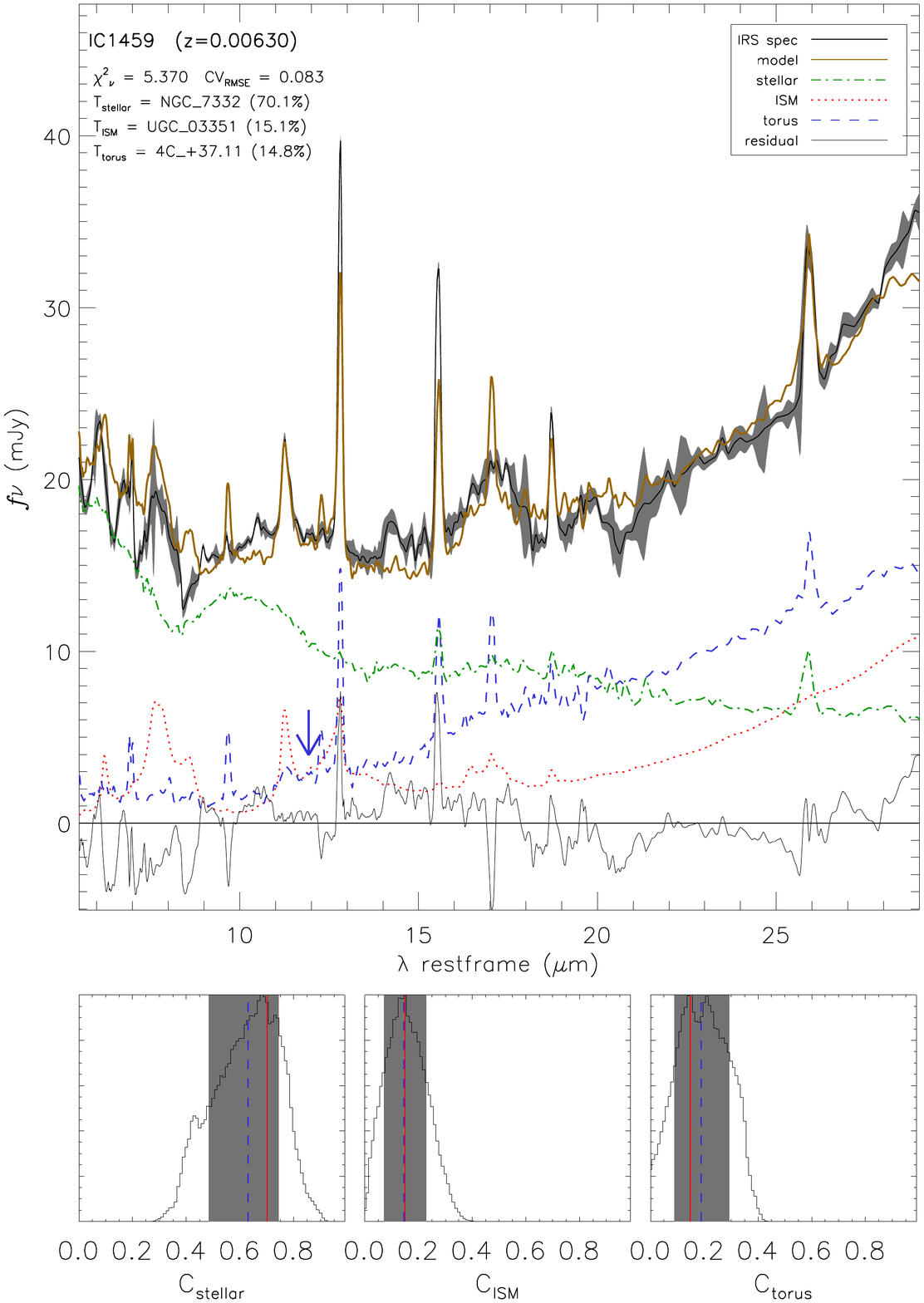}
\includegraphics[width=0.45\columnwidth]{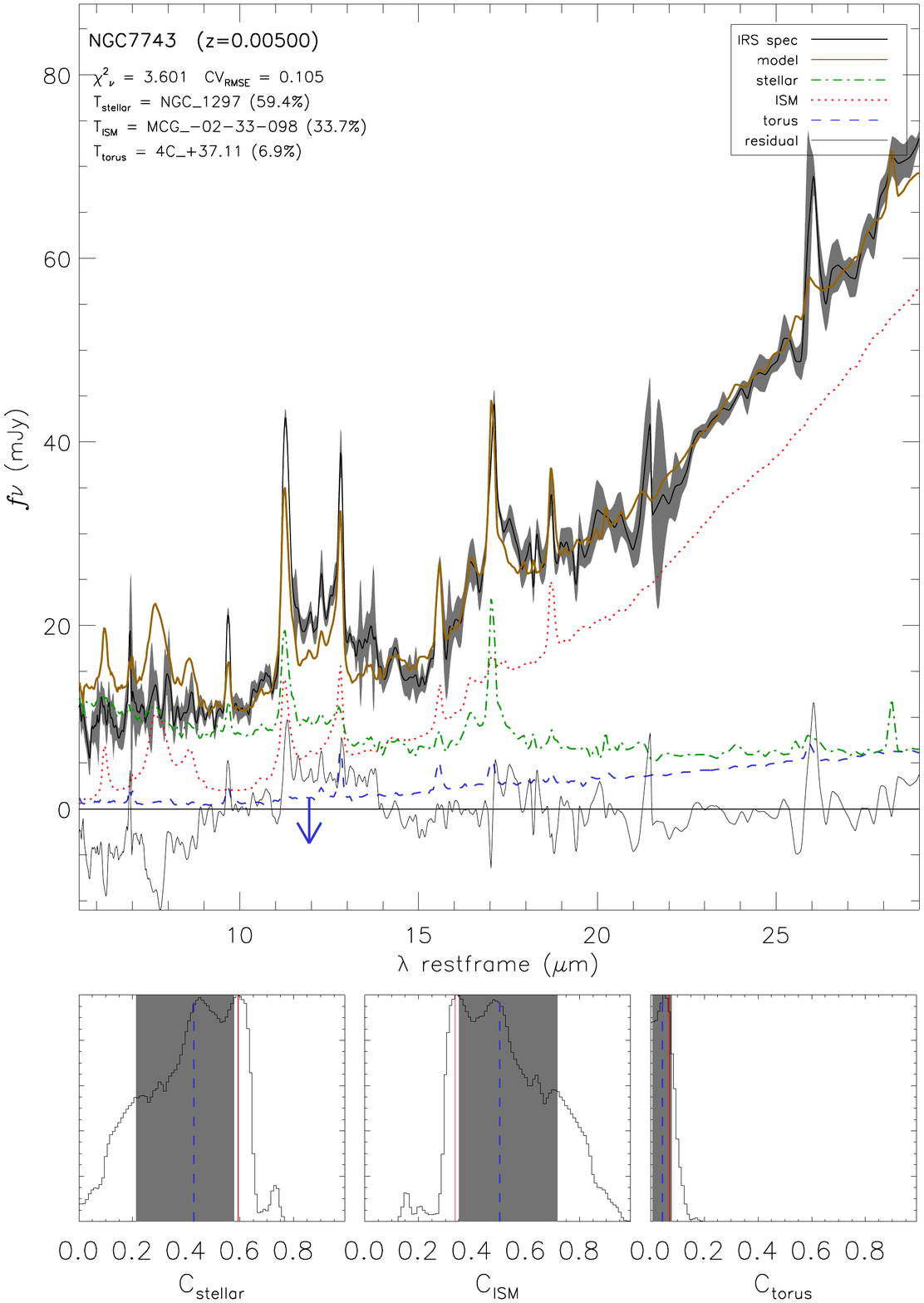}
\includegraphics[width=0.45\columnwidth]{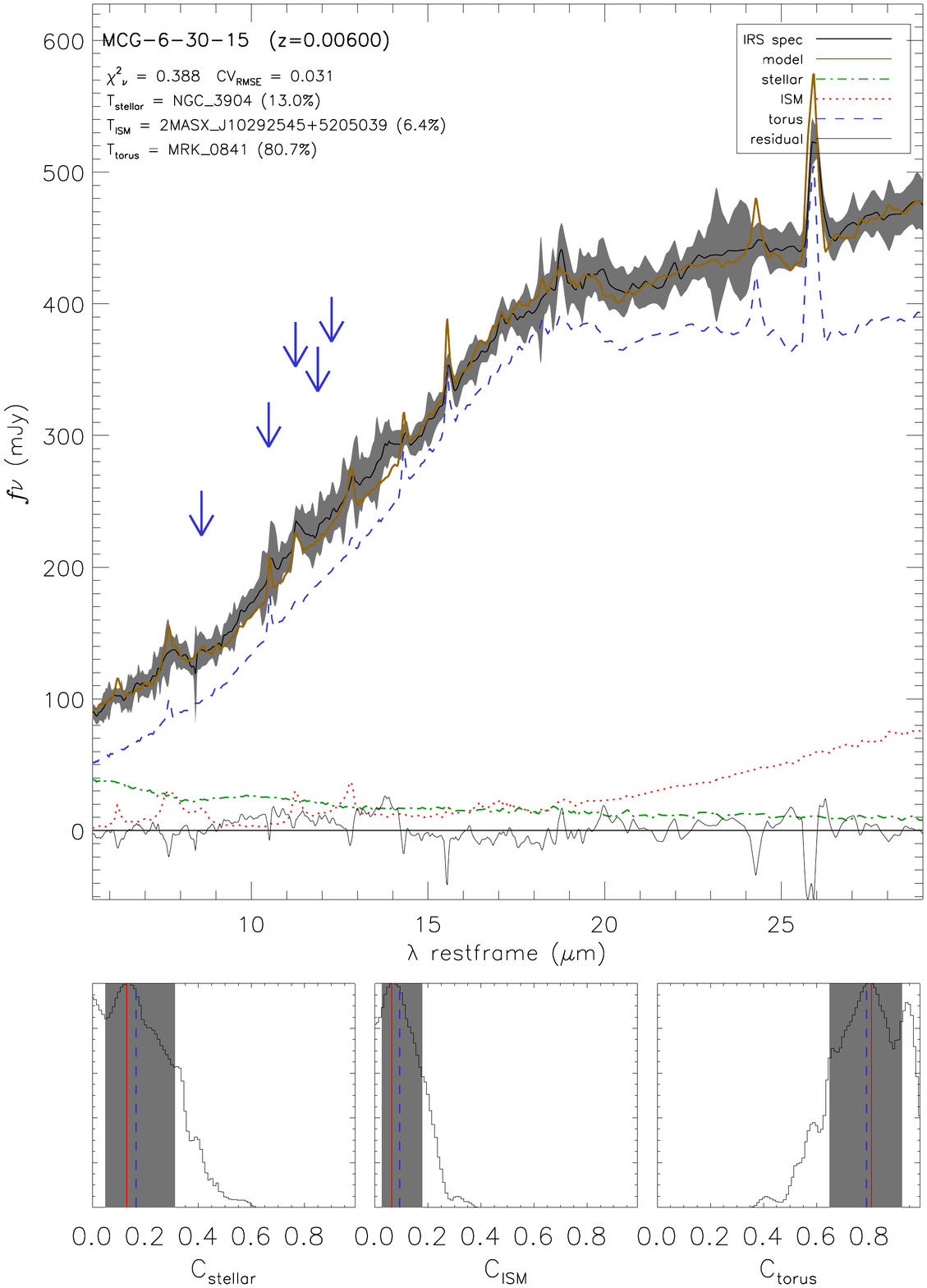}
\caption{...continued.}
\label{fig:CatSpectra}
\end{center}
\end{figure*}

\begin{figure*}
\begin{center}
\includegraphics[width=0.45\columnwidth]{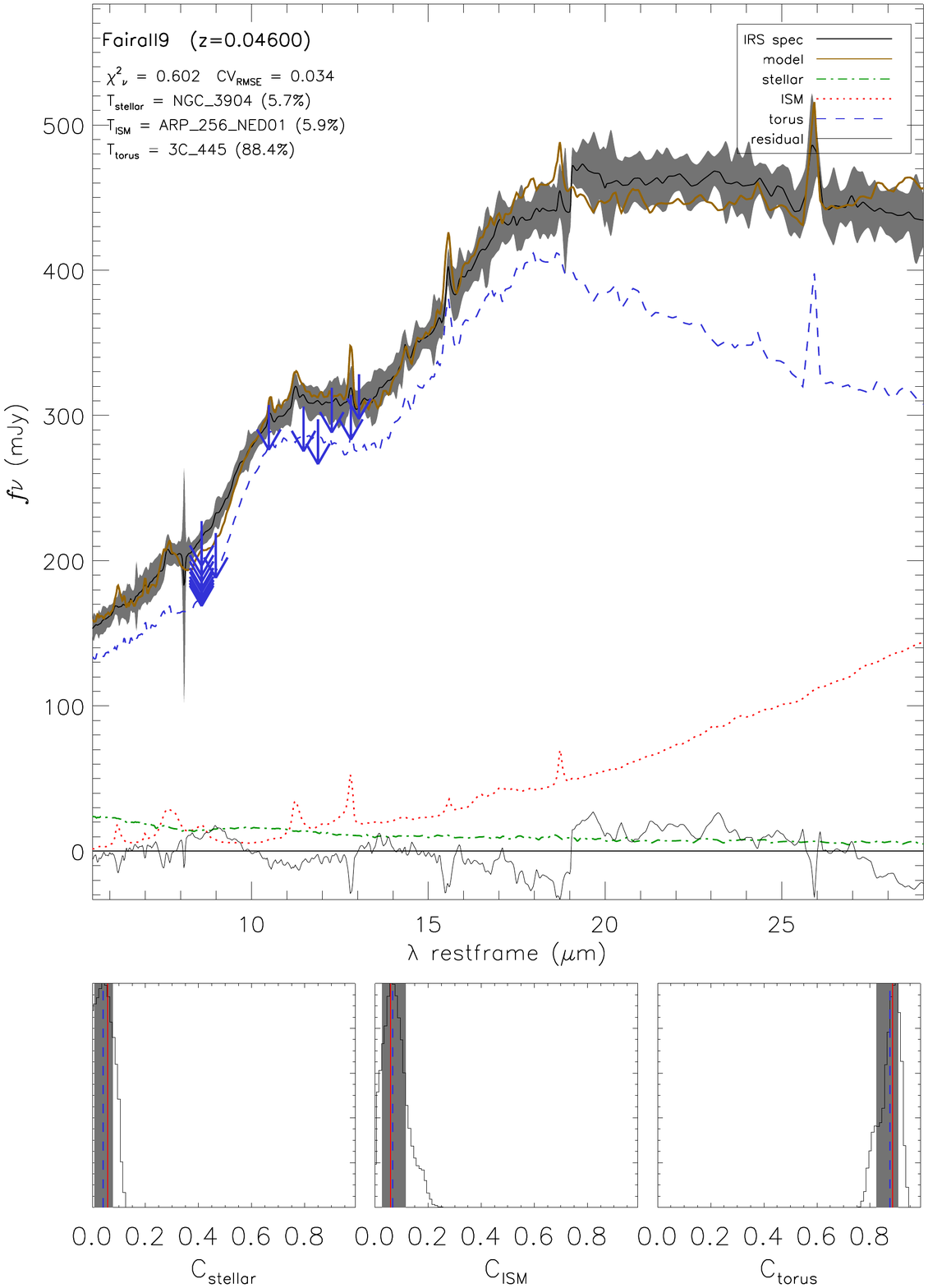}
\includegraphics[width=0.45\columnwidth]{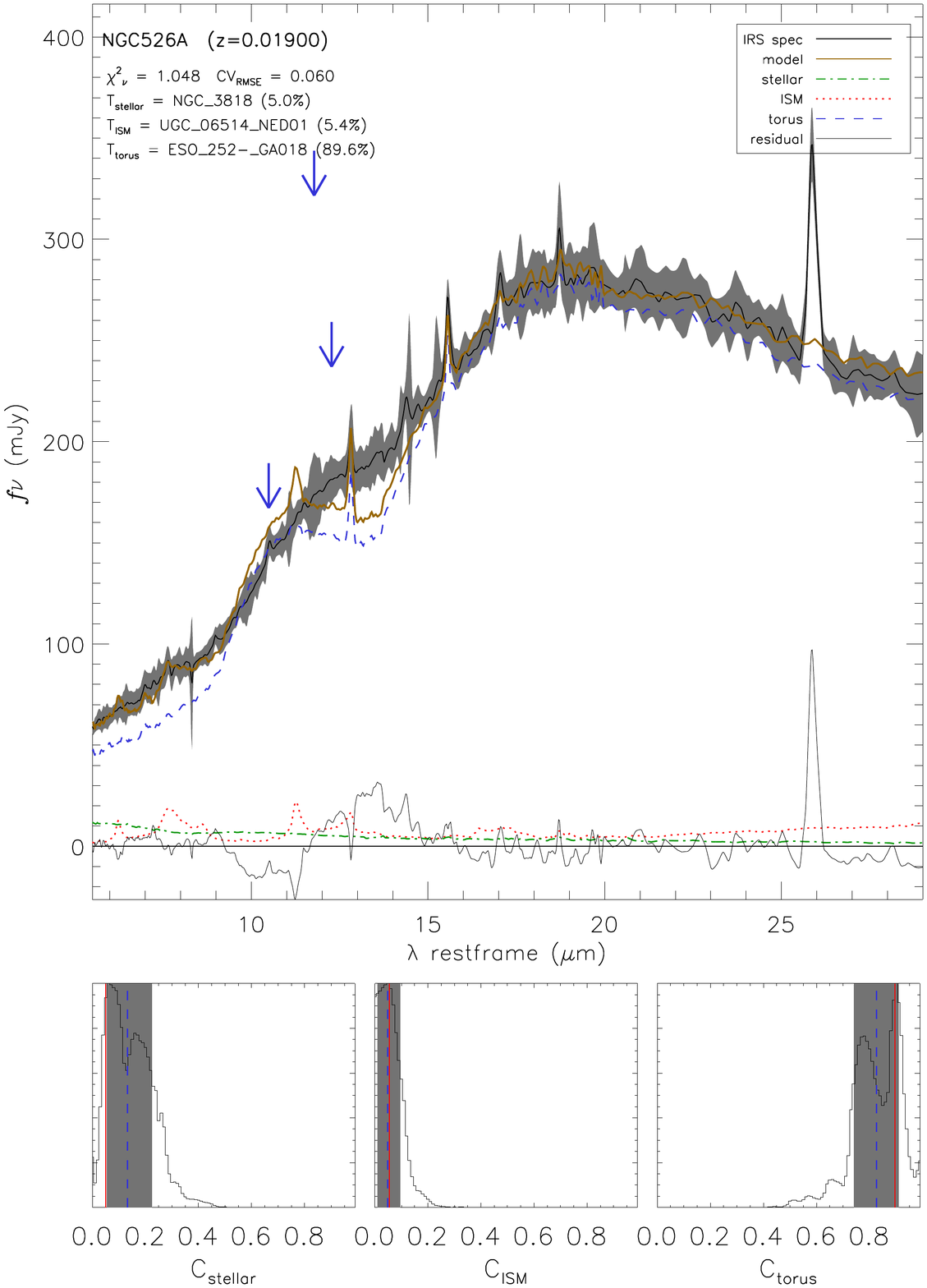}
\includegraphics[width=0.45\columnwidth]{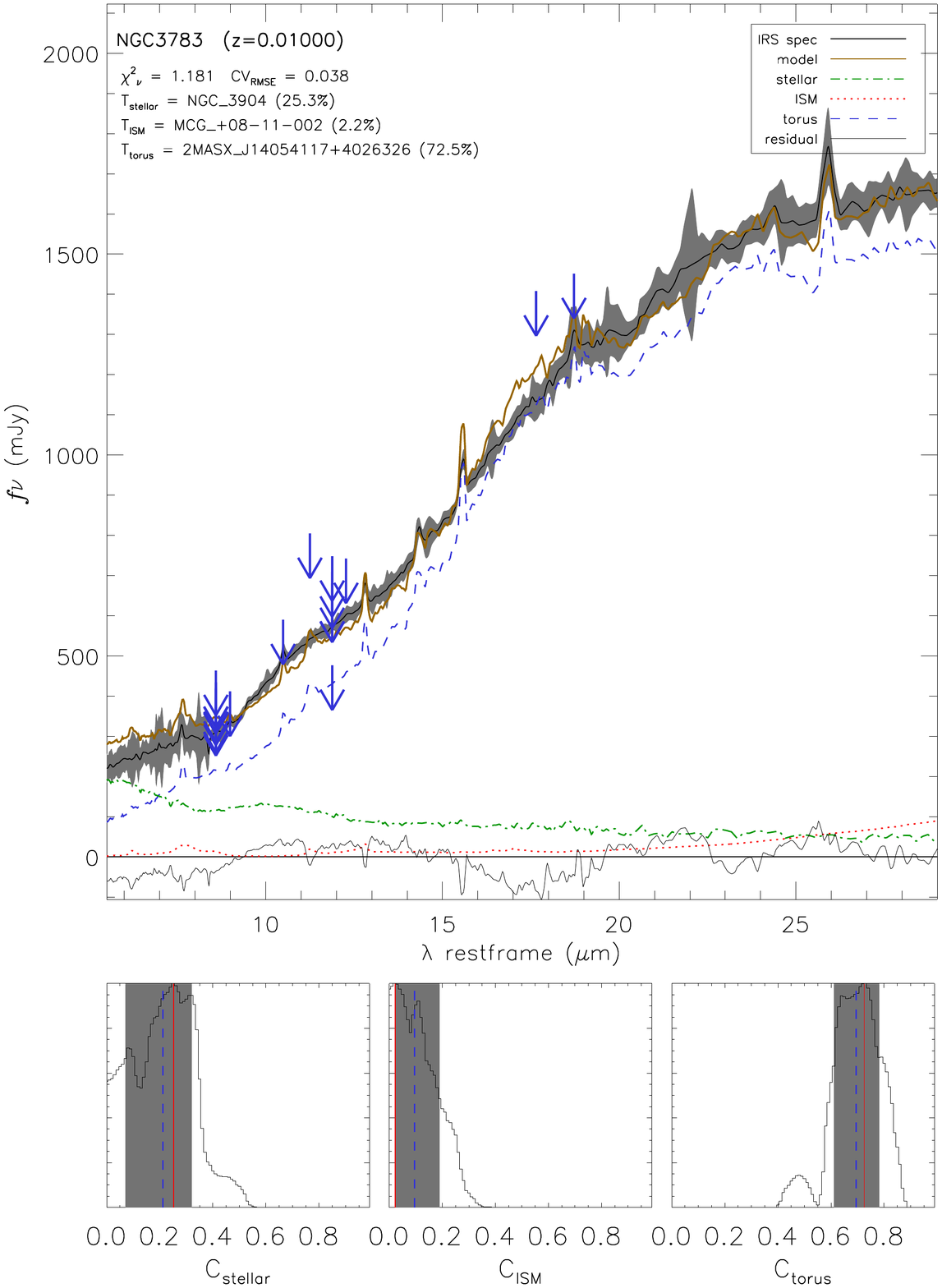}
\includegraphics[width=0.45\columnwidth]{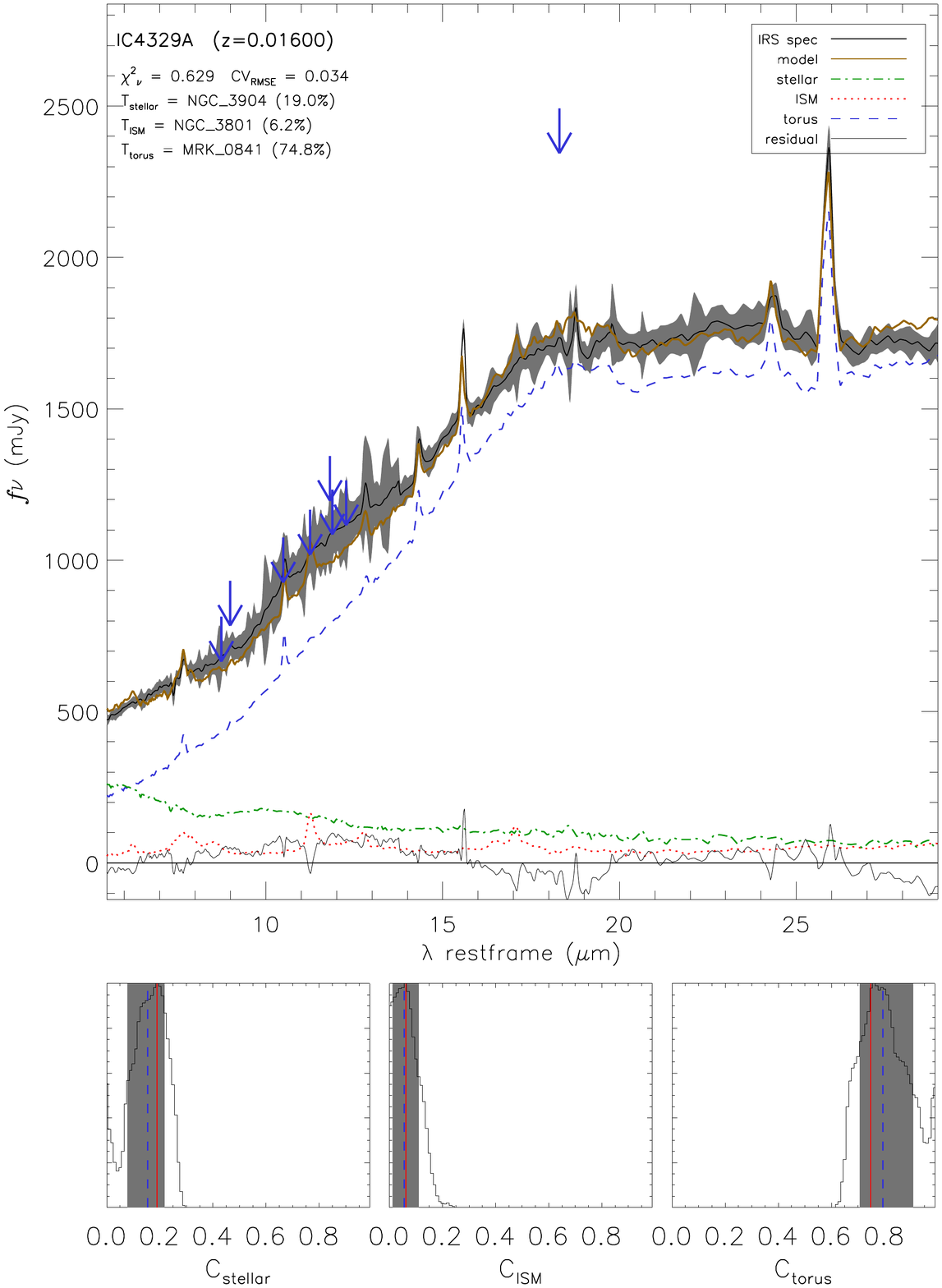}
\caption{...continued.}
\label{fig:CatSpectra}
\end{center}
\end{figure*}

\begin{figure*}
\begin{center}
\includegraphics[width=0.45\columnwidth]{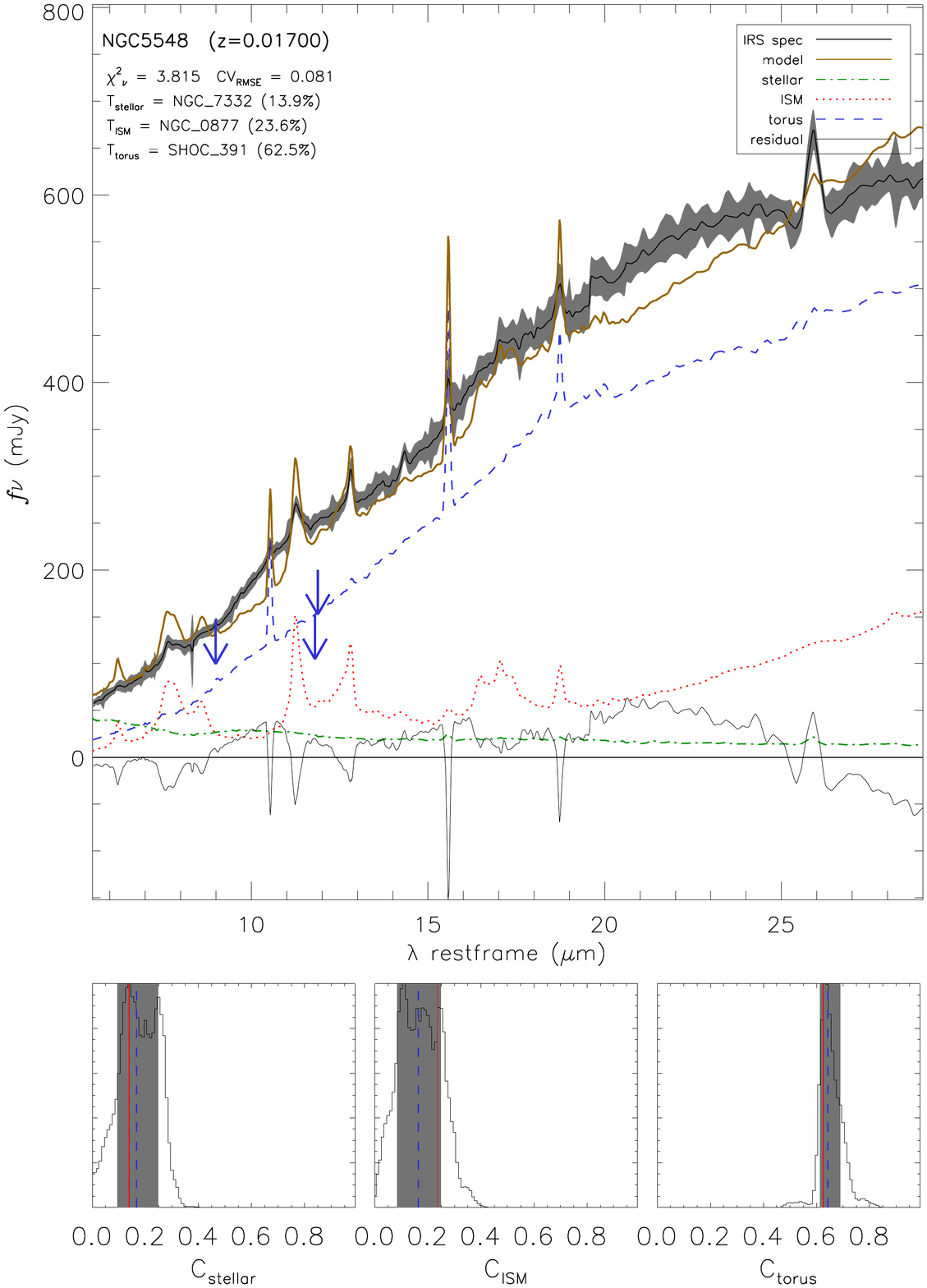}
\includegraphics[width=0.45\columnwidth]{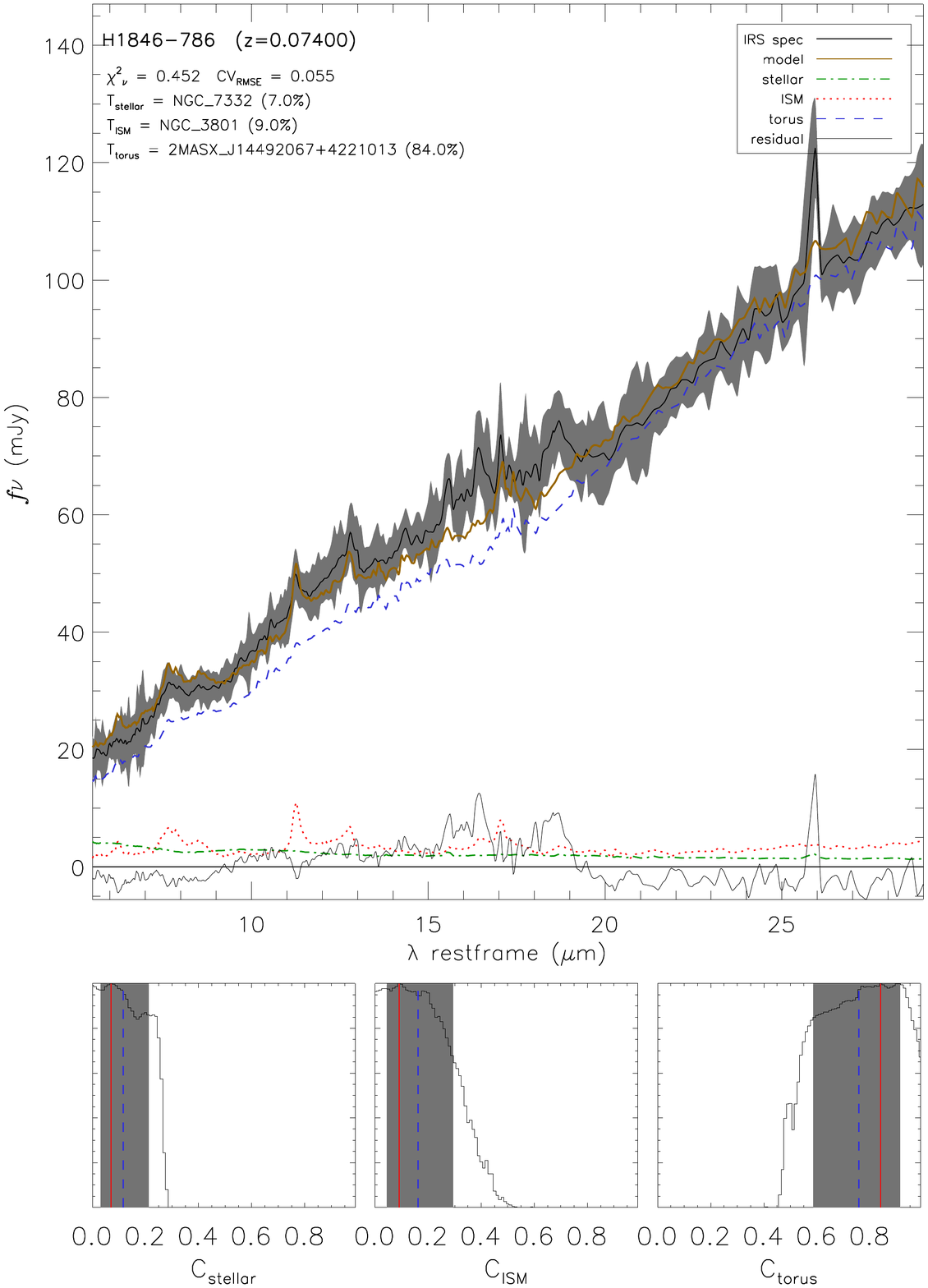}
\includegraphics[width=0.45\columnwidth]{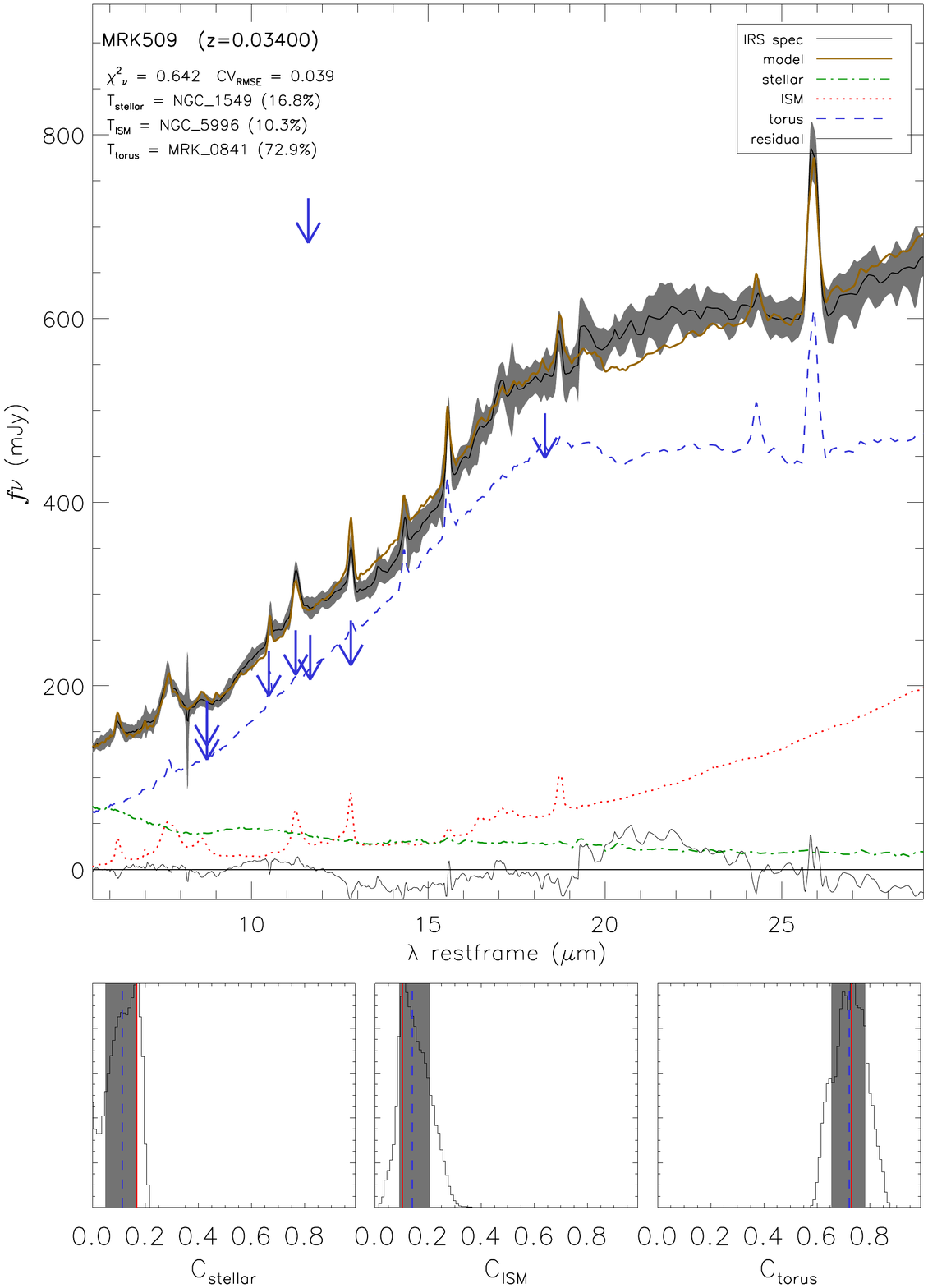}
\includegraphics[width=0.45\columnwidth]{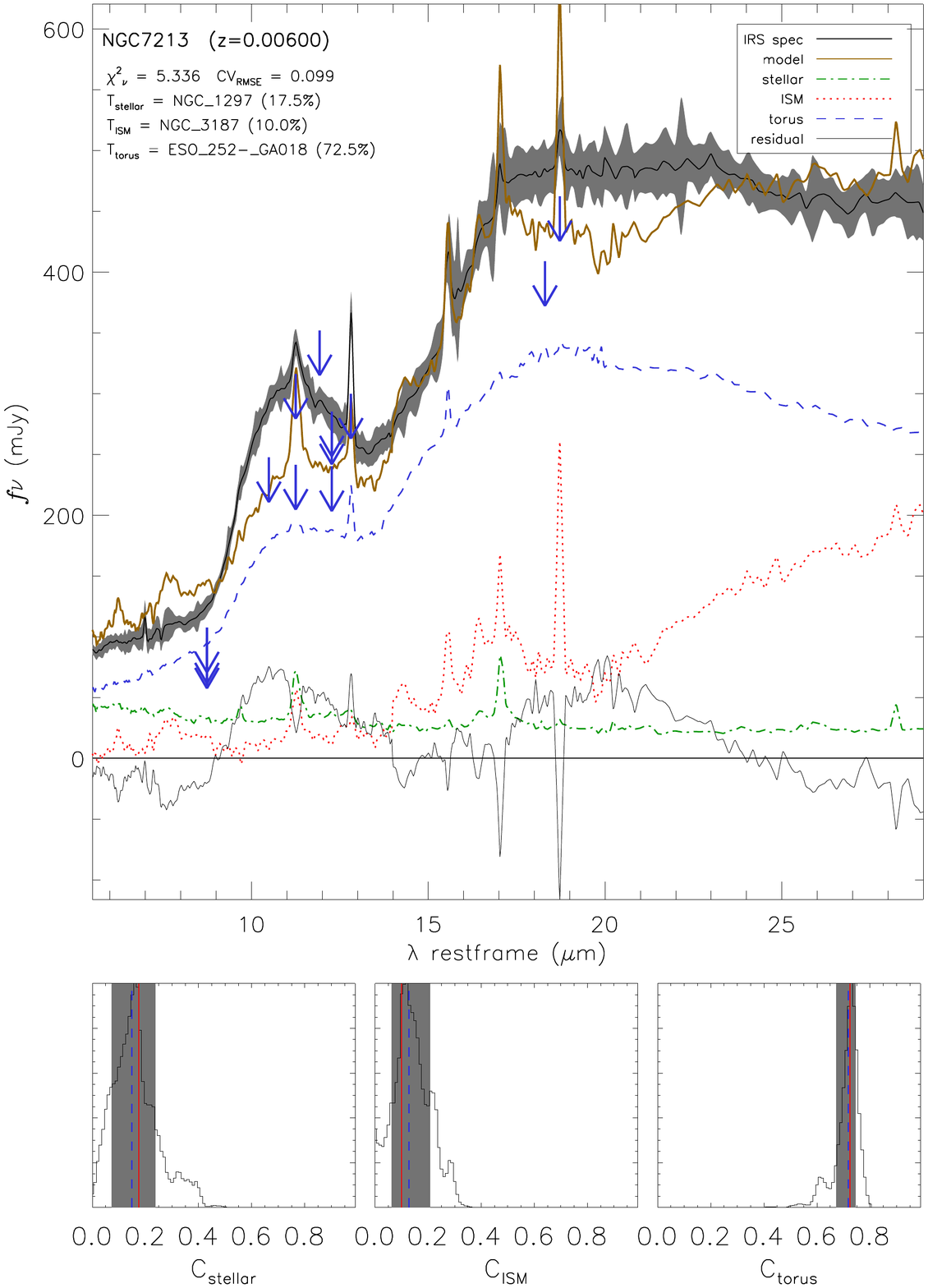}
\caption{...continued.}
\label{fig:CatSpectra}
\end{center}
\end{figure*}

\begin{figure*}
\begin{center}
\includegraphics[width=0.45\columnwidth]{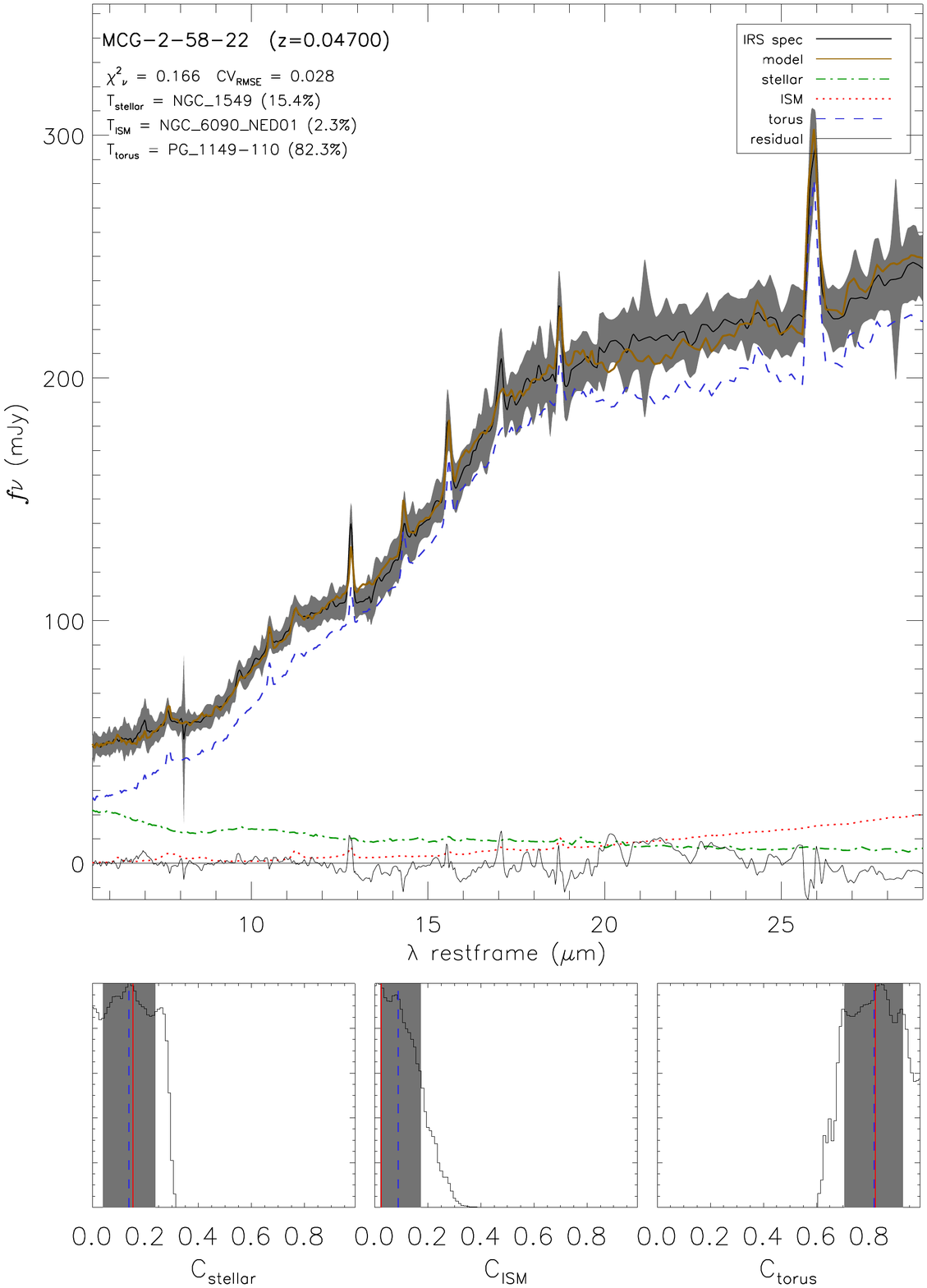}
\includegraphics[width=0.45\columnwidth]{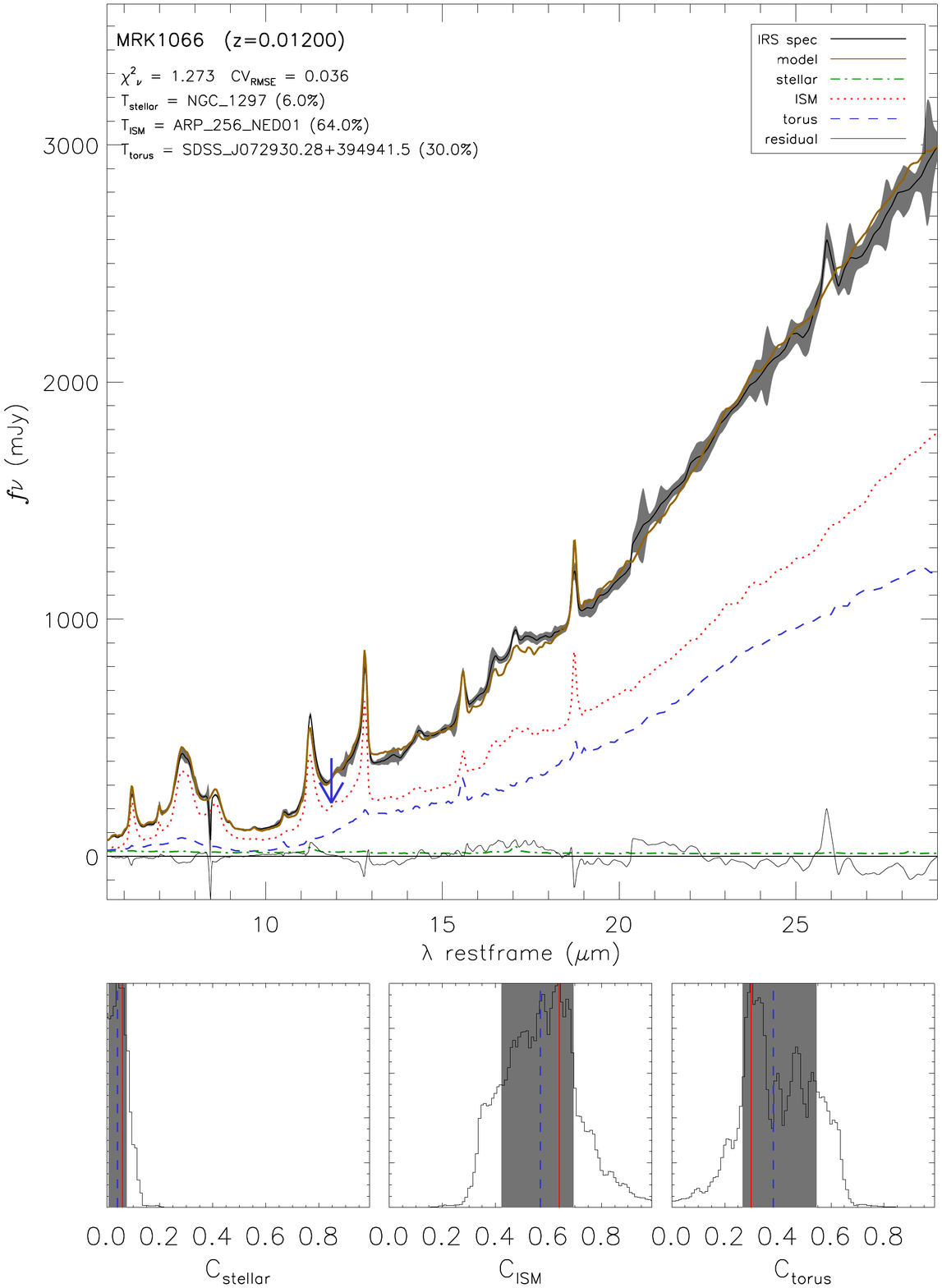}
\includegraphics[width=0.45\columnwidth]{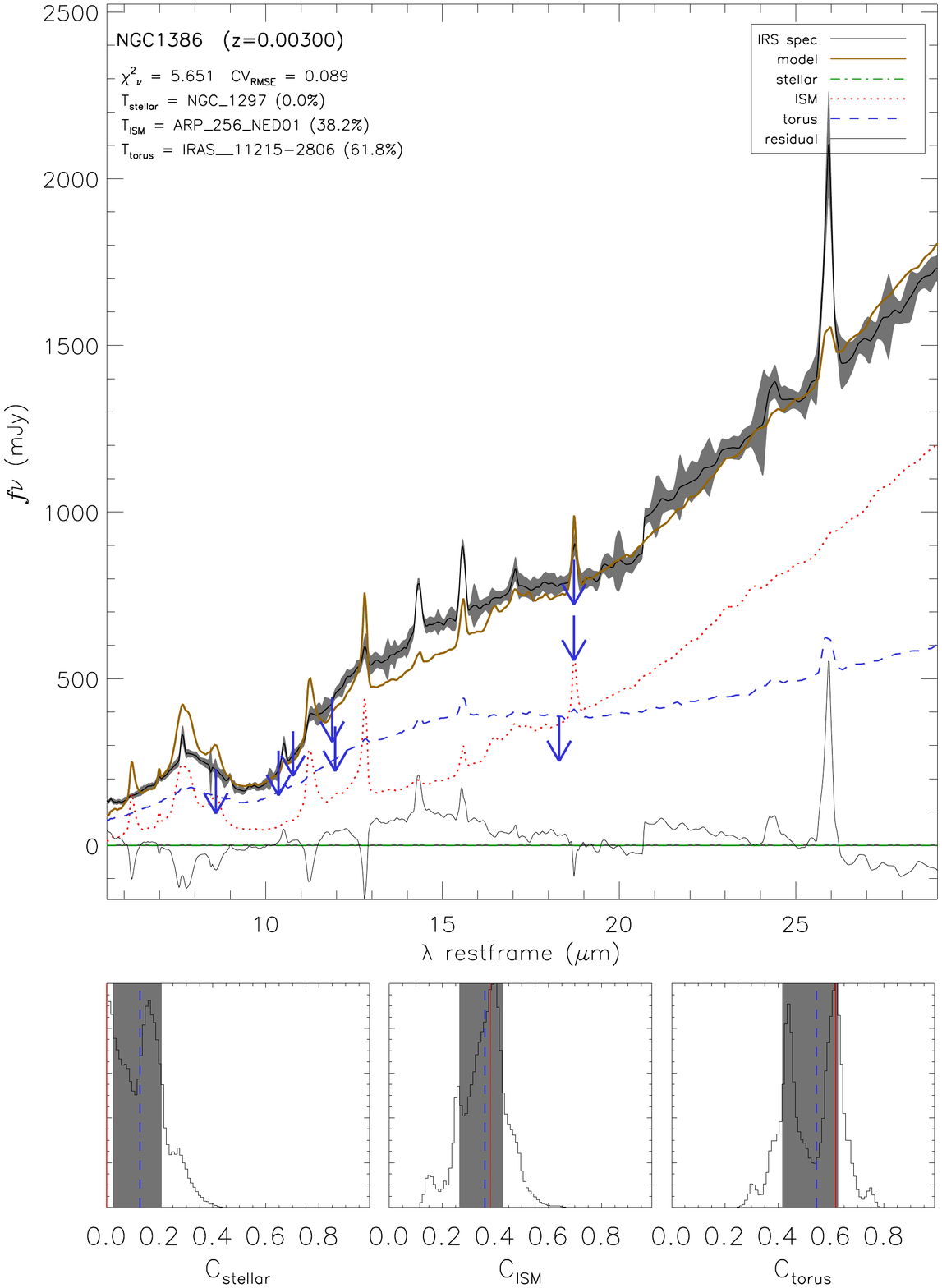}
\includegraphics[width=0.45\columnwidth]{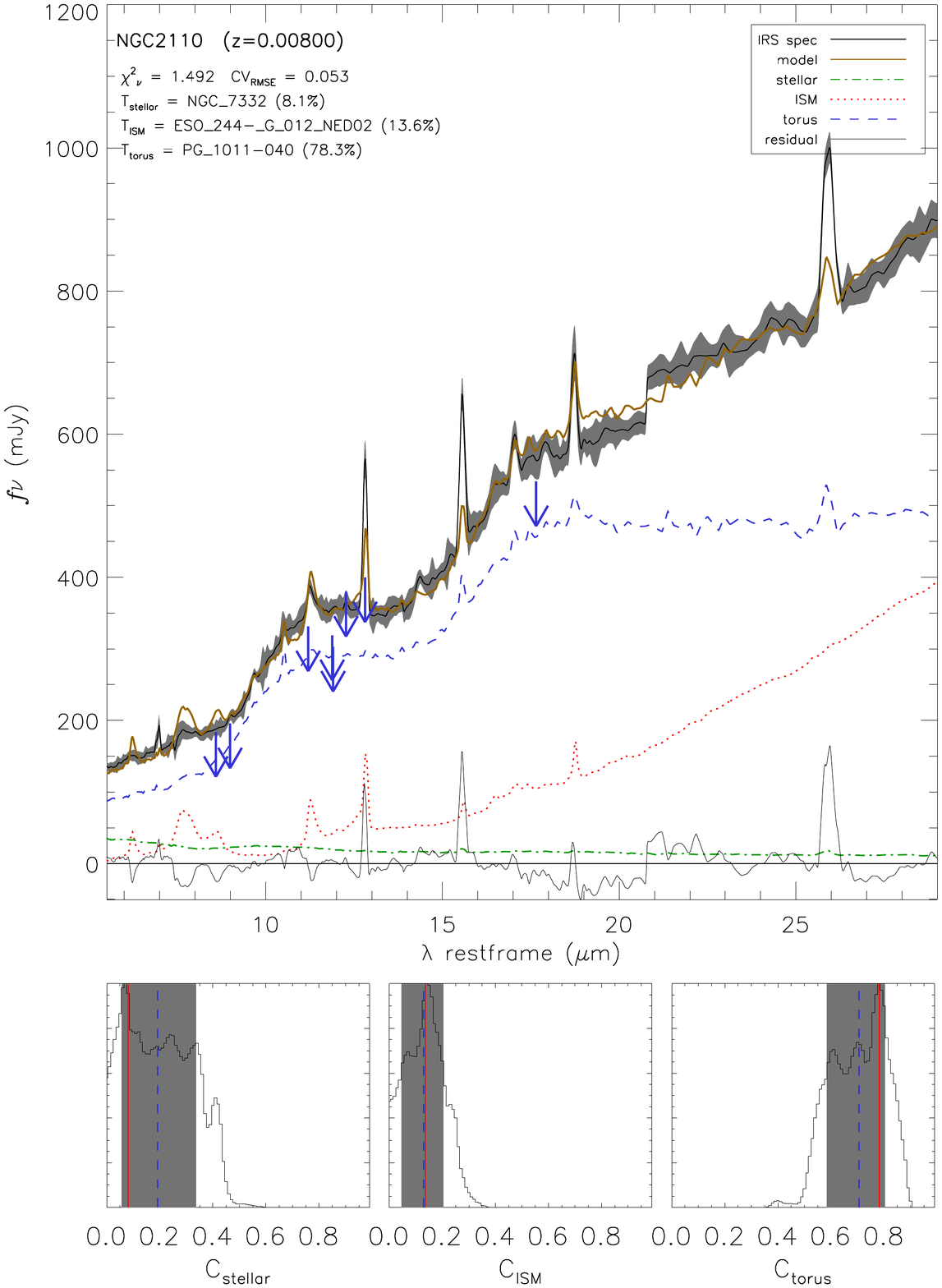}
\caption{...continued.}
\label{fig:CatSpectra}
\end{center}
\end{figure*}

\begin{figure*}
\begin{center}
\includegraphics[width=0.45\columnwidth]{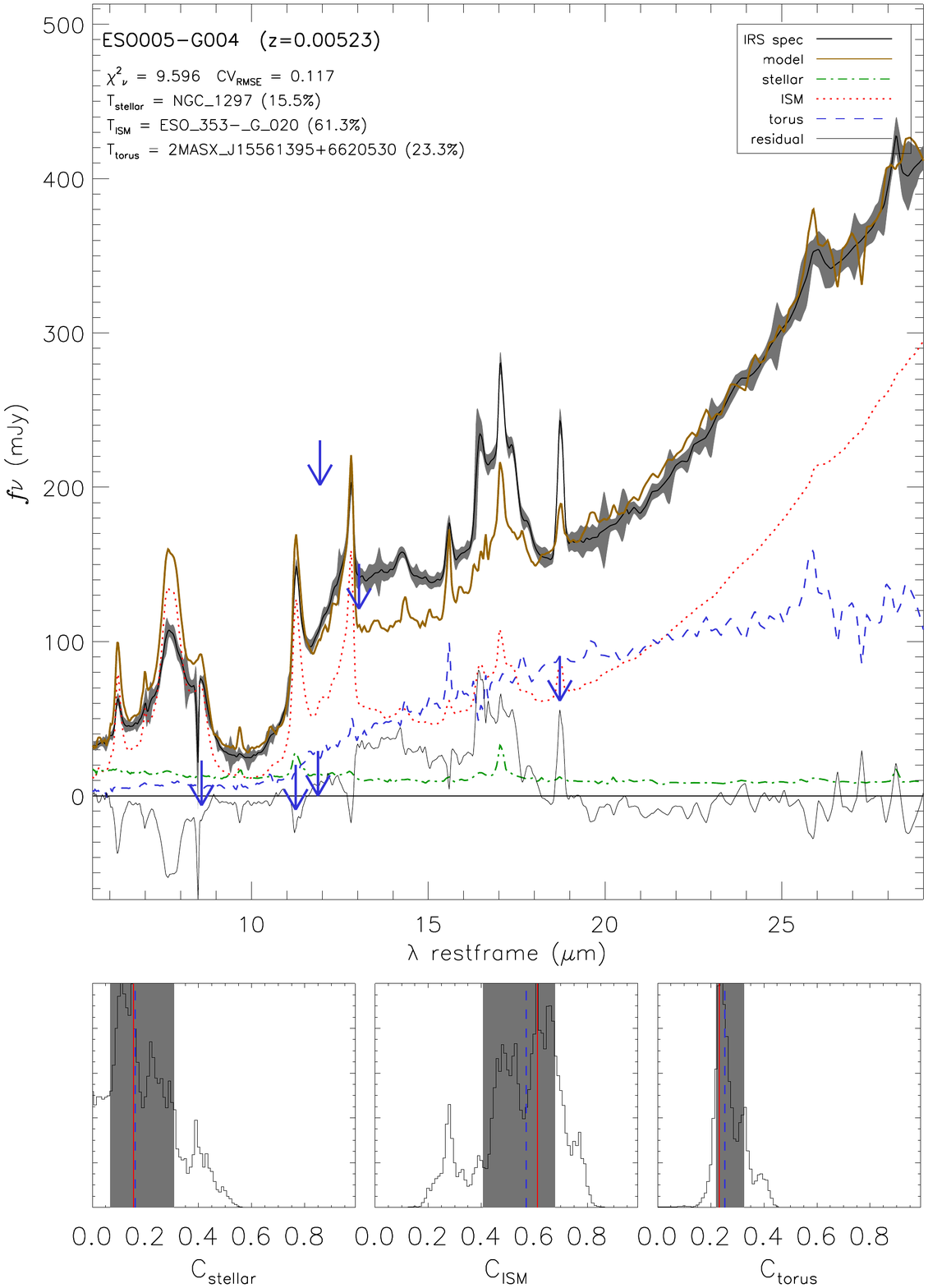}
\includegraphics[width=0.45\columnwidth]{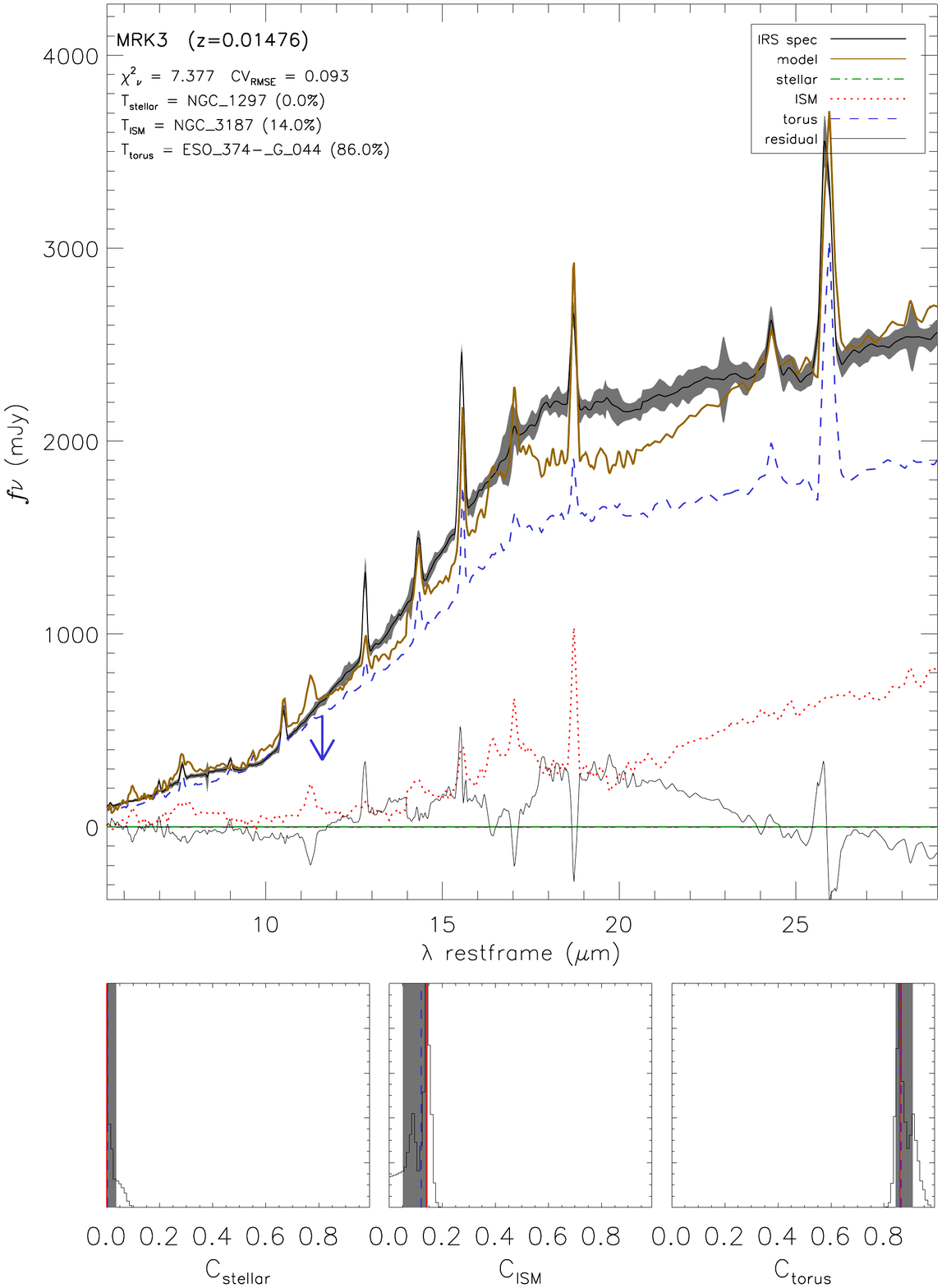}
\includegraphics[width=0.45\columnwidth]{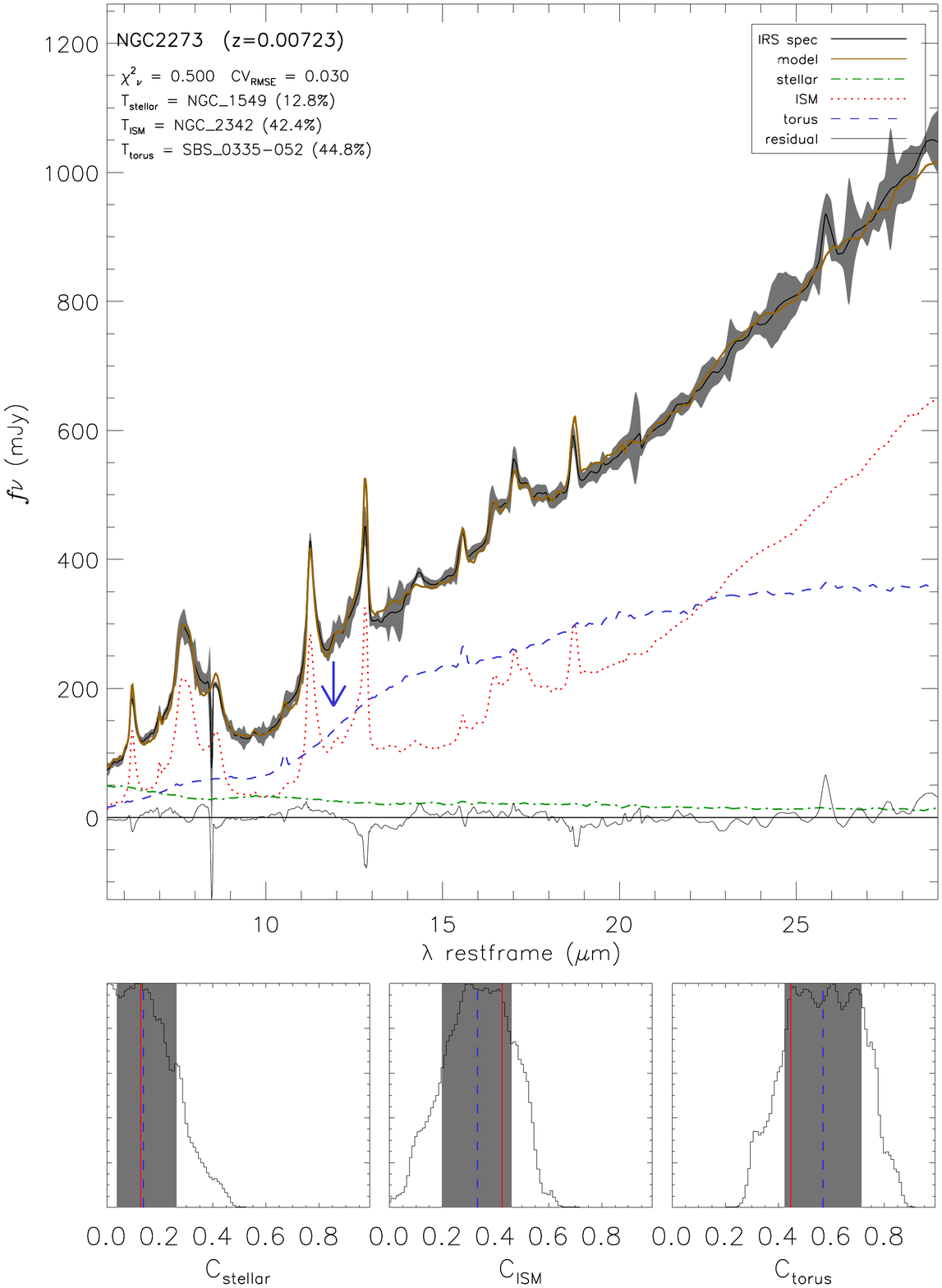}
\includegraphics[width=0.45\columnwidth]{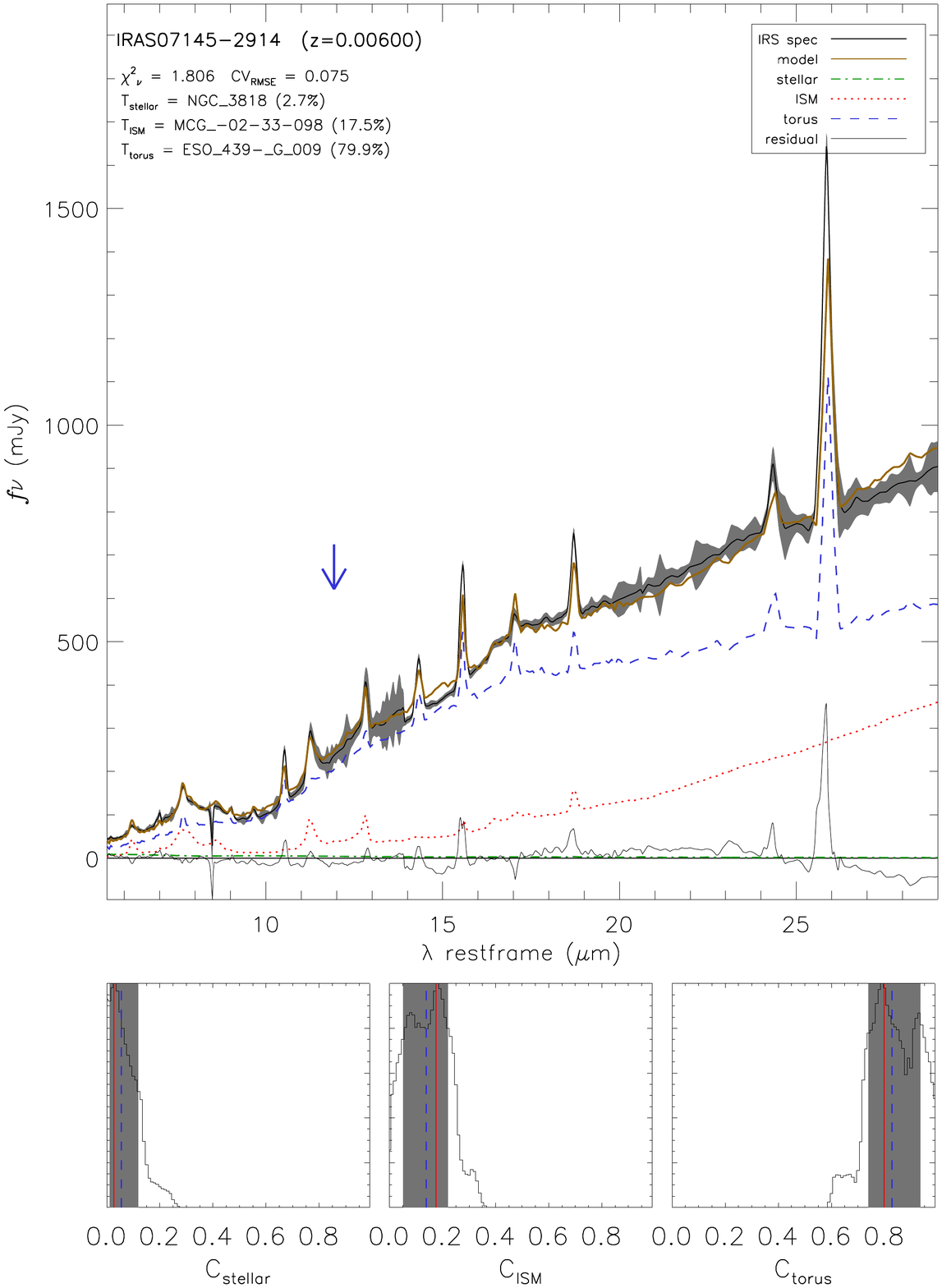}
\caption{...continued.}
\label{fig:CatSpectra}
\end{center}
\end{figure*}

\begin{figure*}
\begin{center}
\includegraphics[width=0.45\columnwidth]{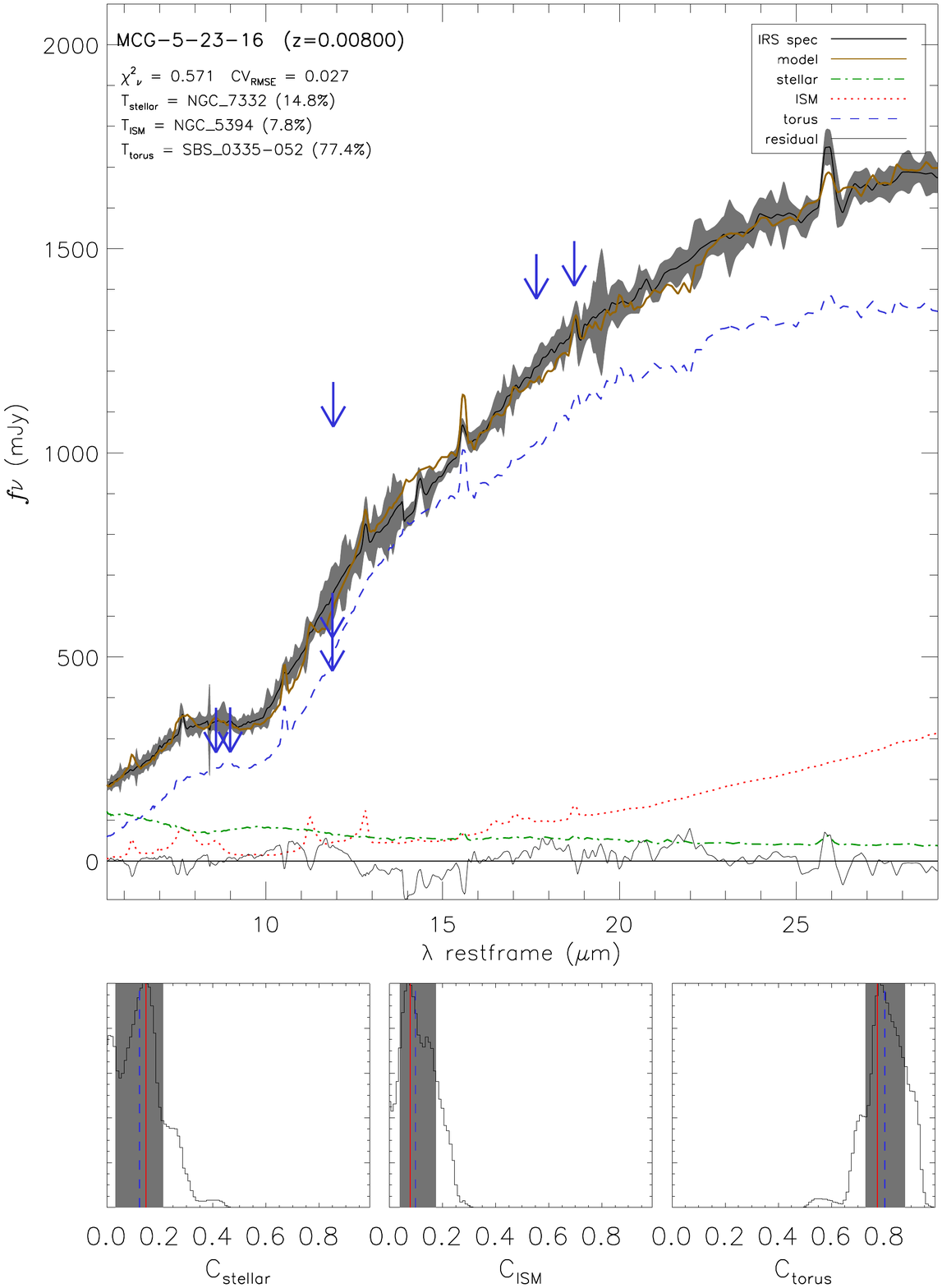}
\includegraphics[width=0.45\columnwidth]{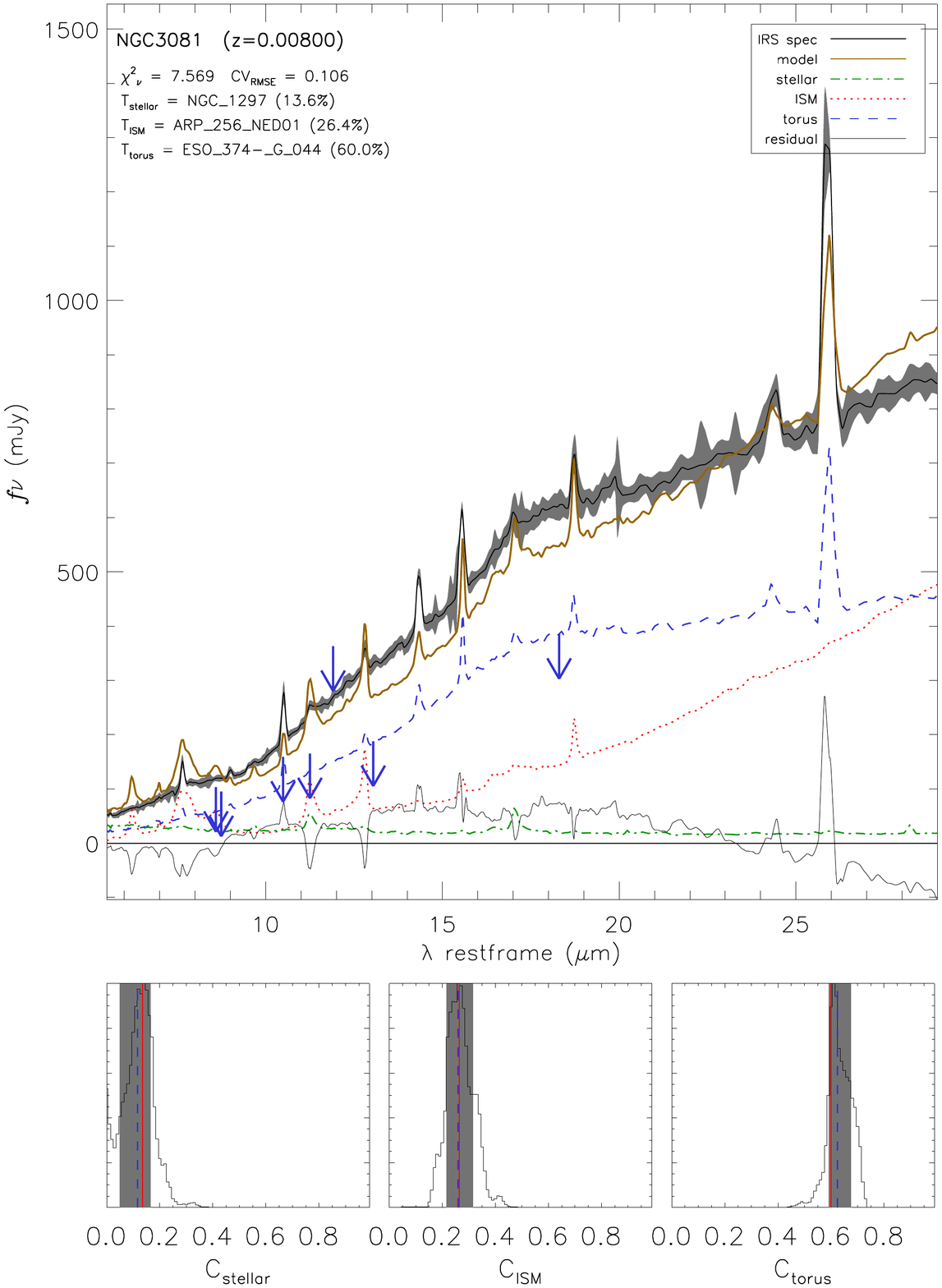}
\includegraphics[width=0.45\columnwidth]{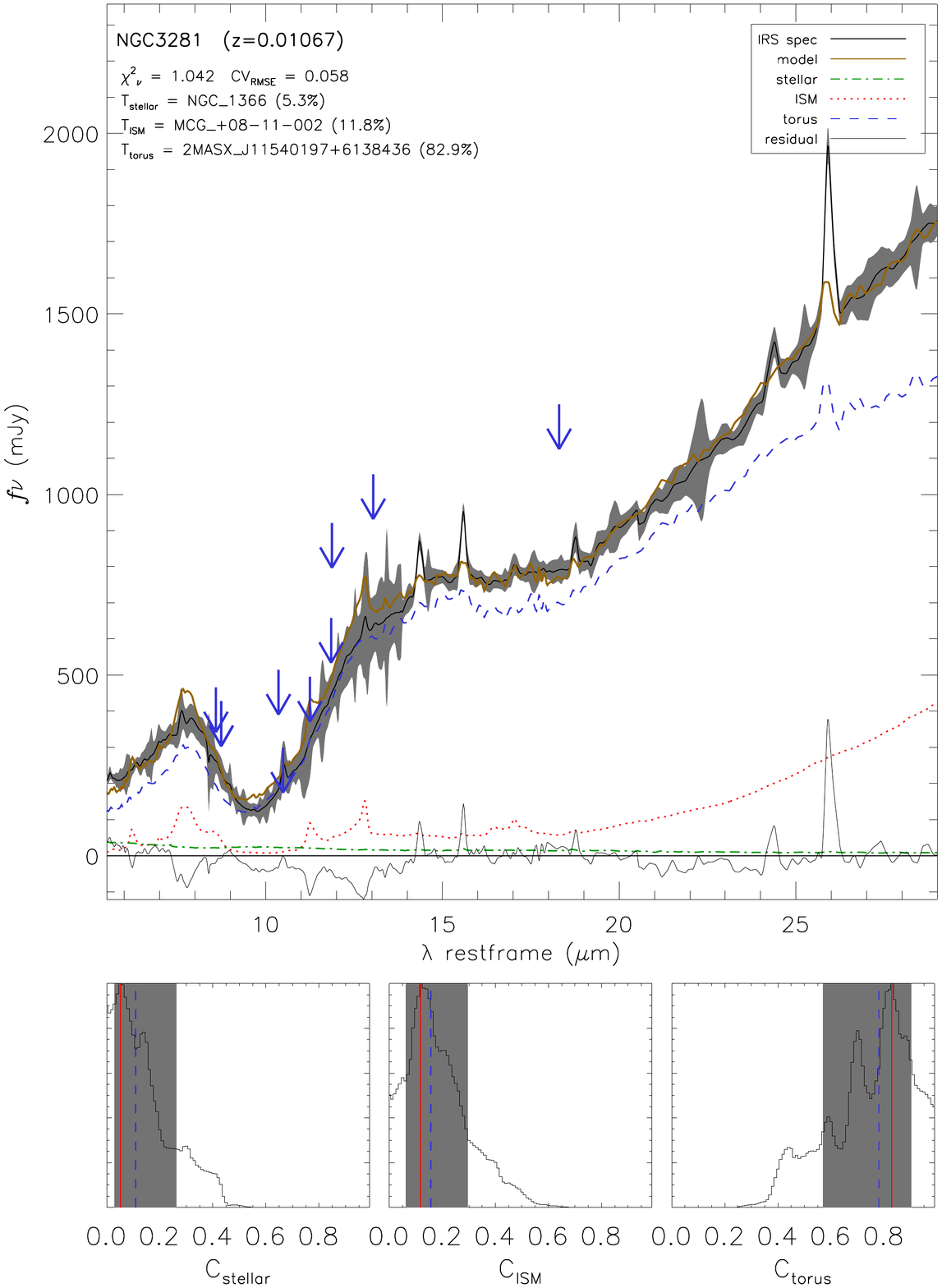}
\includegraphics[width=0.45\columnwidth]{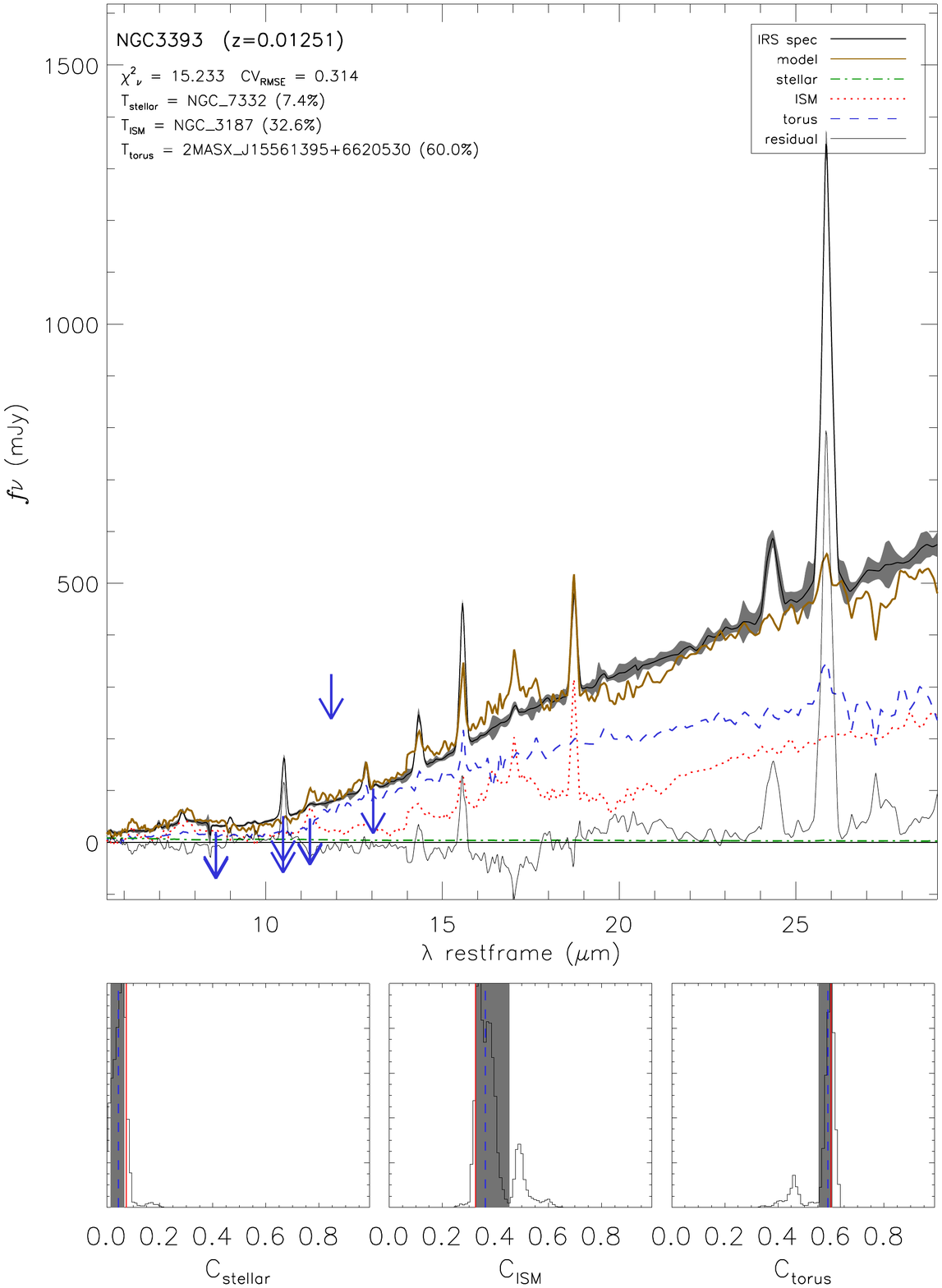}
\caption{...continued.}
\label{fig:CatSpectra}
\end{center}
\end{figure*}

\begin{figure*}
\begin{center}
\includegraphics[width=0.45\columnwidth]{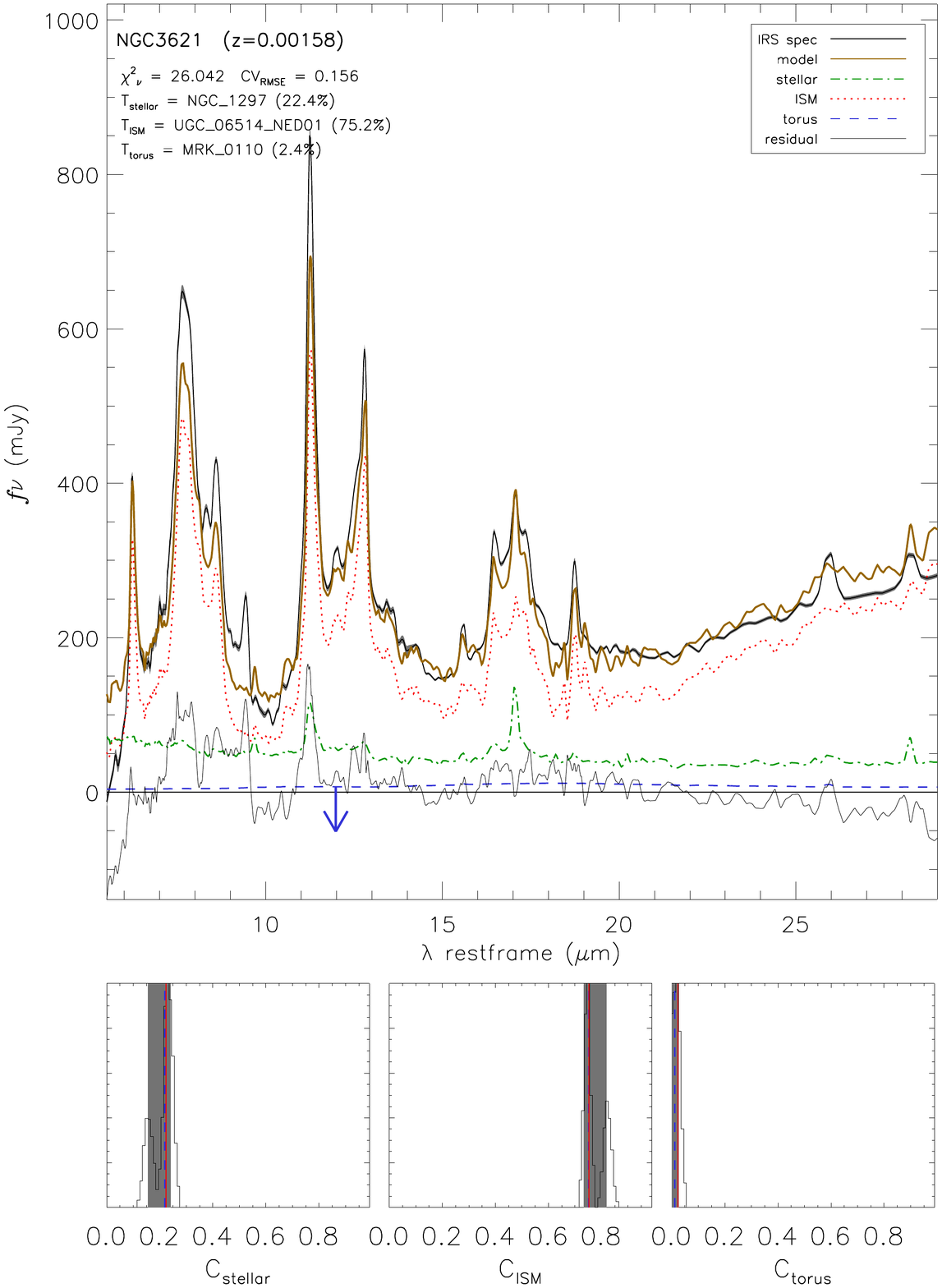}
\includegraphics[width=0.45\columnwidth]{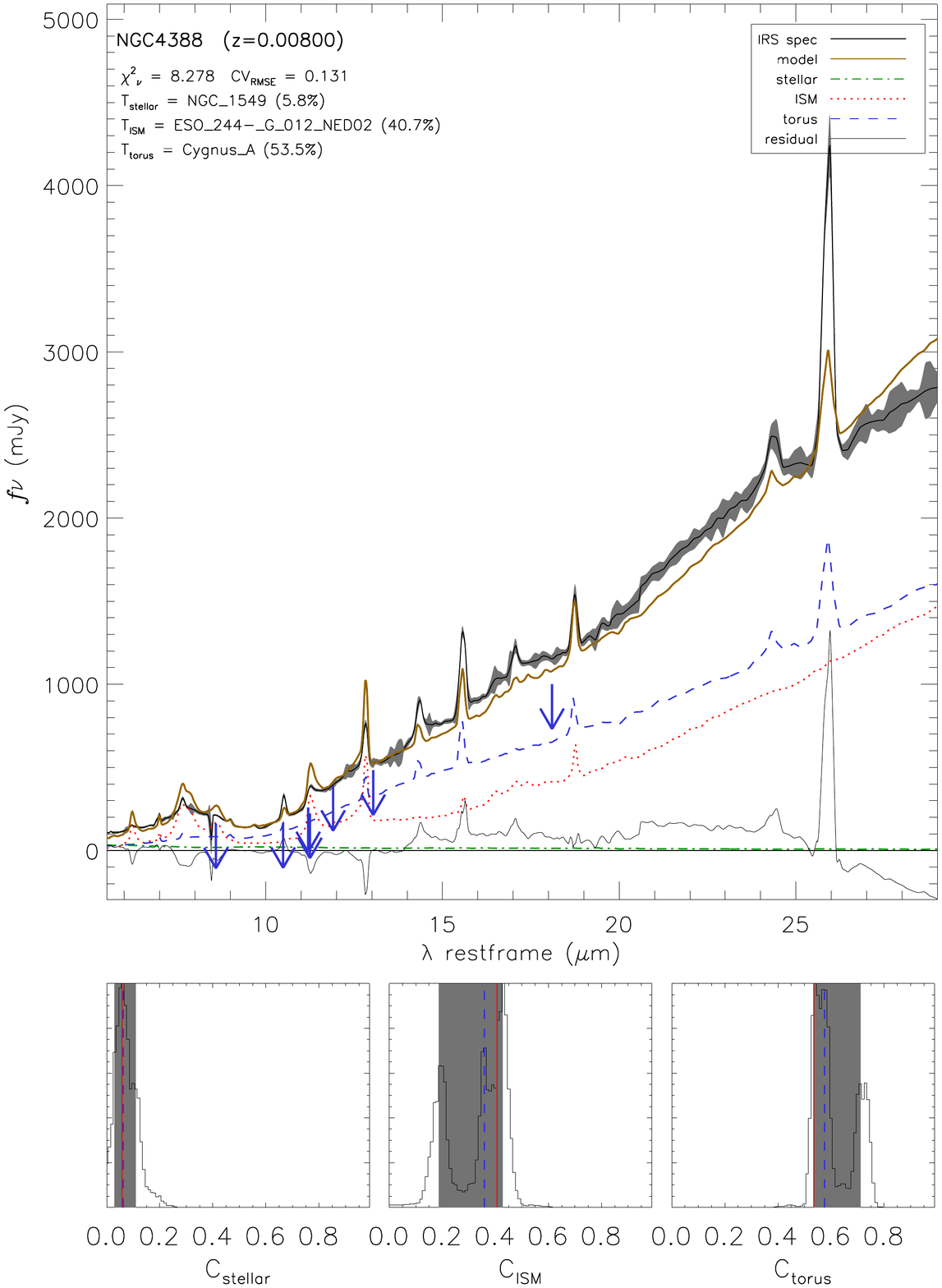}
\includegraphics[width=0.45\columnwidth]{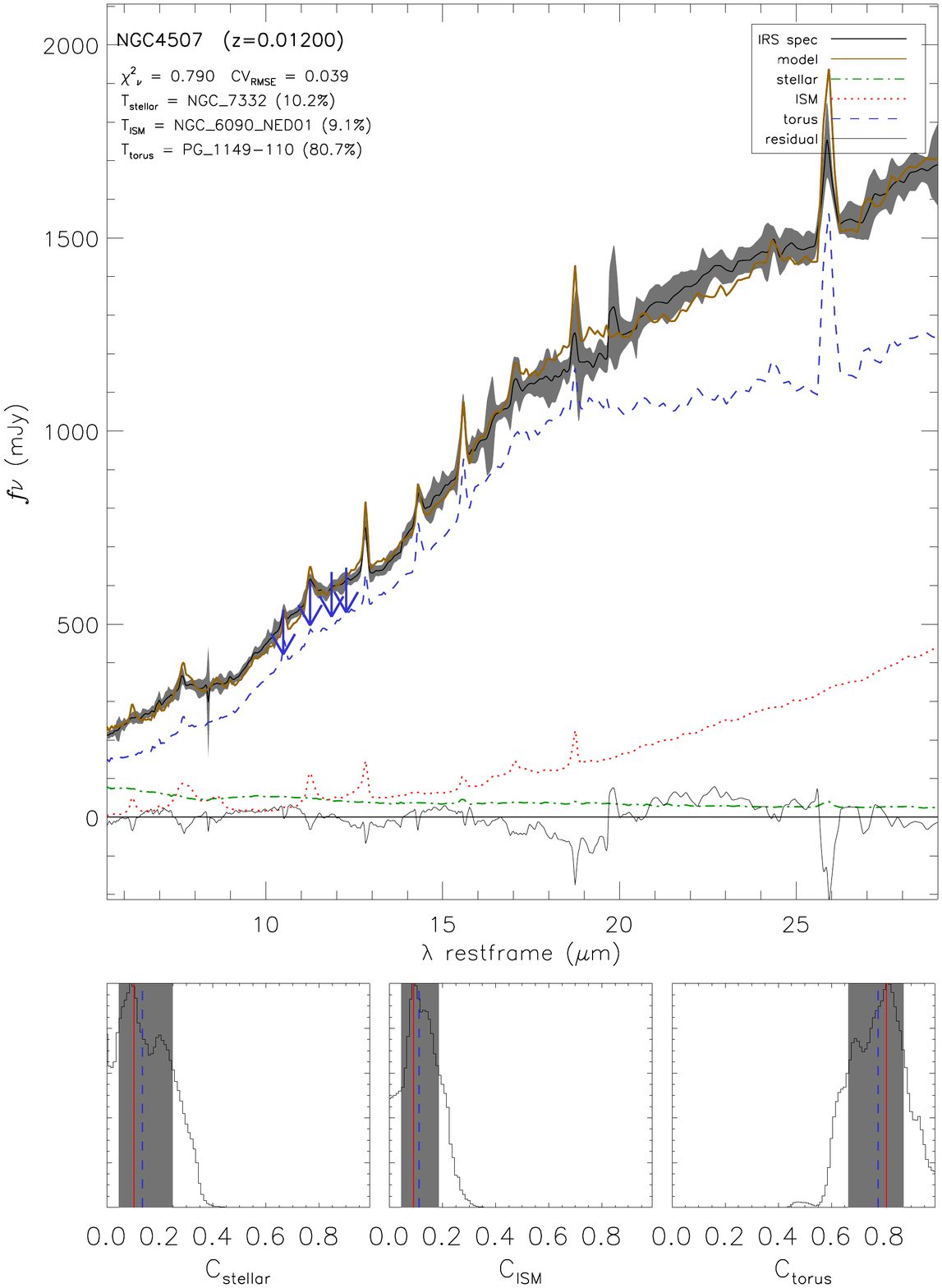}
\includegraphics[width=0.45\columnwidth]{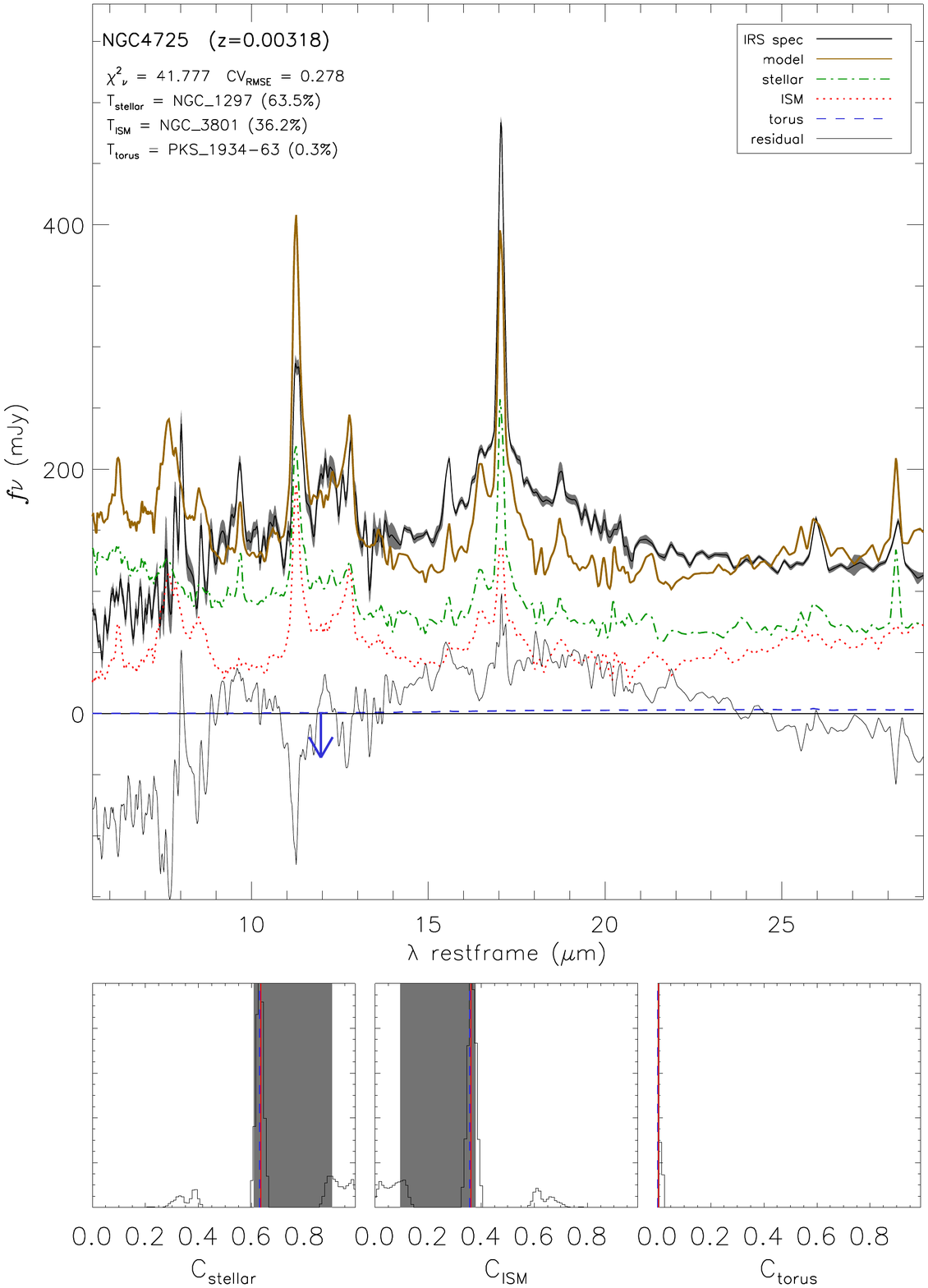}
\caption{...continued.}
\label{fig:CatSpectra}
\end{center}
\end{figure*}
\clearpage

\begin{figure*}
\begin{center}
\includegraphics[width=0.45\columnwidth]{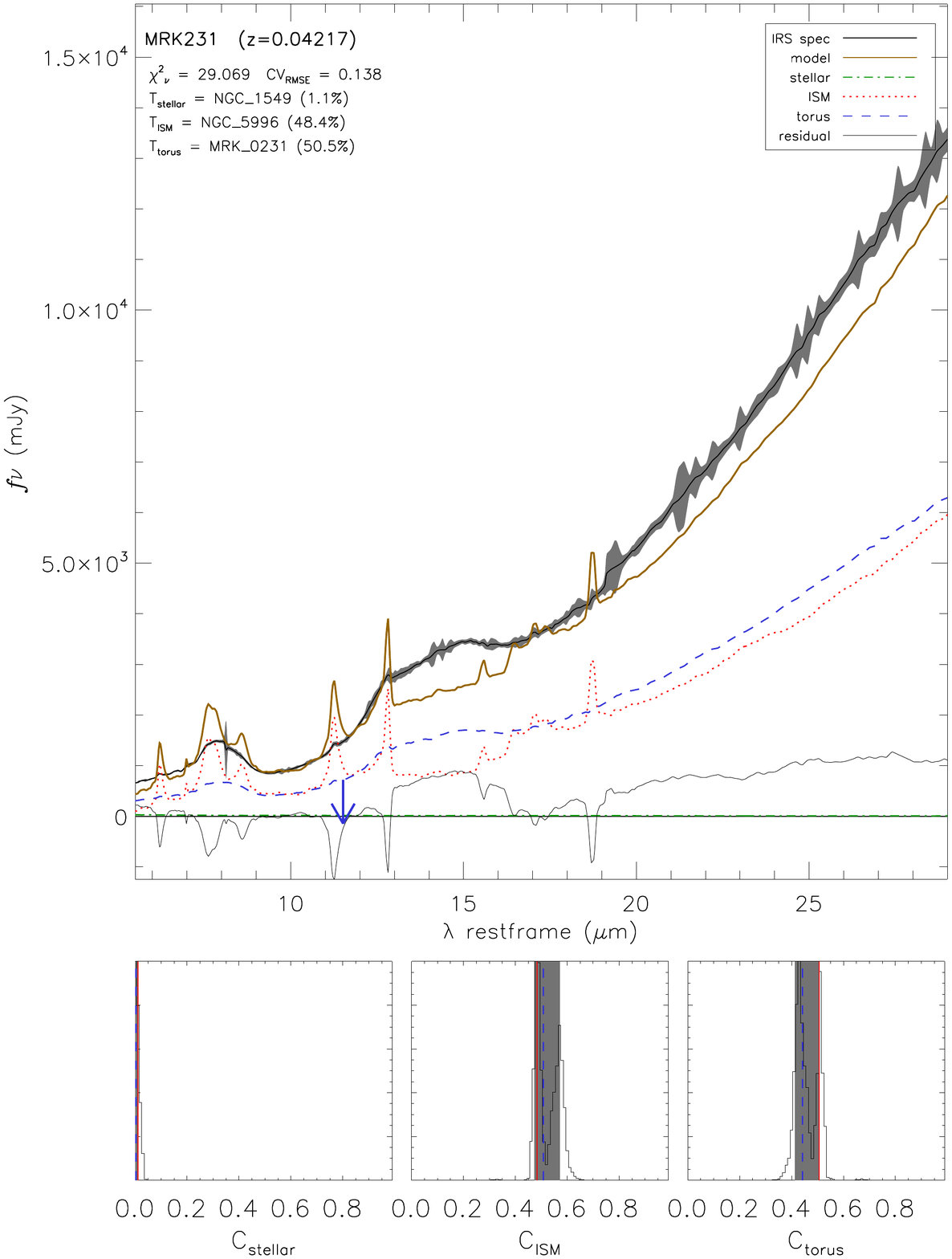}
\includegraphics[width=0.45\columnwidth]{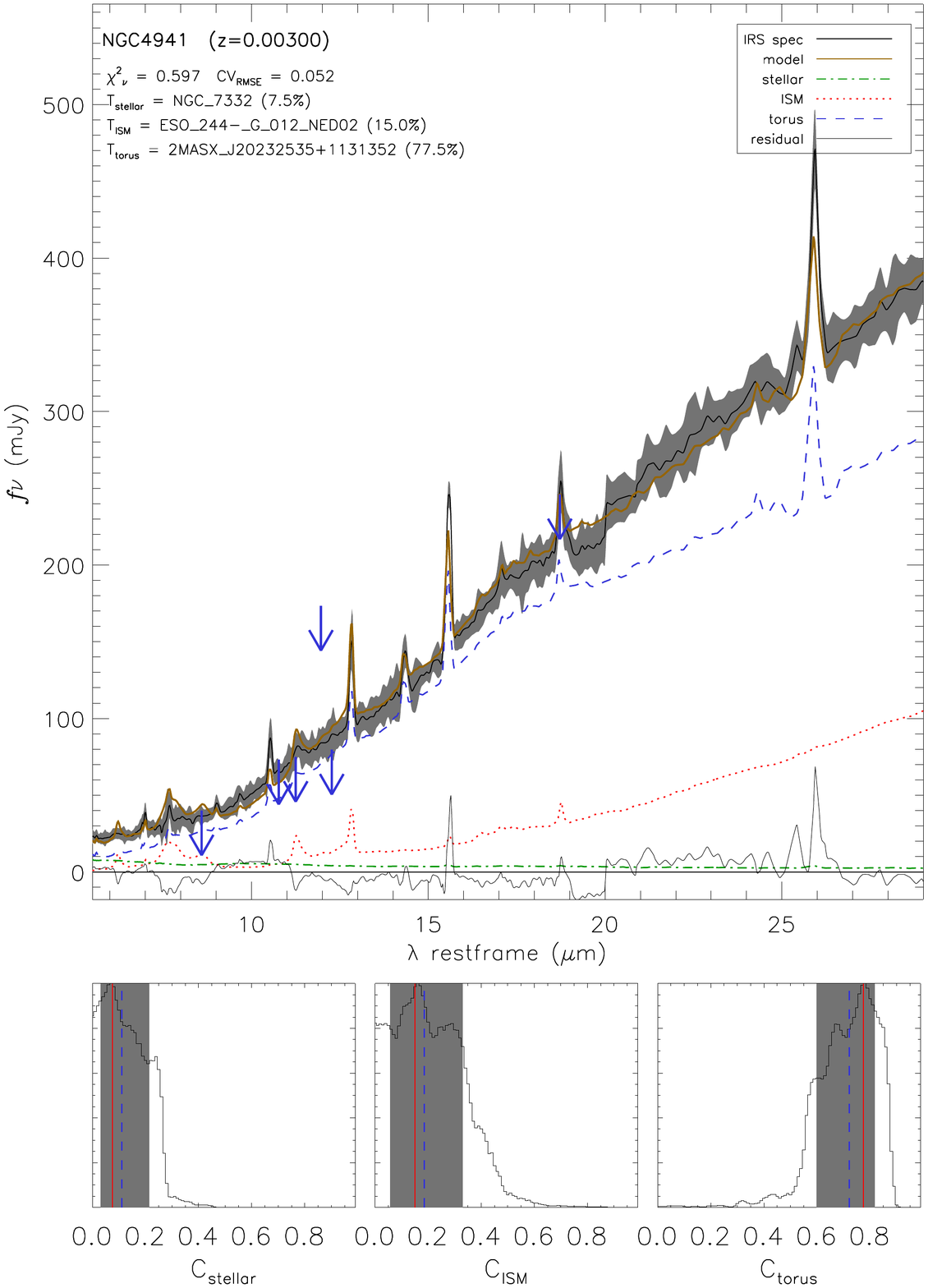}
\includegraphics[width=0.45\columnwidth]{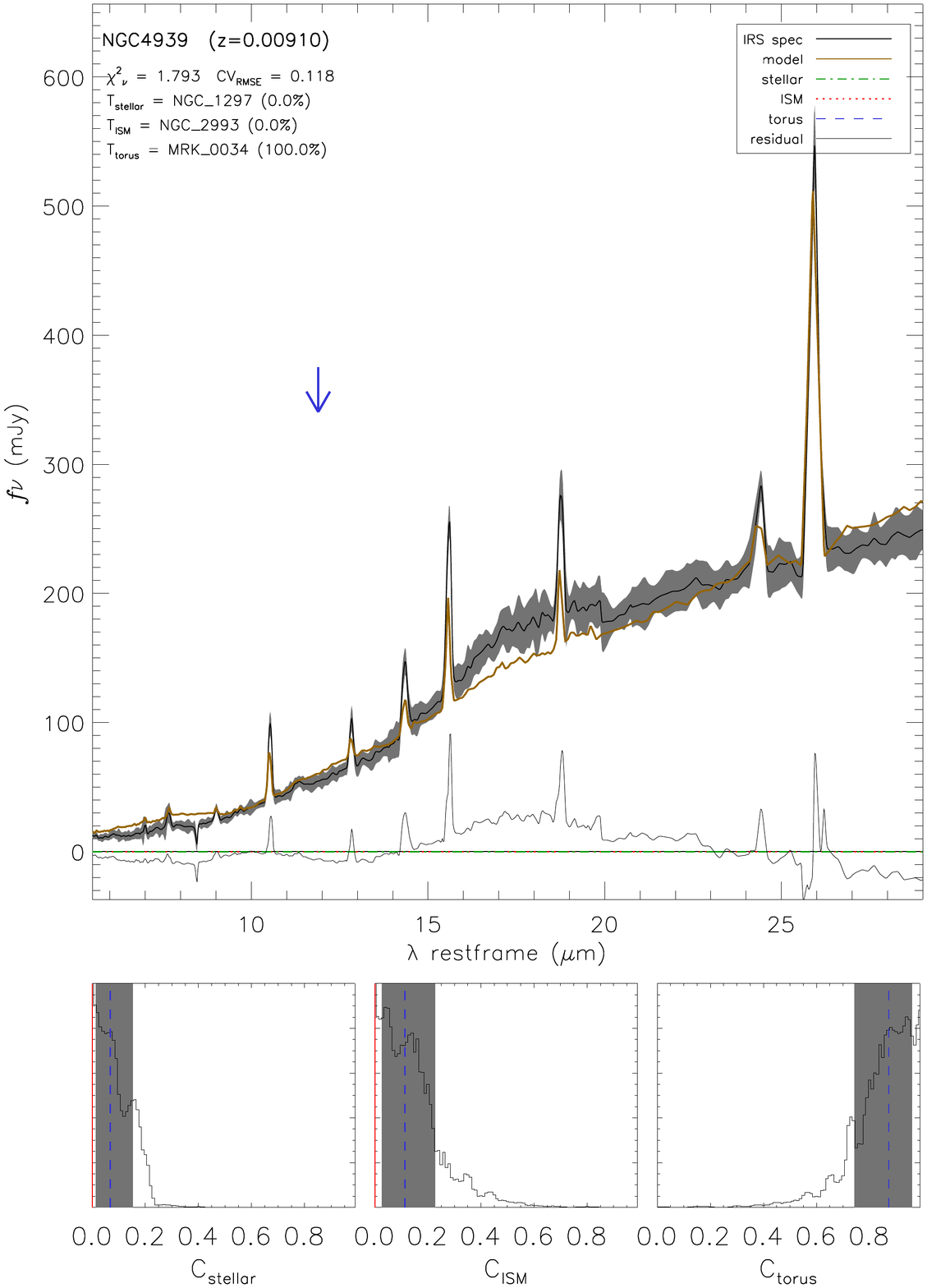}
\includegraphics[width=0.45\columnwidth]{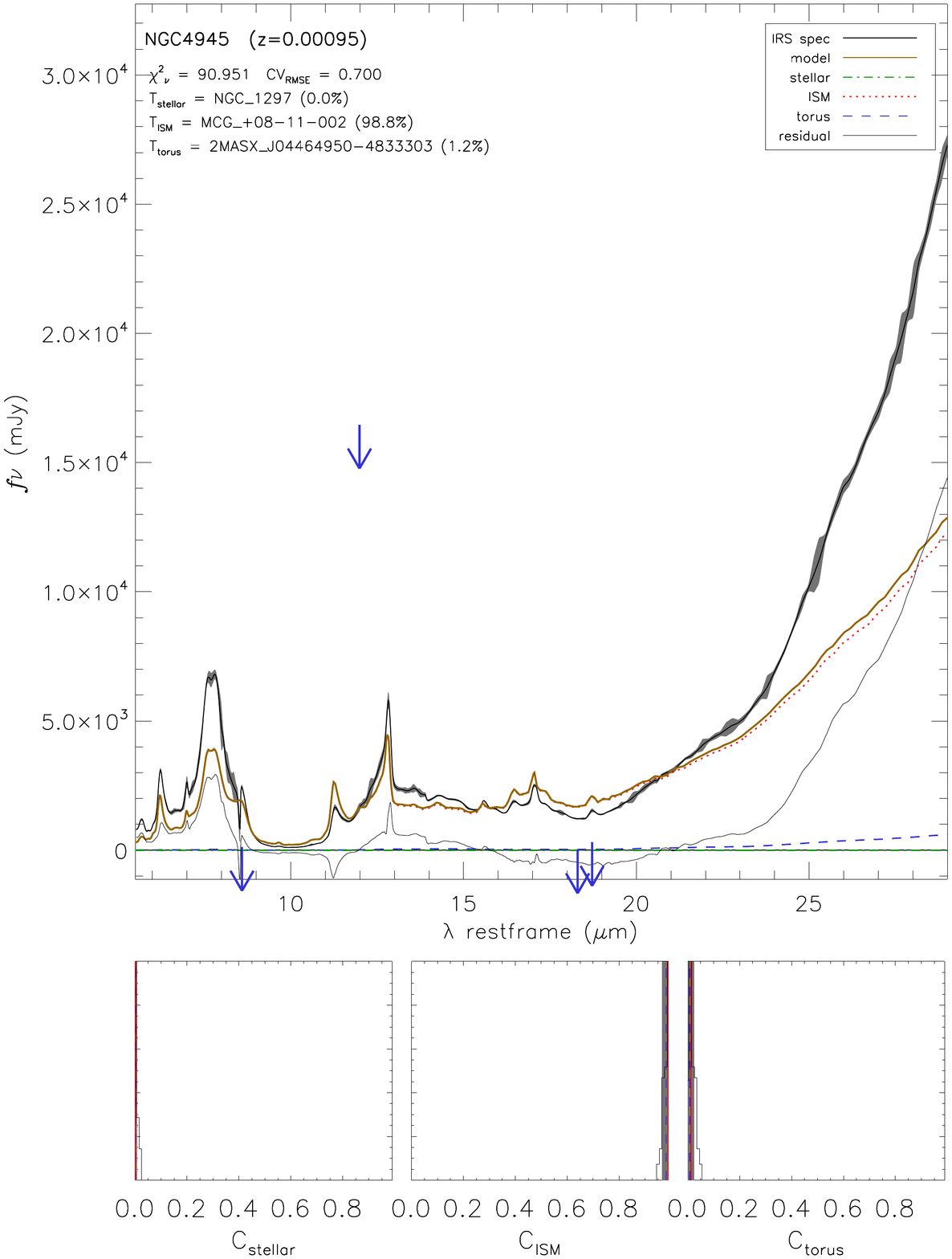}
\caption{...continued.}
\label{fig:CatSpectra}
\end{center}
\end{figure*}

\begin{figure*}
\begin{center}
\includegraphics[width=0.45\columnwidth]{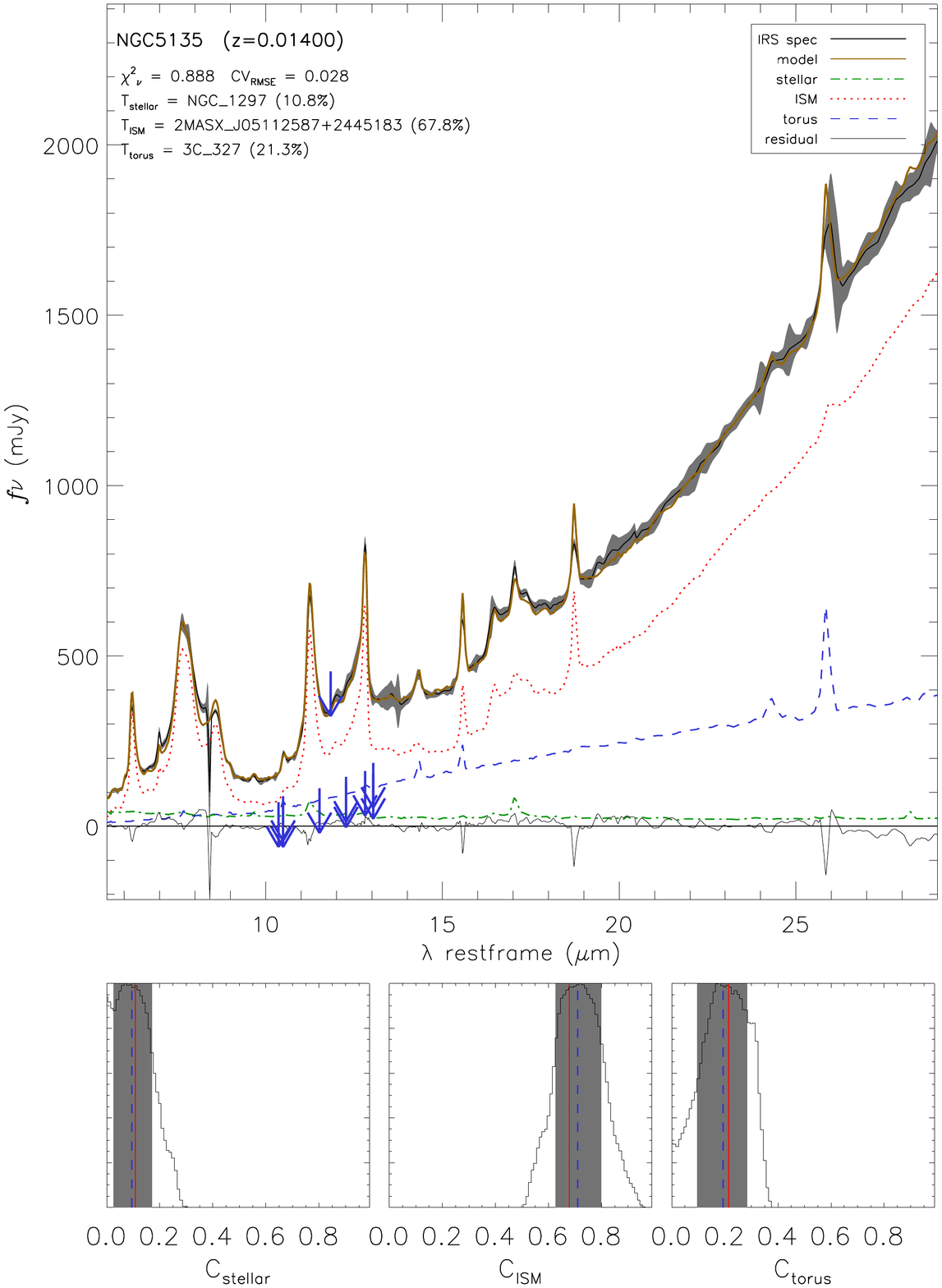}
\includegraphics[width=0.45\columnwidth]{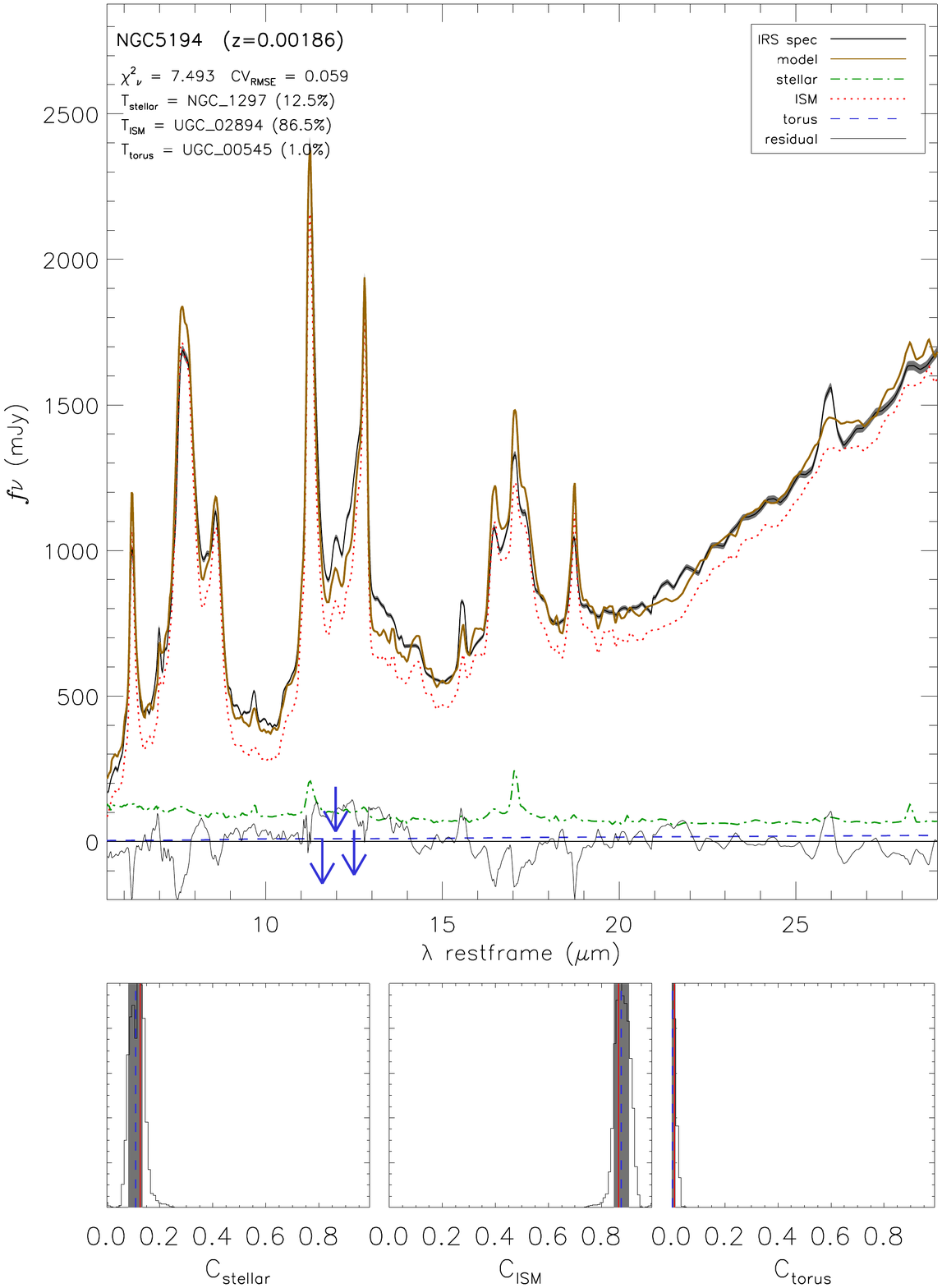}
\includegraphics[width=0.45\columnwidth]{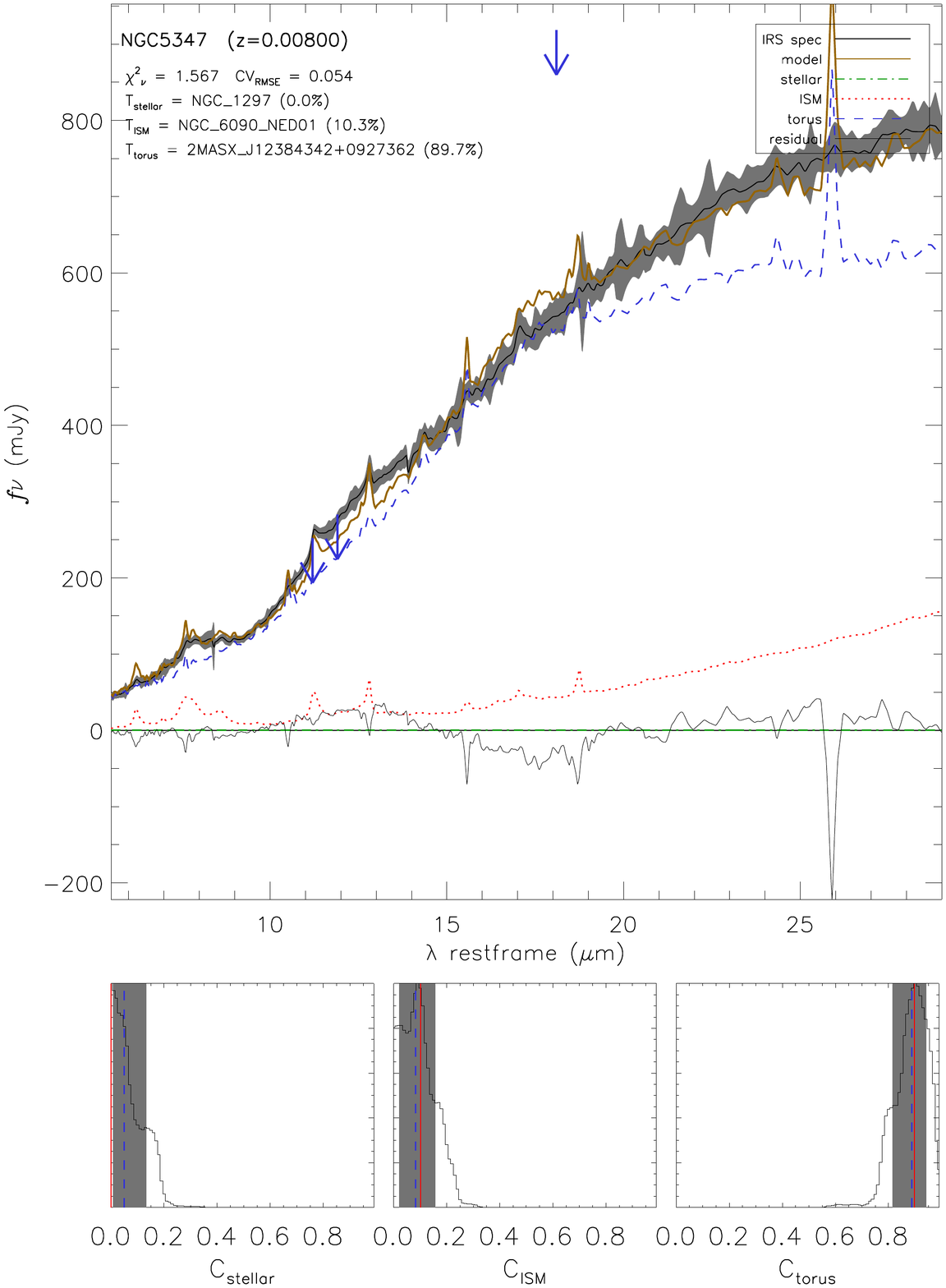}
\includegraphics[width=0.45\columnwidth]{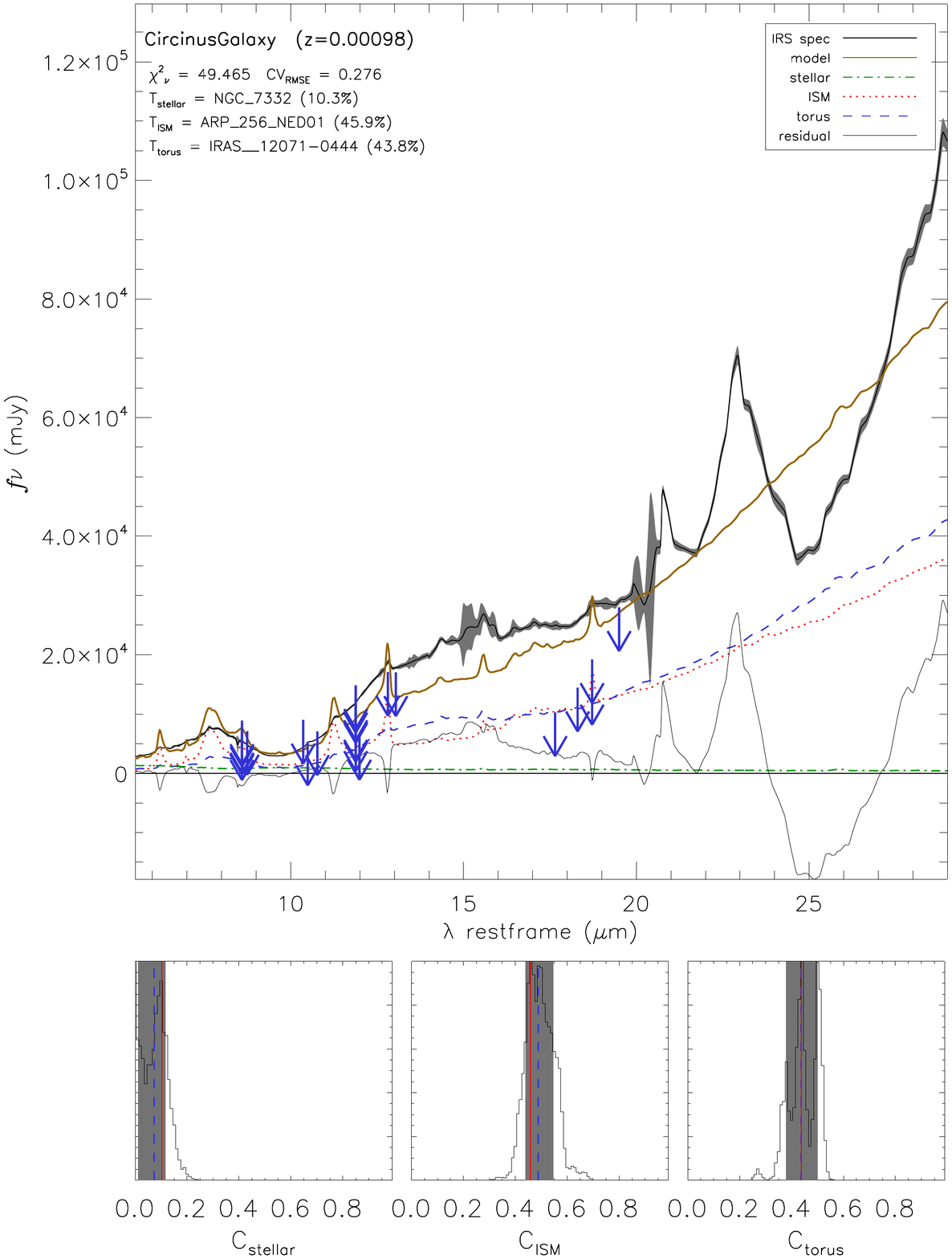}
\caption{...continued.}
\label{fig:CatSpectra}
\end{center}
\end{figure*}

\begin{figure*}
\begin{center}
\includegraphics[width=0.45\columnwidth]{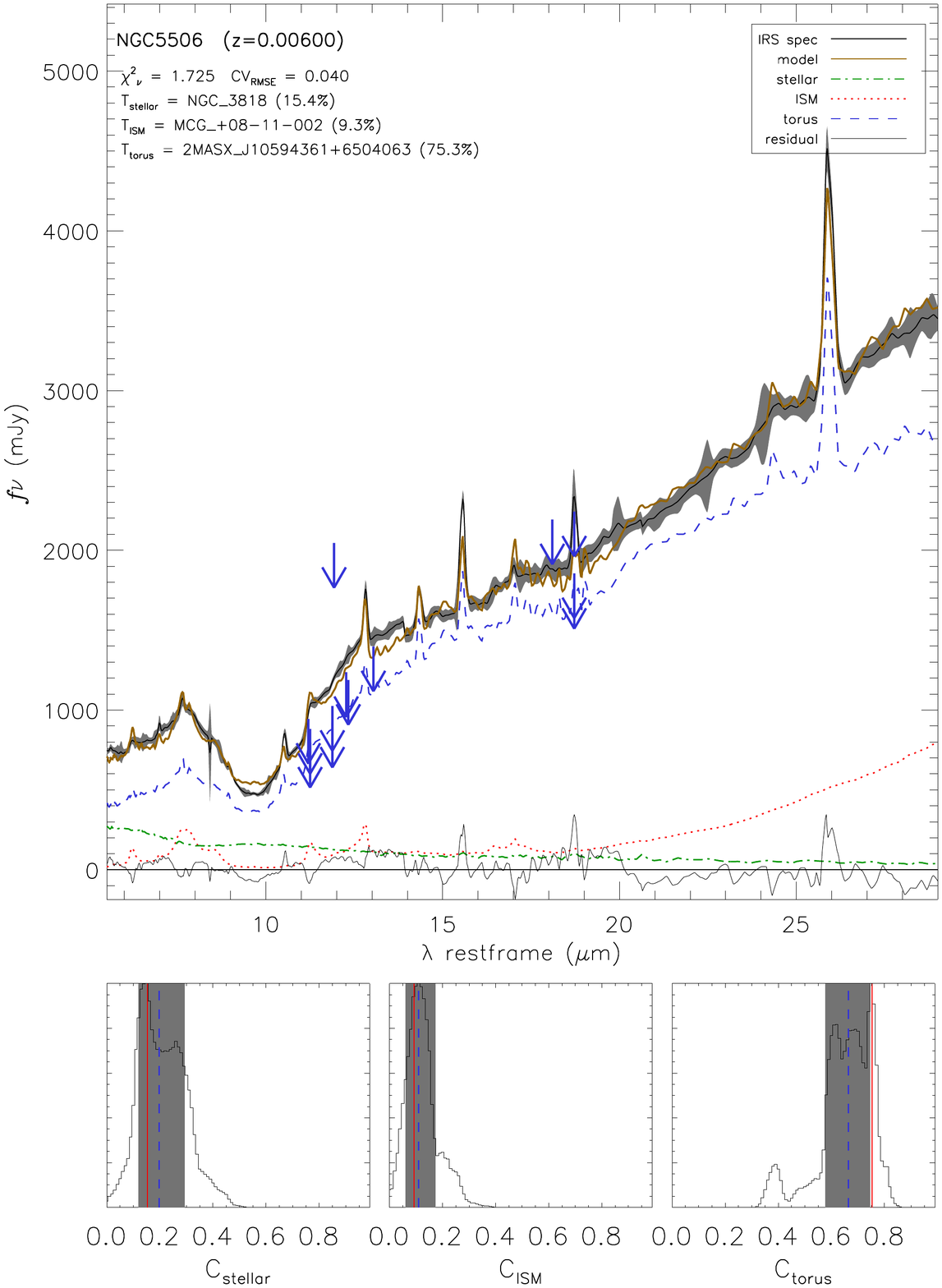}
\includegraphics[width=0.45\columnwidth]{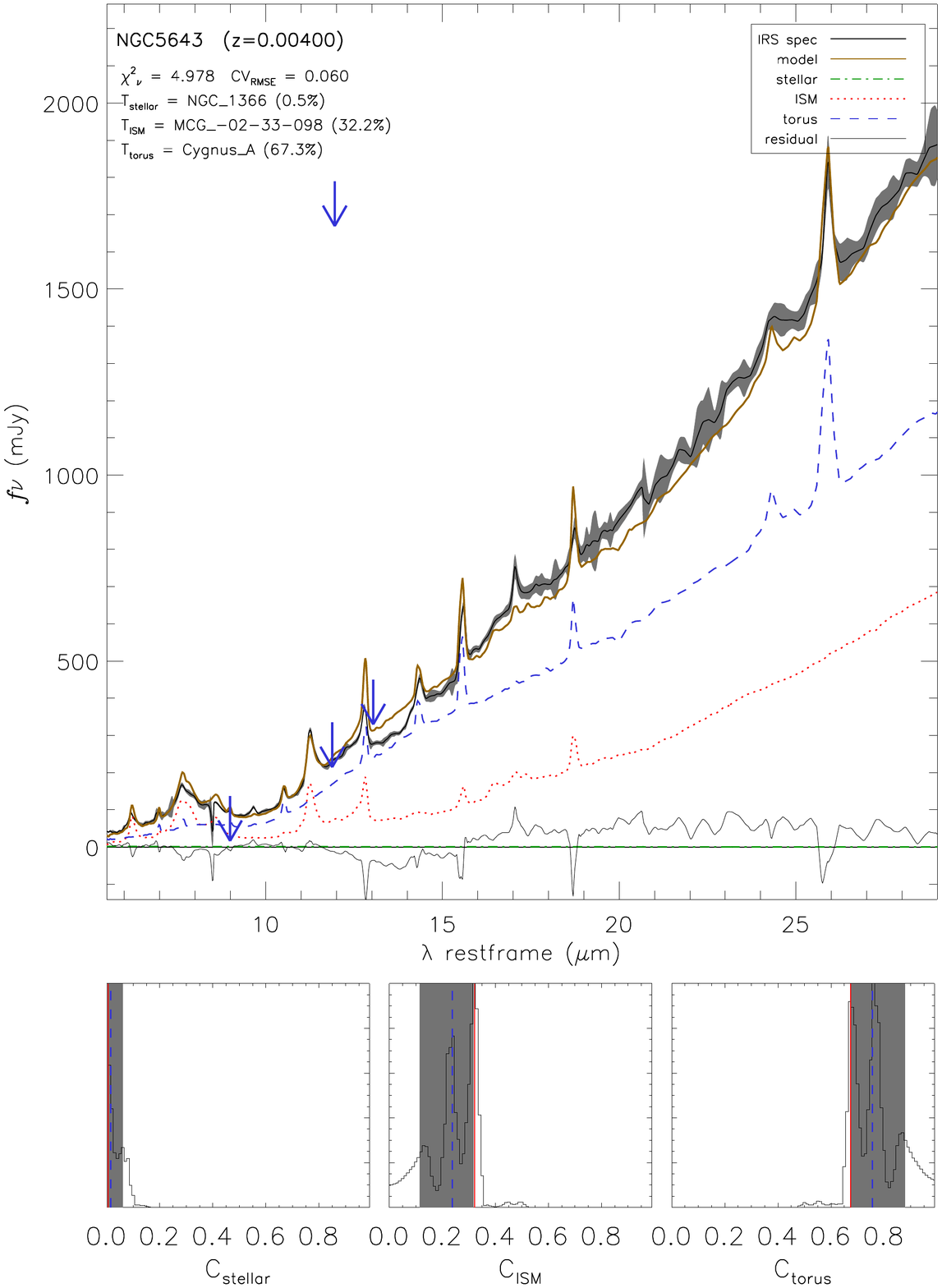}
\includegraphics[width=0.45\columnwidth]{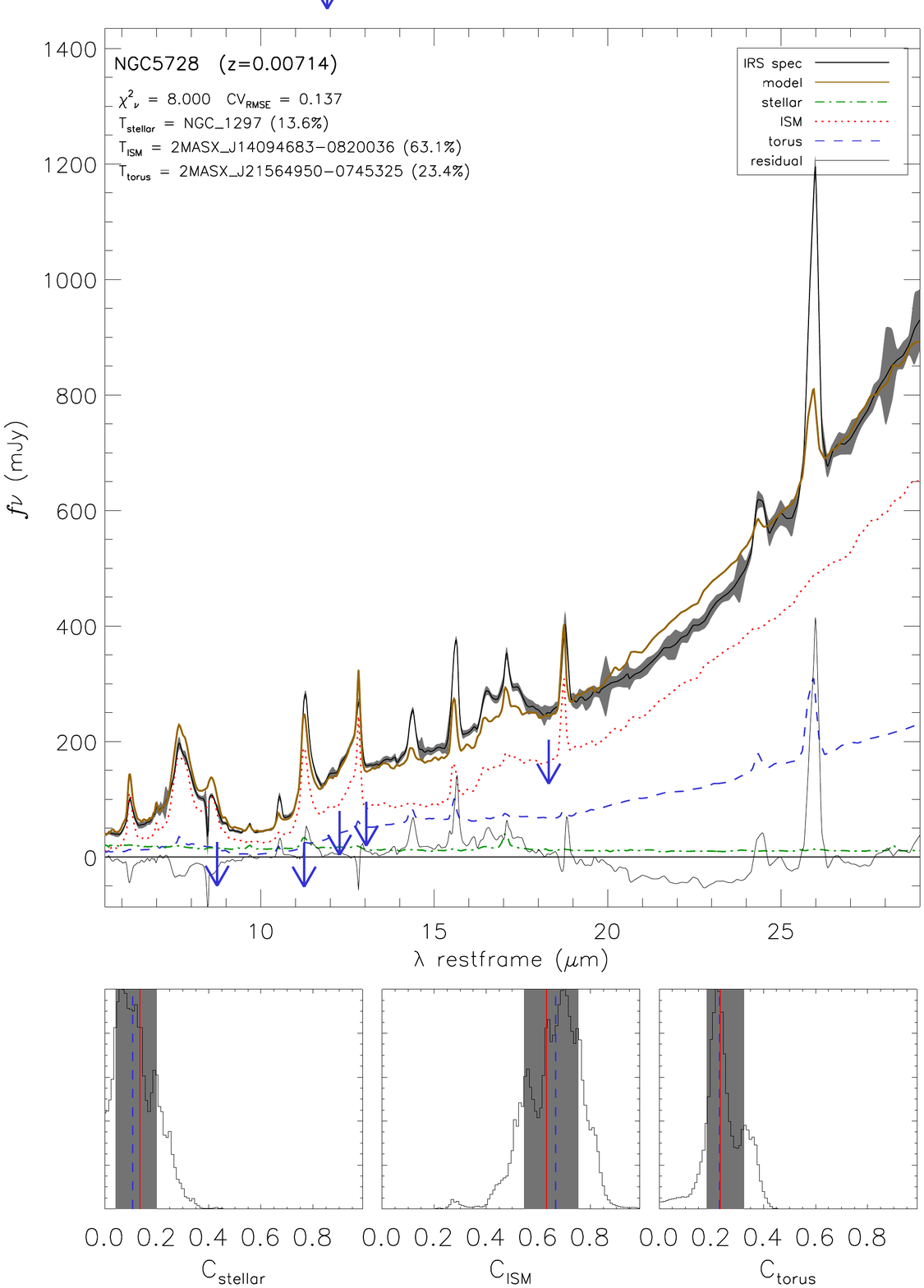}
\includegraphics[width=0.45\columnwidth]{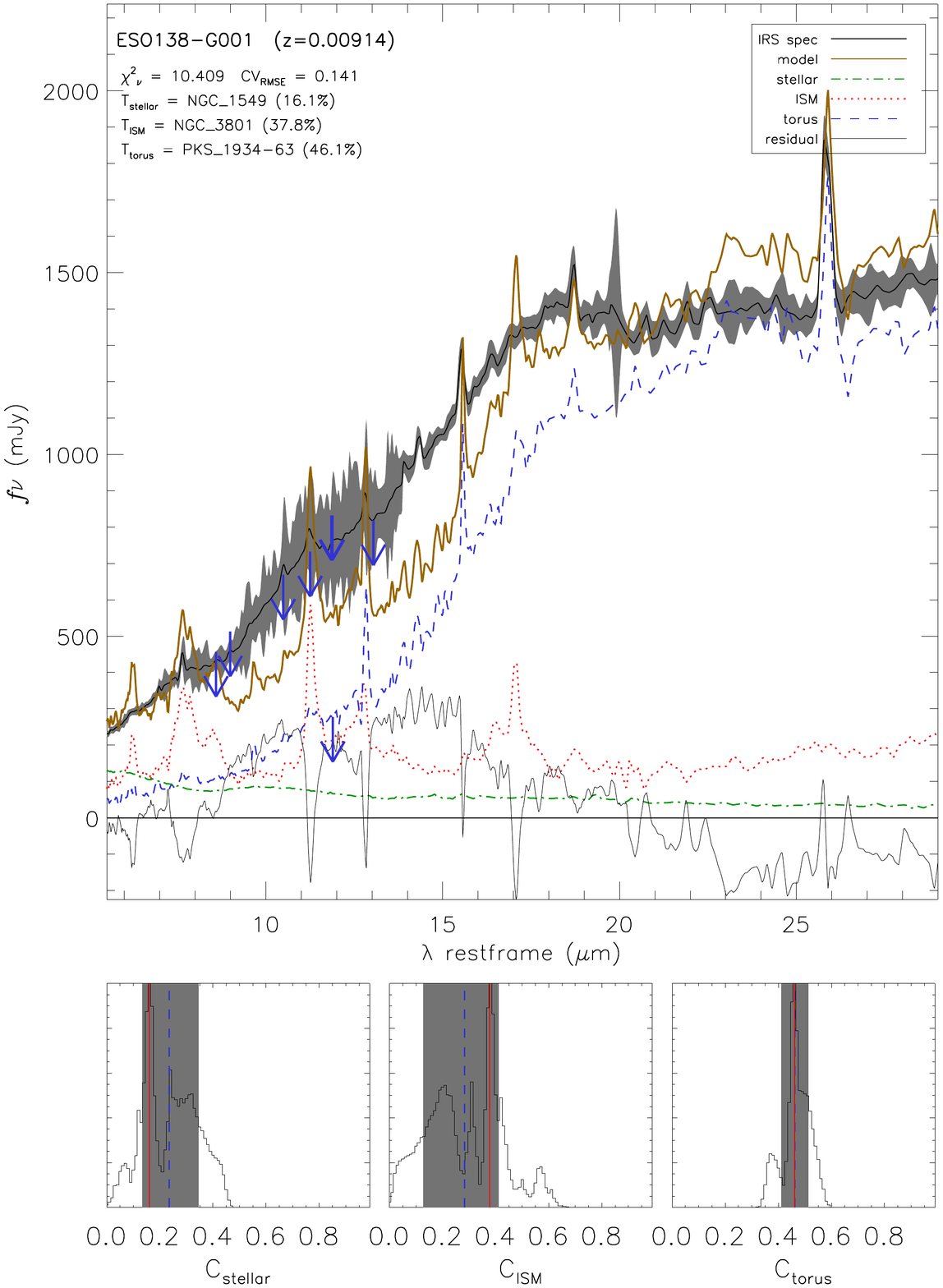}
\caption{...continued.}
\label{fig:CatSpectra}
\end{center}
\end{figure*}

\begin{figure*}
\begin{center}
\includegraphics[width=0.45\columnwidth]{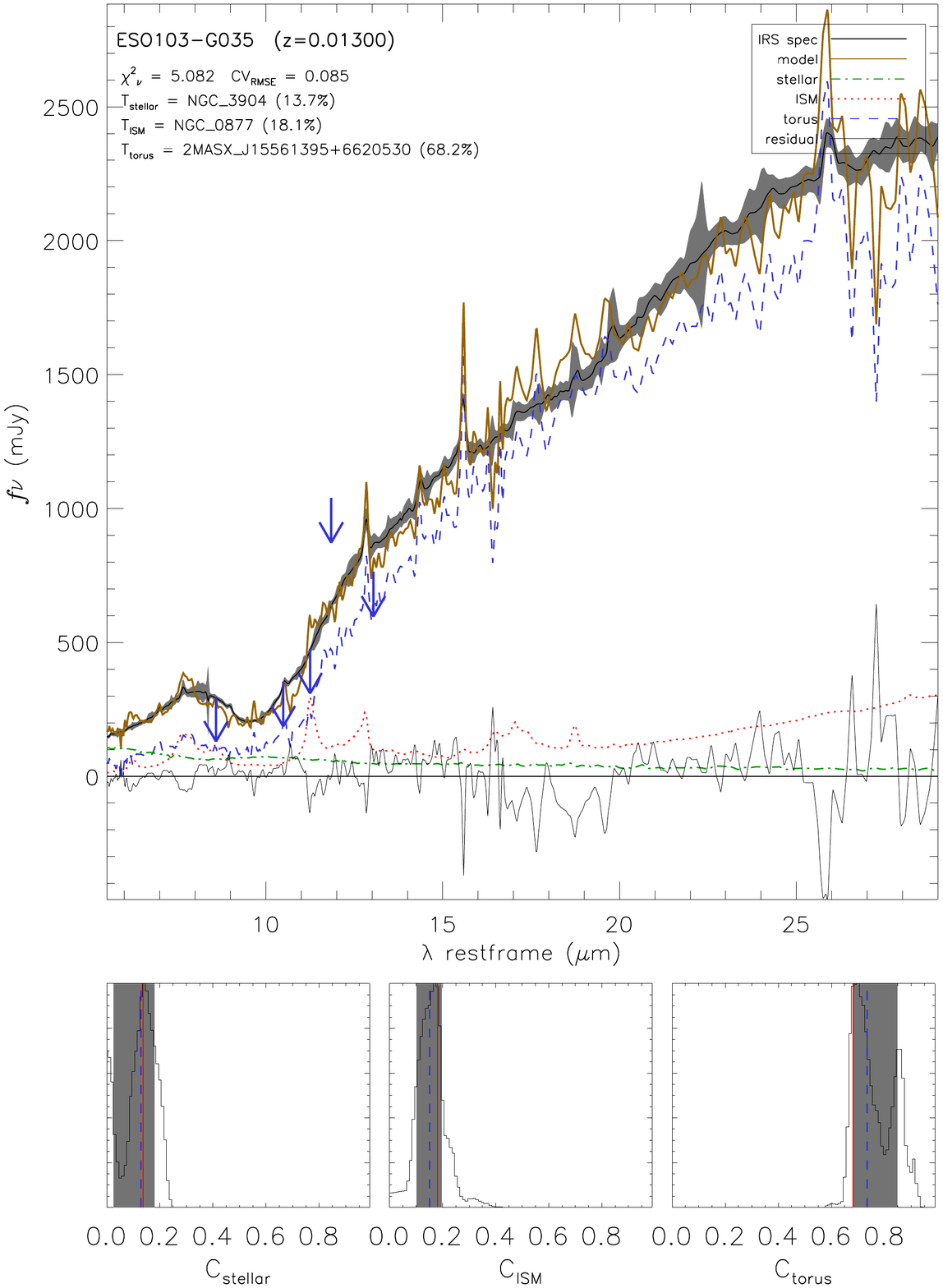}
\includegraphics[width=0.45\columnwidth]{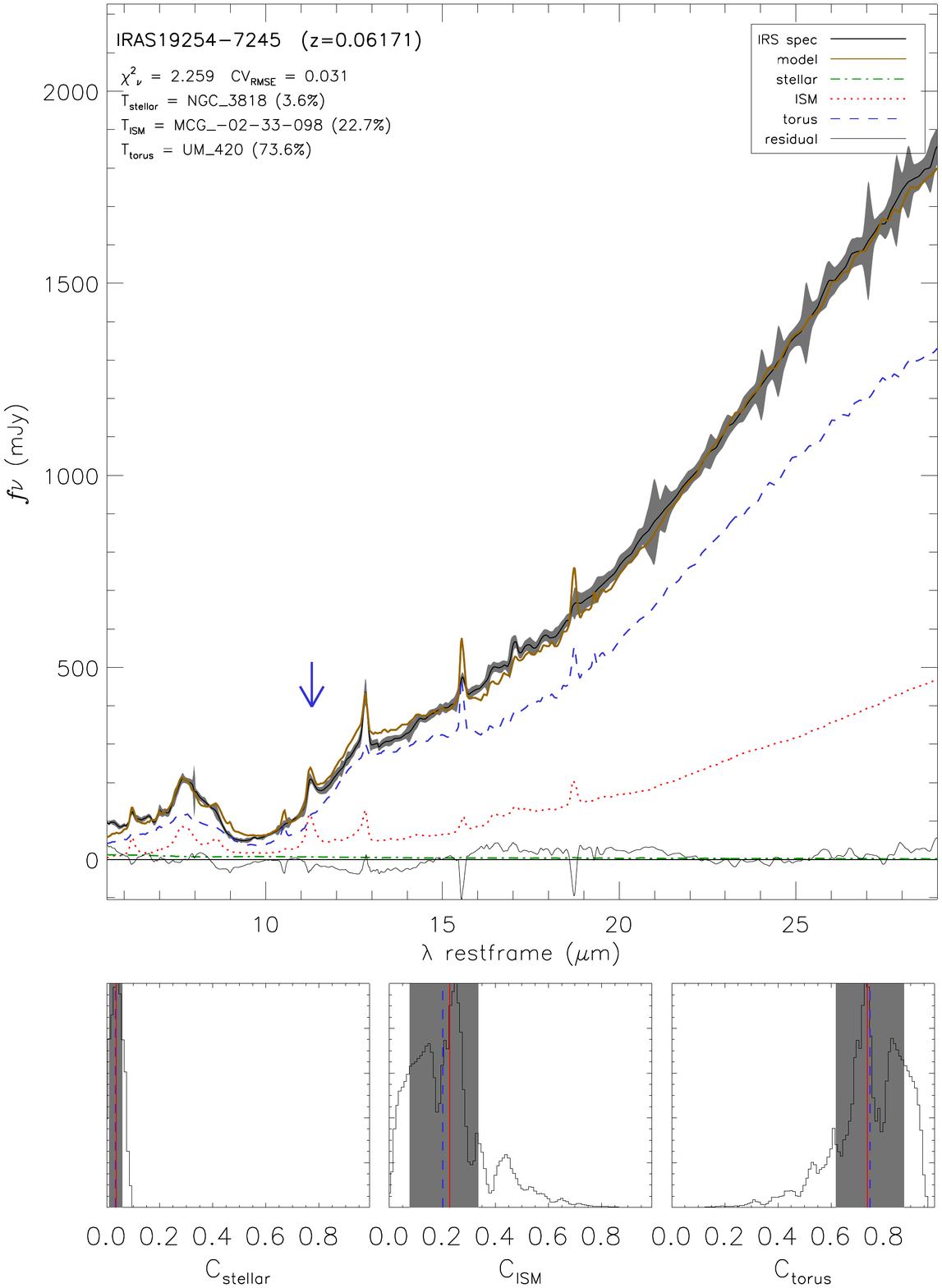}
\includegraphics[width=0.45\columnwidth]{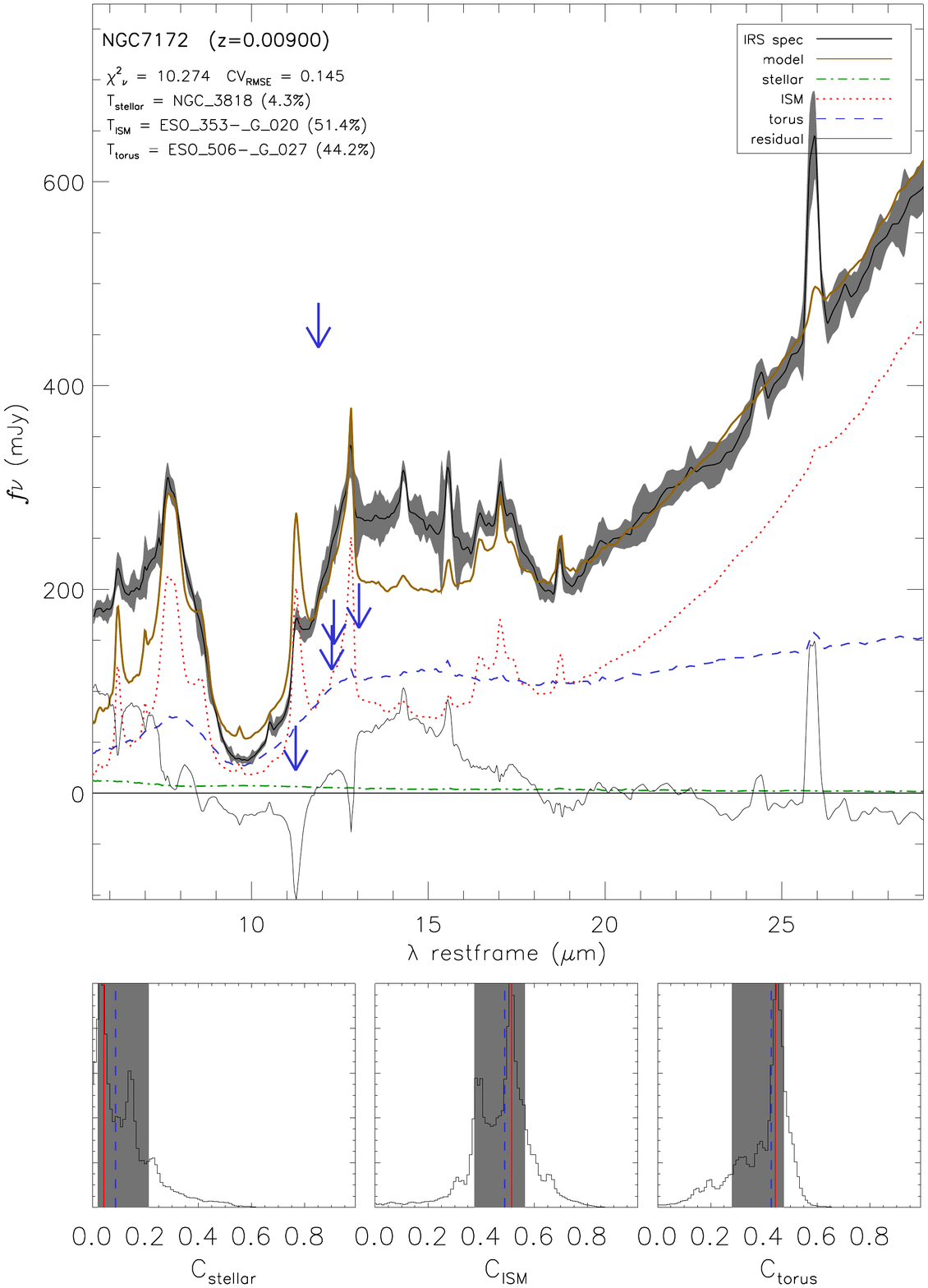}
\includegraphics[width=0.45\columnwidth]{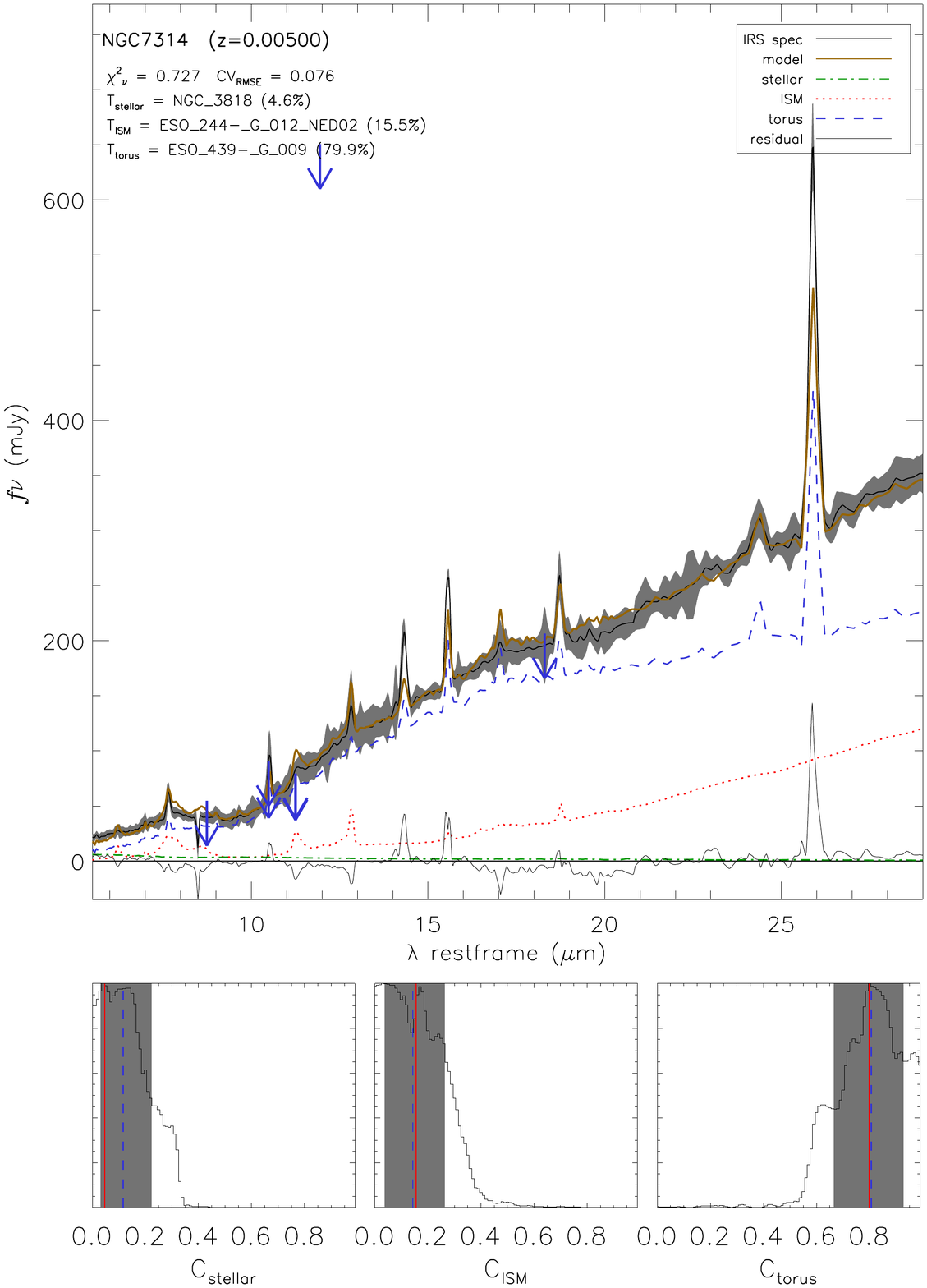}
\caption{...continued.}
\label{fig:CatSpectra}
\end{center}
\end{figure*}

\begin{figure*}
\begin{center}
\includegraphics[width=0.45\columnwidth]{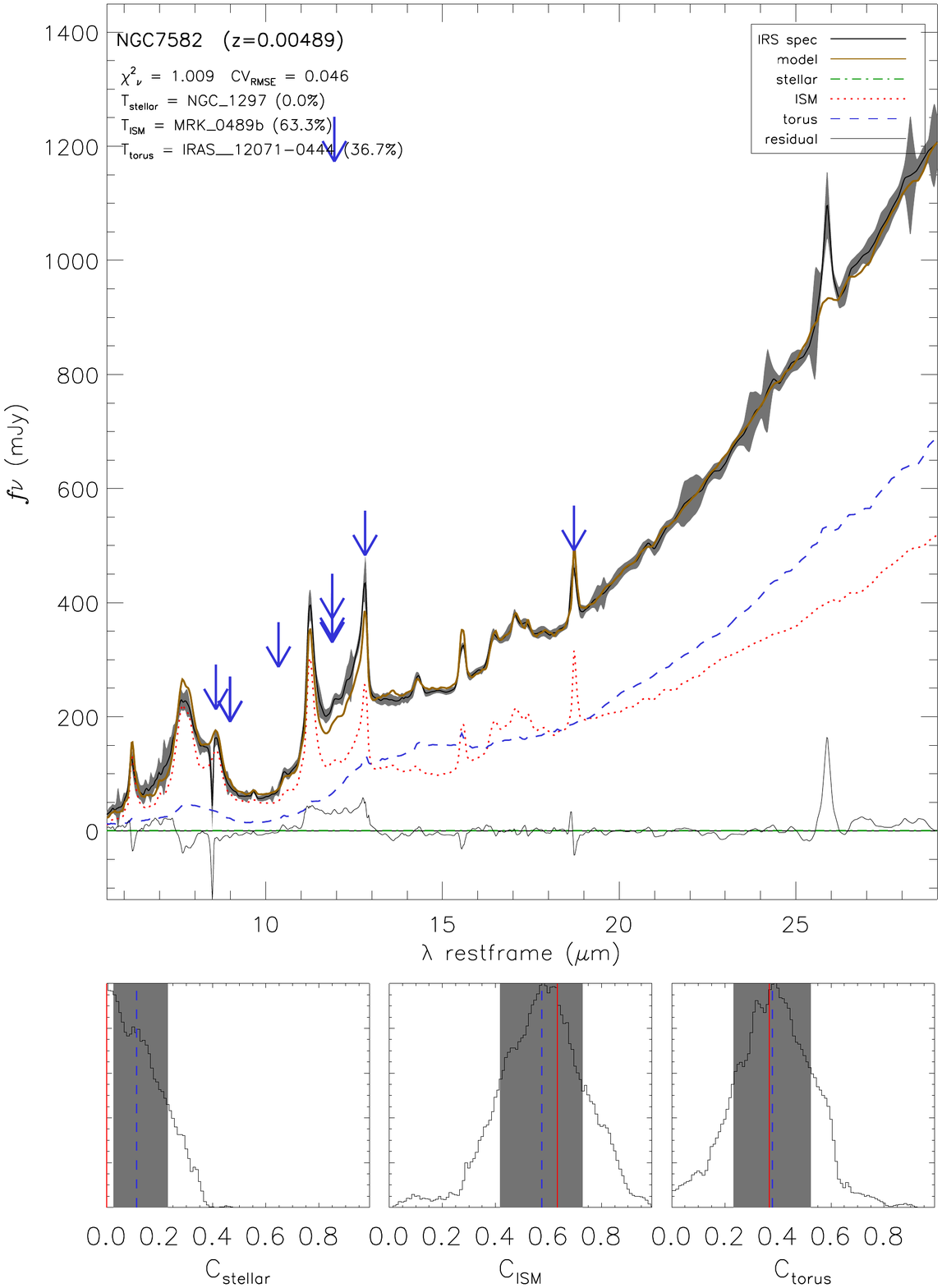}
\includegraphics[width=0.45\columnwidth]{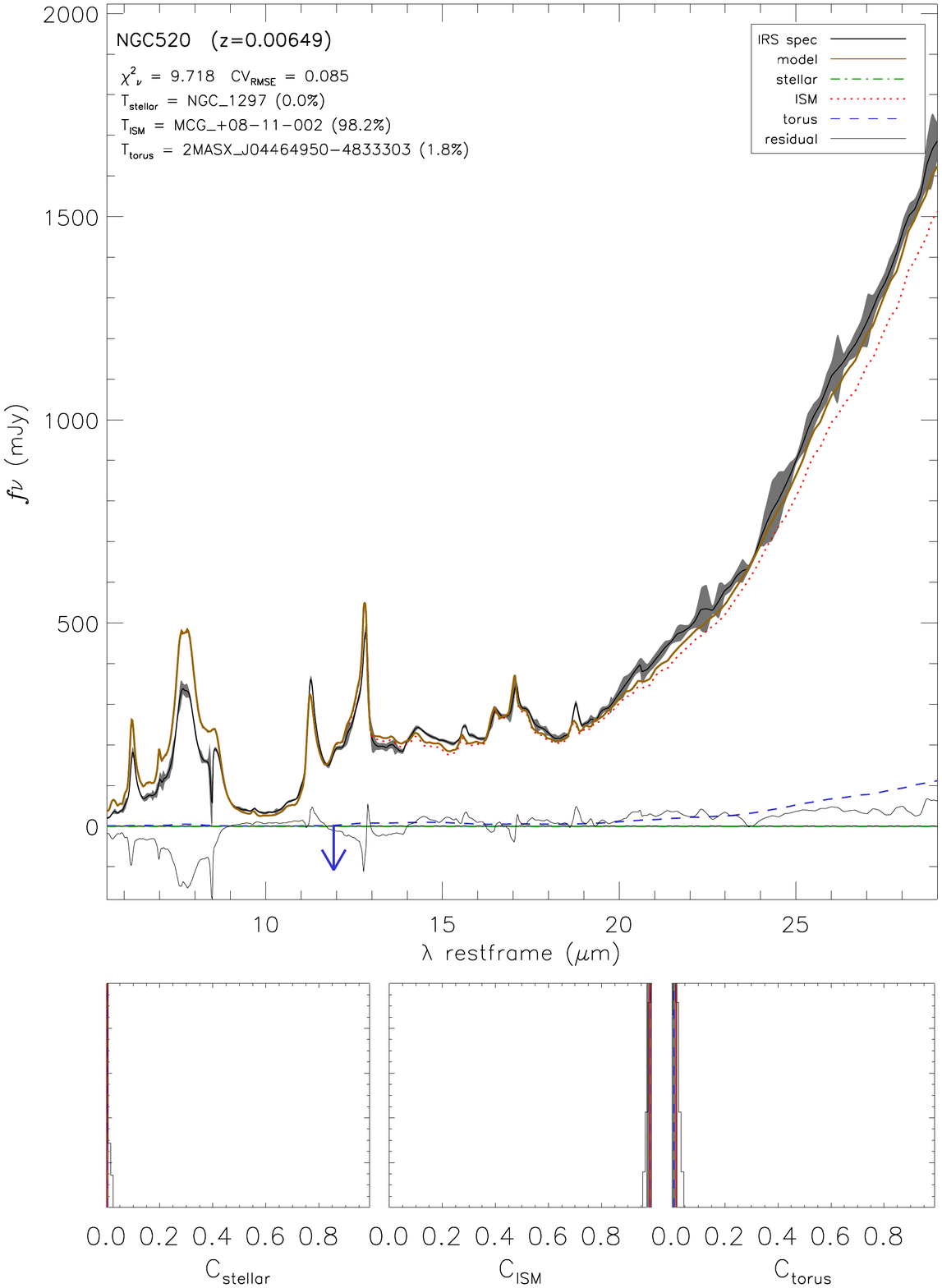}
\includegraphics[width=0.45\columnwidth]{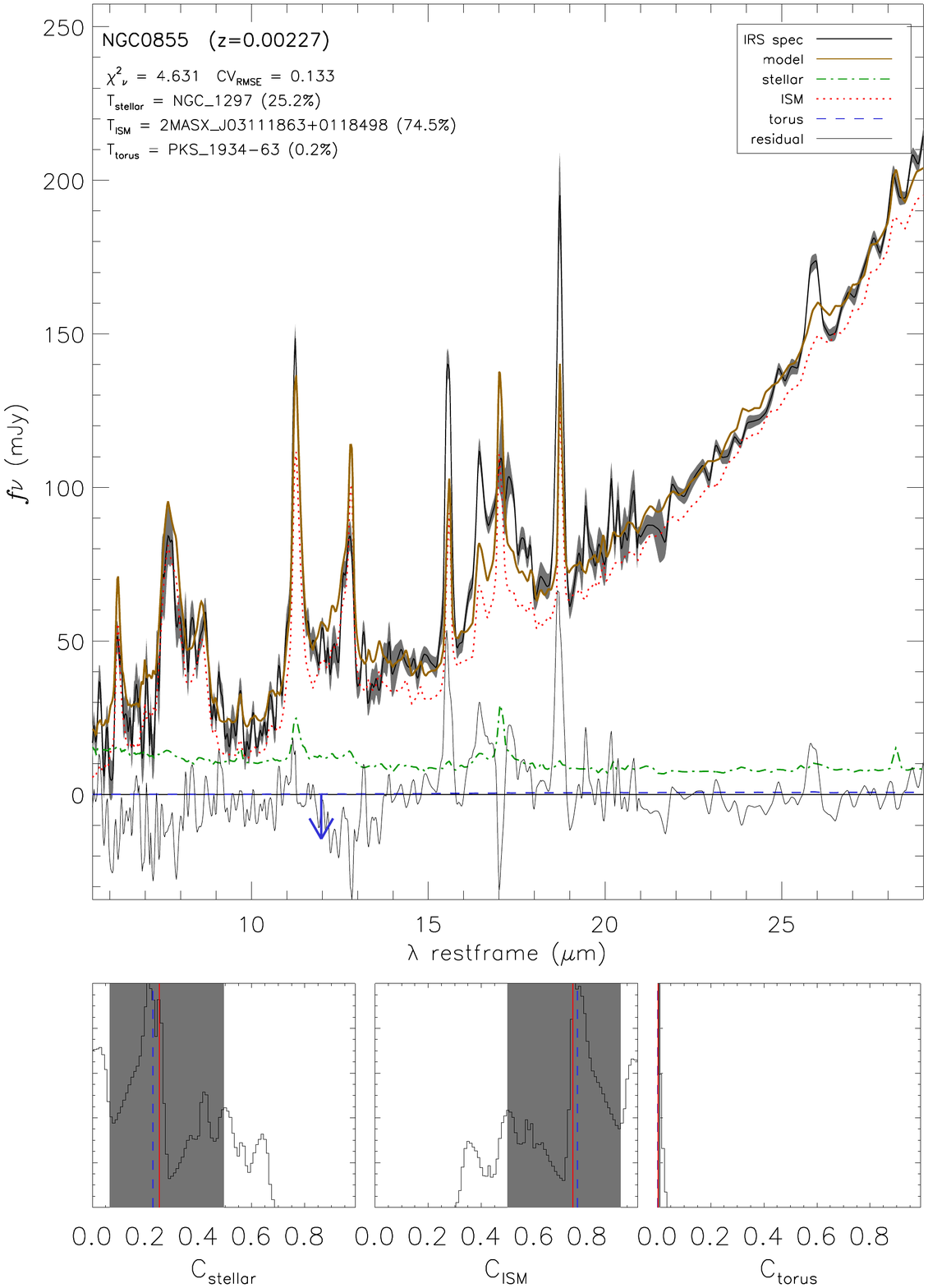}
\includegraphics[width=0.45\columnwidth]{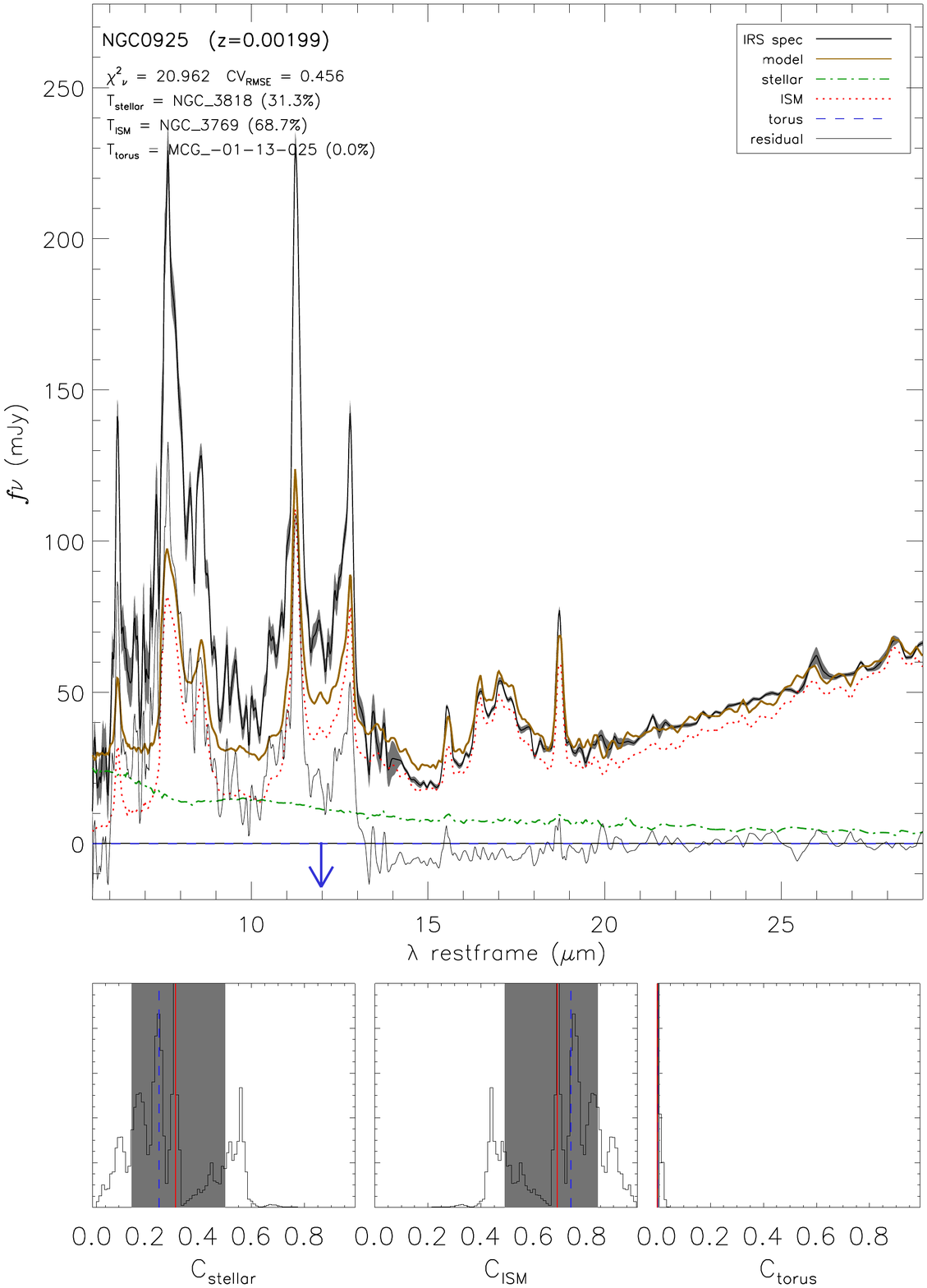}
\caption{...continued.}
\label{fig:CatSpectra}
\end{center}
\end{figure*}

\begin{figure*}
\begin{center}
\includegraphics[width=0.45\columnwidth]{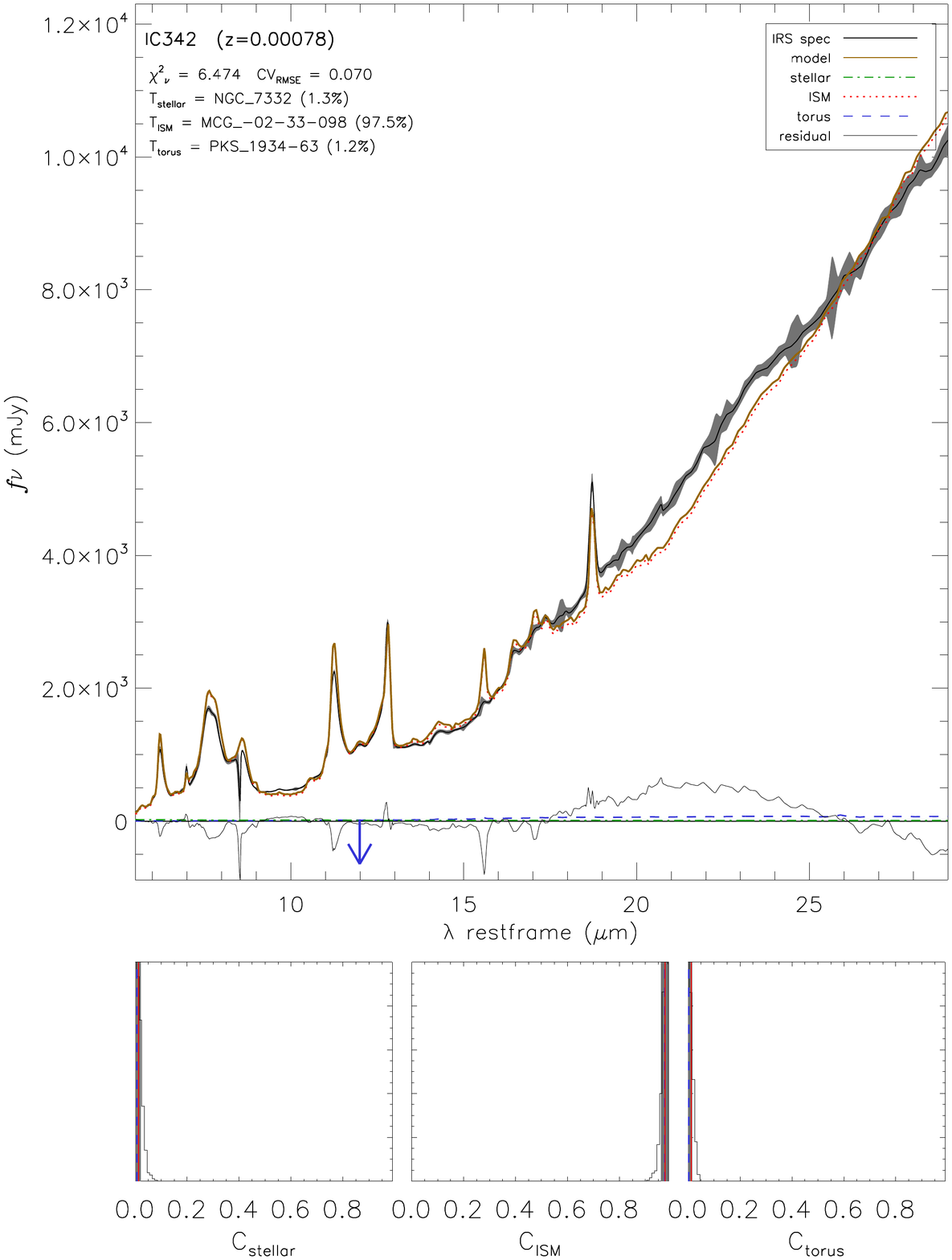}
\includegraphics[width=0.45\columnwidth]{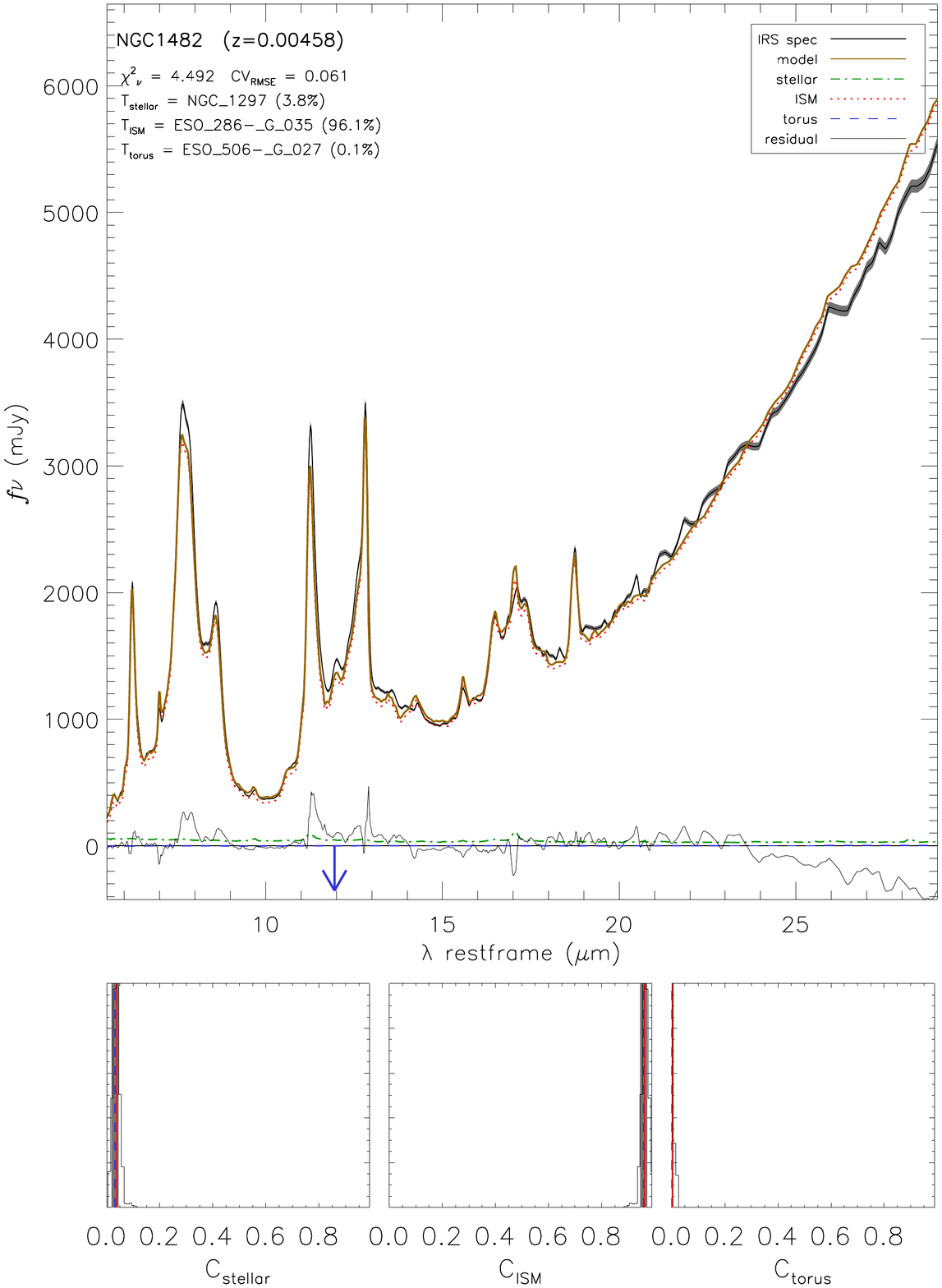}
\includegraphics[width=0.45\columnwidth]{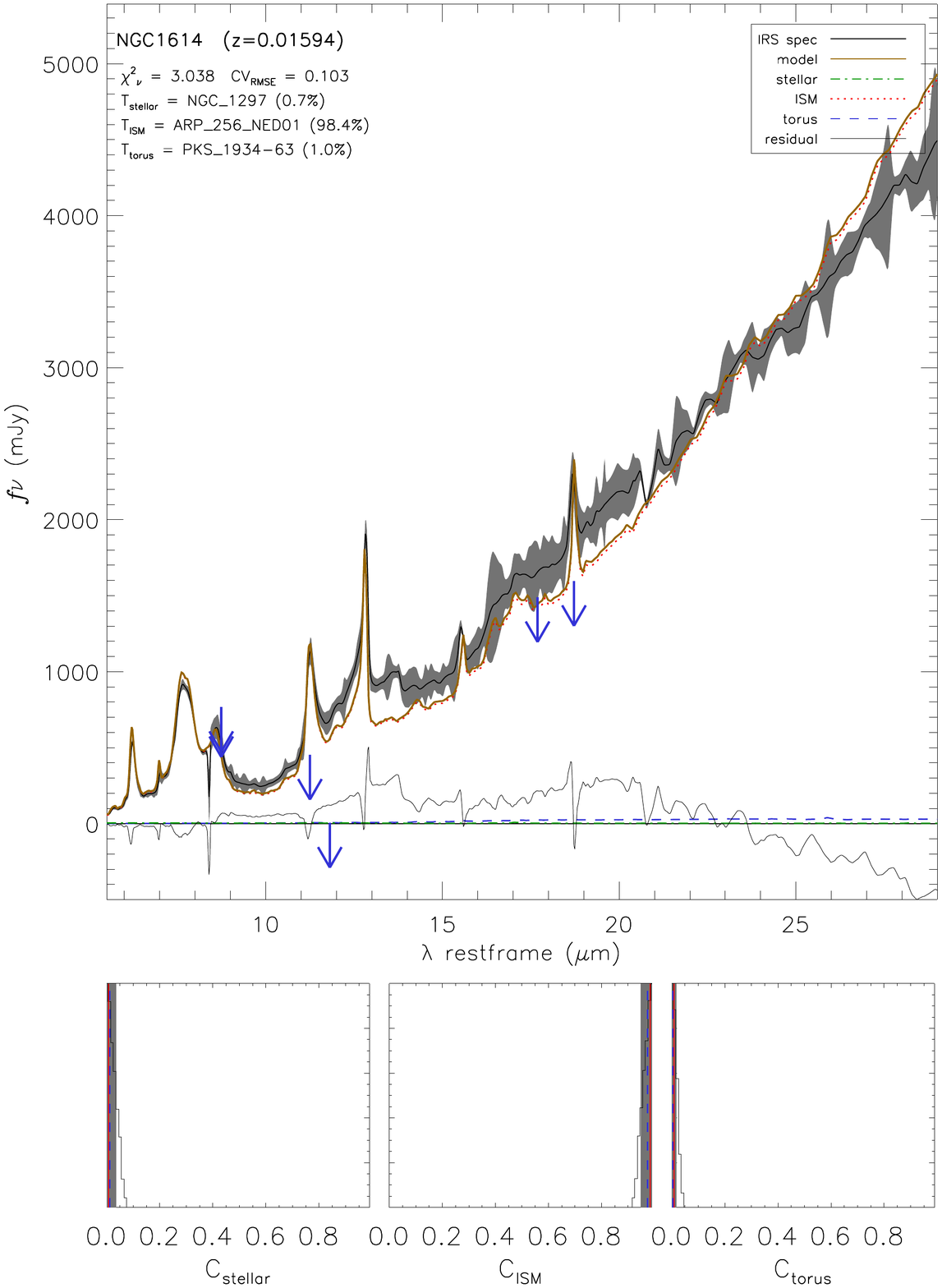}
\includegraphics[width=0.45\columnwidth]{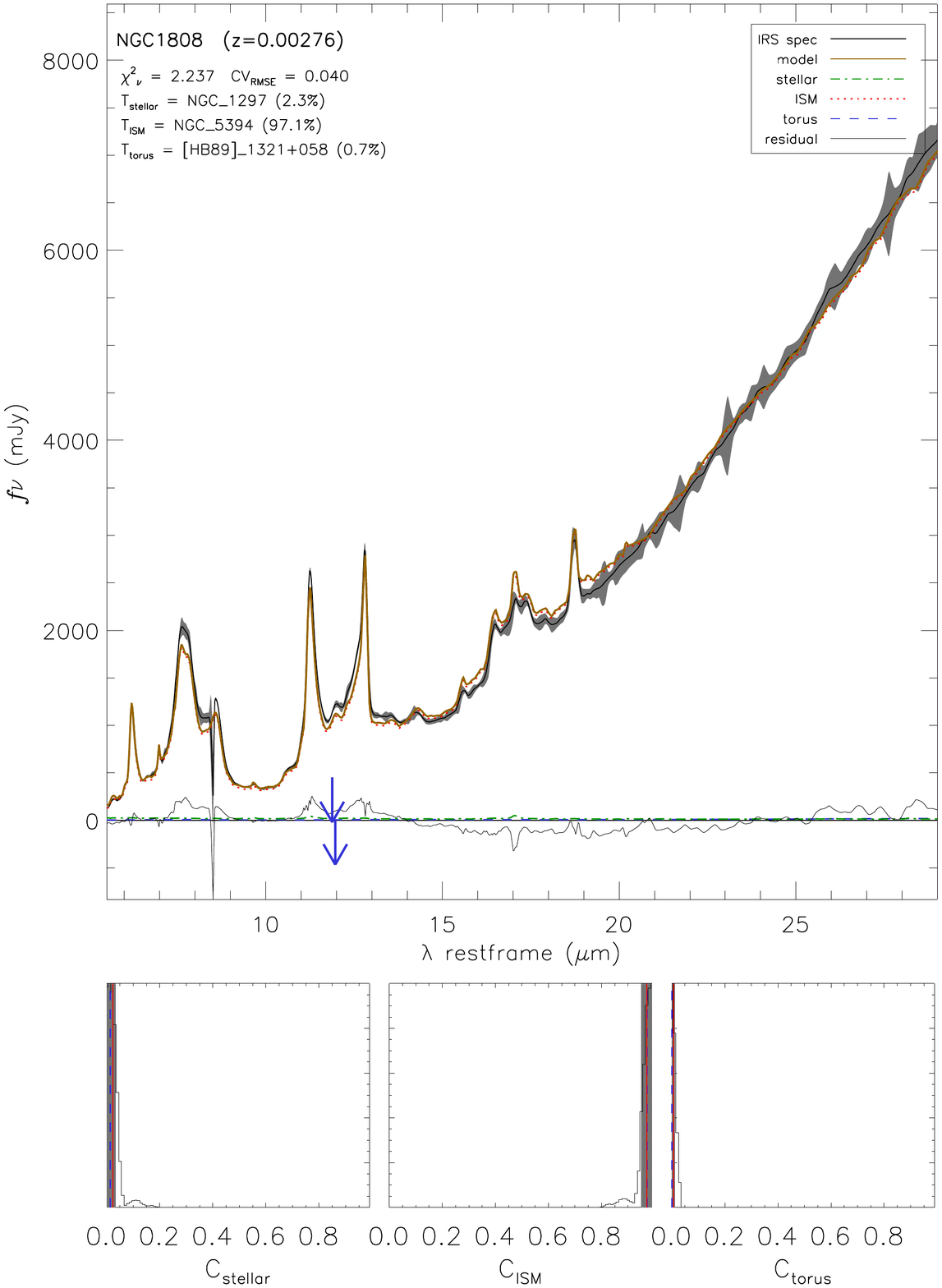}
\caption{...continued.}
\label{fig:CatSpectra}
\end{center}
\end{figure*}

\begin{figure*}
\begin{center}
\includegraphics[width=0.45\columnwidth]{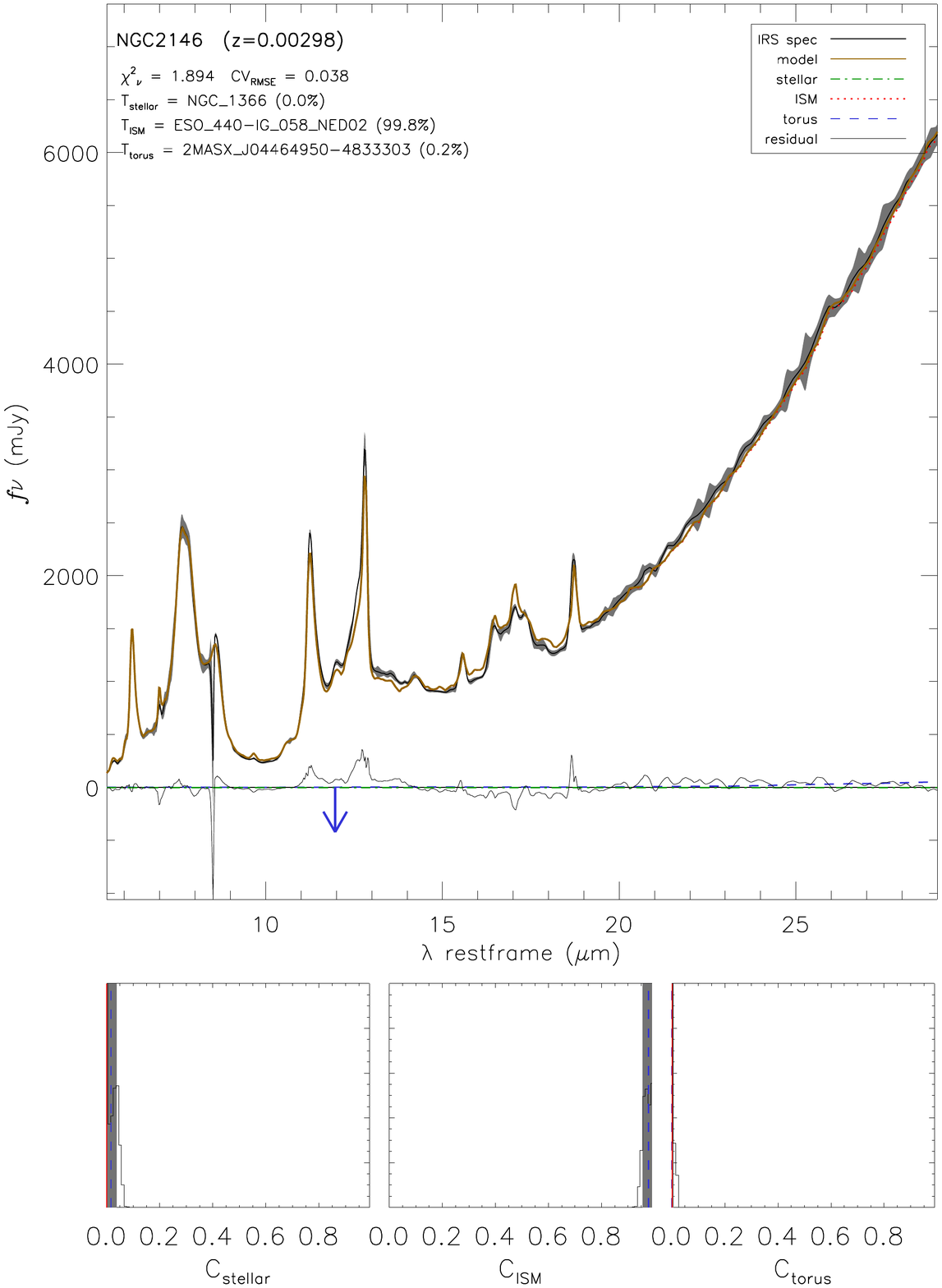}
\includegraphics[width=0.45\columnwidth]{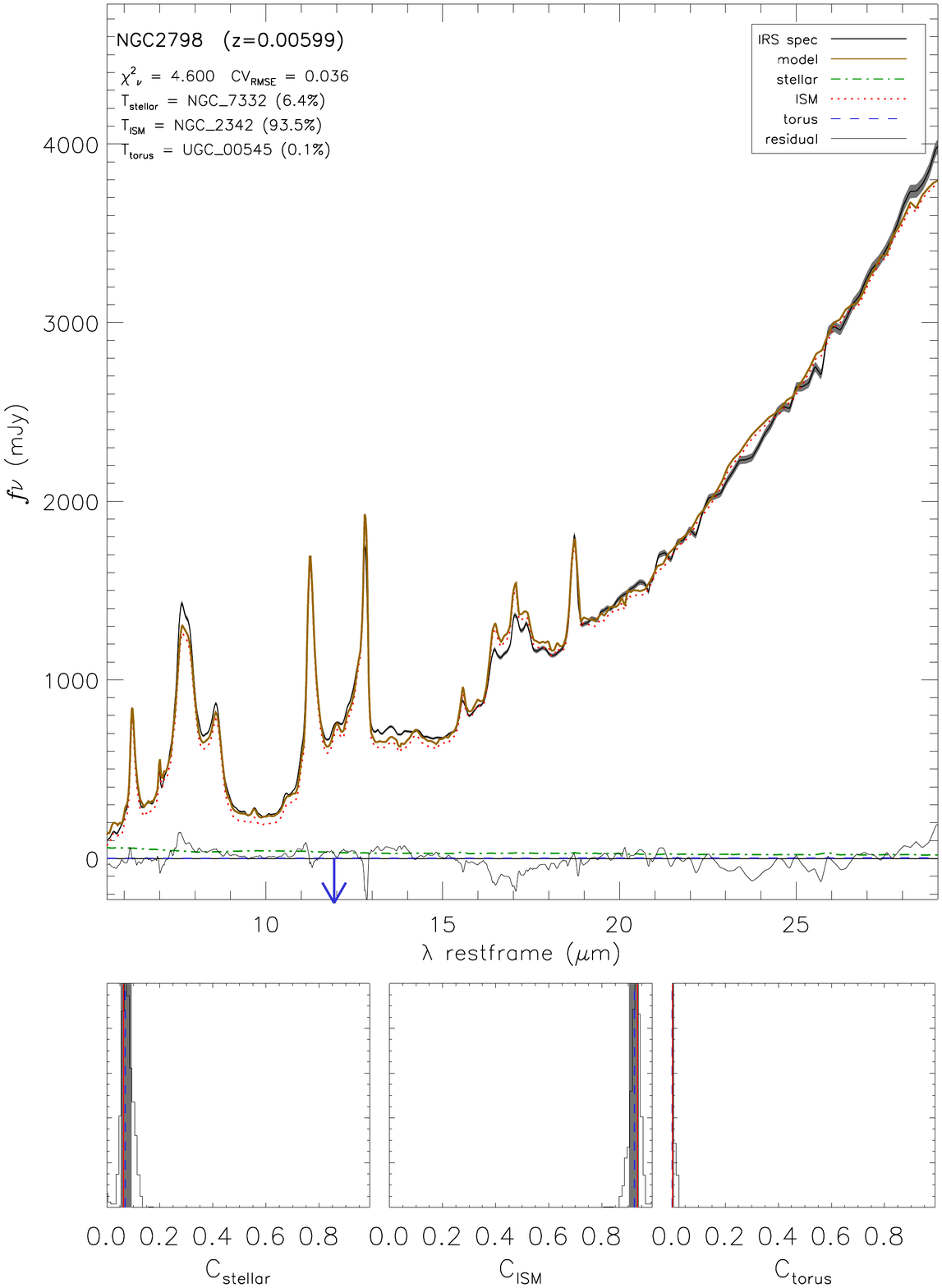}
\includegraphics[width=0.45\columnwidth]{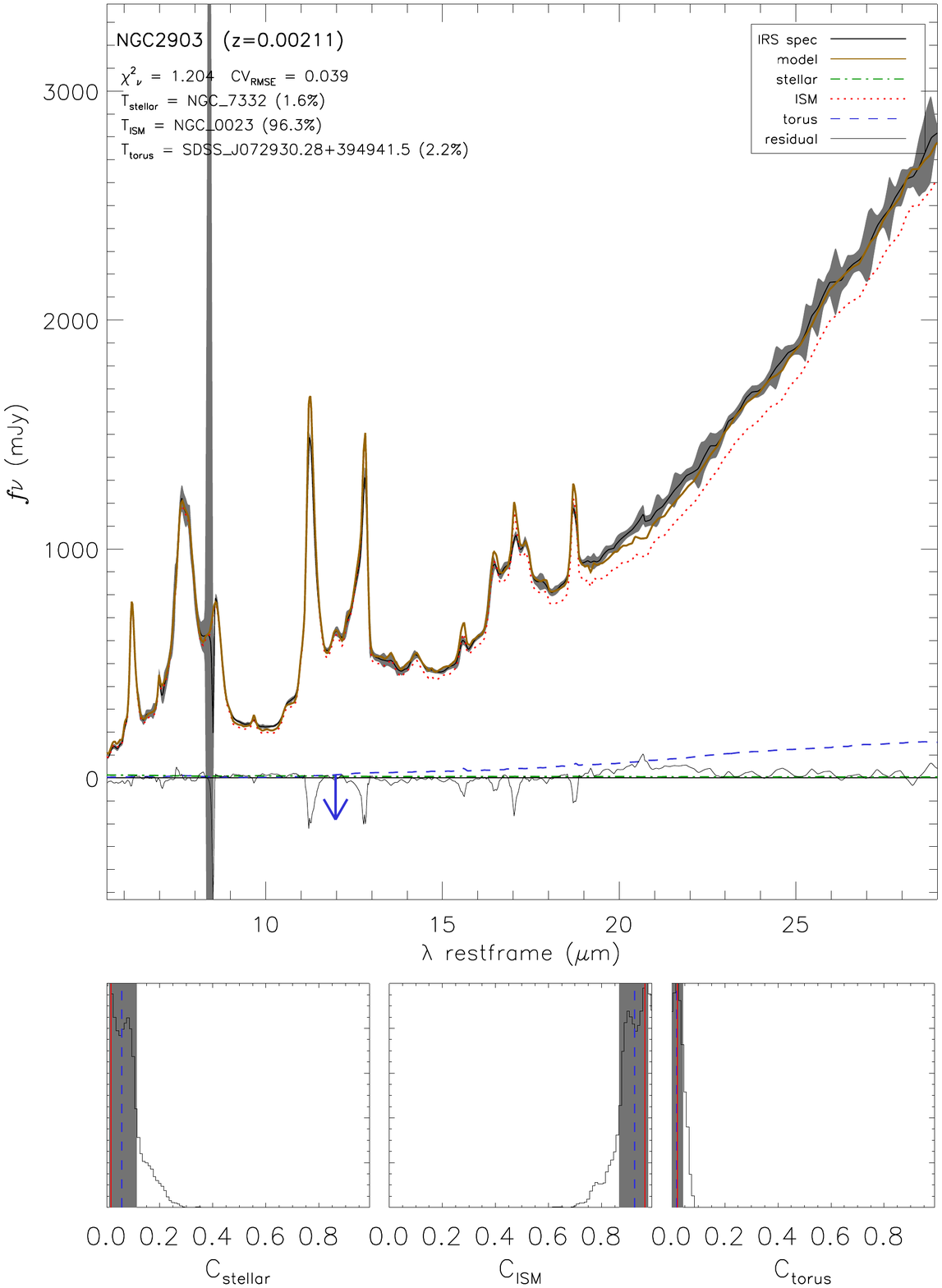}
\includegraphics[width=0.45\columnwidth]{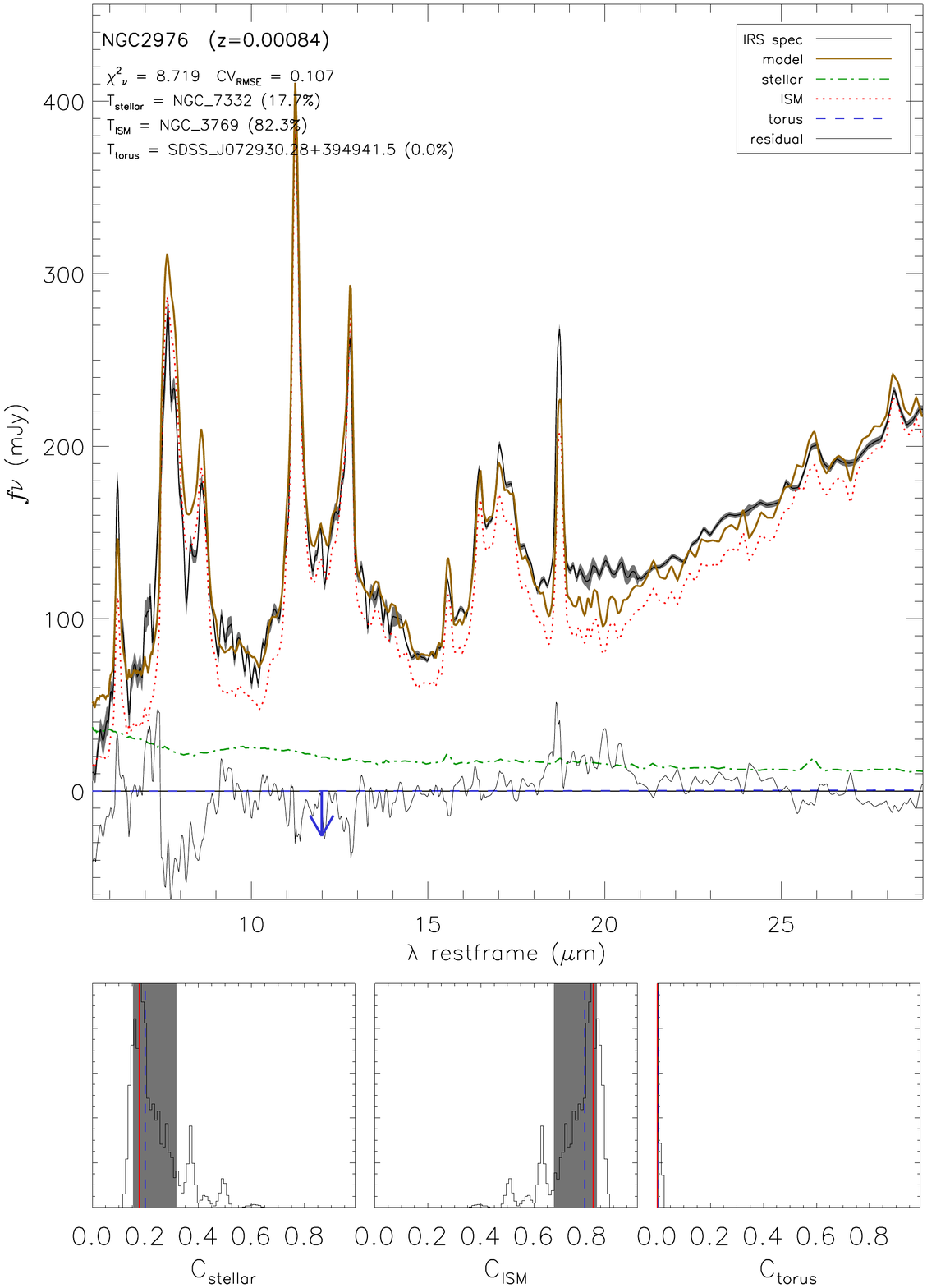}
\caption{...continued.}
\label{fig:CatSpectra}
\end{center}
\end{figure*}

\begin{figure*}
\begin{center}
\includegraphics[width=0.45\columnwidth]{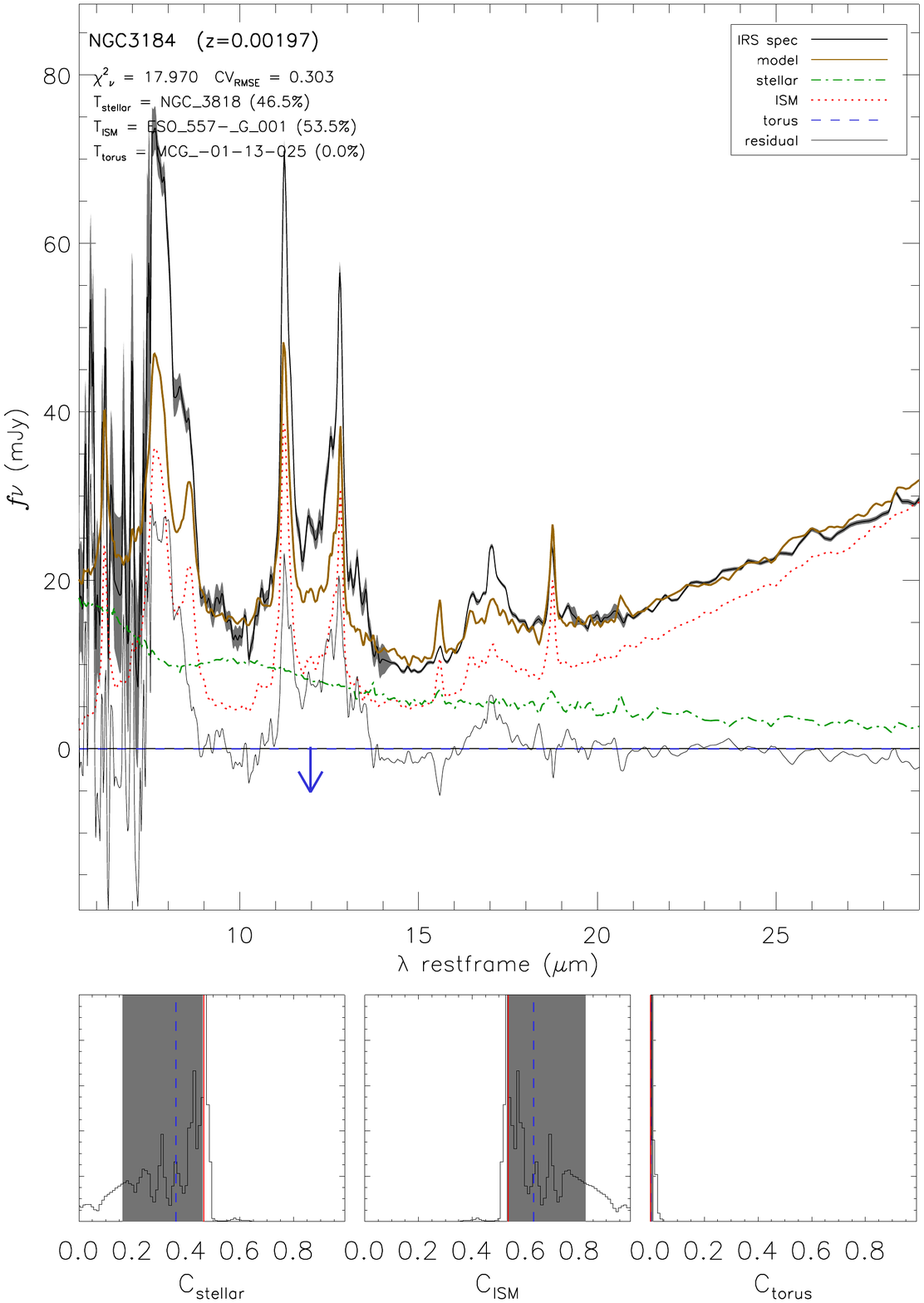}
\includegraphics[width=0.45\columnwidth]{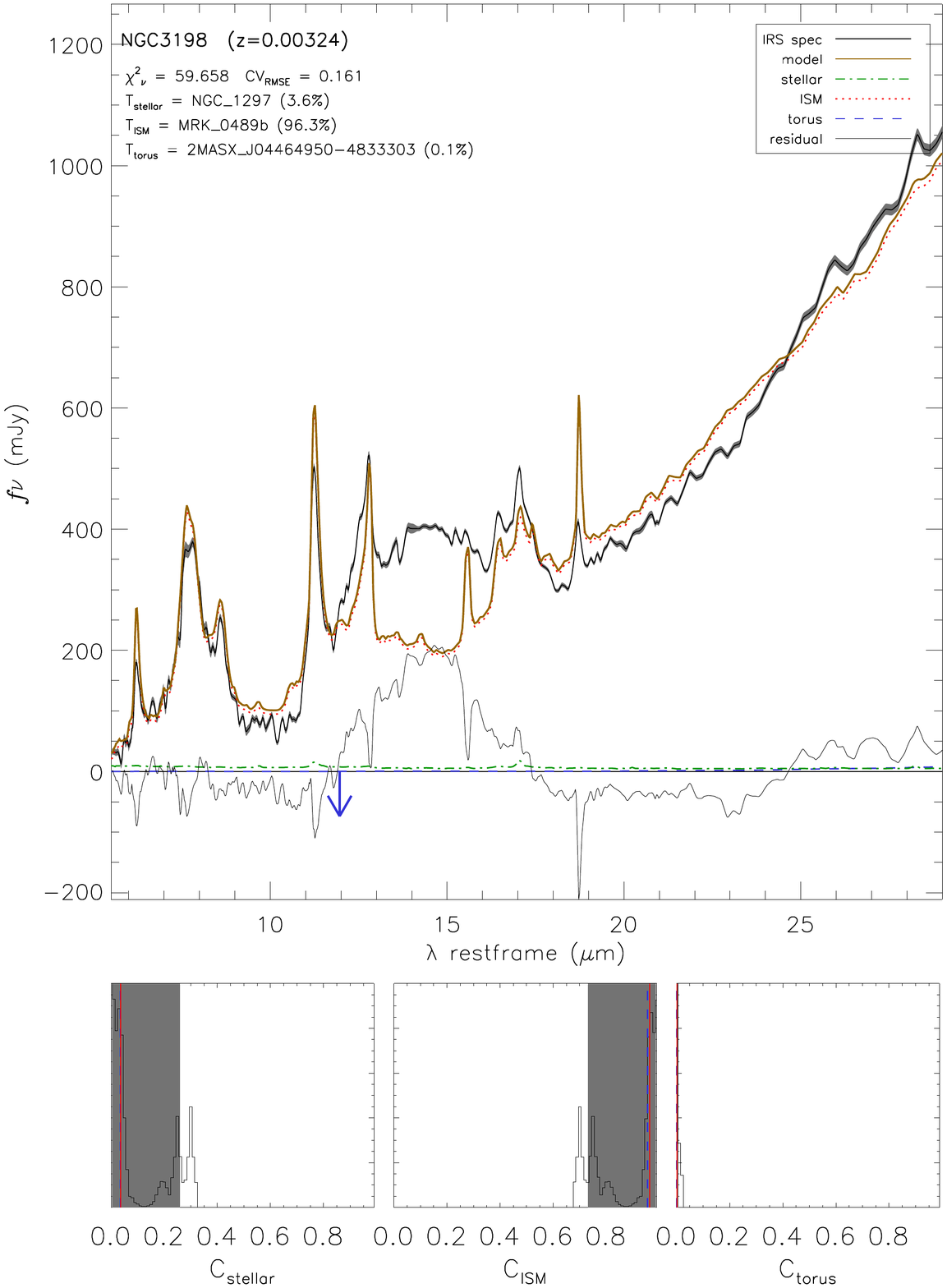}
\includegraphics[width=0.45\columnwidth]{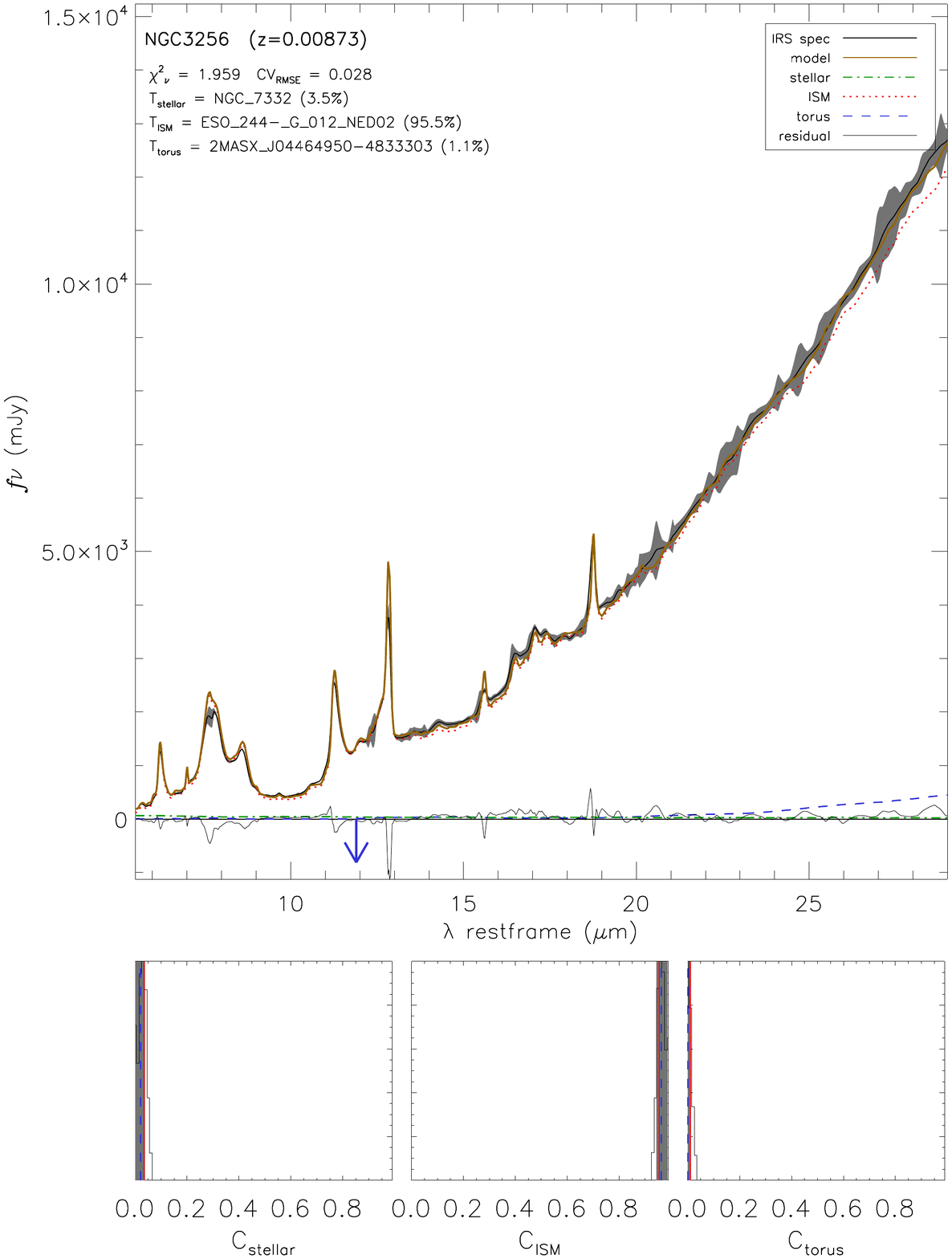}
\includegraphics[width=0.45\columnwidth]{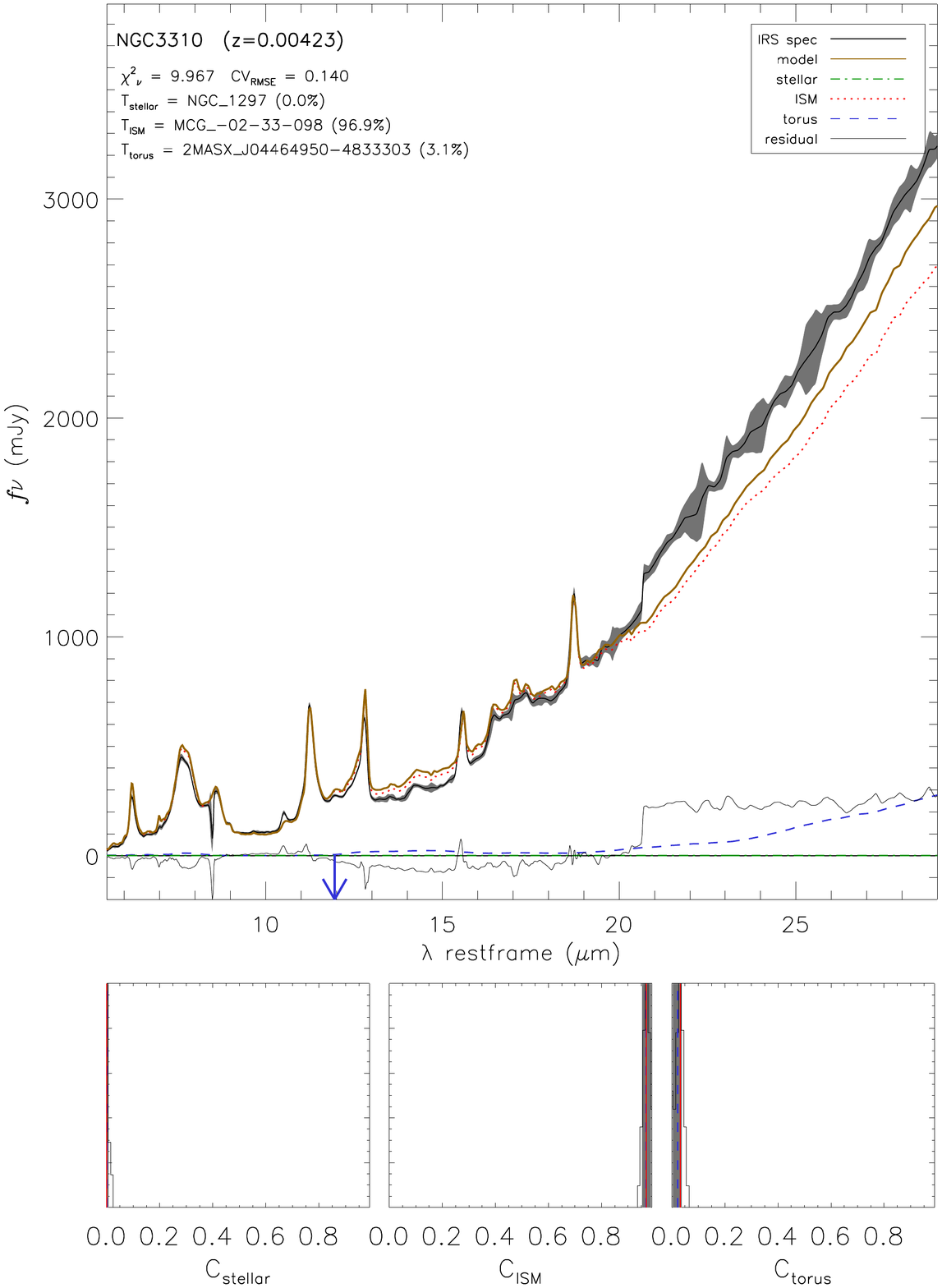}
\caption{...continued.}
\label{fig:CatSpectra}
\end{center}
\end{figure*}

\begin{figure*}
\begin{center}
\includegraphics[width=0.45\columnwidth]{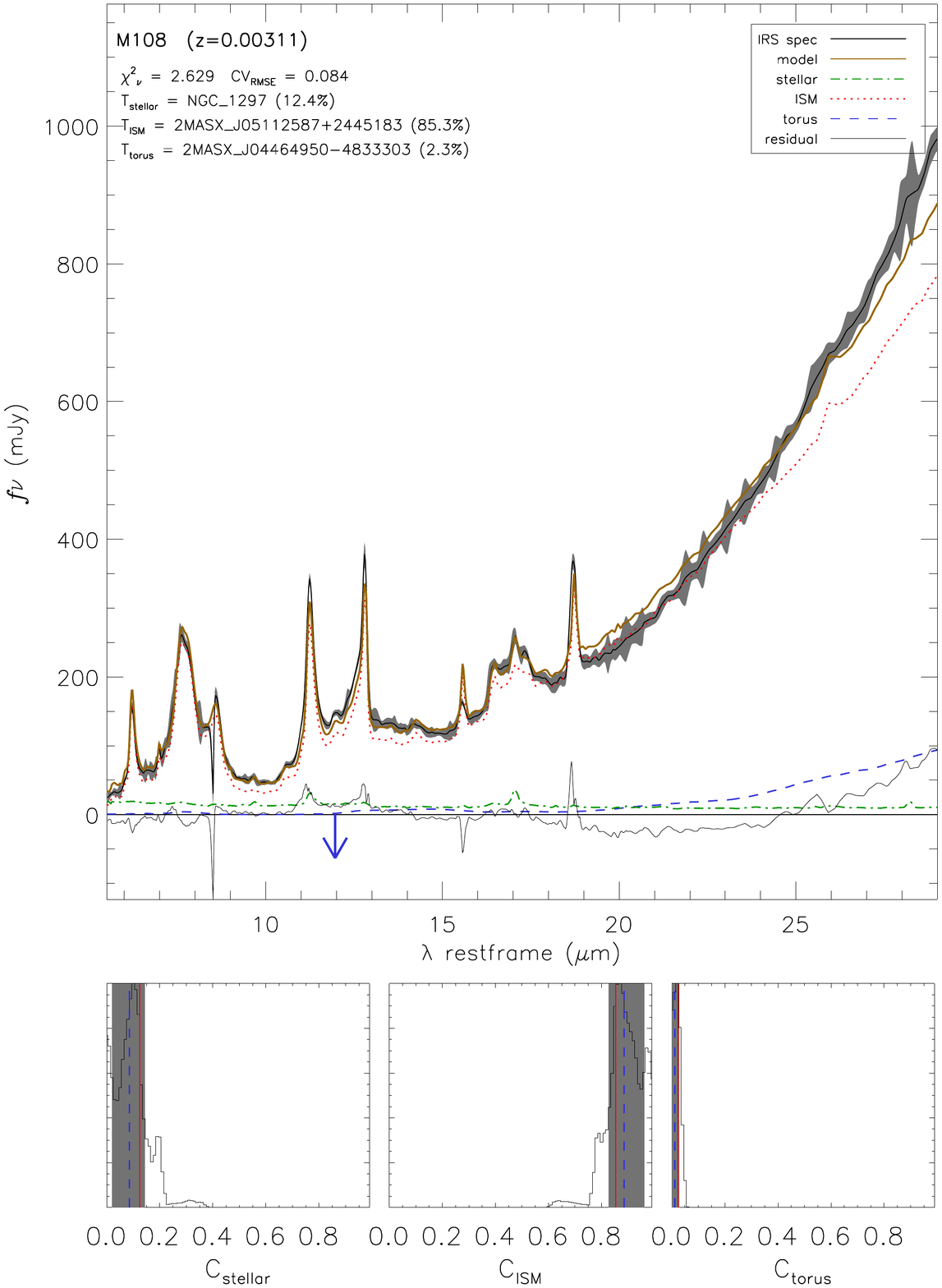}
\includegraphics[width=0.45\columnwidth]{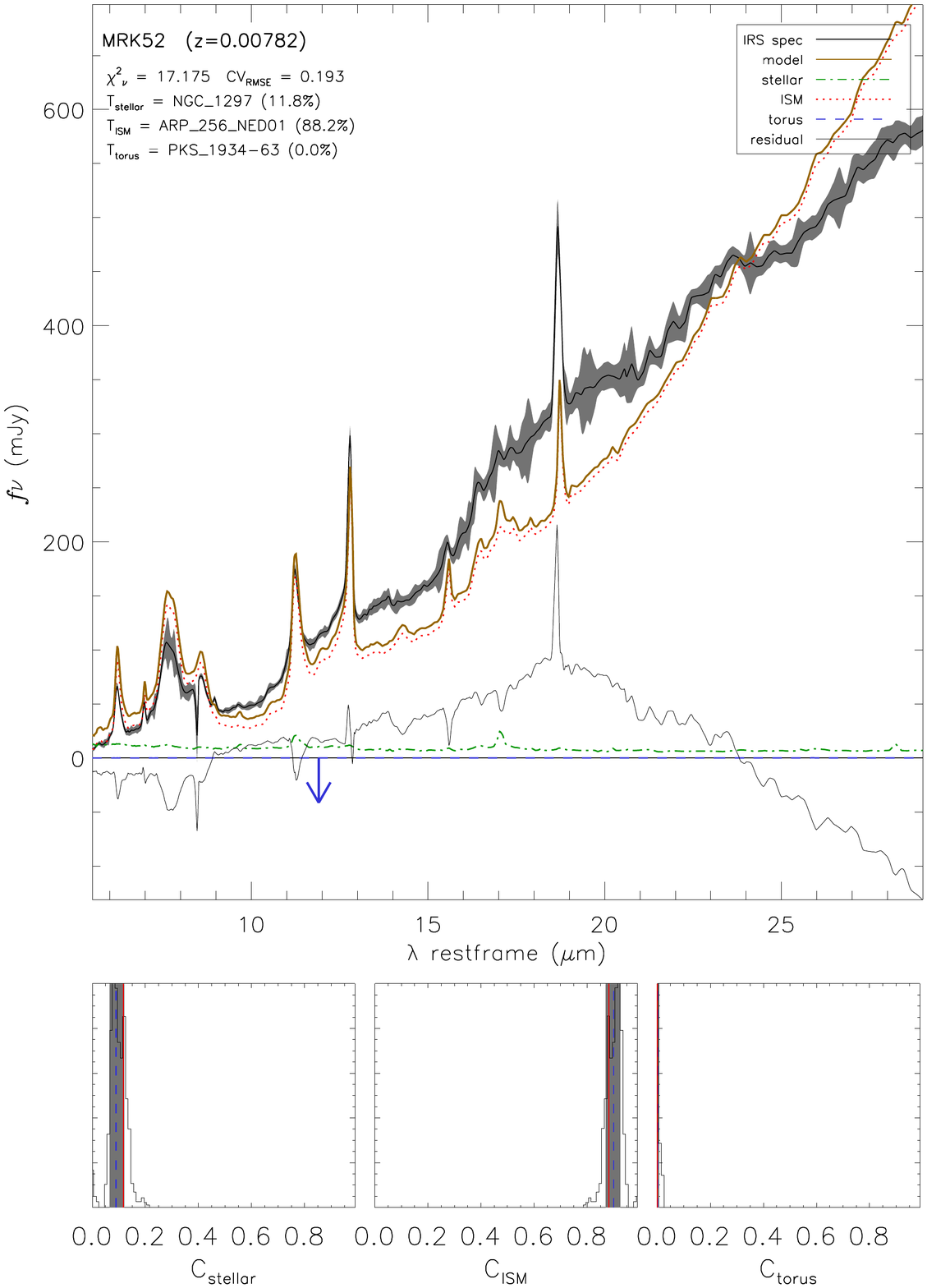}
\includegraphics[width=0.45\columnwidth]{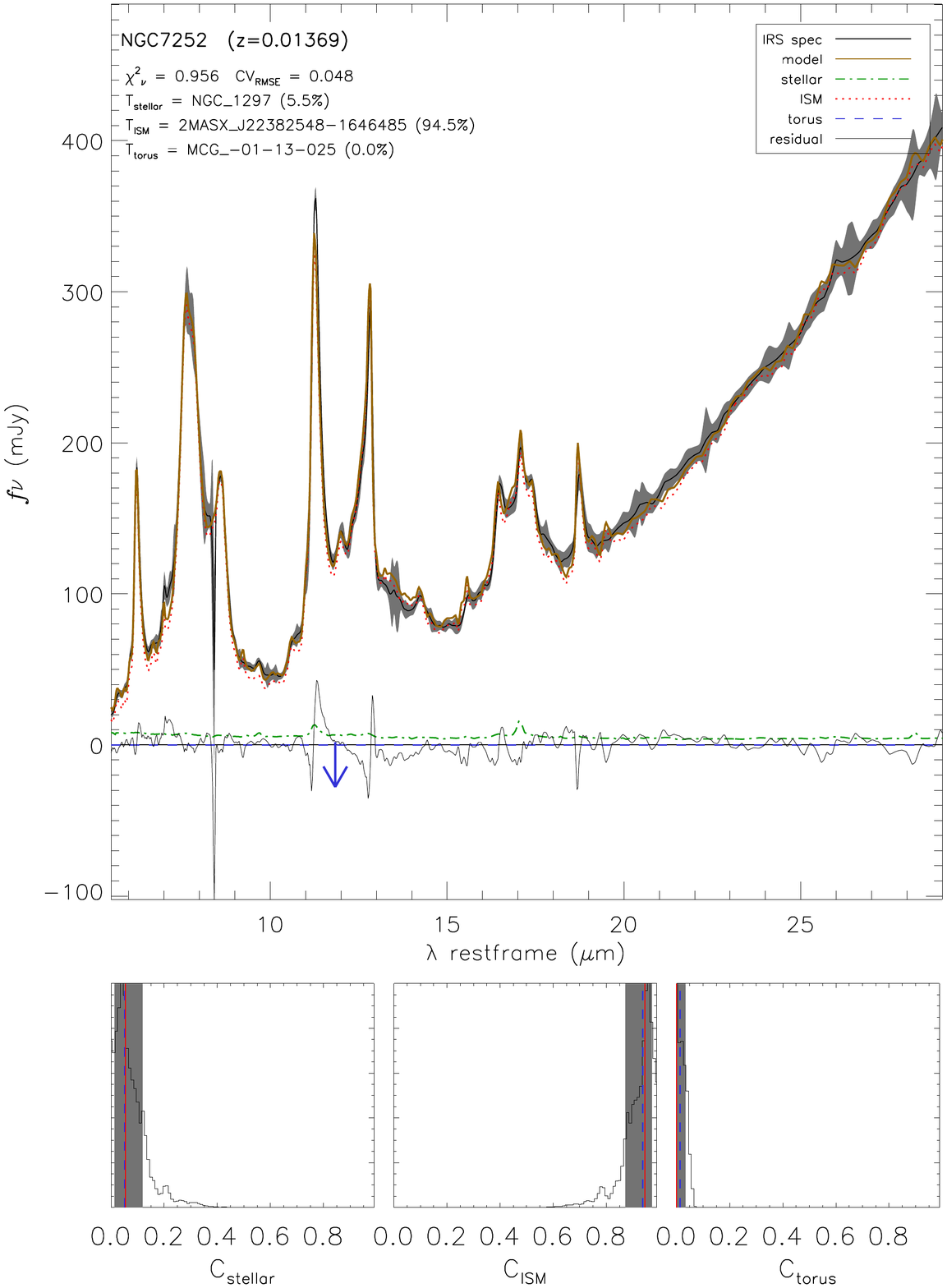}
\caption{...continued.}
\label{fig:CatSpectra}
\end{center}
\end{figure*}

\end{appendix}

%% file: mOGM2017.bbl
\begin{thebibliography}{}
\bibitem[Alonso-Herrero et al.(2006)]{Alonso-Herrero06} Alonso-Herrero, A., Colina, L., Packham, C., et al.\ 2006, \apjl, 652, L83 
\bibitem[Alonso-Herrero et al.(2011)]{Alonso-Herrero11} Alonso-Herrero, A., Ramos Almeida, C., Mason, R., et al.\ 2011, \apj, 736, 82 
\bibitem[Alonso-Herrero et al.(2016)]{Alonso-Herrero16} Alonso-Herrero, A., Esquej, P., Roche, P.~F., et al.\ 2016, \mnras, 455, 563 
\bibitem[Antonucci(1993)]{Antonucci93} Antonucci, R.\ 1993, \araa, 31, 473 
\bibitem[Arshakian(2005)]{Arshakian05} Arshakian, T.~G.\ 2005, \aap, 436, 817 
\bibitem[Asensio Ramos \& Ramos Almeida(2009)]{Asensio-Ramos09} Asensio Ramos, A., \& Ramos Almeida, C.\ 2009, \apj, 696, 2075 
\bibitem[Asmus et al.(2011)]{Asmus11} Asmus, D., Gandhi, P., Smette, A., H{\"o}nig, S.~F., \& Duschl, W.~J.\ 2011, \aap, 536, A36 
\bibitem[Asmus et al.(2014)]{Asmus14} Asmus, D., H{\"o}nig, S.~F., Gandhi, P., Smette, A., \& Duschl, W.~J.\ 2014, \mnras, 439, 1648 
\bibitem[Asmus et al.(2015)]{Asmus15} Asmus, D., Gandhi, P., H{\"o}nig, S.~F., Smette, A., \& Duschl, W.~J.\ 2015, \mnras, 454, 766 
\bibitem[Asmus et al.(2016)]{Asmus16} Asmus, D., H{\"o}nig, S.~F., \& Gandhi, P.\ 2016, \apj, 822, 109 
\bibitem[Assef et al.(2013)]{Assef13} Assef, R.~J., Stern, D., Kochanek, C.~S., et al.\ 2013, \apj, 772, 26 
\bibitem[Barvainis(1987)]{Barvainis87} Barvainis, R.\ 1987, \apj, 320, 537 
\bibitem[Belloni et al.(2002)]{Belloni02} Belloni, T., Psaltis, D., \& van der Klis, M.\ 2002, \apj, 572, 392 
\bibitem[Bentz et al.(2009)]{Bentz09} Bentz, M.~C., Walsh, J.~L., Barth, A.~J., et al.\ 2009, \apj, 705, 199 
\bibitem[Bentz \& Katz(2015)]{Bentz15} Bentz, M.~C., \& Katz, S.\ 2015, \pasp, 127, 67
\bibitem[Black \& van Dishoeck(1987)]{Black87} Black, J.~H., \& van Dishoeck, E.~F.\ 1987, \apj, 322, 412 
\bibitem[Brandl et al.(2006)]{Brandl06} Brandl, B.~R., Bernard-Salas, J., Spoon, H.~W.~W., et al.\ 2006, \apj, 653, 1129 
\bibitem[Burtscher et al.(2013)]{Burtscher13} Burtscher, L., Meisenheimer, K., Tristram, K.~R.~W., et al.\ 2013, \aap, 558, A149 
\bibitem[Cameron et al.(2012)]{Cameron12} Cameron, D.~T., McHardy, I., Dwelly, T., et al.\ 2012, \mnras, 422, 902 
\bibitem[Cao(2010)]{Cao10} Cao, X.\ 2010, \apj, 724, 855 
\bibitem[Carrillo et al.(1999)]{Carrillo99} Carrillo, R., Masegosa, J., Dultzin-Hacyan, D., \& Ordo{\~n}ez, R.\ 1999, \rmxaa, 35, 187 
\bibitem[Cid Fernandes et al.(2004)]{Cid-Fernandes04} Cid Fernandes, R., Gonz{\'a}lez Delgado, R.~M., Schmitt, H., et al.\ 2004, \apj, 605, 105 
\bibitem[D{\'{\i}}az-Santos et al.(2010)]{Diaz-Santos10} D{\'{\i}}az-Santos, T., Alonso-Herrero, A., Colina, L., et al.\ 2010, \apj, 711, 328 
\bibitem[Elitzur \& Shlosman(2006)]{Elitzur06} Elitzur, M., \& Shlosman, I.\ 2006, \apjl, 648, L101 
\bibitem[Elitzur \& Ho(2009)]{Elitzur09} Elitzur, M., \& Ho, L.~C.\ 2009, \apjl, 701, L91 
\bibitem[Elitzur(2012)]{Elitzur12} Elitzur, M.\ 2012, \apjl, 747, L33 
\bibitem[Elitzur et al.(2014)]{Elitzur14} Elitzur, M., Ho, L.~C., \& Trump, J.~R.\ 2014, \mnras, 438, 3340 
\bibitem[Elitzur \& Netzer(2016)]{Elitzur16} Elitzur, M., \& Netzer, H.\ 2016, \mnras, 459, 585 
\bibitem[Emmering et al.(1992)]{Emmering92} Emmering, R.~T., Blandford, R.~D., \& Shlosman, I.\ 1992, \apj, 385, 460 
\bibitem[Frey \& Dueck(2007)]{Frey07} Frey, B.~J., \& Dueck, D.\ 2007, Science, 315, 972 
\bibitem[Fritz et al.(2006)]{Fritz06} Fritz, J., Franceschini, A., \& Hatziminaoglou, E.\ 2006, \mnras, 366, 767 
\bibitem[Gandhi et al.(2009)]{Gandhi09} Gandhi, P., Horst, H., Smette, A., et al.\ 2009, \aap, 502, 457 
\bibitem[Garc{\'{\i}}a-Burillo et al.(2016)]{Garcia-Burillo16} Garc{\'{\i}}a-Burillo, S., Combes, F., Ramos Almeida, C., et al.\ 2016, \apjl, 823, L12 
\bibitem[Garc{\'{\i}}a-Bernete et al.(2016)]{Garcia-Bernete16} Garc{\'{\i}}a-Bernete, I., Ramos Almeida, C., Acosta-Pulido, J.~A., et al.\ 2016, \mnras, 463, 3531 
\bibitem[Gonz{\'a}lez Delgado et al.(2004)]{Gonzalez-Delgado04} Gonz{\'a}lez Delgado, R.~M., Cid Fernandes, R., P{\'e}rez, E., et al.\ 2004, \apj, 605, 127 
\bibitem[Gonz{\'a}lez Delgado et al.(2008)]{Gonzalez-Delgado08} Gonz{\'a}lez Delgado, R.~M., P{\'e}rez, E., Cid Fernandes, R., \& Schmitt, H.\ 2008, \aj, 135, 747 
\bibitem[Gonz{\'a}lez-Mart{\'{\i}}n et al.(2009A)]{Gonzalez-Martin09A} Gonz{\'a}lez-Mart{\'{\i}}n, O., Masegosa, J., M{\'a}rquez, I., Guainazzi, M., \& Jim{\'e}nez-Bail{\'o}n, E.\ 2009, \aap, 506, 1107 
\bibitem[Gonz{\'a}lez-Mart{\'{\i}}n et al.(2009B)]{Gonzalez-Martin09B} Gonz{\'a}lez-Mart{\'{\i}}n, O., Masegosa, J., M{\'a}rquez, I., \& Guainazzi, M.\ 2009, \apj, 704, 1570 
\bibitem[Gonz{\'a}lez-Mart{\'{\i}}n \& Vaughan(2012)]{Gonzalez-Martin12} Gonz{\'a}lez-Mart{\'{\i}}n, O., \& Vaughan, S.\ 2012, \aap, 544, A80 
\bibitem[Gonz{\'a}lez-Mart{\'{\i}}n et al.(2013)]{Gonzalez-Martin13} Gonz{\'a}lez-Mart{\'{\i}}n, O., Rodr{\'{\i}}guez-Espinosa, J.~M., D{\'{\i}}az-Santos, T., et al.\ 2013, \aap, 553, A35 
\bibitem[Gonz{\'a}lez-Mart{\'{\i}}n et al.(2014)]{Gonzalez-Martin14} Gonz{\'a}lez-Mart{\'{\i}}n, O., D{\'{\i}}az-Gonz{\'a}lez, D., Acosta-Pulido, J.~A., et al.\ 2014, \aap, 567, A92 
\bibitem[Gonz{\'a}lez-Mart{\'{\i}}n et al.(2015)]{Gonzalez-Martin15} Gonz{\'a}lez-Mart{\'{\i}}n, O., Masegosa, J., M{\'a}rquez, I., et al.\ 2015, \aap, 578, A74 
\bibitem[Gonz{\'a}lez-Mart{\'{\i}}n et al.(2016)]{Gonzalez-Martin16} Gonz{\'a}lez-Mart{\'{\i}}n, O., Hern{\'a}ndez-Garc{\'{\i}}a, L., Masegosa, J., et al.\ 2016, \aap, 587, A1 
\bibitem[Gopal-Krishna et al.(1996)]{Gopal-Krishna96} Gopal-Krishna, Kulkarni, V.~K., \& Wiita, P.~J.\ 1996, \apjl, 463, L1 
\bibitem[Goulding et al.(2012)]{Goulding12} Goulding, A.~D., Alexander, D.~M., Bauer, F.~E., et al.\ 2012, \apj, 755, 5 
\bibitem[Grier et al.(2011)]{Grier11} Grier, C.~J., Mathur, S., Ghosh, H., \& Ferrarese, L.\ 2011, \apj, 731, 60 
\bibitem[Hao et al.(2005)]{Hao05} Hao, L., Strauss, M.~A., Fan, X., et al.\ 2005, \aj, 129, 1795 
\bibitem[Hasinger(2008)]{Hasinger08} Hasinger, G.\ 2008, \aap, 490, 905 
\bibitem[Hatziminaoglou et al.(2015)]{Hatziminaoglou15} Hatziminaoglou, E., Hern{\'a}n-Caballero, A., Feltre, A., \& Pi{\~n}ol Ferrer, N.\ 2015, \apj, 803, 110 
\bibitem[Heckman(1980)]{Heckman80} Heckman, T.~M.\ 1980, \aap, 87, 152 
\bibitem[Hern{\'a}n-Caballero et al.(2015)]{Hernan-Caballero15} Hern{\'a}n-Caballero, A., Alonso-Herrero, A., Hatziminaoglou, E., et al.\ 2015, \apj, 803, 109 
\bibitem[Hern{\'a}ndez-Garc{\'{\i}}a et al.(2013)]{Hernandez-Garcia13} Hern{\'a}ndez-Garc{\'{\i}}a, L., Gonz{\'a}lez-Mart{\'{\i}}n, O., M{\'a}rquez, I., \& Masegosa, J.\ 2013, \aap, 556, A47 
\bibitem[Hern{\'a}ndez-Garc{\'{\i}}a et al.(2014)]{Hernandez-Garcia14} Hern{\'a}ndez-Garc{\'{\i}}a, L., Gonz{\'a}lez-Mart{\'{\i}}n, O., Masegosa, J., \& M{\'a}rquez, I.\ 2014, \aap, 569, A26 
\bibitem[Hern{\'a}ndez-Garc{\'{\i}}a et al.(2015)]{Hernandez-Garcia15} Hern{\'a}ndez-Garc{\'{\i}}a, L., Masegosa, J., Gonz{\'a}lez-Mart{\'{\i}}n, O., \& M{\'a}rquez, I.\ 2015, \aap, 579, A90 
\bibitem[Hern{\'a}ndez-Garc{\'{\i}}a et al.(2016)]{Hernandez-Garcia16} Hern{\'a}ndez-Garc{\'{\i}}a, L., Masegosa, J., Gonz{\'a}lez-Mart{\'{\i}}n, O., M{\'a}rquez, I., \& Perea, J.\ 2016, \apj, 824, 7 
\bibitem[Ho et al.(1997)]{Ho97} Ho, L.~C., Filippenko, A.~V., Sargent, W.~L.~W., \& Peng, C.~Y.\ 1997, \apjs, 112, 391 
\bibitem[Ho et al.(2003)]{Ho03} Ho, L.~C., Filippenko, A.~V., \& Sargent, W.~L.~W.\ 2003, \apj, 583, 159 
\bibitem[Ho(2008)]{Ho08} Ho, L.~C.\ 2008, \araa, 46, 475 
\bibitem[Ho(2009)]{Ho09} Ho, L.~C.\ 2009, \apj, 699, 626 
\bibitem[Hollenbach \& McKee(1989)]{Hollenbach89} Hollenbach, D., \& McKee, C.~F.\ 1989, \apj, 342, 306 
\bibitem[H{\"o}nig et al.(2006)]{Honig06} H{\"o}nig, S.~F., Beckert, T., Ohnaka, K., \& Weigelt, G.\ 2006, \aap, 452, 459 
\bibitem[H{\"o}nig \& Beckert(2007)]{Honig07} H{\"o}nig, S.~F., \& Beckert, T.\ 2007, \mnras, 380, 1172 
\bibitem[H{\"o}nig \& Kishimoto(2010)]{Honig10} H{\"o}nig, S.~F., \& Kishimoto, M.\ 2010, \aap, 523, A27 
\bibitem[H{\"o}nig et al.(2013)]{Honig13} H{\"o}nig, S.~F., Kishimoto, M., Tristram, K.~R.~W., et al.\ 2013, \apj, 771, 87 
\bibitem[Horst et al.(2006)]{Horst06} Horst, H., Smette, A., Gandhi, P., \& Duschl, W.~J.\ 2006, \aap, 457, L17 
\bibitem[Horst et al.(2008)]{Horst08} Horst, H., Gandhi, P., Smette, A., \& Duschl, W.~J.\ 2008, \aap, 479, 389 
\bibitem[Iwasawa \& Taniguchi(1993)]{Iwasawa93} Iwasawa, K., \& Taniguchi, Y.\ 1993, \apjl, 413, L15 
\bibitem[Kennicutt et al.(2003)]{Kennicutt03} Kennicutt, R.~C., Jr., Armus, L., Bendo, G., et al.\ 2003, \pasp, 115, 928 
\bibitem[Kishimoto et al.(2009)]{Kishimoto09} Kishimoto, M., H{\"o}nig, S.~F., Tristram, K.~R.~W., \& Weigelt, G.\ 2009, \aap, 493, L57 
\bibitem[Kishimoto et al.(2011)]{Kishimoto11} Kishimoto, M., H{\"o}nig, S.~F., Antonucci, R., et al.\ 2011, \aap, 536, A78 
\bibitem[Kormendy \& Richstone(1995)]{Kormendy95} Kormendy, J., \& Richstone, D.\ 1995, \araa, 33, 581 
\bibitem[Krabbe et al.(2001)]{Krabbe01} Krabbe, A., B{\"o}ker, T., \& Maiolino, R.\ 2001, \apj, 557, 626 
\bibitem[Krolik \& Begelman(1988)]{Krolik88} Krolik, J.~H., \& Begelman, M.~C.\ 1988, \apj, 329, 702 
\bibitem[Krongold et al.(2003)]{Krongold03} Krongold, Y., Dultzin-Hacyan, D., Marziani, P., \& de Diego, J.~A.\ 2003, \rmxaa, 39, 225 
\bibitem[Lawrence(1991)]{Lawrence91} Lawrence, A.\ 1991, \mnras, 252, 586 
\bibitem[Lebouteiller et al.(2011)]{Lebouteiller11} Lebouteiller, V., Barry, D.~J., Spoon, H.~W.~W., et al.\ 2011, \apjs, 196, 8 
\bibitem[Levenson et al.(2009)]{Levenson09} Levenson, N.~A., Radomski, J.~T., Packham, C., et al.\ 2009, \apj, 703, 390 
\bibitem[L{\'o}pez-Gonzaga et al.(2016)]{Lopez-Gonzaga16} L{\'o}pez-Gonzaga, N., Burtscher, L., Tristram, K.~R.~W., Meisenheimer, K., \& Schartmann, M.\ 2016, \aap, 591, A47 
\bibitem[Lutz et al.(2004)]{Lutz04} Lutz, D., Maiolino, R., Spoon, H.~W.~W., \& Moorwood, A.~F.~M.\ 2004, \aap, 418, 465 
\bibitem[Maiolino et al.(2007)]{Maiolino07} Maiolino, R., Shemmer, O., Imanishi, M., et al.\ 2007, \aap, 468, 979 
\bibitem[Maloney et al.(1996)]{Maloney96} Maloney, P.~R., Hollenbach, D.~J., \& Tielens, A.~G.~G.~M.\ 1996, \apj, 466, 561 
\bibitem[Marconi et al.(2004)]{Marconi04} Marconi, A., Risaliti, G., Gilli, R., et al.\ 2004, \mnras, 351, 169 
\bibitem[Masegosa et al.(2013)]{Masegosa13} Masegosa, J., M{\'a}rquez, I., Gonz{\'a}lez-Mart{\'{\i}}n, O., et al.\ 2013, Revista Mexicana de Astronomia y Astrofisica Conference Series, 42, 51 
\bibitem[Mason et al.(2012)]{Mason12} Mason, R.~E., Lopez-Rodriguez, E., Packham, C., et al.\ 2012, \aj, 144, 11 
\bibitem[Mason et al.(2013)]{Mason13} Mason, R.~E., Ramos Almeida, C., Levenson, N.~A., Nemmen, R., \& Alonso-Herrero, A.\ 2013, \apj, 777, 164 
\bibitem[Maoz et al.(2005)]{Maoz05} Maoz, D., Nagar, N.~M., Falcke, H., \& Wilson, A.~S.\ 2005, \apj, 625, 699 
\bibitem[Mateos et al.(2016)]{Mateos16} Mateos, S., Carrera, F.~J., Alonso-Herrero, A., et al.\ 2016, \apj, 819, 166 
\bibitem[Mayo \& Lawrence(2013)]{Mayo13} Mayo, J.~H., \& Lawrence, A.\ 2013, \mnras, 434, 1593 
\bibitem[McKernan et al.(2010)]{McKernan10} McKernan, B., Ford, K.~E.~S., \& Reynolds, C.~S.\ 2010, \mnras, 407, 2399 
\bibitem[Mendoza-Castrej{\'o}n et al.(2015)]{Mendoza-Castrejon15} Mendoza-Castrej{\'o}n, S., Dultzin, D., Krongold, Y., Gonz{\'a}lez, J.~J., \& Elitzur, M.\ 2015, \mnras, 447, 2437 
\bibitem[M{\"u}ller-S{\'a}nchez et al.(2013)]{Muller-Sanchez13} M{\"u}ller-S{\'a}nchez, F., Prieto, M.~A., Mezcua, M., et al.\ 2013, \apjl, 763, L1 
\bibitem[Mor et al.(2009)]{Mor09} Mor, R., Netzer, H., \& Elitzur, M.\ 2009, \apj, 705, 298 
\bibitem[Mor \& Netzer(2012)]{Mor12} Mor, R., \& Netzer, H.\ 2012, \mnras, 420, 526 
\bibitem[Nenkova et al.(2002)]{Nenkova02} Nenkova, M., Ivezi{\'c}, {\v Z}., \& Elitzur, M.\ 2002, \apjl, 570, L9 
\bibitem[Nenkova et al.(2008A)]{Nenkova08A} Nenkova, M., Sirocky, M.~M., Ivezi{\'c}, {\v Z}., \& Elitzur, M.\ 2008, \apj, 685, 147-159 
\bibitem[Nenkova et al.(2008B)]{Nenkova08B} Nenkova, M., Sirocky, M.~M., Nikutta, R., Ivezi{\'c}, {\v Z}., \& Elitzur, M.\ 2008, \apj, 685, 160-180 
\bibitem[Netzer \& Maoz(1990)]{Netzer90} Netzer, H., \& Maoz, D.\ 1990, \apjl, 365, L5 
\bibitem[Netzer(2015)]{Netzer15} Netzer, H.\ 2015, \araa, 53, 365 
\bibitem[Nicastro(2000)]{Nicastro00} Nicastro, F.\ 2000, \apjl, 530, L65 
\bibitem[Panuzzo et al.(2011)]{Panuzzo11} Panuzzo, P., Rampazzo, R., Bressan, A., et al.\ 2011, \aap, 528, A10 
\bibitem[Ranalli et al.(2003)]{Ranalli03} Ranalli, P., Comastri, A., \& Setti, G.\ 2003, \aap, 399, 39 
\bibitem[Ramos Almeida et al.(2007)]{Ramos-Almeida07} Ramos Almeida, C., P{\'e}rez Garc{\'{\i}}a, A.~M., Acosta-Pulido, J.~A., \& Rodr{\'{\i}}guez Espinosa, J.~M.\ 2007, \aj, 134, 2006 
\bibitem[Ramos Almeida et al.(2009)]{Ramos-Almeida09} Ramos Almeida, C., Levenson, N.~A., Rodr{\'{\i}}guez Espinosa, J.~M., et al.\ 2009, \apj, 702, 1127 
\bibitem[Ramos Almeida et al.(2011)]{Ramos-Almeida11} Ramos Almeida, C., Levenson, N.~A., Alonso-Herrero, A., et al.\ 2011, \apj, 731, 92 
\bibitem[Ramos Almeida et al.(2014)]{Ramos-Almeida14} Ramos Almeida, C., Alonso-Herrero, A., Levenson, N.~A., et al.\ 2014, \mnras, 439, 3847 
\bibitem[Ramos Almeida et al.(2016)]{Ramos-Almeida16} Ramos Almeida, C., Mart{\'{\i}}nez Gonzalez, M.~J., Asensio Ramos, A., et al.\ 2016, arXiv:1606.02204 
\bibitem[Ricci et al.(2011)]{Ricci11} Ricci, C., Walter, R., Courvoisier, T.~J.-L., \& Paltani, S.\ 2011, \aap, 532, A102 
\bibitem[Ricci et al.(2013)]{Ricci13} Ricci, C., Paltani, S., Awaki, H., et al.\ 2013, \aap, 553, A29 
\bibitem[Rigopoulou et al.(2002)]{Rigopoulou02} Rigopoulou, D., Kunze, D., Lutz, D., Genzel, R., \& Moorwood, A.~F.~M.\ 2002, \aap, 389, 374 
\bibitem[Risaliti et al.(2002)]{Risaliti02} Risaliti, G., Elvis, M., \& Nicastro, F.\ 2002, \apj, 571, 234 
\bibitem[Risaliti(2010)]{Risaliti10} Risaliti, G.\ 2010, X-ray Astronomy 2009; Present Status, Multi-Wavelength Approach and Future Perspectives, 1248, 351 
\bibitem[Roussel et al.(2007)]{Roussel07} Roussel, H., Helou, G., Hollenbach, D.~J., et al.\ 2007, \apj, 669, 959 
\bibitem[Sarzi et al.(2005)]{Sarzi05} Sarzi, M., Rix, H.-W., Shields, J.~C., et al.\ 2005, \apj, 628, 169 
\bibitem[Schartmann et al.(2008)]{Schartmann08} Schartmann, M., Meisenheimer, K., Camenzind, M., et al.\ 2008, \aap, 482, 67 
\bibitem[Shi et al.(2006)]{Shi06} Shi, Y., Rieke, G.~H., Hines, D.~C., et al.\ 2006, \apj, 653, 127 
\bibitem[Siebenmorgen et al.(2015)]{Siebenmorgen15} Siebenmorgen, R., Heymann, F., \& Efstathiou, A.\ 2015, \aap, 583, A120 
\bibitem[Simpson(1998)]{Simpson98} Simpson, C.\ 1998, \mnras, 297, L39 
\bibitem[Simpson(2003)]{Simpson03} Simpson, C.\ 2003, \nar, 47, 211 
\bibitem[Simpson(2005)]{Simpson05} Simpson, C.\ 2005, \mnras, 360, 565 
\bibitem[Sirocky et al.(2008)]{Sirocky08} Sirocky, M.~M., Levenson, N.~A., Elitzur, M., Spoon, H.~W.~W., \& Armus, L.\ 2008, \apj, 678, 729-743 
\bibitem[Stalevski et al.(2012)]{Stalevski12} Stalevski, M., Fritz, J., Baes, M., Nakos, T., \& Popovi{\'c}, L.~{\v C}.\ 2012, \mnras, 420, 2756 
\bibitem[Stalevski et al.(2016)]{Stalevski16} Stalevski, M., Ricci, C., Ueda, Y., et al.\ 2016, \mnras, 458, 2288 
\bibitem[Stern(2015)]{Stern15} Stern, D.\ 2015, \apj, 807, 129 
\bibitem[Sturm et al.(2006)]{Sturm06} Sturm, E., Rupke, D., Contursi, A., et al.\ 2006, \apjl, 653, L13 
\bibitem[Suganuma et al.(2006)]{Suganuma06} Suganuma, M., Yoshii, Y., Kobayashi, Y., et al.\ 2006, \apj, 639, 46 
\bibitem[Tristram et al.(2007)]{Tristram07} Tristram, K.~R.~W., Meisenheimer, K., Jaffe, W., et al.\ 2007, \aap, 474, 837 
\bibitem[Tristram et al.(2009)]{Tristram09} Tristram, K.~R.~W., Raban, D., Meisenheimer, K., et al.\ 2009, \aap, 502, 67 
\bibitem[Tristram \& Schartmann(2011)]{Tristram11} Tristram, K.~R.~W., \& Schartmann, M.\ 2011, \aap, 531, A99 
\bibitem[Urry \& Padovani(1995)]{Urry95} Urry, C.~M., \& Padovani, P.\ 1995, \pasp, 107, 803 
\bibitem[V{\'e}ron-Cetty \& V{\'e}ron(2010)]{Veron10} V{\'e}ron-Cetty, M.-P., \& V{\'e}ron, P.\ 2010, \aap, 518, A10 
\bibitem[Wada(2012)]{Wada12} Wada, K.\ 2012, \apj, 758, 66 
\bibitem[Willott et al.(2000)]{Willott00} Willott, C.~J., Rawlings, S., Blundell, K.~M., \& Lacy, M.\ 2000, \mnras, 316, 449 
\bibitem[Woo \& Urry(2002)]{Woo02} Woo, J.-H., \& Urry, C.~M.\ 2002, \apj, 579, 530 
\bibitem[Zoghbi et al.(2014)]{Zoghbi14} Zoghbi, A., Cackett, E.~M., Reynolds, C., et al.\ 2014, \apj, 789, 56 




\end{thebibliography}
